\documentclass[a4paper,11pt]{article}
\pdfoutput=1 
\usepackage{jheppub} 
\usepackage[T1]{fontenc}
\usepackage{graphicx}
\usepackage{dcolumn}
\usepackage{bm}
\usepackage{tabu}
\usepackage{color}
\usepackage{float} 
\allowdisplaybreaks
\newcommand{\hs}{\hspace*{0.5cm}}

\newcommand{\be}{\begin{equation}}
\newcommand{\ee}{\end{equation}}
\newcommand{\bea}{\begin{eqnarray}}
\newcommand{\eea}{\end{eqnarray}}
\newcommand{\ben}{\begin{enumerate}}
\newcommand{\een}{\end{enumerate}}
\newcommand{\bde}{\begin{widetext}}
\newcommand{\ede}{\end{widetext}}
\newcommand{\nn}{\nonumber}
\newcommand{\crn}{\nonumber \\}
\newcommand{\Tr}{\mathrm{Tr}}

\newcommand{\al}{\alpha}
\newcommand{\la}{\lambda}

\newcommand{\ga}{\gamma}

\newcommand{\pa}{\partial}

\newcommand{\fr}{\frac}
\newcommand{\bc}{\begin{center}}
\newcommand{\ec}{\end{center}}

\newcommand{\ep}{\epsilon}

\newcommand{\La}{\Lambda}

\newcommand{\Om}{\Omega}
\newcommand{\overlr}{\overleftrightarrow} 
\newcommand{\AdrHEPC}{Phenikaa Institute for Advanced Study and Faculty of Basic Science, Phenikaa University, Yen Nghia, Ha Dong, Hanoi 100000, Vietnam}
\title{\boldmath Multicomponent dark matter in noncommutative $B-L$ gauge theory} 
\author[a]{Cao H. Nam,}\affiliation[a]{\AdrHEPC}
\emailAdd{nam.caohoang@phenikaa-uni.edu.vn}
\author[a]{Duong Van Loi,} 
\emailAdd{loi.duongvan@phenikaa-uni.edu.vn}
\author[a,b]{Le Xuan Thuy}\affiliation[b]{Faculty of Basic Science, Vinh Long University of Technology Education, Ward 2, Vinh Long City, Vinh Long 890000, Vietnam}
\emailAdd{thuylx@vlute.edu.vn}
\author[a,1]{Phung Van Dong,\note{Corresponding author at Phenikaa University, Hanoi 100000, Vietnam.}} 
\emailAdd{dong.phungvan@phenikaa-uni.edu.vn}

\abstract{It is shown that for a higher weak isospin symmetry, $SU(P)_L$ with $P\geq 3$, the baryon minus lepton charge $B-L$ neither commutes nor closes algebraically with $SU(P)_L$ similar to the electric charge $Q$, which all lead to a $SU(3)_C\otimes SU(P)_L\otimes U(1)_X\otimes U(1)_N$ gauge completion, where $X$ and $N$ determine $Q$ and $B-L$, respectively. As a direct result, the neutrinos obtain appropriate masses via a canonical seesaw. While the version with $P=3$ supplies the schemes of single-component dark matter well established in the literature, we prove in this work that the models with $P\geq 4$ provide the novel scenarios of multicomponent dark matter, which contain simultaneously at least $P-2$ stable candidates, respectively. In this setup, the multicomponet dark matter is nontrivially unified with normal matter by gauge multiplets, and their stability is ensured by a residual gauge symmetry which is a remnant of the gauge symmetry after spontaneous symmetry breaking. The three versions with $P=4$ according to the new lepton electric charges are detailedly investigated. The mass spectrum of the scalar sector is diagonalized when the scale of the $U(1)_N$ breaking is much higher than that of the usual 3-4-1 symmetry breaking. All the interactions of gauge bosons with fermions and scalars are obtained. We figure out viable parameter regimes given that the multicomponent dark matter satisfies the Planck and (in)direct detection experiments.}


\begin{document} 
\maketitle

\section{Introduction}

The standard model of fundamental particles and interactions has been extremely successful in describing observed phenomena, especially predicting the existence of the Higgs boson. However, it leaves a number of striking physics features of our world unexplained. The experimental evidences of neutrino oscillation have indicated that the neutrinos have non-zero small masses and flavor mixing, which cannot be solved within the framework of the standard model \cite{Kajita:2016cak,McDonald:2016ixn}. Additionally, the standard model fails to account for the cosmological issues relevant to particle physics, such as the matter-antimatter asymmetry of the universe and the fact that the standard model addresses only about 5\% matter content of the universe~\cite{Tanabashi:2018oca}. The rest includes 26.5\% dark matter and 68.5\% dark energy, which all lie beyond the standard model \cite{Hinshaw:2012aka,Ade:2015xua}. In this work, we concentrate on the dark matter issue, finding its implication to the other puzzles. 

The most widely studied dark matter candidate in particle physics and cosmology is a new kind of colorless, electrically-neutral, and weakly-interacting massive particles, called WIMPs \cite{Jungman:1995df,Bertone:2004pz}. Such particles arise naturally in many extensions of the standard model from the supersymmetric models \cite{PhysRevLett.50.1419,Ellis:1983ew,Kane:1993td,Edsjo:1997bg} to models with universal extra dimensions \cite{Kolb:1983fm,Appelquist:2000nn,PhysRevD.66.036005,PhysRevLett.93.231805}, little Higgs models \cite{ArkaniHamed:2002pa,ArkaniHamed:2002qy,Low:2004xc,PhysRevD.71.035016}, and other interesting scenarios \cite{Deshpande:1977rw,Silveira:1985rk,Chacko:2005un,Cirelli:2005uq,Ma:2006km,Barbieri:2006dq,PhysRevLett.96.231802,Mizukoshi:2010ky,Goudelis:2013uca,Dong:2013ioa,Dong:2013wca,Dong:2014esa,Dong:2014wsa,Dong:2015yra,Dong:2016sat,Dong:2016gxl,Huong:2016ybt,Alves:2016fqe,Dong:2017zxo,Kownacki:2017uyq,Ma:2017qad,Kownacki:2018lkj,Huong:2018ytz,Dong:2018aak,VanLoi:2019xud,VanLoi:2019eax,Huong:2019vej,VanDong:2020nwb,Leite:2020bnb}. Despite severe constraints from stability condition on cosmological timescales, relic density \cite{Hinshaw:2012aka,Ade:2015xua}, direct \cite{Akerib:2016vxi,Tan:2016zwf,Cui:2017nnn,Aprile:2017iyp,Aprile:2018dbl} and indirect \cite{Abdallah:2018qtu,Ahnen:2016qkx,Atwood_2009,Boyarsky:2014jta,Bulbul:2014sua} searches, and particle colliders \cite{Belyaev:2018pqr,Abercrombie:2015wmb}, the dark matter candidate can be viable to be a fermion, a vector, or a scalar, with mass scales ranging from a few GeV to several TeV.\footnote{See \cite{Arcadi:2017kky,Roszkowski:2017nbc} for recent reviews and updates.} The stability of the dark matter candidate is usually ensured by an unbroken discrete symmetry, such as $R$-parity in supersymmetry, KK-parity in universal extra dimension, $T$-parity in the little Higgs theory, matter parity in $B-L$ extensions \cite{Kadastik:2009dj,Dong:2013wca,VanDong:2020cjf}, or lepton parity in neutrino mass generation schemes \cite{Ma:2015xla}. Generally, all of the standard model particles are even, whereas the relevant new particles are odd, such that the lightest odd particle is stabilized, contributing to dark matter. A lot of such discrete symmetries must be imposed by hand, assumed to be exact or appropriately violated. The possibility of discrete symmetry that arises as a residual gauge symmetry is compelling, because the gauge symmetry not only determines and stabilizes dark matter candidates but also sets dark matter interactions and observables. 

The mentioned experiments on relic density, direct and indirect detections, and particle colliders have not yet provided the particle picture of dark matter. Obviously, the mentioned theories often assume dark matter to be composed of a kind of a single particle---the lightest particle that is odd under the discrete symmetry. Since the constituent of dark matter is still an open question, there is no reason why dark matter comes from such a single particle kind. The scenario of multicomponent dark matter seems to be naturally in view of the rich structure of stable normal matter---the atoms. Furthermore, they have been phenomenologically and/or theoretically motivated \cite{Boehm:2003ha,Chialva:2012rq,Aoki:2013gzs,Kajiyama:2013rla,Bhattacharya:2013hva,Karam:2016rsz,Bhattacharya:2016ysw,Arcadi:2016kmk,Borah:2017xgm,Ahmed:2017dbb,Biswas:2013nn,Bhattacharya:2018cgx,Chakraborti:2018lso,Borah:2019epq,Borah:2019aeq,Bhattacharya:2019fgs,PhysRevD.79.115002,Fukuoka:2010kx,Bernal:2018aon,Biswas:2019ygr} and revealed interesting consequences for galaxy structure \cite{Fan:2013yva,Fan:2013tia}. The schemes of multicomponent dark matter have been especially to accommodate the multiple gamma-ray line and boosted dark matter signals \cite{Agashe:2014yua,Kong:2014mia,Alhazmi:2016qcs,Kim:2016zjx,Giudice:2017zke,Chatterjee:2018mej,Kim:2018veo} as well as dark matter self-interactions \cite{Elbert:2014bma,Tulin:2017ara}. Theoretically, to have a multiple dark matter scenario, the simplest way adds to the standard model an exact discrete symmetry $Z_2\otimes Z_2$. One can also add an exact $Z_2$ symmetry to supersymmetric models, or to universal extra dimension models, or to $U(1)_{B-L}$ models. Besides, there are other interesting approaches \cite{Heeck:2012bz,Bian:2013wna,Bian:2014cja,Esch:2014jpa,Karam:2015jta,DiFranzo:2016uzc,DuttaBanik:2016jzv,Bhattacharya:2017fid,Bhattacharya:2018cqu,Aoki:2018gjf,DuttaBanik:2018emv,Barman:2018esi,YaserAyazi:2018lrv,Chakraborti:2018aae,Elahi:2019jeo,Bhattacharya:2019tqq}. Since the dark matter structure is enriched, such scenarios possess interesting phenomenological consequences, attracting current research.

Among the standard model extensions, the models that include $B-L$ as a gauge charge have intriguing features. They can explain small neutrino masses through the exchange of heavy right-handed neutrinos, which arise as a result of $B-L$ anomaly cancelation, while the right-handed mass scale is induced by $B-L$ breaking \cite{Minkowski:1977sc,GellMann:1980vs,Yanagida:1979as,Glashow:1979nm,Mohapatra:1979ia,Mohapatra:1980yp,Lazarides:1980nt,Schechter:1980gr,Schechter:1981cv}. The theories that contain noncommutative $B-L$ charge define the dark sector to be nontrivially unified with the normal sector in gauge multiplets, while the residual $B-L$ charge stabilizes dark matter candidates \cite{Dong:2013wca,Dong:2014wsa,Huong:2016ybt,Alves:2016fqe,Dong:2015yra,Dong:2016sat,Dong:2016gxl,Dong:2017zxo,Huong:2018ytz,Huong:2019vej}. Since dark matter takes part in gauge multiplets, the gauge symmetry would govern the dark matter observables, providing very predictive signals. The inflation and leptogenesis can be derived by the $B-L$ symmetry when its breaking scale is high enough \cite{Huong:2015dwa,Huong:2016ybt,Dong:2018aak,VanDong:2020nwb}. In this article, we develop a gauge theory that contains noncommutative $B-L$ charge. Starting from a higher weak isospin symmetry $SU(P)_L$, we prove that the complete gauge symmetry must be $SU(3)_C\otimes SU(P)_L\otimes U(1)_X\otimes U(1)_N$, called 3-$P$-1-1, where the last two factors determine the electric charge and $B-L$, respectively. We show that this theory provides multicomponent dark matter naturally for $P\geq 4$. Whereas, the model with $P=3$ yield single component dark matter, which has been well established in the literature \cite{Dong:2013wca,Dong:2014wsa,Dong:2015yra,Huong:2016ybt,Alves:2016fqe,Dong:2018aak,Huong:2019vej}.

The rest of this work is organized as follows: In Sec. \ref{model}, we construct a general noncommutative $B-L$ gauge theory for multicomponent dark matter, discussing the spontaneous symmetry breaking, residual symmetries, dark matter stability, and fermion masses. In Sec. \ref{scalar}, we study the mass spectra of the scalar and gauge boson sectors according to the minimal model of multicomponent dark matter, i.e. the 3-4-1-1 model. All the interactions of gauge bosons with fermions and scalars are obtained in Sec. \ref{interaction}. In Sec. \ref{DMP}, we consider the three different scenarios of multicomponent dark matter and examine the dark matter observables. We summarize the results and make conclusions in Sec. \ref{conl}.

\section{\label{model} Noncommutative $B-L$ gauge theory}

The purpose of this section proposes a general $B-L$ gauge model responsible for multicomponent dark matter. The minimal realization of the model is presented. 

\subsection{General setup}

Apart from the QCD group, let the $SU(2)_L$ symmetry of weak isospin be enlarged to $SU(P)_L$ for $P=3,4,5,\cdots$, a higher weak isospin symmetry. Correspondingly, let each standard model fermion doublet be enlarged to either the defining representation ($P$-plet) or the complex conjugate of defining representation (anti-$P$-plet) of $SU(P)_L$. The $[SU(P)_L]^3$ anomaly cancellation demands that the number of fermion $P$-plets equals that of fermion anti-$P$-plets, since a representation and its conjugate have opposite anomaly contributions. It follows that the number of generations is a multiple of fundamental color number, 3. Further, the QCD asymptotic freedom condition implies that the number of generations is not larger than $[33/(2P)]=5,4,3$ for $P=3,4,5$, respectively.       
Hence the generation number is just three, matching that of colors, and $P\leq 5$. That property disappears in the standard model due to vanishing $[SU(2)_L]^3$ anomaly for every representation, unlike that of $SU(P)_L$. The higher weak isospin extension is thus motivated.   

Therefore, the fermion content under $SU(P)_L$ is arranged as
\bea \psi_{aL} &=& \left(\begin{array}{l} \nu^{0,-1}\\ e^{-1,-1}\\ E^{q_1,n_1}_1\\ E^{q_2,n_2}_2\\ \vdots \\ E^{q_{P-2},n_{P-2}}_{P-2}\end{array}\right)_{aL}\sim P,\label{plepton}\\
Q_{\al L}&=& \left(\begin{array}{l} d^{-1/3,1/3}\\ -u^{2/3,1/3}\\ J^{-q_1-1/3,-n_1-2/3}_1\\ J^{-q_2-1/3,-n_2-2/3}_2\\ \vdots\\ J^{-q_{P-2}-1/3,-n_{P-2}-2/3}_{P-2}\end{array}\right)_{\al L}\sim P^*,\label{pquark1}\\
Q_{3 L}&=& \left(\begin{array}{l} u^{2/3,1/3}\\ d^{-1/3,1/3}\\ J^{q_1+2/3,n_1+4/3}_1\\ J^{q_2+2/3,n_2+4/3}_2\\ \vdots\\ J^{q_{P-2}+2/3,n_{P-2}+4/3}_{P-2}\end{array}\right)_{3 L}\sim P,\label{pquark2}\eea plus the corresponding right-handed components transforming as $SU(P)_L$ singlets. The generation indices are $a=1,2,3$ and $\al =1,2$. The new fields $E$'s and $J$'s are necessarily included to complete the representations. Their subscripts are $SU(P)_L$ indices, while the superscripts of all fields are clarified below.  

Without loss of generality, take a lepton $P$-plet into account. Since the new fields $E$'s are unknown, let their electric charge $Q$ and baryon minus lepton charge $B-L$ be $q$'s and $n$'s, respectively. Thus, each lepton field ($\nu$, $e$, $E$'s) possesses a pair of the characteristic charges $(Q,B-L)$ as superscripted, respectively. Suppose that $Q$ and $B-L$ are conserved, which are all approved by the standard model and the experiment. Both $Q$ and $B-L$ neither commute nor close algebraically with $SU(P)_L$. Indeed, we have $Q=\mathrm{diag}(0,-1,q_1,q_2,\cdots,q_{P-2})$ and $B-L=\mathrm{diag}(-1,-1,n_1,n_2,\cdots,n_{P-2})$ for $\psi_{aL}$, which are generally not commuted with the $SU(P)_L$ weight raising/lowering generators: 
\bea && [Q,T_1\pm i T_2]=\pm(T_1\pm i T_2),\\ 
&&[Q,T_4\pm i T_5]=\mp q_1 (T_4\pm i T_5),\\ 
&& [Q,T_6\pm i T_7]=\mp(1+q_1)(T_6\pm i T_7),\\
&& [Q,T_9\pm i T_{10}]=\mp q_2(T_9\pm i T_{10}),\\ 
&& [Q,T_{11}\pm i T_{12}]=\mp(1+q_2)(T_{11}\pm i T_{12}),\\
&& [Q,T_{13}\pm i T_{14}]=\mp(q_2-q_1)(T_{13}\pm i T_{14}),\\
&&\cdots\cdots\cdots,\crn
&& [Q,T_{P^2-3}\pm i T_{P^2-2}]=\mp(q_{P-2}-q_{P-3})(T_{P^2-3}\pm i T_{P^2-2}),\eea
for the electric charge, and \bea
&& [B-L,T_4\pm i T_5]=\mp (1+n_1) (T_4\pm i T_5),\\
&& [B-L,T_6\pm i T_7]=\mp(1+n_1)(T_6\pm i T_7),\\ 
&& [B-L,T_9\pm i T_{10}]=\mp (1+n_2)(T_9\pm i T_{10}),\\ 
&& [B-L,T_{11}\pm i T_{12}]=\mp(1+n_2)(T_{11}\pm i T_{12}),\\
&& [B-L,T_{13}\pm i T_{14}]=\mp(n_2-n_1)(T_{13}\pm i T_{14}),\\
&&\cdots\cdots\cdots,\crn
&& [B-L,T_{P^2-3}\pm i T_{P^2-2}]=\mp(n_{P-2}-n_{P-3})(T_{P^2-3}\pm i T_{P^2-2}),\eea for the baryon minus lepton charge, where $T_i=\fr 1 2 \la_i$ ($i=1,2,3,\cdots,P^2-1$) are the $SU(P)_L$ charges, given in the basis of the generalized Gell-Mann matrices. The nonclosure is due to the fact that if $Q$ and $B-L$ are some generators of $SU(P)_L$, they must be linearly combined of $Q=y_i T_i$ and $B-L=x_i T_i$, implying $\Tr Q=0$ and $\Tr(B-L)=0$, respectively. The last ones are not valid for the general $E$'s charges, especially for the ordinary right-handed lepton. In other words, $Q$, $B-L$, and $T_i$ do not form a symmetry by themselves. 

To close the symmetries, two Abelian charges must be imposed, yielding a complete gauge symmetry,
\be SU(3)_C\otimes SU(P)_L\otimes U(1)_X\otimes U(1)_N,\ee called 3-$P$-1-1, where the color group is also included, and $X,N$ determines $Q$ and $B-L$,
\bea Q=\sum^{P-2}_{k=0}\beta_k H_k + X,\hs B-L = \sum^{P-2}_{k=0} b_k H_k + N,\eea respectively.\footnote{They are given in terms of diagonal generators due to the conservation and additive nature.} Here $H_k=T_{(k+2)^2-1}=T_3,T_8,T_{15},\cdots, T_{P^2-1}$ according to $k=0,1,2,\cdots,P-2$ are the $SU(P)_L$ Cartan generators. Note that $X$ and $N$ are linearly independent as $Q$ and $B-L$ are. All the charges $Q$, $X$, $B-L$, and $N$ must be gauged, since $H_k$ are gauged charges, a consequence of the noncommutation. The coefficients $\beta$'s and $b$'s can be obtained by acting $Q$ and $B-L$ on $\psi_{aL}$, which obey              
\bea \beta_{k} &=&\sqrt{\fr{k}{k+2}}\beta_{k-1} +\sqrt{\fr{2(k+1)}{k+2}}(q_{k-1}-q_k),\\
b_{k} &=&\sqrt{\fr{k}{k+2}}b_{k-1} +\sqrt{\fr{2(k+1)}{k+2}}(n_{k-1}-n_k),\eea where the initial conditions are ($\beta_0=1$, $q_0=-1$) and ($b_0=0,n_0=-1$), respectively. For $P\leq 5$, we find $\beta_1=-(1+2q_1)/\sqrt{3}$, $\beta_2=-(1-q_1+3q_2)/\sqrt{6}$, $\beta_3=-(1-q_1-q_2+4q_3)/\sqrt{10}$, $b_1=-2(1+n_1)/\sqrt{3}$, $b_2=-(2-n_1+3n_2)/\sqrt{6}$, and $b_3=-(2-n_1+7n_2-4n_3)/\sqrt{10}$. Last, but not least, acting $Q$ and $B-L$ on the quark multiplets, $Q_{\al L}$ and $Q_{3L}$, we obtain the pairs of the corresponding ($Q,B-L$) charges for component quark fields, as already superscripted in (\ref{pquark1}) and (\ref{pquark2}).       

The important remark is that the higher weak isospin symmetry $SU(P)_L$ contains two noncommutative charges $Q$ and $B-L$, and the algebraic closure among them yields the gauge model 3-$P$-1-1, where $B-L$ is nontrivially unified with the weak interaction in the same manner the electroweak theory does so for the electric charge. We will shortly prove that the new fermions $E$'s and $J$'s including new scalar and gauge bosons transform nontrivially under a residual gauge symmetry survived after spontaneous breaking, which contribute to dark matter. This theory determines dark matter nontrivially unified with normal matter in gauge multiplets by the gauge principle, a consequence of the noncommutative $B-L$ charge. The multicomponent dark matter arises due to a nontrivial structure of the residual gauge symmetry relevant to the existence of multi $B-L$ charges $n_1,n_2,\cdots,n_{P-2}$, or in other words, the 3-$P$-1-1 extension for $P\geq 4$. 

To summarize the full fermion content transforms under the 3-$P$-1-1 symmetry as 
\bea &&\psi_{aL}\sim \left(1,P,\fr{q-1}{P},\fr{n-2}{P}\right),\\
&& Q_{\al L}\sim \left(3,P^*,-\fr 1 3 +\fr{1-q}{P},-\fr 2 3 +\fr{2-n}{P}\right),\\
&&  Q_{3 L}\sim \left(3,P,\fr 2 3 +\fr{q-1}{P},\fr 4 3 +\fr{n-2}{P}\right),\\
&& \nu_{aR}\sim (1,1,0,-1),\hs e_{aR}\sim (1,1,-1,-1),\hs E_{k aR}\sim (1,1,q_k,n_k),\\
&& u_{aR}\sim (3,1,2/3,1/3),\hs d_{aR}\sim (3,1,-1/3,1/3),\\
&& J_{k \al R}\sim (3,1, -q_k-1/3,-n_k-2/3),\hs J_{k 3 R}\sim (3,1, q_k+2/3,n_k+4/3),\eea where we denote $k=1,2,3,\cdots,P-2$, $q\equiv q_1+q_2+\cdots+q_{P-2}$, and $n\equiv n_1+n_2+\cdots+n_{P-2}$. Of course, the right-handed fermions have the same $(Q,B-L)$ charges as those of the left-handed fermions in (\ref{plepton}), (\ref{pquark1}), and (\ref{pquark2}), which were not supperscripted (i.e. omitted) without confusion. It is easily verified that all the anomalies vanish, as indicated in Appendix \ref{anomaly}. Especially, $\nu_{aR}$ must be presented to cancel the $U(1)_N$ anomalies which are relevant to $B-L$ charge.

To break the gauge symmetry and generate the particle masses properly, we introduce $P$ scalar $P$-plets plus a scalar singlet, 
\bea &&\left(\begin{array}{l} 
\varphi^{0,0}_{11}\\
\varphi^{-1,0}_{21}\\
\varphi^{q_1,n_1+1}_{31}\\
\varphi^{q_2,n_2+1}_{41}\\
\vdots\\
\varphi^{q_{P-2},n_{P-2}+1}_{P1}
\end{array}\right),\hs
\left(\begin{array}{l} 
\varphi^{1,0}_{12}\\
\varphi^{0,0}_{22}\\
\varphi^{q_1+1,n_1+1}_{32}\\
\varphi^{q_2+1,n_2+1}_{42}\\
\vdots\\
\varphi^{q_{P-2}+1,n_{P-2}+1}_{P2}
\end{array}\right),\hs
\left(\begin{array}{l} 
\varphi^{-q_1,-1-n_1}_{13}\\
\varphi^{-1-q_1,-1-n_1}_{23}\\
\varphi^{0,0}_{33}\\
\varphi^{q_2-q_1,n_2-n_1}_{43}\\
\vdots\\
\varphi^{q_{P-2}-q_1,n_{P-2}-n_1}_{P3}
\end{array}\right),\label{scalarm1}\\
&& 
\left(\begin{array}{l} 
\varphi^{-q_2,-1-n_2}_{14}\\
\varphi^{-1-q_2,-1-n_2}_{24}\\
\varphi^{q_1-q_2,n_1-n_2}_{34}\\
\varphi^{0,0}_{44}\\
\vdots\\
\varphi^{q_{P-2}-q_2,n_{P-2}-n_2}_{P4}
\end{array}\right),
\cdots\cdots\cdots,
\left(\begin{array}{l} 
\varphi^{-q_{P-2},-1-n_{P-2}}_{1P}\\
\varphi^{-1-q_{P-2},-1-n_{P-2}}_{2P}\\
\varphi^{q_1-q_{P-2},n_1-n_{P-2}}_{3P}\\
\varphi^{q_2-q_{P-2},n_2-n_{P-2}}_{4P}\\
\vdots\\
\varphi^{0,0}_{PP}
\end{array}\right),\hs \phi\sim (1,1,0,2).\label{scalarm2}
 \eea Here $\phi$ owning such quantum numbers is given in order to break $U(1)_N$ and produce right-handed neutrino masses through the couplings $\nu_{aR} \nu_{bR} \phi$, when it develops a vacuum expectation value (vev), $\langle \phi\rangle \sim \La$. The scalar $P$-plets correspondingly possess the components $\varphi_{11},\varphi_{22},\varphi_{33},\cdots,\varphi_{PP}$ which have both $Q=0$ and $B-L=0$, hence possibly developing vevs, such as $v_1,v_2,v_3,\cdots,v_P$, respectively. The remaining scalar fields have vanishing vev due to the electric charge conservation. The first two vevs $v_{1,2}$ break the standard model symmetry and give mass for ordinary particles, while $v_{3,4,\cdots,P}$ including $\La$ break the extended symmetry and provide mass for new particles. To be consistent with the standard model, we impose $v_{1,2}\ll v_{3,4,5,\cdots,P},\La$. The scheme of the gauge symmetry breaking is therefore summarized as
\bc
\begin{tabular}{c}
$SU(3)_C\otimes SU(P)_L\otimes U(1)_X\otimes U(1)_N$\\
$\downarrow v_{3,4,\cdots,P},\La$\\
$SU(3)_C\otimes SU(2)_L\otimes U(1)_Y\otimes P$\\
$\downarrow v_{1,2}$\\
$SU(3)_C\otimes U(1)_Q\otimes P$
\end{tabular} 
\ec
where $Y=\sum^{P-2}_{k=1}\beta_k H_k+X$ is the hypercharge, while $P$ is a residual symmetry of $B-L$ obtained below. To achieve the appropriate scenario of multicomponent dark matter, one must take $v_{3,4,\cdots,P}$ at TeV scale, while $\La$ that would define both the seesaw scale and the multiple matter parity $P$ can take values from TeV to the inflation scale $\sim 10^{15}$ GeV. 

As indicated before, $B-L=\sum_{k=1}^{P-2} b_k H_k +N$ is a nonfactorized charge of $SU(P)_L\otimes U(1)_N$. It transforms a general field as $\Phi\rightarrow \Phi'=U(\al)\Phi$, where $U(\al)=e^{i\al (B-L)}$, and $\al$ is a transforming parameter. $B-L$ annihilates the vevs $v_{1,2,\cdots,P}$ since the corresponding scalar fields all have $B-L=0$; in other words $(B-L)\langle \varphi\rangle=0$ for every scalar $P$-plet. $B-L$ is survived after the gauge symmetry breaking by $v_{1,2,\cdots,P}$. However, $B-L$ is broken by $\La$, i.e. $(B-L)\La\neq 0$, since $B-L=N=2$ for $\phi$. A residual symmetry of $B-L$ satisfies $U(\al)\La=\La$, thus $e^{i\al 2}=1$, leading to $\al=z\pi$ for $z=0,\pm 1,\pm 2,\cdots$. The final survival symmetry includes $U(z\pi)=(-1)^{z(B-L)}$, which is actually larger than a $Z_2$ (at least homomorphic to $Z_6$) for $z=0,\pm1,\pm2,\cdots$. We consider an invariant (or normal) subgroup of it that is generated by the survival transformation $P\equiv U(3\pi)=(-1)^{3(B-L)}$ according to $z=3$. We further redefine \be P=(-1)^{3(B-L)+2s},\ee by multiplying the spin parity $(-1)^{2s}$, which is always conserved by the Lorentz symmetry. The transformation $P$ is similar to $R$-parity in supersymmetry, but in our model it arises from the gauge symmetry and means a multiple matter parity.  

Indeed, let us calculate all $P$ values for fields, as collected in Table \ref{mmp}. 
\begin{table}[h]
\bc
\begin{tabular}{c|ccccccccc}
\hline\hline
Field & $\nu_a$ & $e_a$ & $u_a$ & $d_a$ & $E_{ka}$ & $J_{k\al}$ & $J_{k3}$ & $\varphi_{11}$ & $\varphi_{12}$  \\
\hline
$P$ & 1 & 1 & 1 & 1 & $P^+_k$ & $P^-_k$ & $P^+_k$ & 1 & 1  \\
\hline\hline
Field & $\varphi_{21}$ & $\varphi_{22}$ & $\varphi_{k+2,1}$ & $\varphi_{k+2,2}$ & $\varphi_{1,l+2}$ & $\varphi_{2,l+2}$ & $\varphi_{k+2,l+2}$ & $\phi$ & $G$ \\
$P$ & 1& 1& $P^+_k$ & $P^+_k$ & $P^-_l$ & $P^-_l$ & $P^+_k P^-_l$ & 1 & 1 \\
\hline\hline
Field  & $B$ & $C$ & $A$ & $W_{12}$ & $W_{k+2,1}$ & $W_{k+2,2}$ & $W_{1,l+2}$ & $W_{2,l+2}$ & $W_{k+2,l+2}$\\
\hline
$P$  & 1& 1& 1 & 1 & $P^+_k$ & $P^+_k$ & $P^-_l$ & $P^-_l$ & $P^+_k P^-_l$\\
\hline\hline
\end{tabular}
\caption[]{\label{mmp} The multiple matter parity $P$ value of the model particles, where $P^{\pm}_{k,l}\equiv (-1)^{\pm(3 n_{k,l}+1)}$ and $k,l=1,2,3,\cdots,P-2$. Additionally, $G$, $B$, $C$, $A$ and $W$'s denote the gauge bosons associated with the color, $X$, $N$, the Cartan and weight raising/lowering generators.}
\ec
\end{table}        
First note that $(P^\pm_k)^\dagger = P^\mp_k$ and $(P^\pm_k)^2\neq 1$ which generally differ from a parity.\footnote{Only if all $3n_k+1$ are integer, $P$ reduces to a parity.} It is clear that the fields that have a $B-L$ charge dependent on $n_k$ or $n_k-n_l$ transform as $P^\pm_k$ or $P^\pm_k P^\mp_l$, respectively, while the other fields including the standard model ones have $P=1$, called {\it normal fields}. $P^\pm_k$ and $P^\pm_k P^\mp_l$ are nontrivial ($\neq 1$) if $n_k\neq (2z-1)/3=\pm 1/3,\pm 1,\pm 5/3,\cdots$ and $n_k-n_l\neq 2z/3=0,\pm 2/3,\mp 4/3,\cdots$ for every $z$ integer, respectively. Such conditions generally apply since $n_k,n_l$ can in principle be arbitrary. Hence, the fields that have a nontrivial $P$ value are called {\it wrong fields}, since they possess an abnormal $B-L$ charge, opposite to the standard model definition of $B-L$ for normal fields ($B-L=-1$, $1/3$, and 0 for leptons, quarks, and bosons, respectively).  

Because our theory conserves the $P$-transformation, there is no single wrong field in interactions. Hence, the wrong fields are only coupled in pairs or self-interacted in interactions. Further, if an interaction has $r_k$ of $P^+_k$ fields and $s_l$ of $P^-_l$ fields, where $r_k,s_l$ are integer, the $P$ conservation implies $\sum_k r_k (3n_k+1) - \sum_l s_l (3n_l+1)=2 z $ for $z$ integer. This is valid for arbitrary $n_k,n_l$ charge parameters if and only if $k=l$ and $r_k=s_l$, i.e. $P^+_k$ and $P^-_k$ always appear in pairs. If an interaction has $t_{kl}$ of $P^+_k P^-_l$ fields for $t_{kl}$ integer, the $P$ conservation leads to $\sum_{kl} t_{kl} (3n_k-3n_l)=2 z$ for $z$ integer. This happens for arbitrary $n_k,n_l$ charge parameters if $t_{kl}=t_{lk}$, i.e. $P^+_k P^-_l$ and its conjugate always appear in pairs. Last, but not least, if an interaction contains both kinds of the above wrong fields, then the $P^+_k P^-_l$ field can self-interact with two $P^-_k$ and $P^+_l$ fields, such that $P$ is conserved. 

The above analysis yields that $P^\pm_k$ is separately conserved, for each $k$. Thus the invariant subgroup that includes $P$ must span \be P=\bigotimes_{k=1}^{P-2} P_k=P_1\otimes P_2\otimes \cdots \otimes P_{P-2},\ee where the operator $P_k$ has values $P^\pm_k$ or 1 when acting on a field. Each field would possess a multiple $P$ value such as $P=(P_1,P_2,\cdots,P_{P-2})$. The normal fields have $P=(+,+,\cdots,+)$, while the wrong fields have at least a $P_k=P^\pm_k\neq 1$ in $P$. For the latter, we call singly-, doubly-, triply-, etc. wrong fields if they contain one, two, three, etc. nontrivial $P_k$'s in $P$, respectively. Considering the special case, $n_k=2z/3$ for each $k$ and any $z$ integer, we have $P^\pm_k=-1$ for every $k$. In this case, each $P_k$ is a $Z_2$, and the invariant subgroup that contains \be P=Z_2\otimes Z_2\otimes \cdots \otimes Z_2\ee closes by itself, called multiple matter parity, as stated before. This structure will be used for investigating realistic models, where $P$ always possesses odd/even values when acting on fields, likely $P=(\cdots \pm,\cdots,\pm,\cdots,\pm\cdots )$. It is noteworthy that even if the number of odd values are even, the corresponding field always belongs to the class of wrong fields, despite having a product of partial parities $P=1$.

Three remarks are in order 
\ben
\item It is sufficiently to consider $P$, since the theory automatically conserves the quotient group of the residual symmetry by $P$.\footnote{We can construct a theory such that the residual symmetry coincides with $P$, assuming a heavy scalar singlet $\phi'\sim (1,1,0,2/3)$ for $B-L$ breaking and then integrating it out. But, this is not necessary.}  
\item The wrong scalar fields always have vanishing vev, even if electrically neutral, due to the $P$ conservation. This validates the choice of vevs from the outset.   
\item The lightest $P^\pm_k$ field is stabilized responsible for dark matter due to the $P_k$ conservation, for each $k$. Hence, there are simultaneously $P-2$ stable dark matter candidates corresponding to the $P_1,P_2,\cdots,P_{P-2}$ transformations, called multicomponent dark matter, as stated before. The 3-4-1-1 and 3-5-1-1 models contain two-component dark matter and three-component dark matter, respectively, whereas the well-established 3-3-1-1 model has only single-component dark matter.     
\een                    

Furthermore, dark matter must be colorless and electrically neutral. We have various schemes for dark matter candidates,   
\ben \item $q_k=0$: The candidate is either a lepton $E_k$, a non-Hermitian gauge boson that couples $\nu E_k$, or a scalar combination of $\varphi_{k+2,1}$ and $\varphi_{1,k+2}$.\item $q_l=-1$: The candidate is either a non-Hermitian gauge boson that couples $e E_l$ or a scalar combination of $\varphi_{l+2,2}$ and $\varphi_{2,l+2}$.
\item $q_{r}=q_{s}$: The candidate is either a non-Hermitian gauge boson that couples $E_r E_s$ or a scalar combination of $\varphi_{r+2,s+2}$ and $\varphi_{s+2,r+2}$. \een 
The multicomponent dark matter scenarios are given by composing the above $P-2$ conditions. For instance, $q_k=0$ for all $k$, the multicomponent dark matter may contain either only lepton candidates $E_1,E_2,\cdots, E_{P-2}$, only scalar candidates as combinations of $\varphi_{k+2,1}$ and $\varphi_{1,k+2}$ for $k=1,2,3,\cdots,P-2$, or composition of those $N$ lepton plus $P-2-N$ scalar candidates. Alternatively, $q_l=-1$ for all $l$, the multicomponent dark matter may include only scalar candidates as combinations of $\varphi_{l+2,2}$ and $\varphi_{2,l+2}$ for $l=1,2,3,\cdots,P-2$. Similarly, $q_k=0$ for $k=1,2,\cdots,N$ and $q_l=-1$ for $l=N+1,N+2,\cdots,P-2$, this case may consist of $N$ lepton and $P-2-N$ scalar candidates. 

\begin{figure}[h]
\bc
\includegraphics[scale=0.8]{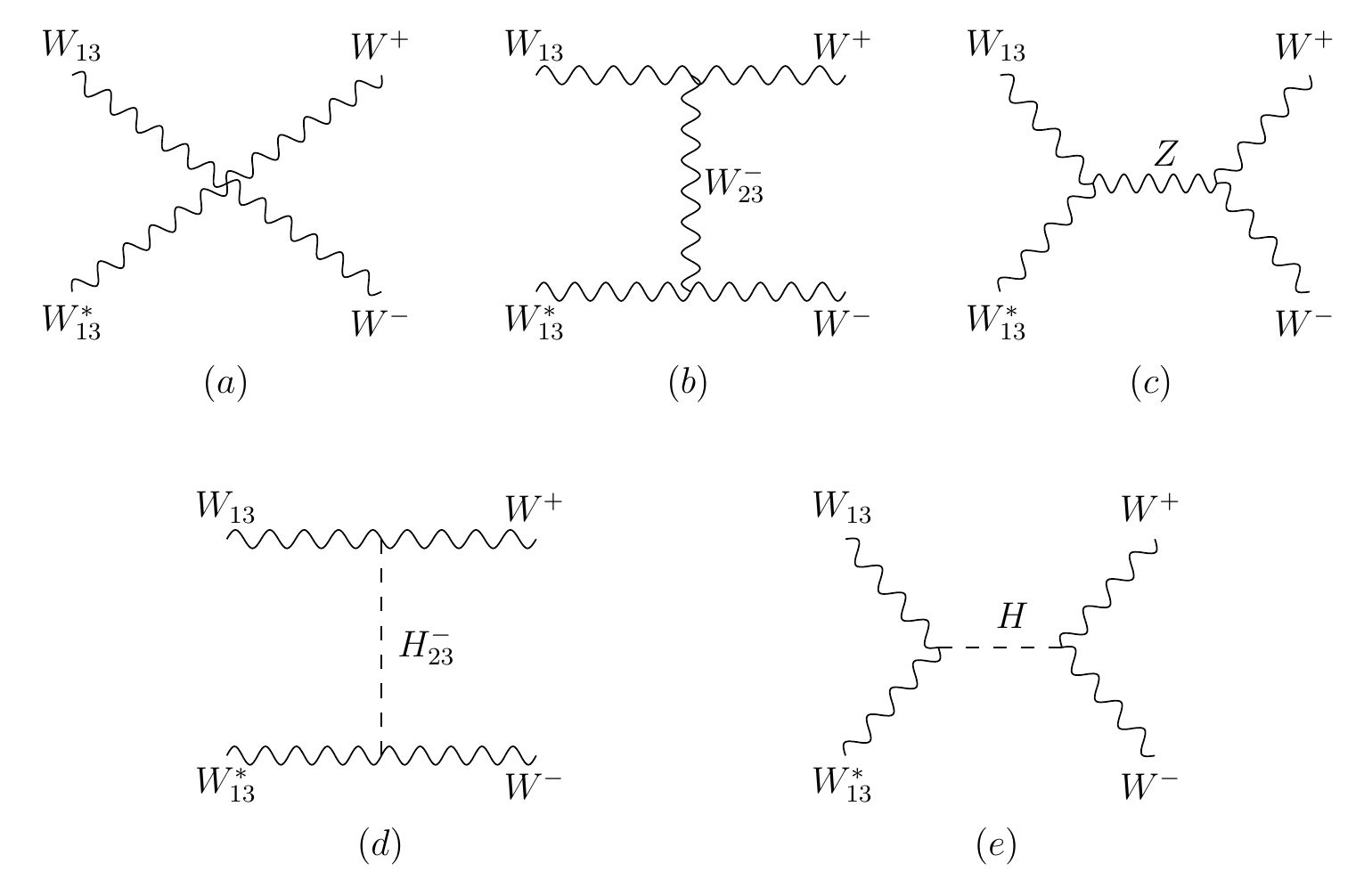}
\caption[]{\label{ffigddtta0} Dominant contributions to the process $W_{13}W^*_{13}\rightarrow W^+W^-$, where $H^-_{23}$ is the physical charged wrong Higgs field combined of $\varphi^-_{32}$ and $\varphi^-_{23}$, while the remaining combination orthogonal to $H^-_{23}$ is the Goldstone boson eaten by $W^-_{23}$. For a reference, the fields $Z$, $H$, and $H^-_{23}$ are identical to $Z_1$, $H_1$, and $\mathcal{H}^-_4$ in the following 3-4-1-1 model, respectively.}
\ec
\end{figure}

Above, we do not interpret the vector candidates, since they can be imposed to be heavier than the scalar or fermion candidates that belong to the same kind of wrong particles with the vector fields, for simplicity. Otherwise, the multicomponent dark matter scenarios may contain (stable) vector candidates. In this case, let us remind the reader that a particular study of the 3-3-1-1 model presented in \cite{Dong:2013wca} by one of us argued that the vector candidates annihilated entirely before freeze-out due to the gauge interactions with the standard model $W,Z$, which did not contribute to the present dark matter relic. But, such conclusion is not true, presenting a mistake, since the cross-section obtained therein enhanced with the energy scaling would break the unitarity bound at high energy. Hence, we would like to revisit the question of vector dark matter annihilation here for the 3-$P$-1-1 model in general. Consider the candidate $W_{13}$, thus $q_1=0$ and $\beta_1=-1/\sqrt{3}$, without loss of generality. The annihilation of $W_{13}$ to the standard model particles is proceeded through processes, $W_{13} W^*_{13}\rightarrow W^+ W^-, ZZ, HH, f\bar{f}$, where $f$ denotes ordinary leptons and quarks. The last two processes to $HH, f\bar{f}$ yield cross-sections proportional to $1/m^2_{W_{13}}$ as the normal WIMP, which should be skipped. For the first two processes to gauge boson pairs, let us particularly consider the one $W_{13} W^*_{13}\rightarrow W^+ W^-$, which is induced by five tree-level diagrams as despited in Fig. \ref{ffigddtta0}, where the diagrams in the first raw ($a,b,c$) are similar to Fig. 1 in \cite{Dong:2013wca}. 
The diagram ($a$) is via the contact interaction $\fr{g^2}{2}(g_{\mu\nu}g_{\al\beta}+g_{\mu\beta}g_{\al\nu}-2g_{\mu\al}g_{\nu\beta})W_{13\mu} W^*_{13\nu}W^+_{\al} W^-_{\beta}$. The diagram ($b$) is by $t$-channel $W_{23}$ exchange via couplings $\fr{g}{\sqrt{2}}L^{\mu\nu\al} W^+_{\mu} W^-_{23\nu}W^*_{13\al} +H.c.$, while the diagram ($c$) is by $s$-channel $Z$ exchange via couplings $\fr{g}{2c_W}L^{\mu\nu\al}Z_\mu W_{13\nu}W^*_{13\al}-g c_{W} L^{\mu\nu\al} Z_\mu W^+_\nu W^-_\al $, where $L^{\mu\nu\al}=g^{\mu\nu}(p_1-p_2)^\al+g^{\nu\al}(p_2-p_3)^\mu+g^{\al\mu}(p_3-p_1)^\nu$ depends on the momenta of $V_{1\mu}(p_1), V_{2\nu}(p_2), V_{3\al}(p_3)$ vector fields that go into the vertex. The diagram ($d$) is via couplings $(g^2 v_2/2\sqrt{2}) H^-_{23} W^{*\mu}_{13} W^+_\mu+H.c.$, while the diagram ($e$) is via couplings $\fr{g^2}{2}(v^2_1/\sqrt{v^2_1+v_2^2}) H W^\mu_{13}W^*_{13\mu}+\fr{g^2}{2}\sqrt{v^2_1+v^2_2} H W^{+\mu}W^-_{\mu}$. Notice that the only $Z,H$ contribute to the $s$-channel diagram, while the photon $\ga$ does not since it does not couple to $W_{13}W^*_{13}$. Additionally, the new heavy neutral gauge and Higgs bosons do not contribute to the $s$-channel diagram since they do not couple to $W^+W^-$ at the effective limit $v_{1,2}\ll v_{3,4,...,P},\La$. We perform calculation in the unitary gauge, so the Goldstone bosons are not presented. Denote the Lorentz indices and momenta for the incoming and outgoing particles such as \be W_{13\mu}(p_2)+W^*_{13\nu}(p_1)\rightarrow W^+_\beta (k_2)+W^-_\al (k_1).\ee To justify the enhancement of the scattering amplitudes with energy scale, it is conveniently to work in the center-of-mass frame of two incoming dark matter, such that $p_1=(E,\vec{p})$ and $p_2=(E,-\vec{p})$, thus $k_1=(E,\vec{k})$ and $k_2=(E,-\vec{k})$ due to the energy-momentum conservation. Because the dark matter is nonrelativistic ($v\sim 10^{-3}$), we approximate $E\simeq m_{W_{13}}$ and $|\vec{p}|\simeq m_{W_{13}}v$, thus the scattering energy $\sqrt{s}=2E\simeq 2m_{W_{13}}$ and $|\vec{k}|=(E^2-m^2_W)^{1/2}\simeq m_{W_{13}}$. Because of $m_{W_{13}}\gg m_W$, the $W$ bosons are strongly boosted in dark matter annihilation and their longitudinal polarization components dominate over the amplitudes, for which one has \be \ep_\al (k_1)=\fr{k_{1\al}}{m_{W}}+\mathcal{O}\left(\fr{m_W}{m_{W_{13}}}\right),\hs \ep_\beta (k_2)=\fr{k_{2\beta}}{m_{W}}+\mathcal{O}\left(\fr{m_W}{m_{W_{13}}}\right).\ee Note that the polarization vectors of dark matter are proportional to unity and satisfy the Lorentz conditions, $p^\nu_1 \ep_\nu (p_1)=0$ and $p^\mu_2 \ep_{\mu }(p_2)=0$. Further, we use $p_1+p_2=k_1+k_2$, $p_1k_1=p_2 k_2$, $(m_{Z},m_W, m_H) \ll m_{W_{13}}$, $m_{H_{23}}\sim m_{W_{13}}$, and $m_{W_{13}}\simeq m_{W_{23}}$, since $v_{1,2}\ll v_3$. With these at hand, we obtain the amplitudes for producing the longitudinal $W$'s corresponding to the given diagrams to be 
\bea i M(a) &=& \fr{ig^2}{2 m^2_W}\ep_\mu (p_2)\ep_\nu (p_1) \left(g^{\mu\nu} k_1 k_2 + k_2^\mu k_1^\nu - 2 k_1^\mu k_2^\nu \right)+\mathcal{O}\left(\fr{m^2_W}{m^2_{W_{13}}}\right),\\
i M(b) &=& \fr{ig^2}{2 m^2_W}\ep_\mu (p_2)\ep_\nu (p_1)\left[k^\mu_2 k^\nu_1+g^{\mu\nu}(k^2_1-2p_1 k_1)\right]+\mathcal{O}\left(\fr{m^2_W}{m^2_{W_{13}}}\right),\\ 
i M(c)&=& \fr{ig^2}{2 m^2_W}\ep_\mu (p_2)\ep_\nu (p_1) \left[g^{\mu\nu} p_1 (k_1-k_2)-p^\mu_1(k_1-k_2)^\nu \right.\crn
&&\left. + (k_1+k_2)^\nu (k_1-k_2)^\mu\right]+\mathcal{O}(1),\\
i M (d)&=&\fr{-i g^4 v^2_2}{8 m^2_W} \ep_\mu (p_2)\ep_\nu (p_1) \fr{k^\mu_2 k^\nu_1}{m^2_{W_{13}}-m^2_{H_{23}}} + \mathcal{O}\left(\fr{m^4_W}{m^4_{W_{13}}}\right)\sim 1,\\
iM(e)&=&\fr{-i g^2}{4}\ep_\mu (p_2)\ep_\nu (p_1) \fr{k^\mu_2 k^\nu_1}{m^2_{W_{13}}} + \mathcal{O}\left(\fr{m^4_W}{m^4_{W_{13}}}\right)\sim 1. \eea The last two amplitudes are proportional to 1 due to $k_{1,2}\sim m_{W_{13}}$, preserving the unitarity condition, whereas the first three amplitudes are proportional to $m^2_{W_{13}}/m^2_W$, i.e. to $s/m^2_W$ at high energy. In other words, $M(a,b,c)$ separately violate the unitarity bound, since the amplitude must be smaller than a constant. However, it is clear that \be M(a)+M(b)+M(c)\sim \mathcal{O}(1),\ee i.e. their leading terms are added up to zero. Hence, the unitarity is satisfied by the process. And, the average cross-section times relative velocity  $\langle \sigma v_{\mathrm{rel}}\rangle_{W_{13}W^*_{13}\rightarrow W^+W^-} \simeq |M(a)+M(b)+M(c)+M(d)+M(e)|^2/(32\pi m^2_{W_{13}})$ is proportional to $1/m^2_{W_{13}}$ as expected. The cross-section for $W_{13}W^*_{13}\rightarrow ZZ$ is similarly suppressed by $1/m^2_{W_{13}}$ due to the unitarity condition, although it has a contact interaction $W_{13\mu}W^*_{13\nu}Z_\al Z_\beta$ and other channels as the previous process. Therefore, the total cross-section satisfies $\langle \sigma v_{\mathrm{rel}}\rangle =\langle \sigma v_{\mathrm{rel}}\rangle_{W_{13}W^*_{13}\rightarrow W^+W^-}+\langle \sigma v_{\mathrm{rel}}\rangle_{W_{13}W^*_{13}\rightarrow ZZ}+\langle \sigma v_{\mathrm{rel}}\rangle_{W_{13}W^*_{13}\rightarrow HH}+\langle \sigma v_{\mathrm{rel}}\rangle_{W_{13}W^*_{13}\rightarrow f\bar{f}}\sim 1/m^2_{W_{13}}$, suppressed by the dark matter mass squared as in the normal WIMP. The relic density of $W_{13}$ gets the correct value, $\Om_{W_{13}}h^2\simeq 0.1\ \mathrm{pb}/\langle \sigma v_{\mathrm{rel}}\rangle \simeq 0.12$, if $m_{W_{13}}$ is large enough, in TeV regime; that said, $W_{13}$ may be present-day viable, in contradiction to \cite{Dong:2013wca}. The above proof can apply for other wrong vector fields, such as $W_{1,l+2}$, $W_{2,l+2}$, $W_{k+2,l+2}$, and their conjugation, as a result of the unitarity. In other words, although the $SU(P)_L$ adjoint component fields are hierarchical in mass, their annihilation obeys the unitarity bound, yielding viable dark matter relic. Extra comment is that the doubly-wrong fields $W_{k+2,l+2}$ not only couple to relevant singly-wrong fields, e.g. $W_{1,l+2}$, $W_{k+2,2}$, and so on, through the gauge self-interactions, but also are always heavier than such singly-wrong fields (comparing $W_{34}$ with the remaining wrong vector-fields in the following 3-4-1-1 model, for clarity). Hence, in the early universe, $W_{k+2,l+2}$ also annihilate to the singly-wrong vectors via such gauge interaction. In summary, the possibility of vector dark matter existence in this kind of the model is worth exploring, a task to be taken up elsewhere.

In Appendix \ref{fermionmass} we present the fermion mass generation. There, the neutrino masses are given along with the determination of the seesaw scale $\La\sim [(h^\nu)^2/f^\nu]\times 10^{14}$ GeV, which is proportional to the inflation scale. Therefore, it is naturally to impose \be 100\ \mathrm{GeV}\sim v_{1,2}\ll v_{3,4,...,P-2}\ll \La\sim 10^{14}\ \mathrm{GeV},\ee where the intermediate physical regime $v_{3,4,...,P-2}\sim$ 5--10 TeV sets the multicomponent dark matter observables.    

The 3-3-1 model might encounter a low Landau pole around 4--5 TeV as shown in \cite{Dias:2004dc,Dias:2004wk} and such issue is potentially presented in the current model, as a result of the weak isospin extension to $SU(P)_L$ for $P=3,4,5$. Notice that the gauge couplings commonly denoted by $g=\{g_s,g,g_X,g_N\}$ generally change with the renormalization scale $M$, satisfying the RG equation, \be M\fr{\pa g}{\pa M}=\beta(g)=-\fr{g^3}{16\pi^2}b_g,\ee where the beta function is given at 1-loop level as \be b_g=\fr{11}{3} C_V-\fr 2 3 \sum_L C_L-\fr 2 3 \sum_R C_R -\fr 1 3 \sum_S C_S,\ee where $V,L,R,S$ denote the vector, left-chiral fermion, right-chiral fermion, and scalar representations of the gauge group. For $SU(\mathcal{N})$ group, one has $C_V=\mathcal{N}$ and $C_L=C_R=C_S=1/2$ for fundamental (anti)multiplets, while for $U(1)_{\mathcal{X}}$ group, one obtains $C_V=0$ and $C_{L,R,S}=\mathcal{X}^2$ for each $L,R,S$ field with corresponding $\mathcal{X}$-charge. For the 3-$P$-1-1 gauge groups, we compute $b_{g_s}=11-2P>0$ and $b_g=(7P-8)/2>0$, since $P=3,4,5$. Hence, $g_s$ and $g$ decrease when $M$ increases and obviously $g,g_s\rightarrow 0$ when $M\rightarrow \infty$. There is no Landau pole associate to $g_s,g$. However, since $C_V=0$ and $C_{L,R,S}=\mathcal{X}^2>0$ for the $U(1)$ groups, we have $b_{g_X}<0$ and $b_{g_N}<0$. Namely, $g_X$ and $g_N$ increase when $M$ increases. In this case, a Landau pole $M$ may be present at which $g_X(M)=\infty$ or $g_N(M)=\infty$. It is easily realized that $g_N$ is finite when $M$ tends to the Planck scale, while $g_X$ is different because of $U(1)_X$ along with $SU(P)_L$ embedded in $U(1)_Q$. Indeed, from the electric charge operator $Q=\sum_k \beta_k H_k +X$, we can directly deduce the photon field, $A$, that couples to it, as \be 
\fr{A}{e}=\fr{1}{g}\sum_k \beta_k A_k+\fr{B}{g_X},\ee where $A_k$ and $B$ are the gauge fields respectively associated with $H_k$ and $X$, as mentioned. Note that $e=g s_W$ required to match the standard model and thus the field normalization condition leads to \be s^2_W=\fr{g^2_X}{g^2+g^2_X\sum_k \beta^2_k}<\fr{1}{\sum_k \beta^2_k}.\label{adttn1}\ee Since $ \sum_k\beta_k^2=1+\sum_{k=1}^{P-2}\beta_k^2>1$ and $g^2$ decreases but does not vanish, $s^2_W$ approaches the r.h.s of inequality when $g^2_X=\infty$ at some finite energy scale $M$. If the 3-$P$-1-1 model has the embedding coefficients of $Q$ such that $ \sum_k\beta_k^2\rightarrow 1/s^2_W$, i.e. $\sum_k \beta^2_k$ is close to $1/s^2_W=4.329$ given at the weak scale, the Landau pole $M$ is presented above the weak scale, at which $s^2_W(M)=1/\sum_k\beta_k^2$ or $g_X(M)=\infty$. The choice of the basic charges $q_{1,2,...,P-2}$ responsible for a dark matter scheme is subjected to such condition necessarily justified, since the new physics is only valid below the Landau pole.

\subsection{Minimal realization}

The minimal realization of multicomponent dark matter corresponds to $P=4$, i.e. the 3-4-1-1 model. We now provides the necessary features of the model as well as obtaining the two-component dark matter scenarios that the model contains. 

The gauge symmetry is given by
\be SU(3)_C\otimes SU(4)_L\otimes U(1)_X\otimes U(1)_N.\ee
The $Q$ and $B-L$ charges are embedded as
\be Q=T_3+\beta T_8 + \ga T_{15}+X,\hs B-L= b T_8 +c T_{15}+N,\ee
where we redefine the coefficients $\beta\equiv \beta_1$, $\ga \equiv \beta_2$, $b\equiv b_1$, and $c\equiv b_2$ for brevity, and note that the $SU(4)_L$ charges are $T_i$ for $i = 1,2,3,\dots,15$.  

The fermion and scalar contents under the gauge symmetry are expressed in Table~\ref{tab22}, where we relabel the new fermions $E\equiv E_1$, $F\equiv E_2$, $J\equiv J_1$, $K\equiv J_2$, their $Q,B-L$ charges $q\equiv q_1$, $p\equiv q_2$, $n\equiv n_1$, $m\equiv n_2$, which are identical to the $X,N$ charges of the corresponding right-handed fermions, and the scalar multiplets $\eta\equiv \varphi_1$, $\rho\equiv \varphi_2$, $\chi\equiv \varphi_3$, $\Xi\equiv \varphi_4$, for clarity.  
\begin{table}[h]
\bc
\begin{tabular}{lcccc}
\hline\hline
Multiplet & $SU(3)_C$ & $SU(4)_L$ & $U(1)_X$ & $U(1)_N$\\
\hline
$\psi_{aL}=\left(\nu \ e \ E \ F\right)^T_{aL}$ & 1 & 4 & $\fr{q+p-1}{4}$   & $\fr{n+m-2}{4}$\\
$Q_{\al L}= \left( d\  -u \ J \ K \right)^T_{\al L}$ & 3 & $4^*$ & $-\fr{q+p+1/3}{4}$ & $-\fr{n+m+2/3}{4}$\\
$Q_{3 L}= \left( u \ d \ J \ K\right)^T_{3L}$ & 3 & 4 & $\fr{q+p+5/3}{4}$ & $\fr{n+m+10/3}{4}$\\ 
$\nu_{aR}$ & 1 & 1 & 0 & $-1$\\
$e_{aR}$ & 1 & 1 & $-1$ & $-1$\\
$u_{a R}$ & 3 & 1 & $\fr 2 3$ & $\fr 1 3$\\
$d_{aR}$ & 3 & 1 & $-\fr 1 3$ & $\fr 1 3 $\\ 
$E_{aR}$ & 1 & 1 & $q$ & $n$\\ 
$F_{aR}$ & 1 & 1 & $p$ & $m$\\
$J_{\al R}$&3 & 1 & $-q-\fr 1 3$ & $-n-\fr 2 3$\\ 
$J_{3R}$ & 3 & 1 & $q+\fr 2 3$ & $n+\fr 4 3$\\
$K_{\al R}$ & 3 & 1 & $-p-\fr 1 3$ & $-m-\fr 2 3$ \\
$K_{3R}$ & 3 & 1 & $p+\fr 2 3$ & $m+\fr 4 3$ \\
$\eta =\left( \eta_1 \ \eta_2 \ \eta_3 \ \eta_4 \right)^T$ & 1 & 4 & $\fr{q+p-1}{4}$ & $\fr{n+m+2}{4}$ \\
$\rho=\left( \rho_1 \ \rho_2 \ \rho_3 \ \rho_4 \right)^T$ &  1 & 4 & $\fr{q+p+3}{4}$ & $\fr{n+m+2}{4}$ \\
$\chi =
\left( \chi_1\ \chi_2\ \chi_3\ \chi_4\right)^T$ & 1 & 4 & $\fr{p-3q-1}{4}$ & $\fr{m-3n-2}{4}$ \\
$ \Xi =
\left( \Xi_1 \ \Xi_2 \ \Xi_3 \ \Xi_4\right)^T$ & 1 & 4 & $\fr{q-3p-1}{4}$ & $\fr{n-3m-2}{4}$ \\
$\phi$ & 1 & 1 & 0 & 2\\
\hline\hline
\end{tabular}
\caption[]{\label{tab22} Field representation content of the model}
\ec
\end{table}
The charge parameters $q,p,n,m$ are related to the embedding coefficients as
\bea \beta&=&-\frac{1}{\sqrt3}(2q+1),\hs \gamma=\frac{1}{\sqrt6}(q-3p-1),\label{dtttn137} \\
 b&=&-\frac{2}{\sqrt3}(n+1),\hs c=\frac{1}{\sqrt6}(n-3m-2). \eea
 
The vevs of the scalar multiplets are written as  
\bea \langle \eta \rangle &=& \fr{1}{\sqrt{2}}\left(
\begin{array}{c} u \\ 0 \\ 0 \\ 0 \end{array}\right),\hs
\langle \rho\rangle = \fr{1}{\sqrt{2}} \left( \begin{array}{c} 0 \\ v \\ 0 \\ 0 \end{array}\right),\\ 
\langle \chi\rangle &=&
\fr{1}{\sqrt{2}} \left(\begin{array}{c} 0 \\ 0 \\ w \\ 0 \end{array}\right),\hs
\langle \Xi\rangle =
\fr{1}{\sqrt{2}} \left(\begin{array}{c} 0 \\ 0 \\ 0 \\ V \end{array}\right),\hs
\langle \phi\rangle = \fr{1}{\sqrt{2}} \La,\eea where we redenote $u\equiv v_1$, $v\equiv v_2$, $w\equiv v_3$, and $V\equiv v_4$, for convenience. As mentioned, $u,v$ define the electroweak scale, while $w,V$ determine the intermediate new physics scale relevant to dark matter and the seesaw scale $\La$ prevents small neutrino masses. The consistent condition is $\La\gg w, V \gg u,v$.

The multiple matter parity takes the form \be P=P_n\otimes P_m,\ee where the partial parities $P_n$ and $P_m$ have values to be either 1 or $P_n^\pm = (-1)^{\pm (3n+1)}=-1$ and $P_m^\pm = (-1)^{\pm (3m+1)}=-1$, provided that $n,m=2z/3=0,\pm2/3,\pm 4/3,\pm2,\cdots$, respectively. $P$ classifies the particles, such as \ben \item Normal particles for $P=(+,+)$, which include the standard model particles; \item Wrong particles for $P=(+,-)$, $(-,+)$, or $(-,-)$, containing the most new particles.\een They are all collected in Table \ref{BLCNF}, along with the corresponding $Q,B-L$ charges and $P$ multiple matter parity.  
\begin{table}[h]
\bc
\begin{tabular}{lccc}
\hline\hline
Particle & $Q$ & $B-L$ & $P$\\
\hline
$\nu_a$ & 0 & $-1$ & $(+,+)$ \\ 
$e_a$ & $-1$ & $-1$ & $(+,+)$\\
$E_a$ & $q$ & $n$ & $(-,+)$\\
$F_a$ & $p$ & $m$ & $(+,-)$\\
$u_a$ & $2/3$ & $1/3$ & $(+,+)$\\
$d_a$ & $-1/3$ & $1/3$ & $(+,+)$\\
$J_\alpha$ & $-q-1/3$ & $-n-2/3$ & $(-,+)$\\
 $K_\alpha$ & $-p-1/3$ & $-m-2/3$ & $(+,-)$\\
 $J_3$ & $q+2/3$ & $n+4/3$ & $(-,+)$\\
 $K_3$ & $p+2/3$ & $m+4/3$ & $(+,-)$\\
$\eta_1,\rho_2,\chi_3,\Xi_4$ & 0 & 0 & $(+,+)$\\
$\rho_1,\eta^*_2$ & 1 & 0 & $(+,+)$\\
$\chi_1,\eta^*_3$ & $-q$ & $-n-1$ & $(-,+)$\\
$\chi_2,\rho^*_3$ & $-q-1$ & $-n-1$ & $(-,+)$\\
$\Xi_1,\eta^*_4$ & $-p$ &$-m-1$ & $(+,-)$\\
$\Xi_2,\rho^*_4$ & $-p-1$& $-m-1$ & $(+,-)$ \\
$\Xi_3,\chi^*_4$  & $q-p$ & $n-m$ & $(-,-)$\\
$\phi$ & 0 & 2 & $(+,+)$\\  
$G,\ga,Z_{1,2,3,4}$ & 0 & 0 & $(+,+)$\\
$W$& $1$ & 0 & $(+,+)$\\
 $W_{13}$ & $-q$ & $-n-1$ & $(-,+)$\\ 
$W_{14}$ &  $-p$ & $-m-1$ & $(+,-)$\\
$W_{23}$ & $-q-1$ & $-n-1$ & $(-,+)$\\
$W_{24}$ & $-p-1$ & $-m-1$ & $(+,-)$\\
$W_{34}$ & $q-p$ & $n-m$ & $(-,-)$\\
 \hline\hline
\end{tabular}
\caption[]{\label{BLCNF} $Q$, $B-L$ charges and $P$ multiple matter parity of the model particles.}
\ec
\end{table}

The two-component dark matter scenarios can be extracted when imposing color and electric neutrality conditions for both kinds of the candidates (i.e., $P_n$ and $P_m$ odd fields). With the aid of the physical scalar and gauge boson fields identified in the next section, we derive such dark matter schemes as given in Table \ref{twocdts}.
\begin{table}[h]
\bc
\begin{tabular}{lcccc}
\hline\hline
Model & $\sum_k \beta^2_k$ & $(-,+)$ candidate & $(+,-)$ candidate & $(-,-)$ candidate \\
\hline
$q=p=0$ & 1.5 & $E_{1,2,3}$, $\mathcal{H}_{2}$, $W_{13}$ & $F_{1,2,3}$, $\mathcal{H}_{3}$, $W_{14}$ & $\mathcal{H}_{6}$, $W_{34}$\\
\hline
$q=0,p=-1$ & 2 & $E_{1,2,3}$, $\mathcal{H}_{2}$, $W_{13}$ & $\mathcal{H}_{5}$, $W_{24}$ & Non \\
\hline
$q=-1,p=0$ & 2 & $\mathcal{H}_{4}$, $W_{23}$ & $F_{1,2,3}$, $\mathcal{H}_{3}$, $W_{14}$ & Non \\
\hline
$q=p=-1$ & 1.5 & $\mathcal{H}_{4}$, $W_{23}$ & $\mathcal{H}_{5}$, $W_{24}$ & $\mathcal{H}_{6}$, $W_{34}$ \\
\hline
$q=p\neq 0,-1$ & $1+\fr 1 2 (2q+1)^2$ & Non & Non & $\mathcal{H}_{6}$, $W_{34}$\\
\hline \hline
\end{tabular}
\caption[]{\label{twocdts} Schemes of two-component dark matter.}
\ec
\end{table}
As mentioned, since the wrong gauge bosons do not contribute to dark matter, the model in the last raw is not appropriate for multicomponent dark matter, whereas the first four models do. Additionally, from the second column, the interested models have $\sum_k \beta^2_k$ to be equal to 1.5 or 2, which are very far from $1/s^2_W=4.329$ at the weak scale. A Landau pole if presented in these models is much beyond the TeV scale (perhaps beyond the GUT scale, since at that scale typically $1/s^2_W=8/3\simeq 2.67$ as in $SU(5)$ is still larger than those of our model). The model with $q=p=0$ has a rich two-component dark matter structure, to be further investigated in this work.     
 
The total Lagrangian is given, up to the gauge fixing and ghost terms, by 
\be \mathcal{L}= \bar{F}i\ga^\mu D_\mu F+(D^\mu S)^\dagger (D_\mu S) -\fr 1 4 A_{\mu\nu} A^{\mu\nu} + \mathcal{L}_{\mathrm{Yukawa}}-V_{\mathrm{Higgs}},\ee where $F$, $S$, and $A$ run over the fermion, scalar, and gauge-boson multiplets, respectively. The covariant derivative and field strength tensors are
\bea D_\mu &=& \pa_\mu + i g_s t_r G_{r\mu}+ i g T_i A_{i \mu}+ i g_X X B_\mu + i g_N N C_\mu,\label{code}\\
G_{r\mu\nu}&=&\pa_\mu G_{r\nu}-\pa_\nu G_{r\mu}-g_s f_{rst} G_{s\mu} G_{t\nu},\\
A_{i\mu\nu}&=&\pa_\mu A_{i\nu}-\pa_\nu A_{i\mu}-g f'_{ijk} A_{j\mu} A_{k\nu},\\
B_{\mu\nu}&=&\pa_\mu B_\nu-\pa_\nu B_\mu, \hs C_{\mu\nu}=\pa_\mu C_\nu-\pa_\nu C_\mu, \eea where $(g_s,g,g_X,g_N)$, $(t_r, T_i,X,N)$, and $(G_r,A_i,B,C)$ are the coupling constants, generators, and gauge bosons according to the 3-4-1-1 subgroups, respectively. $f_{rst}$ and $f'_{ijk}$ are the $SU(3)$ and $SU(4)$ structure constants, respectively.   

The Yukawa interactions can be extracted from Appendix \ref{fermionmass}, such as 
\bea \mathcal{L}_{\mathrm{Yukawa}}&=&\fr 1 2 f^\nu_{ab}\bar{\nu}^c_{aR}\nu_{bR}\phi+h^\nu_{ab}\bar{\psi}_{aL}\eta\nu_{bR}+ h^e_{ab}\bar{\psi}_{aL}\rho e_{bR} + h^E_{ab}\bar{\psi}_{aL}\chi E_{bR}+ h^{F}_{ab}\bar{\psi}_{aL}\Xi F_{bR}\crn
&&+ h^u_{3a} \bar{Q}_{3L}\eta u_{aR}+h^u_{\al a } \bar{Q}_{\al L}\rho^* u_{aR} +h^d_{3a}\bar{Q}_{3L}\rho d_{aR} + h^d_{\al a} \bar{Q}_{\al L}\eta^* d_{aR} \crn
&& + h^J_{33}\bar{Q}_{3L}\chi J_{3R}+ h^{K}_{33}\bar{Q}_{3L}\Xi K_{3R} + h^J_{\al \beta}\bar{Q}_{\al L} \chi^* J_{\beta R}+ h^{K}_{\al \beta}\bar{Q}_{\al L} \Xi^* K_{\beta R} \crn 
&&+ H.c.\eea 
The scalar potential is given by 
\bea
V_{\mathrm{Higgs}} &=& \mu^2_1\eta^\dagger \eta + \mu^2_2 \rho^\dagger \rho + \mu^2_3 \chi^\dagger \chi + \mu^2_4 \Xi^\dagger \Xi +\la_1 (\eta^\dagger \eta)^2 + \la_2 (\rho^\dagger \rho)^2 + \la_3 (\chi^\dagger \chi)^2 + \la_4 (\Xi^\dagger \Xi)^2\crn
&& +(\eta^\dagger \eta) (\la_5 \rho^\dagger \rho +\la_6 \chi^\dagger \chi +\la_7\Xi^\dagger \Xi)+(\rho^\dagger \rho)(\la_8 \chi^\dagger \chi +\la_9 \Xi^\dagger \Xi) +\la_{10}(\chi^\dagger \chi)(\Xi^\dagger \Xi)\crn
&&+\la_{11} (\eta^\dagger \rho)(\rho^\dagger \eta) +\la_{12} (\eta^\dagger \chi)(\chi^\dagger \eta)+\la_{13} (\eta^\dagger \Xi)(\Xi^\dagger \eta)+\la_{14} (\rho^\dagger \chi)(\chi^\dagger \rho)\crn
&& +\la_{15} (\rho^\dagger \Xi)(\Xi^\dagger \rho)+\la_{16} (\chi^\dagger \Xi)(\Xi^\dagger \chi)+ (\la_{17}\eta \rho \chi \Xi+H.c.) +V(\phi), 
\eea where the last term is the potential of $\phi$ plus the interactions of $\phi$ with $\eta$, $\rho$, $\chi$, and $\Xi$,
\be V(\phi)=\mu^2 \phi^*\phi +\la (\phi^*\phi)^2+ (\phi^*\phi) (\la_{18}\eta^\dagger \eta +\la_{19} \rho^\dagger \rho +\la_{20}\chi^\dagger \chi +\la_{21}\Xi^\dagger \Xi). \ee

We will identify the scalar mass spectrum and calculate the gauge interactions of scalars and fermions, which were all skipped in \cite{VanLoi:2019xud}.

\section{\label{scalar}Scalar and gauge sectors}

The necessary conditions for the scalar potential to be bounded from below and to induce the gauge symmetry breaking properly are
\be \mu^2, \mu^2_{1,2,3,4}<0,\hs \la,\la_{1,2,3,4}>0. \ee Here the conditions for the $\la$'s couplings could be determined for $V_{\mathrm{Higgs}}>0$ when $\phi$, $\eta$, $\rho$, $\chi$, and $\Xi$ separately tend to infinity, respectively. Further, additional conditions are required for $V_{\mathrm{Higgs}}>0$ when every two of those fields simultaneously tend to infinity, which obey \bea && \la_5 + \la_{11}\Theta(-\la_{11}) > -2\sqrt{\la_1\la_2},\\ 
&&\la_6 + \la_{12}\Theta(-\la_{12})> -2\sqrt{\la_1\la_3},\\ 
&&\la_7 + \la_{13}\Theta(-\la_{13}) > -2\sqrt{\la_1\la_4},\\
&& \la_8 + \la_{14}\Theta(-\la_{14}) > -2\sqrt{\la_2\la_3},\\
&& \la_9 + \la_{15}\Theta(-\la_{15}) > -2\sqrt{\la_2\la_4},\\
&& \la_{10} + \la_{16}\Theta(-\la_{16}) > -2\sqrt{\la_3\la_4},\\ 
&& \la_{18}  > -2\sqrt{\la\la_1},\hs \la_{19}>-2\sqrt{\la \la_2},\\
&& \la_{20}> -2\sqrt{\la \la_3},\hs \la_{21}> -2\sqrt{\la \la_4}, \eea where $\Theta(x)$ is the Heaviside step function. Notice that the total potential required to be positive for every three (and every four, every give) of the scalar fields simultaneously tending to infinity would supply extra conditions for the vacuum stability. Additionally, the constraints of physical scalar masses squared to be positive may also be existed, but most of them should be equivalent to the given conditions.
On the other hand, the hierarchies $u,v\ll w,V\ll \La$ can be obtained by requiring \be |\mu_{1,2}|\ll |\mu_{3,4}|\ll |\mu|.\ee 

We consider the large hierarchy case $|\mu| \gg |\mu_{1,2,3,4}|$, such that $\phi$ is decoupled. The $\phi$ field obtains a large vev, $\La^2\simeq -\mu^2/\la$, 
implied by $V(\phi)$. Let us expand \be \phi = \fr{1}{\sqrt2}(\Lambda+H_N+iG_{Z_N}),\ee where $H_N$ and $G_{Z_N}$ are heavy Higgs and massless Goldstone bosons associated with the $U(1)_N$ breaking, with the gauge boson $Z_N\equiv C$. The $U(1)_N$ gauge and Higgs bosons have masses, $m_{Z_N}\simeq 2g_N\La$ and $m_{H_N}\simeq \sqrt{2}\Lambda$, which are proportional to $\La$ scale and decoupled from the particle spectrum.  

Below $\La$, integrating $\phi$ out, we find that the effective potential at the leading order coincides with $V_{\mathrm{Higgs}}$ when omitting $V(\phi)$, \be V^{\mathrm{eff}}_{\mathrm{Higgs}}\simeq V_{\mathrm{Higgs}}-V(\phi).\ee We expand the scalar quadruplets,
\bea \eta &=&\left(\begin{array}{cccc}
\frac{1}{\sqrt2}(u+S_1+iA_1) & \eta^-_2 & \eta^q_3 & \eta^p_4
\end{array}\right)^T,\\
\rho &=&\left(\begin{array}{cccc}
\rho^+_1 & \frac{1}{\sqrt2}(v+S_2+iA_2) & \rho^{q+1}_3 & \rho^{p+1}_4
\end{array}\right)^T,\\
\chi &=&\left(\begin{array}{cccc}
\chi^{-q}_1 & \chi^{-q-1}_2 & \frac{1}{\sqrt2}(w+S_3+iA_3) & \chi^{p-q}_4
\end{array}\right)^T, \\
\Xi &=&\left(\begin{array}{cccc}
\Xi^{-p}_1 & \Xi^{-p-1}_2 & \Xi^{q-p}_3 & \frac{1}{\sqrt2}(V+S_4+iA_4)\\
\end{array}\right)^T.\eea
Substituting them to the effective potential, we derive 
\be V^{\mathrm{eff}}_{\mathrm{Higgs}} = V_{\mathrm{min}}+V_{\mathrm{linear}}+V_{\mathrm{mass}}+V_{\mathrm{interaction}}, \ee
which includes vacuum energy, linear, mass, interaction terms, respectively. 

Because of the gauge invariance, the coefficients of $V_{\mathrm{linear}}$ vanish,
\bea
(2\mu_1^2 +2\la_1 u^2+\la_5 v^2+\la_6 w^2+\la_7 V^2)u+\la_{17} vwV &=& 0,\label{minconf}\\
(2\mu_2^2 +2\la_2 v^2+\la_5 u^2+\la_8 w^2+\la_9 V^2)v+\la_{17} uwV &=& 0,\label{minconf1}\\
(2\mu_3^2 +2\la_3 w^2+\la_6 u^2+\la_8 v^2+\la_{10} V^2)w+\la_{17} uvV &=& 0,\label{minconf2}\\
(2\mu_4^2 +2\la_4 V^2+\la_7 u^2+\la_9 v^2+\la_{10} w^2)V+\la_{17} uvw &=& 0,\label{minconl}
\eea which provide a solution for $u,v,w,V$ related to the potential parameters. 

The mass terms can be separated into, 
\be V_{\mathrm{mass}} = V_{\mathrm{mass}}^S+V_{\mathrm{mass}}^A+V_{\mathrm{mass}}^{\mathrm{charged}}, \ee
where the first two terms contain the mass terms of the CP-even and CP-odd scalar fields, respectively, while the last one consists of the mass terms of the charged scalar fields. Using the potential minimum conditions,(\ref{minconf}), (\ref{minconf1}), (\ref{minconf2}), and (\ref{minconl}), $V_{\mathrm{mass}}^S$ is given by
\be V_{\mathrm{mass}}^S=\fr 1 2 \left(\begin{array}{cccc}
S_1 & S_2 & S_3 & S_4 \end{array}\right)M^2_S \left(\begin{array}{cccc}
S_1 & S_2 & S_3 & S_4 \end{array}\right)^T, \ee
where
\be M^2_S=\left(\begin{array}{cccc}
2\la_1 u^2-\frac{\la_{17} v}{2u}wV &\la_5 uv +\frac{\la_{17}}{2}wV & \la_6 uw +\frac{\la_{17}}{2}vV & \la_7 uV+\frac{\la_{17}}{2}vw \\
\la_5 uv+\frac{\la_{17}}{2} wV & 2\la_2 v^2 -\frac{\la_{17} u}{2v}wV & \la_8 vw +\frac{\la_{17}}{2}uV & \la_9 vV+\frac{\la_{17}}{2}uw \\
\la_6 uw+\frac{\la_{17}}{2} vV & \la_8 vw +\frac{\la_{17}}{2}uV & 2\la_3 w^2 -\frac{\la_{17} V}{2w}uv & \la_{10} wV+\frac{\la_{17}}{2}uv \\
\la_7 uV+\frac{\la_{17}}{2} vw & \la_9 vV +\frac{\la_{17}}{2}uw & \la_{10} wV+\frac{\la_{17}}{2}uv & 2\la_4 V^2 -\frac{\la_{17} w}{2V}uv \end{array}\right). \ee

Due to the condition $u,v\ll w,V$, the above mass matrix would provide a small eigenvalue identical to the standard model Higgs boson ($H_1$) mass and three large eigenvalues corresponding to new neutral Higgs bosons ($H_{2,3,4}$). Indeed, at the leading order, we obtain
\bea
H_1 &=& \frac{uS_1+vS_2}{\sqrt{u^2+v^2}}, \hs m^2_{H_1} = 0,\\
H_2 &=& \frac{vS_1-uS_2}{\sqrt{u^2+v^2}}, \hs m^2_{H_2} = -\frac{\la_{17}(u^2+v^2)}{2uv}wV,\\
H_3 &=& c_{\al_1}S_3 - s_{\al_1}S_4,\hs m^2_{H_3} = \la_4V^2 + \la_3w^2 -\sqrt{(\la_4V^2 - \la_3w^2)^2+ 4\la_{10}^2w^2V^2},\\
H_4 &=& s_{\al_1}S_3 + c_{\al_1}S_4,\hs m^2_{H_4} = \la_4V^2 + \la_3w^2 +\sqrt{(\la_4V^2 - \la_3w^2)^2+ 4\la_{10}^2w^2V^2},
\eea
where
\be \tan (2\al_1) = \frac{\la_{10}wV}{\la_4V^2-\la_3w^2} \ee is finite, given that $w\sim V$. 

In the new basis, $(H_1,H_2,H_3,H_4)$, the mass matrix $M^2_S$ has a type I and II seesaw form, which includes $m^2_{H_1}\sim (u,v)^2$ plus the mixing terms between the Higgs bosons $H_{1,2,3,4}$ proportional to $(u,v)(w,V)$, while the mass terms of $H_{2,3,4}$ depend on $(w,V)^2$. Hence, at the next-to-leading order, the standard model Higgs boson $H_1$ gains a mass via the seesaw formula, approximated to be 
\be m^2_{H_1}\simeq \frac{2}{u^2+v^2}(\la_1u^4+\la_2v^4+\la_5u^2v^2), \ee while the heavy Higgs bosons $H_{2,3,4}$ have the masses as retained. The mixings between the Higgs bosons are suppressed by $(u,v)/(w,V)\ll 1$, as neglected.   

The second term $V_{\mathrm{mass}}^A$ is given by
\be V_{\mathrm{mass}}^A=\fr 1 2 \left(\begin{array}{cccc}
A_1 & A_2 & A_3 & A_4 \end{array}\right)M^2_A \left(\begin{array}{cccc}
A_1 & A_2 & A_3 & A_4 \end{array}\right)^T, \ee
where
\be M^2_A=-\fr{\la_{17}}{2} \left(\begin{array}{cccc}
\frac{v}{u} wV & wV & vV & vw  \\
wV & \frac{u}{v} wV & uV & uw  \\
vV & uV &  \frac{V}{w}uv & uv  \\
vw & uw & uv &  \frac{w}{V}uv 
\end{array}\right). \ee

This mass matrix provides a physical pseudoscalar with a corresponding mass,
\bea && \mathcal{A}=\frac{(v A_1 + u A_2)wV + (V A_3 + w A_4)uv}{\sqrt{v^2w^2V^2 + u^2 [w^2V^2 + v^2(w^2+V^2)]}}, \\
&& m^2_\mathcal{A}=-\frac{\la_{17}}{2}\left(\frac{u^2+v^2}{uv} wV + \frac{w^2+V^2}{wV}uv\right). \eea
The remaining eigenstates are three massless pseudoscalars identical to the Goldstone bosons of the corresponding neutral gauge bosons, $Z_{1,2,3}$, such that 
\bea G_{Z_1}&=&\frac{u A_1 - v A_2}{\sqrt{u^2 + v^2}}, \hs G_{Z_2}=\frac{w A_3 - V A_4}{\sqrt{w^2 + V^2}},\\
G_{Z_3}&=&\frac{(v A_1 + u A_2)(w^2 + V^2)uv - (V A_3 + w A_4)(u^2 + v^2) wV}{\sqrt{(u^2 + v^2)(w^2 + V^2)[(u^2+v^2)w^2V^2+(w^2+V^2)u^2v^2]}},\eea where $Z_1$ and $G_{Z_1}$ are identical to those of the standard model, while the others $Z_{2,3}$ and $G_{Z_{2,3}}$ are new physical states. 

Concerning the charged scalar term, we obtain
\bea V_{\mathrm{mass}}^{\mathrm{charged}} &=& \left(\begin{array}{cc}
\rho_1^+ & \eta_2^+ \end{array}\right)M^2_{1} \left(\begin{array}{c}
\rho_1^- \\ \eta_2^- \end{array}\right) + \left(\begin{array}{cc}
\chi_1^q & \eta_3^q \end{array}\right)M^2_{q} \left(\begin{array}{c}
\chi_1^{-q} \\ \eta_3^{-q} \end{array}\right)\crn
&& + \left(\begin{array}{cc}
\Xi_1^p & \eta_4^p \end{array}\right)M^2_{p} \left(\begin{array}{c}
\Xi_1^{-p} \\ \eta_4^{-p} \end{array}\right) + \left(\begin{array}{cc}
\chi_2^{q+1} & \rho_3^{q+1} \end{array}\right)M^2_{q+1} \left(\begin{array}{c}
\chi_2^{-(q+1)} \\ \rho_3^{-(q+1)} \end{array}\right) \crn
&& + \left(\begin{array}{cc}
\Xi_2^{p+1} & \rho_4^{p+1} \end{array}\right)M^2_{p+1} \left(\begin{array}{c}
\Xi_2^{-(p+1)} \\ \rho_4^{-(p+1)} \end{array}\right)+ \left(\begin{array}{cc}
\Xi_3^{(q-p)} & \chi_4^{(q-p)} \end{array}\right)M^2_{q-p} \left(\begin{array}{c}
\Xi_3^{-(q-p)} \\ \chi_4^{-(q-p)} \end{array}\right).\nn \eea

Each of the mass matrices $M^2$'s of the charged scalars yields a massless eigenstate identical to the Goldstone boson of a corresponding non-Hermitian gauge boson and a massive eigenstate with a mass at $w,V$ scale. They are summarized as 
\bea M^2_{1}&=&\fr 1 2 \left(\begin{array}{cc}\frac{u}{v}(\la_{11}uv-\la_{17} wV)& \la_{11} uv - \la_{17} wV\\
\la_{11} uv - \la_{17} wV & \frac{v}{u}(\la_{11}uv-\la_{17} wV)\end{array}\right),\\
G^\pm_W &=& \frac{v \rho_1^\pm - u \eta_2^\pm}{\sqrt{u^2+v^2}},\
\mathcal{H}^\pm_1 = \frac{u \rho_1^\pm + v \eta_2^\pm}{\sqrt{u^2+v^2}},\\
 m^2_{\mathcal{H}^\pm_1} &=& \frac{u^2+v^2}{2uv}(\la_{11} uv - \la_{17} wV),
\\ 
M^2_{q}&=&\fr 1 2 \left(\begin{array}{cc}\left(\la_{12}u-\la_{17} v\frac{V}{w}\right)u & \la_{12} uw - \la_{17} vV\\
\la_{12} uw - \la_{17} vV & \left(\la_{12}w-\la_{17} V\frac{v}{u}\right)w \end{array}\right), \\
G^{\pm q}_{W_{13}} &=& \frac{w\chi_1^{\pm q} - u\eta_3^{\pm q}}{\sqrt{u^2+w^2}},\
\mathcal{H}^{\pm q}_2 = \frac{u\chi_1^{\pm q} + w\eta_3^{\pm q}}{\sqrt{u^2+w^2}},\\
 m^2_{\mathcal{H}^{\pm q}_2}&=&\fr 1 2 \left(\la_{12}-\la_{17}\frac{vV}{uw}\right)(u^2+w^2),
\\ 
M^2_{p}&=&\fr 1 2 \left(\begin{array}{cc}\left(\la_{13}u-\la_{17} v\frac{w}{V}\right)u & \la_{13} uV - \la_{17} vw\\
\la_{13} uV - \la_{17} vw & \left(\la_{13}V-\la_{17} w\frac{v}{u}\right)V \end{array}\right), \\
G^{\pm p}_{W_{14}} &=& \frac{V\Xi_1^{\pm p} - u\eta_4^{\pm p}}{\sqrt{u^2+V^2}},\
\mathcal{H}^{\pm p}_3 = \frac{u\Xi_1^{\pm p} + V\eta_4^{\pm p}}{\sqrt{u^2+V^2}},\\
 m^2_{\mathcal{H}^{\pm p}_3}&=&\fr 1 2 \left(\la_{13}-\la_{17}\frac{vw}{uV}\right)(u^2+V^2),
\\ 
M^2_{q+1}&=&\fr 1 2 \left(\begin{array}{cc}\left(\la_{14}v-\la_{17} u\frac{V}{w}\right)v & \la_{14} vw - \la_{17} uV\\
\la_{14} vw - \la_{17} uV & \left(\la_{14}w-\la_{17} V\frac{u}{v}\right)w \end{array}\right), \\
G^{\pm (q+1)}_{W_{23}} &=& \frac{w\chi_2^{\pm (q+1)} - v\rho_3^{\pm (q+1)}}{\sqrt{v^2+w^2}},\
\mathcal{H}^{\pm (q+1)}_4 = \frac{v\chi_2^{\pm (q+1)} + w\rho_3^{\pm (q+1)}}{\sqrt{v^2+w^2}},\\ m^2_{\mathcal{H}^{\pm (q+1)}_4}&=&\fr 1 2 \left(\la_{14}-\la_{17}\frac{uV}{vw}\right)(v^2+w^2),
\\ 
M^2_{p+1}&=&\fr 1 2 \left(\begin{array}{cc}\left(\la_{15}v-\la_{17} u\frac{w}{V}\right)v & \la_{15} vV - \la_{17} uw\\
\la_{15} vV - \la_{17} uw & \left(\la_{15}V-\la_{17} w\frac{u}{v}\right)V \end{array}\right), \\
G^{\pm (p+1)}_{W_{24}} &=& \frac{V\Xi_2^{\pm (p+1)} - v\rho_4^{\pm (p+1)}}{\sqrt{v^2+V^2}},\
\mathcal{H}^{\pm (p+1)}_5 = \frac{v\Xi_2^{\pm (p+1)} + V\rho_4^{\pm (p+1)}}{\sqrt{v^2+V^2}},\\
m^2_{\mathcal{H}^{\pm (p+1)}_5}&=&\fr 1 2 \left(\la_{15}-\la_{17}\frac{uw}{vV}\right)(v^2+V^2),\\
 M^2_{q-p}&=&\fr 1 2 \left(\begin{array}{cc}(\la_{16}wV-\la_{17} uv)\frac{w}{V} & \la_{16} wV - \la_{17} uv\\
\la_{16} wV - \la_{17} uv & (\la_{16}wV-\la_{17} uv)\frac{V}{w} \end{array}\right), \\
G^{\pm (q-p)}_{W_{34}} &=& \frac{V\Xi_3^{\pm(q-p)} - w\chi_4^{\pm(q-p)}}{\sqrt{w^2+V^2}},\
\mathcal{H}^{\pm (q-p)}_6 = \frac{w\Xi_3^{\pm (q-p)} + V\chi_4^{\pm (q-p)}}{\sqrt{w^2+V^2}},\\ m^2_{\mathcal{H}^{\pm (q-p)}_6}&=&\frac{w^2+V^2}{2wV}(\la_{16}wV-\la_{17} uv).
\eea

In summary, at the effective limit $u,v\ll w,V$, the physical scalar states are related to the gauge states as follows:
\bea
\left(\begin{array}{c} H_1 \\ H_2 \end{array}\right) &\simeq & \left(\begin{array}{cc} c_{\al_2} & s_{\al_2} \\ s_{\al_2} & -c_{\al_2} \end{array}\right)\left(\begin{array}{c} S_1 \\ S_2 \end{array}\right),\hs \left(\begin{array}{c} H_3 \\ H_4 \end{array}\right) \simeq \left(\begin{array}{cc} c_{\al_1} & -s_{\al_1} \\ s_{\al_1} & c_{\al_1} \end{array}\right)\left(\begin{array}{c} S_3 \\ S_4 \end{array}\right),\\
\crn
\left(\begin{array}{c} \mathcal{A} \\ G_{Z_1} \end{array}\right) &\simeq & \left(\begin{array}{cc} s_{\al_2} & c_{\al_2} \\ 
c_{\al_2} & -s_{\al_2} \end{array}\right)\left(\begin{array}{c} A_1 \\ A_2 \end{array}\right),\hs 
\left(\begin{array}{c} G_{Z_2} \\ G_{Z_3}  \end{array}\right) \simeq \left(\begin{array}{cc}  
s_{\al_3} & -c_{\al_3} \\ 
-c_{\al_3} & -s_{\al_3}  \end{array}\right)\left(\begin{array}{c} A_3 \\ A_4 \end{array}\right),\\
\crn
\left(\begin{array}{c} G^\pm_W \\ \mathcal{H}^\pm_1 \end{array}\right) &= & \left(\begin{array}{cc} s_{\al_2} & -c_{\al_2} \\ 
c_{\al_2} & s_{\al_2} \end{array}\right)\left(\begin{array}{c} \rho^\pm_1 \\ \eta^\pm_2 \end{array}\right),\hs 
\left(\begin{array}{c} G^{\pm (q-p)}_{W_{34}} \\ \mathcal{H}^{\pm (q-p)}_6  \end{array}\right) = \left(\begin{array}{cc}  
c_{\al_3} & -s_{\al_3} \\ 
s_{\al_3} & c_{\al_3}  \end{array}\right)\left(\begin{array}{c} \Xi^{\pm (q-p)}_3 \\ \chi^{\pm (q-p)}_4 \end{array}\right),\\
\crn
G^{\pm q}_{W_{13}}&\simeq &\chi_1^{\pm q},\hs G^{\pm p}_{W_{14}}\simeq \Xi_1^{\pm p},\hs G^{\pm (q+1)}_{W_{23}}\simeq \chi_2^{\pm (q+1)}, \hs G^{\pm (p+1)}_{W_{24}} \simeq \Xi_2^{\pm (p+1)},\\
\crn
\mathcal{H}^{\pm q}_2&\simeq &\eta_3^{\pm q},\hs \mathcal{H}^{\pm p}_3\simeq \eta_4^{\pm p},\hs \mathcal{H}^{\pm (q+1)}_4 \simeq \rho_3^{\pm (q+1)},\hs \mathcal{H}^{\pm (p+1)}_5 \simeq \rho_4^{\pm (p+1)},
\eea
where the $\alpha_{2,3}$ angles are defined by $\tan({\al_2})= v/u $ and $\tan({\al_3})= w/V $, respectively. 

Let us investigate the mass spectrum of the gauge bosons, given by
\be \mathcal{L}\supset \sum_{S=\eta,\rho,\chi,\Xi}\left(D^\mu\langle S\rangle\right)^\dag\left(D_\mu\langle S\rangle\right),\ee where the $U(1)_N$ gauge and Higgs sectors are decoupled as presented above. Note that in this case, the kinetic mixing between the two $U(1)$ gauge bosons does not contribute. Substituting the vevs of the scalar quadruplets, we get
\bea
\mathcal{L}&\supset&\frac{g^2}{4} \left[(u^2+v^2)W^{\mu +}W^-_\mu +(u^2+w^2)W^{q\mu}_{13}W^{-q}_{13\mu}\right.\crn
&&+(u^2+V^2)W^{p\mu}_{14}W^{-p}_{14\mu} +(v^2+w^2)W^{(q+1)\mu}_{23}W^{-(q+1)}_{23\mu}\crn
&&\left. +(v^2+V^2)W^{(p+1)\mu}_{24}W^{-(p+1)}_{24\mu}+(w^2+V^2)W^{(q-p)\mu}_{34}W^{-(q-p)}_{34\mu}\right]\crn && +\fr 1 2 \left(\begin{array}{cccc} A_3^\mu & A_8^\mu & A_{15}^\mu & B^\mu \end{array}\right)M^2_0 \left(\begin{array}{cccc} A_{3\mu} & A_{8\mu} & A_{15\mu} & B_\mu \end{array}\right)^T,
\eea where we have denoted the non-Hermitian gauge bosons as
\bea
W^\pm_\mu &=& \frac{1}{\sqrt2}(A_{1\mu}\mp i A_{2\mu}),\hs W^{\pm q}_{13\mu} = \frac{1}{\sqrt2}(A_{4\mu}\pm i A_{5\mu}),\label{chgauf}\\
W^{\pm p}_{14\mu} &=& \frac{1}{\sqrt2}(A_{9\mu}\pm i A_{10\mu}),\hs W^{\pm (q+1)}_{23\mu} = \frac{1}{\sqrt2}(A_{6\mu}\pm i A_{7\mu}),\label{addl1}\\
W^{\pm (p+1)}_{24\mu} &=& \frac{1}{\sqrt2}(A_{11\mu}\pm i A_{12\mu}),\hs W^{\pm (q-p)}_{34\mu} = \frac{1}{\sqrt2}(A_{13\mu}\mp i A_{14\mu}).\label{chgaul}
\eea

The non-Hermitian gauge bosons are physical eigenstates by themselves with corresponding masses
\bea
m^2_{W} &=& \frac{g^2}{4}(u^2+v^2),\hs m^2_{W_{13}} = \frac{g^2}{4}(u^2+w^2),\hs m^2_{W_{14}} = \frac{g^2}{4}(u^2+V^2),\\
m^2_{W_{23}} &=& \frac{g^2}{4}(v^2+w^2),\hs m^2_{W_{24}} = \frac{g^2}{4}(v^2+V^2),\hs m^2_{W_{34}} = \frac{g^2}{4}(w^2+V^2).
\eea
The $W$ boson has a mass at the weak scale identified to the standard model $W$ boson, thus $u^2+v^2=(246 \text{ GeV})^2$, as mentioned. Whereas, the remainders are new charged gauge bosons with large masses at $w,V$ scale. 

For the neutral gauge bosons, the mass matrix is given by
\bea
M^2_0=\frac{g^2}{4}\left(\begin{array}{cccc} 
u^2+v^2 & \frac{1}{\sqrt3}(u^2-v^2) & \frac{1}{\sqrt6}(u^2-v^2) & m^2_{14} \\
\frac{1}{\sqrt3}(u^2-v^2) & \fr 1 3 (u^2+v^2+4w^2) & \frac{1}{3\sqrt2}(u^2+v^2-2w^2) & m^2_{24} \\
\frac{1}{\sqrt6}(u^2-v^2) & \frac{1}{3\sqrt2}(u^2+v^2-2w^2) & \fr 1 6 (u^2+v^2+w^2+9V^2) & m^2_{34}  \\
 m^2_{14} &  m^2_{24} &  m^2_{34} & m^2_{44} 
\end{array}\right),
\eea
where
\bea
m^2_{14} &=& 2(u^2X_\eta-v^2X_\rho)t_X, \\
m^2_{24} &=& \frac{2}{\sqrt3}(u^2X_\eta+v^2X_\rho-2w^2X_\chi)t_X,\\
m^2_{34} &=& \sqrt{\fr 2 3}(u^2X_\eta+v^2X_\rho+w^2X_\chi-3V^2X_\Xi)t_X, \\
m^2_{44} &=& 4(u^2X_\eta^2+v^2X_\rho^2+w^2X_\chi^2+V^2X_\Xi^2)t_X^2,
\eea
where we define $t_X=g_X/g$ and the scalar $X$-charges are given above. 

The diagonalization of the mass matrix $M^2_0$ can be read off from \citep{VanLoi:2019xud}, which yields the neutral gauge bosons,
\bea
A_\mu &=& s_W A_{3\mu} + c_W \left(\beta t_W A_{8\mu} +\gamma t_W A_{15\mu} +\frac {t_W}{t_X} B_\mu\right),\label{gaua}\\
Z_{1\mu} &=& c_W A_{3\mu} -s_W \left(\beta t_W A_{8\mu} +\gamma t_W A_{15\mu} +\frac {t_W}{t_X} B_\mu\right), \\
Z'_{2\mu} &=& \frac{1}{\sqrt{1-\beta^2 t_W^2}} \left[(1-\beta^2 t_W^2) A_{8\mu}-\beta\gamma t_W^2 A_{15\mu}-\frac{\beta t_W^2}{t_X}  B_\mu\right], \\
Z'_{3\mu} &=& \frac{1}{\sqrt{1+\gamma^2t_X^2}}\left(A_{15\mu}- \gamma t_X B_\mu\right),\label{gauz} \eea
with the corresponding masses, 
\bea && m_A = 0, \hs m^2_{Z_1} = \frac{g^2}{4c^2_W}(u^2+v^2), \crn 
&&m^2_{Z'_2} \simeq \frac{g^2w^2}{3(1+\gamma^2t_X^2)}[1+(\beta^2+\gamma^2)t_X^2],\\
&&m^2_{Z'_3} \simeq \frac{g^2}{24(1+\gamma^2t_X^2)}\{w^2[\gamma(2\sqrt2\beta-\gamma)t_X^2-1]^2+9V^2(1+\gamma^2t_X^2)^2\},\\
&& m^2_{Z'_2Z'_3} \simeq \frac{g^2w^2[\gamma(2\sqrt2 \beta-\gamma)t^2_X-1]\sqrt{1+(\beta^2+\gamma^2)t^2_X}}{6\sqrt2(1+\gamma^2t_X^2)},
\eea
where $s_W = e/g = t_X/\sqrt{1+(1+\beta^2+\gamma^2)t_X^2}$ is the sine of the Weinberg angle. 

The photon field $A$ is an exact massless eigenstate, decoupled from the other states. The field $Z_{1}$ slightly mix with the two heavy states $Z'_{2,3}$, which at the effective limit, $(u,v)^2/(w,V)^2\ll 1$, the $Z_1$ is identified to the standard model $Z$ boson. There is a finite mixing between $Z'_{2}$ and $Z'_{3}$, which produces two new eigenstates, \be Z_{2\mu} = c_\varphi Z'_{2\mu} - s_\varphi Z'_{3\mu},\hs Z_{3\mu} = s_\varphi Z'_{2\mu} + c_\varphi Z'_{3\mu},\ee with corresponding masses,
\bea
m^2_{Z_2,Z_3} = \fr 1 2\left[m^2_{Z'_2}+m^2_{Z'_3}\mp\sqrt{(m^2_{Z'_2}-m^2_{Z'_3})^2+4m^4_{Z'_2Z'_3}}\right], \eea at $w,V$ scale. 
The mixing angle is given by
\bea t_{2\varphi }&=& \fr{4\sqrt2 w^2[\gamma(2\sqrt2 \beta-\gamma)t_X^2-1]\sqrt{1+(\beta^2+\gamma^2)t_X^2}}{w^2\{\gamma^2(2\sqrt2 \beta-\gamma)^2t_X^4-[(2\sqrt2 \beta+\gamma)^2+5\gamma^2]t_X^2-7\}+9V^2(1+\gamma^2t_X^2)^2}.\label{gaup}\eea

In summary, the physical neutral gauge bosons are related to the beginning states by $(A\ Z_1\ Z_2\ Z_3)^T=U(A_3\ A_8\ A_{15}\ B)^T$, where
\be U= \left(\begin{array}{cccc} 
s_W & \beta s_W & \ga s_W & \frac{s_W}{t_X} \\
c_W & -\beta s_W t_W & -\ga s_Wt_W & -\frac{s_Wt_W}{t_X} \\
0 & c_\varphi\sqrt{1-\beta^2t^2_W} & -\fr{s_\varphi}{\sqrt{1+\ga^2t_X^2}}-\fr{c_\varphi\beta\ga t^2_W}{\sqrt{1-\beta^2t_W^2}} & \fr{s_\varphi\ga t_X}{\sqrt{1+\ga^2t_X^2}}-\fr{c_\varphi\beta t^2_W}{t_X\sqrt{1-\beta^2t_W^2}} \\
0 &  s_\varphi\sqrt{1-\beta^2t^2_W} & \fr{c_\varphi}{\sqrt{1+\ga^2t_X^2}}-\fr{s_\varphi\beta\ga t^2_W}{\sqrt{1-\beta^2t_W^2}} & -\fr{c_\varphi\ga t_X}{\sqrt{1+\ga^2t_X^2}}-\fr{s_\varphi\beta t^2_W}{t_X\sqrt{1-\beta^2t_W^2}}  
\end{array}\right).\label{addng1}\ee 

\section{\label{interaction}Interactions}

\subsection{Gauge interactions for fermions}
The interactions of fermions with gauge bosons are derived from the Lagrangian,
\bea \mathcal{L} &\supset& \sum_F \bar{F}i\ga^\mu D_\mu F = \sum_F [\bar{F}i\ga^\mu \pa_\mu F - g_s\bar{F}\ga^\mu t_r G_{r\mu} F-g\bar{F}\ga^\mu (P^{CC}_{\mu}+P^{NC}_{\mu}) F], \eea
where the charged and neutral currents couple to the gauge bosons by
\bea P^{CC}_\mu &=& \sum_{i=1,2,4,5,6,7,9,10,11,12,13,14} T_iA_{i\mu}, \label{PCC}\\
 P^{NC}_\mu &=&\sum_{i=3,8,15} T_iA_{i\mu} + t_X XB_\mu + t_N NC_\mu.\label{PNC}\eea
 
Substituting the charged gauge bosons from (\ref{chgauf}), (\ref{addl1}), and (\ref{chgaul}) into (\ref{PCC}), we get
\bea P^{CC}_{\mu} &=& \frac{1}{\sqrt2}[(T_1+iT_2)W^+_\mu + (T_4-iT_5) W^{q}_{13\mu} + (T_9-iT_{10})W^{p}_{14\mu} \crn
&& + (T_6-iT_7)W^{q+1}_{23\mu} + (T_{11}-iT_{12})W^{p+1}_{24\mu} + (T_{13}+iT_{14})W^{q-p}_{34\mu} +H.c.].\eea
The interactions of the charged gauge bosons with fermions are
\bea -g\sum_F \bar{F}\ga^\mu P^{CC}_{\mu} F &=& J^{-\mu}_W W^+_\mu + J^{-q\mu}_{W_{13}} W^{q}_{13\mu} + J^{-p\mu}_{W_{14}} W^{p}_{14\mu} \crn
&& + J^{-(q+1)\mu}_{W_{23}} W^{q+1}_{23\mu} + J^{-(p+1)\mu}_{W_{24}} W^{p+1}_{24\mu} + J^{-(q-p)\mu}_{W_{34}} W^{q-p}_{34\mu} + H.c.,\eea
where the corresponding charged currents are determined by
\bea
J^{-\mu}_W &=& -\frac{g}{\sqrt2}(\bar{\nu}_{aL}\ga^\mu e_{aL}+\bar{u}_{aL}\ga^\mu d_{aL}),\\
J^{-q\mu}_{W_{13}} &=& -\frac{g}{\sqrt2}(\bar{E}_{aL}\ga^\mu\nu_{aL}-\bar{d}_{\al L}\ga^\mu J_{\al L}+\bar{J}_{3L}\ga^\mu u_{3L}),\\
J^{-p\mu}_{W_{14}} &=& -\frac{g}{\sqrt2}(\bar{F}_{aL}\ga^\mu\nu_{aL}-\bar{d}_{\al L}\ga^\mu K_{\al L}+\bar{K}_{3L}\ga^\mu u_{3L}),\\
J^{-(q+1)\mu}_{W_{23}} &=& -\frac{g}{\sqrt2}(\bar{E}_{aL}\ga^\mu e_{aL}+\bar{u}_{\al L}\ga^\mu J_{\al L}+\bar{J}_{3L}\ga^\mu d_{3L}),\\
J^{-(p+1)\mu}_{W_{24}} &=& -\frac{g}{\sqrt2}(\bar{F}_{aL}\ga^\mu e_{aL}+\bar{u}_{\al L}\ga^\mu K_{\al L}+\bar{K}_{3L}\ga^\mu d_{3L}),\\
J^{-(q-p)\mu}_{W_{34}} &=& -\frac{g}{\sqrt2}(\bar{E}_{aL}\ga^\mu F_{aL}+\bar{K}_{\al L}\ga^\mu J_{\al L}+\bar{J}_{3L}\ga^\mu K_{3L}).
\eea

Substituting the neutral gauge bosons from (\ref{addng1}) into (\ref{PNC}), we obtain
\bea
P^{NC}_\mu &=& s_W Q A_\mu + \frac{1}{c_W}(T_3-s^2_WQ)Z_{1\mu} \crn
&& + \left\{\frac{c_\varphi}{\sqrt{1-\beta^2 t_W^2}}[T_8-\beta (Q-T_3)t_W^2]-\frac{s_\varphi}{\sqrt{1+\ga^2 t_X^2}}(T_{15}-\ga X t_X^2)\right\}Z_{2\mu}\crn
&& + \left\{\frac{s_\varphi}{\sqrt{1-\beta^2 t_W^2}}[T_8-\beta (Q-T_3)t_W^2]+\frac{c_\varphi}{\sqrt{1+\ga^2 t_X^2}}(T_{15}-\ga X t_X^2)\right\}Z_{3\mu}.
\eea
Hence, the interactions of the neutral gauge bosons with fermions are given by
\bea
-g\sum_F \bar{F}\ga^\mu P^{NC}_{\mu} F &=& -eQ(f)\bar{f}\ga^\mu f A_\mu \crn
&&- \frac{g}{2c_W}\{\bar{\nu}_L\ga^\mu\nu_L + \bar{f}\ga^\mu [g_V^{Z_1}(f)- g_A^{Z_1}(f)\ga_5]f \}Z_{1\mu}\crn
&& - \frac{g}{2}\{ C^{Z_2}_{\nu_L}\bar{\nu}_L\ga^\mu\nu_L+ \bar{f}\ga^\mu [g_V^{Z_2}(f)- g_A^{Z_2}(f)\ga_5]f \} Z_{2\mu} \crn
&& - \frac{g}{2}\{C^{Z_3}_{\nu_L}\bar{\nu}_L\ga^\mu\nu_L + \bar{f}\ga^\mu [g_V^{Z_3}(f)- g_A^{Z_3}(f)\ga_5]f\} Z_{3\mu},\label{gaun}
\eea
where $e=gs_W$ and $f$ indicates every charged fermion of the model. 

The first term in (\ref{gaun}) yields electromagnetic interactions, as usual. Concerning the remaining interactions, for the neutrinos we have
\bea C^{Z_2}_{\nu_L} &=& \frac{c_\varphi (1+\sqrt3\beta t_W^2)}{\sqrt3\sqrt{1-\beta^2t_W^2}}-\frac{s_\varphi[1+\ga(\ga+\sqrt2\beta+\sqrt6)t_X^2]}{\sqrt6\sqrt{1+\ga^2t_X^2}},\\
C_{\nu_L}^{Z_3} &=& C_{\nu_L}^{Z_2}(c_\varphi\to s_\varphi, s_\varphi\to -c_\varphi).
\eea
For the other fermions, the vector and axial-vector couplings can be obtained as 
\bea
g_V^{Z_1}(f) &=& T_3(f_L)-2s_W^2 Q(f), \hs g_A^{Z_1}(f)=T_3(f_L),\\
g_V^{Z_2}(f) &=& \frac{c_\varphi \left\{T_8(f_L)-\beta [2Q(f)-T_3(f_L)]t_W^2\right\}}{\sqrt{1-\beta^2 t_W^2}}\crn
&&-\frac{s_\varphi \left\{T_{15}(f_L)-\ga [2Q(f)-T_3(f_L)-\beta T_8(f_L)-\ga T_{15}(f_L)] t_X^2\right\}}{\sqrt{1+\ga^2 t_X^2}},\\
g_A^{Z_2}(f) &=& \frac{c_\varphi [T_8(f_L)+\beta T_3(f_L)t_W^2]}{\sqrt{1-\beta^2 t_W^2}}\crn
&&-\frac{s_\varphi\left\{T_{15}(f_L)+\ga [T_3(f_L)+\beta T_8(f_L)+\ga T_{15}(f_L)] t_X^2\right\}}{\sqrt{1+\ga^2 t_X^2}},\\
g_{V,A}^{Z_3}(f) &=& g_{V,A}^{Z_2}(f)(c_\varphi\to s_\varphi, s_\varphi\to -c_\varphi).
\eea

In Appendix \ref{gvaf}, we compute the couplings of $Z_1$ with fermions as given in Table \ref{Z1}, which are consistent with the standard model. Additionally, the couplings of $Z_2$ with fermions are derived as collected in Table \ref{Z2}. Here, it is noted that the couplings of $Z_3$ with fermions can be obtained from those of $Z_2$ by replacing, $c_\varphi\to s_\varphi$ and $s_\varphi\to -c_\varphi$, which need not necessarily be determined.

\subsection{Gauge interactions for scalars}

The relevant interactions arise from
\be \mathcal{L} \supset \sum_{S=\eta,\rho,\chi,\Xi,\phi} (D^\mu S)^\dagger (D_\mu S), \ee
where $S=\langle S\rangle + S'$ takes the forms,
\bea
\eta &=& \left(\begin{array}{cccc} \frac{1}{\sqrt2}u & 0 & 0 & 0
\end{array} \right)^T + \left(\begin{array}{cccc} \frac{1}{\sqrt2}(c_{\al_2}H_1+s_{\al_2}H_2+i s_{\al_2}\mathcal{A}) &\,\, s_{\al_2}\mathcal{H}_1^- &\,\, \mathcal{H}_2^q &\,\, \mathcal{H}_3^p \end{array} \right)^T,\\
\rho &=& \left(\begin{array}{cccc} 0 &\frac{1}{\sqrt2}v & 0 & 0
\end{array} \right)^T + \left(\begin{array}{cccc} c_{\al_2}\mathcal{H}_1^+ &\,\, \frac{1}{\sqrt2}(s_{\al_2}H_1-c_{\al_2}H_2+i c_{\al_2}\mathcal{A}) &\,\, \mathcal{H}_4^{q+1} &\,\, \mathcal{H}_5^{p+1} \end{array} \right)^T,\\
\chi &=& \left(\begin{array}{cccc} 0 & 0 &\frac{1}{\sqrt2}w & 0
\end{array} \right)^T + \left(\begin{array}{cccc} 0 &\,\, 0 &\,\, \frac{1}{\sqrt2}(c_{\al_1} H_3 + s_{\al_1} H_4) &\,\, c_{\al_3} \mathcal{H}_6^{p-q} \end{array} \right)^T,\\
\Xi &=& \left(\begin{array}{cccc} 0 & 0 & 0 & \frac{1}{\sqrt2}V
\end{array} \right)^T + \left(\begin{array}{cccc} 0 &\,\, 0 &\,\, s_{\al_3} \mathcal{H}_6^{q-p} &\,\, \frac{1}{\sqrt2}(-s_{\al_1} H_3+c_{\al_1} H_4) \end{array} \right)^T,
\eea with the vevs and physical states explicitly shown. 

The Lagrangian is correspondingly expanded by  
\bea
\mathcal{L} &\supset& g[i(\pa^\mu S')^\dagger (P^{CC}_\mu S') + i (\pa^\mu S')^\dagger (P^{NC}_\mu S')+ \mathrm{H.c.}] \crn
&& + g^2[\langle S\rangle^\dagger P^{CC\mu} P^{CC}_\mu S'+\langle S\rangle^\dagger (P^{CC\mu} P^{NC}_\mu + P^{NC\mu} P^{CC}_\mu) S'+ \langle S\rangle^\dagger P^{NC\mu} P^{NC}_\mu S' +\mathrm{H.c.}] \crn
&& + g^2 [S'^\dag P^{CC\mu} P^{CC}_\mu S'+ S'^\dag (P^{CC\mu} P^{NC}_\mu+P^{NC\mu} P^{CC}_\mu) S'+ S'^\dag P^{NC\mu} P^{NC}_\mu S'].
\eea

In Appendix \ref{sgint}, we calculate all the gauge boson and scalar interactions and express the corresponding couplings from Table \ref{1CG2S} to Table \ref{2NG2ST2}. There, the new labels are, 
\bea \beta_1 &\equiv &\frac{1-(1+2q)t_W^2}{\sqrt{1-\beta^2t_W^2}},\hs \ga_1\equiv \frac{\sqrt6+3\ga (1-q-p)t_X^2}{2\sqrt3\sqrt{1+\ga^2t_X^2}},\\
\beta_2 &\equiv &\frac{1+(1+2q)t_W^2}{\sqrt{1-\beta^2t_W^2}},\hs \ga_2\equiv \frac{\sqrt6-3\ga (3+q+p)t_X^2}{2\sqrt3\sqrt{1+\ga^2t_X^2}},\eea which differ from those in the electric charge operator, without confusion.

\section{\label{DMP} Multicomponent dark matter phenomenology}
We consider the model with $q=p=0$. In this case, the neutral particles that transform nontrivially under the multiple matter parity $P=P_n\otimes P_m$ are $E^0_a$, $F^0_a$, $\mathcal{H}^0_2$, $\mathcal{H}^0_3$, $\mathcal{H}^0_6$, $W^0_{13}$, $W^0_{14}$, and $W^0_{34}$, as explicitly shown in Table \ref{twocdts}. We divide into three possibilities of two-component dark matter existence.

For the discussion and numerical calculation throughout the following dark matter schemes, let us remind the reader that the $U(1)_N$ sector is decoupled by the condition $\La\gg w,V,u,v$, as mentioned. So its gauge coupling $g_N$ (or $t_N$), the scalar coupling $\la$, as well as the relevant fields and masses of this sector, disappear. Additionally, from the given expression of the sine of the Weinberg angle, we derive $t_X=g_X/g=s_W/\sqrt{1-(1+\beta^2+\ga^2)s^2_W}$, where $\beta=-1/\sqrt{3}$ and $\ga=-1/\sqrt{6}$ are obtained with the aid of (\ref{dtttn137}), which will be used hereafter. Note that the Yukawa couplings are proportional to the fermion mass matrices, if the VEVs are fixed. So, we only refer to the fermion masses through the scenarios. Moreover, $E_a$ and $F_a$ are assumed to be flavor diagonal, i.e. they are physical fields by themselves. The mixing effect of the ordinary quarks and leptons negligibly affect to the dark matter observables which would be skipped. Since the dark matter abundance is thermally produced, the dark matter masses are always assumed to be larger than the BBN and CMB bounds $> 0.1$ MeV.  

\subsection{Scenario with two-fermion dark matter}
We assume that $E$ (one of three particles $E^0_a$) and $F$ (one of three particles $F^0_a$), which are singly-wrong particles according to the separately conserved single parities $P_n$ and $P_m$, are the lightest particles within the classes of singly-wrong particles of the same kind ($E_a, \mathcal{H}_2$, $W_{13}$) and ($F_a, \mathcal{H}_3$, $W_{14}$), respectively. Note that $E$ and $F$ are only coupled via the new gauge boson $W_{34}$ and new Higgs scalar $\mathcal{H}_6$, due to the gauge and $P_{n,m}$ invariances. We further assume that the net mass of $E$ and $F$ is smaller than each mass of $W_{34}$ and $\mathcal{H}_6$. Hence, they are stable and can play the role of two-component dark matter candidates. 

The dominant channels of the dark matter pair annihilation into the standard model particles are given by
\bea 
EE^c &\to & \nu\nu^c,l^-l^+,qq^c,Z_1H_1,\\
FF^c &\to & \nu\nu^c,l^-l^+,qq^c,Z_1H_1. 
\eea
Particularly through the neutral gauge boson portals, let us note that there is no channel of fermion dark matter annihilation to $W^-W^+$, since $Z_2$ and $Z_3$ are not directly coupled with $W^-W^+$ at the effective limit, $(u,v)^2/(w,V)^2\ll 1$, although these $Z_{2,3}$ couple to $E,F$. Additionally, despite the fact that $Z_1$ couples to $W^-W^+$, the $Z_1$ does not couples to $E$ and $F$ for the model with $q=p=0$ under consideration, as can be verified from the couplings in Appendix \ref{gvaf}.
In addition, there is the conversion between dark matter, which plays the key role in the multicomponent dark matter scenario, in which the heavier dark matter component would annihilate into the lighter one. In this sense, there adds the annihilation process either
\be
EE^c \to FF^c\ \ \ \ \text{if}\ \ m_E>m_F,\ee
or \be FF^c \to EE^c\ \ \ \ \text{if}\ \ m_F>m_E.
\ee

\begin{figure}[!h]
	\centering
	\includegraphics[scale=0.9]{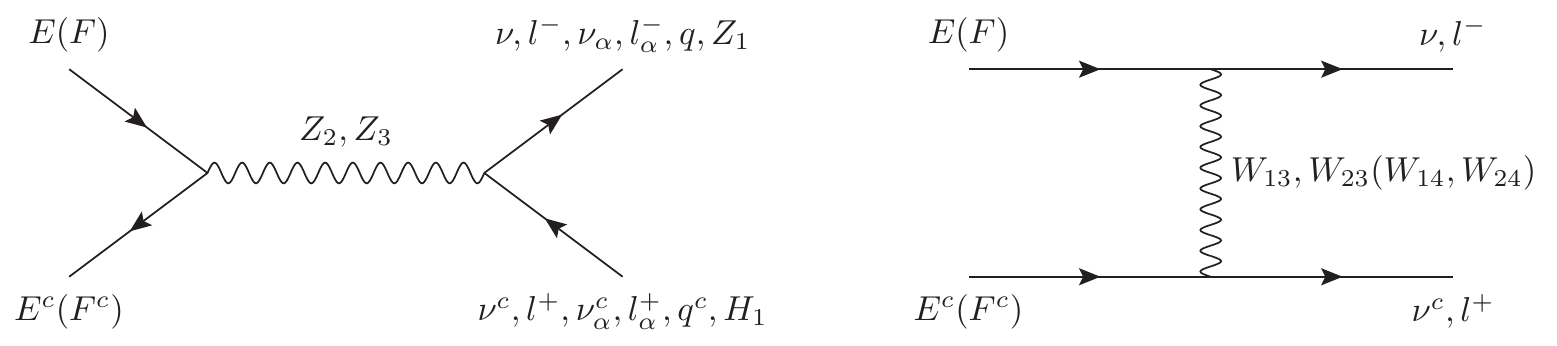}
	\caption{Dominant contributions to annihilation of the two-component fermion dark matter into standard model particles.}	\label{DMSM-1}
\end{figure}

\begin{figure}[!h]
	\centering
	\includegraphics[scale=0.9]{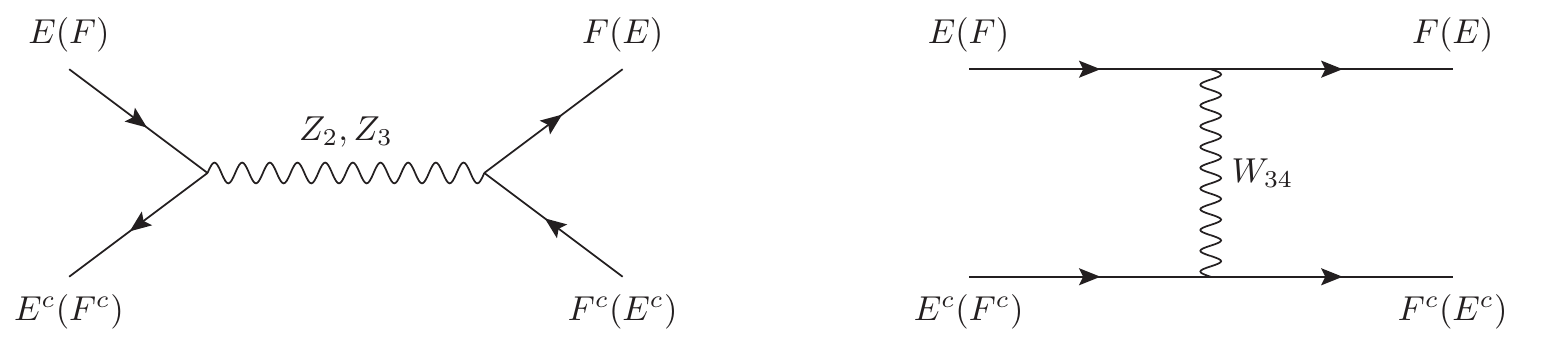}
	\caption{Conversion between fermion dark matter components.}	\label{DMDM-1}
\end{figure}

The relevant Feynman diagrams which describe the dark matter pair annihilation into the standard model particles and the conversion between dark matter components are given in Figures \ref{DMSM-1} and \ref{DMDM-1}, respectively. Hereafter, note that $Z_N$ is superheavy, hence not contributing to the dark matter observables. 

We compute the dark matter relic abundance due to the thermal freeze-out of two dark matter components $E$ and $F$. The dark matter relic abundance is obtained by solving the coupled Boltzmann equations (BEQs), which govern the evolution of $Y_{E(F)}\equiv\frac{n_{E(F)}}{s}$ with $n_{E(F)}$ referring to the number density of the dark matter component $E(F)$ and $s$ to be the entropy density, given by
\bea
\frac{dY_E}{dx}&=&-0.264M_{\text{Pl}}\sqrt{g_*}\frac{\mu}{x^2}\left\{\langle\sigma v\rangle_{EE^c\to\text{SM}\text{SM}}\left[Y^2_E-\left(Y^{\text{EQ}}_E\right)^2\right]\right.\crn
&&\left.+\langle\sigma v\rangle_{EE^c\to FF^c}\left[Y^2_E-\left(\frac{Y^{\text{EQ}}_E}{Y^{\text{EQ}}_F}\right)^2Y^2_F\right]\Theta(m_E-m_F)\right.\crn
&&\left.-\langle\sigma v\rangle_{FF^c\to EE^c}\left[Y^2_F-\left(\frac{Y^{\text{EQ}}_F}{Y^{\text{EQ}}_E}\right)^2Y^2_E\right]\Theta(m_F-m_E)\right\},\label{adta}\\
\frac{dY_F}{dx}&=&-0.264M_{\text{Pl}}\sqrt{g_*}\frac{\mu}{x^2}\left\{\langle\sigma v\rangle_{FF^c\to\text{SM}\text{SM}}\left[Y^2_F-\left(Y^{\text{EQ}}_F\right)^2\right]\right.\crn
&&\left.+\langle\sigma v\rangle_{FF^c\to EE^c}\left[Y^2_F-\left(\frac{Y^{\text{EQ}}_F}{Y^{\text{EQ}}_E}\right)^2Y^2_E\right]\Theta(m_F-m_E)\right.\crn
&&\left.-\langle\sigma v\rangle_{EE^c\to FF^c}\left[Y^2_E-\left(\frac{Y^{\text{EQ}}_E}{Y^{\text{EQ}}_F}\right)^2Y^2_F\right]\Theta(m_E-m_F)\right\},\label{BEQs}
\eea
where $M_{\text{Pl}}=1.22\times10^{19}$ GeV, $g_*=106.75$ is the effective total number of degrees of
freedom, $\mu=\frac{m_Em_F}{m_E+m_F}$, and
\bea
Y^{\text{EQ}}_{E(F)}=0.145\frac{g}{g_*}\left(\frac{m_{E(F)}}{\mu}\right)^{3/2}x^{3/2}e^{-x\frac{m_{E(F)}}{\mu}},
\eea
with $g=2$ being the number of degrees of freedom for dark matter components. Additionally, in the given BEQs, we have used the $\Theta$ functions to describe the dependence on mass hierarchy, either $m_E>m_F$ or $m_F>m_E$, corresponding to either the channel $EE^c\to FF^c$ or the channel $FF^c\to EE^c$ would contribute.

In equations (\ref{adta}) and (\ref{BEQs}), the thermal average annihilation cross-section times the relative velocity for the dark matter components is given in the non-relativistic approximation at the leading order as
\bea
\langle\sigma v\rangle_{EE^c\to\text{SM}\text{SM}}&=&\frac{g^4m^2_E}{\pi}\left\{\left[\frac{1}{(m^2_E+m^2_{W_{13}})^2}-\sum_{i}\frac{g^{Z_i}_V(E)[g^{Z_i}_V(\nu_{L})+g^{Z_i}_A(\nu_{L})]}{4(m^2_E+m^2_{W_{13}})(4m^2_E-m^2_{Z_i})}\right.\right.\crn
&&\left.\left.+\left(m_{W_{13}}\leftrightarrow m_{W_{23}},\nu_{L}\leftrightarrow e\right)\right]+\sum_{i,j}\frac{g_{_{Z_iZ_1H_1}}g_{_{Z_jZ_1H_1}}}{64g^2m^2_{Z_1}}\right.\crn
&&\left.\times\left[g^{Z_i}_V(E)g^{Z_j}_V(E)+g^{Z_i}_A(E)g^{Z_j}_A(E)\right]\frac{(4m^2_E-m^2_{Z_i})^{-1}}{4m^2_E-m^2_{Z_j}}\right.\crn
&&\left.+\sum_{f,i,j}N_C(f)\frac{g^{Z_i}_V(E)g^{Z_j}_V(E)[g^{Z_i}_V(f)g^{Z_j}_V(f)+g^{Z_i}_A(f)g^{Z_j}_A(f)]}{16(4m^2_E-m^2_{Z_i})(4m^2_E-m^2_{Z_j})}\right\},\\
\langle\sigma v\rangle_{EE^c\to FF^c}&=&\frac{g^4\sqrt{m^2_E-m^2_F}}{2\pi m_E}\left\{\frac{2m^2_E-m^2_F}{(m^2_E-m^2_F+m^2_{W_{34}})^2}-\sum_{i}\frac{(4m^2_E-m^2_{Z_i})^{-1}}{4(m^2_E-m^2_F+m^2_{W_{34}})}\right.\crn
&&\left.\times\left[2m^2_Eg^{Z_i}_V(E)[g^{Z_i}_V(F)+g^{Z_i}_A(F)]-m^2_Fg^{Z_i}_V(E)[g^{Z_i}_V(F)-g^{Z_i}_A(F)]\right.\right.\crn
&&\left.\left.+m^2_Fg^{Z_i}_A(F)[g^{Z_i}_V(E)-g^{Z_i}_A(E)]\right]+\sum_{i,j}\frac{(4m^2_E-m^2_{Z_j})^{-1}}{16(4m^2_E-m^2_{Z_i})}\right.\crn
&&\left.\times\left[2m^2_Eg^{Z_i}_V(E)g^{Z_j}_V(E)[g^{Z_i}_V(F)g^{Z_j}_V(F)+g^{Z_i}_A(F)g^{Z_j}_A(F)]\right.\right.\crn
&&\left.\left.+m^2_Fg^{Z_i}_V(E)g^{Z_j}_V(E)[g^{Z_i}_V(F)g^{Z_j}_V(F)-g^{Z_i}_A(F)g^{Z_j}_A(F)]\right.\right.\crn
&&\left.\left.-m^2_Fg^{Z_i}_A(F)g^{Z_j}_A(F)[g^{Z_i}_V(E)g^{Z_j}_V(E)-g^{Z_i}_A(E)g^{Z_j}_A(E)]\right]\right\},\\
\langle\sigma v\rangle_{FF^c\to\text{SM}\text{SM}}&=&\langle\sigma v\rangle_{EE^c\to\text{SM}\text{SM}}\left(E\leftrightarrow F,m_{W_{13}}\leftrightarrow m_{W_{14}},m_{W_{23}}\leftrightarrow m_{W_{24}}\right),\\
\langle\sigma v\rangle_{FF^c\to EE^c}&=&\langle\sigma v\rangle_{EE^c\to FF^c}\left(E\leftrightarrow F\right),
\eea
where $i,j=2,3$, and $f$ refers to every fermion of the standard model. Above, we have assumed that the masses of the new gauge bosons are much larger than the masses of the standard model fermions.

By solving numerically these equations with the following boundary condition
\bea
Y_E(1)&=Y^{\text{EQ}}_E(1),\\
Y_F(1)&=Y^{\text{EQ}}_F(1),
\eea
corresponding to that the dark matter species are in equilibrium with the thermal bath at start, one can obtain the individual relic abundance of each dark matter component as
\bea
\Omega_Eh^2&=&2.752 \frac{m_E}{\text{GeV}}Y_E(x_\infty)\times10^8,\\
\Omega_Fh^2&=&2.752\frac{m_F}{\text{GeV}}Y_F(x_\infty)\times10^8,
\eea
where $x_\infty$ refers to a very large value of $x$ after the thermal freeze-out. In particular, when the production of the lighter dark matter component from the heavier dark matter component is less significant compared to its annihilation to the standard model particles, an approximate analytic solution of BEQs is given by \cite{Bhattacharya:2016ysw}
\begin{eqnarray}
\Omega_{E}h^2&=&\frac{1.07 \times10^9 x_E}{\sqrt{g_*}M_{\text{Pl}}\langle\sigma v\rangle^T_E}\ \text{GeV}^{-1},\\
\Omega_{F}h^2&=&\frac{1.07 \times10^9 x_F}{\sqrt{g_*}M_{\text{Pl}}\langle\sigma v\rangle^T_F} \text{GeV}^{-1},
\end{eqnarray}
where $x_E=m_E/T_E$, $x_F=m_F/T_F$, and
\begin{eqnarray}
\langle\sigma v\rangle^T_E&=&\langle\sigma v\rangle_{EE^c\to\text{SM}\text{SM}}+\langle\sigma v\rangle_{EE^c\to FF^c},\\
\langle\sigma v\rangle^T_F&=&\langle\sigma v\rangle_{FF^c\to\text{SM}\text{SM}},
\end{eqnarray}
for the case $m_E>m_F$, or 
\bea
\langle\sigma v\rangle^T_E&=&\langle\sigma v\rangle_{EE^c\to\text{SM}\text{SM}},\\
\langle\sigma v\rangle^T_F&=&\langle\sigma v\rangle_{FF^c\to\text{SM}\text{SM}}+\langle\sigma v\rangle_{FF^c\to EE^c},
\eea
for the opposite case $m_E<m_F$. The dark matter relic abundance is a sum of the individual contributions as
\bea
\Omega_{\text{DM}}h^2=\Omega_Eh^2+\Omega_Fh^2.\label{tdmef8}
\eea

\begin{figure}[!h]
\begin{center}
	\includegraphics[scale=0.25]{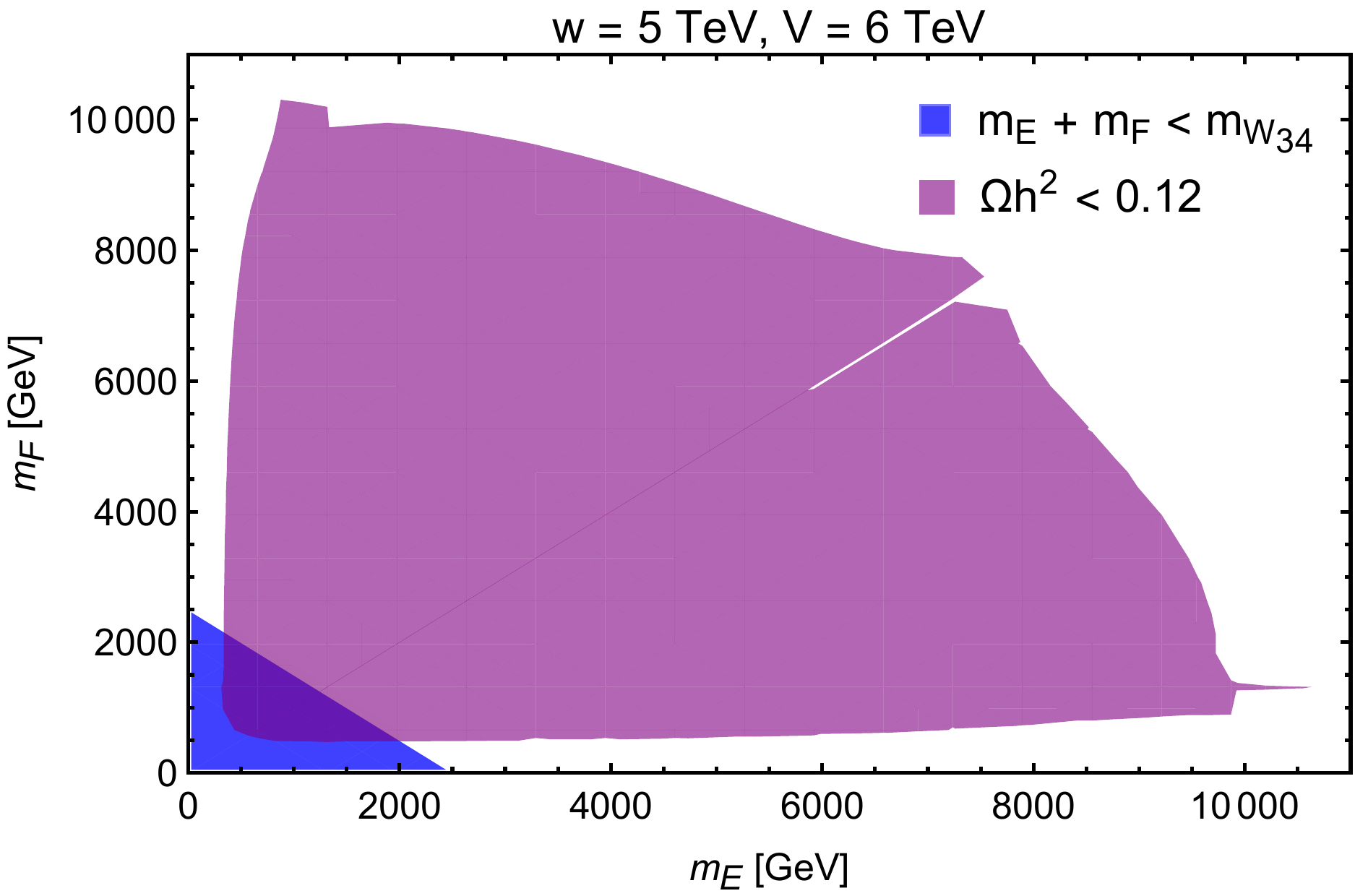}
	\includegraphics[scale=0.25]{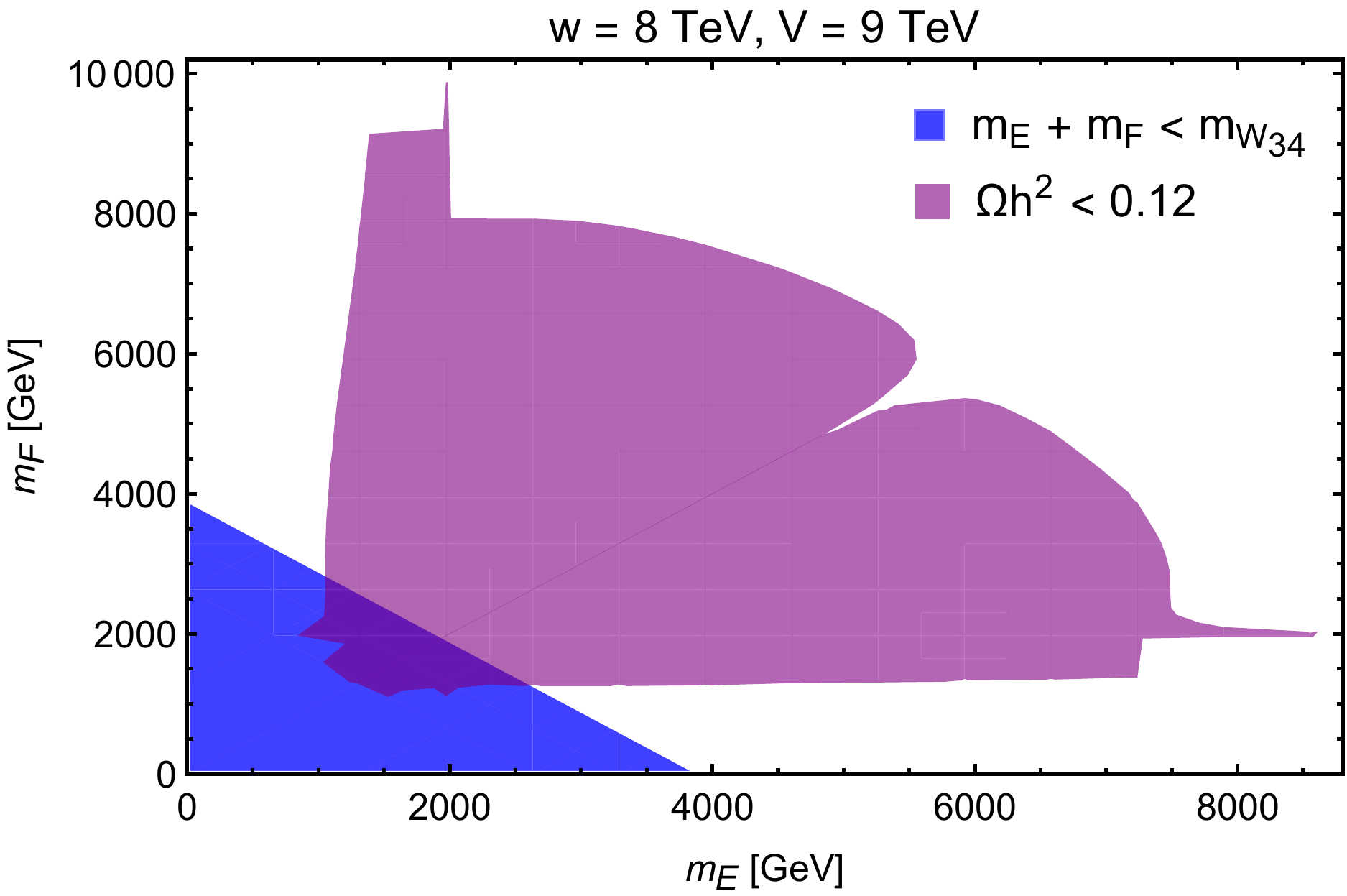}
	\includegraphics[scale=0.25]{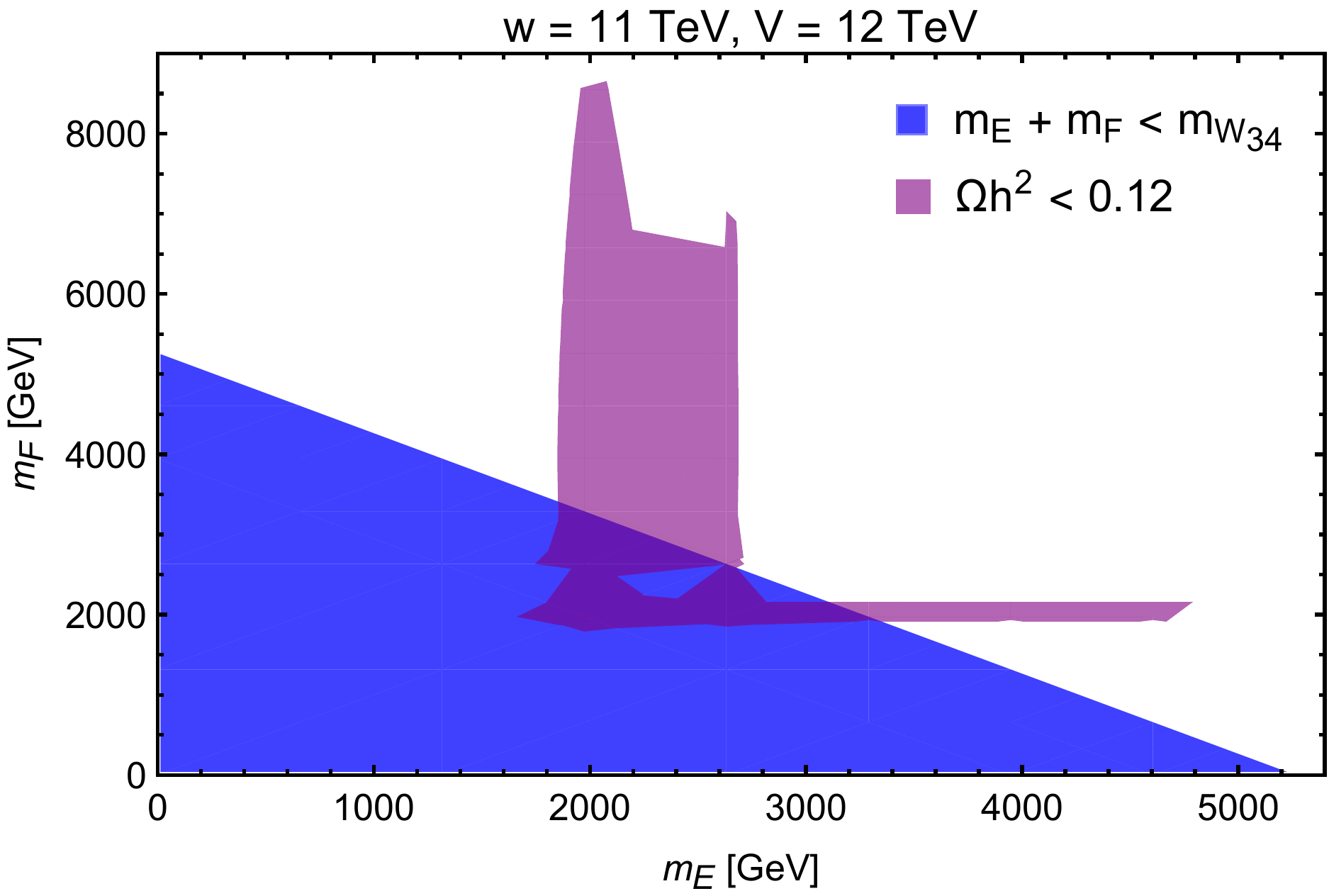}
	\caption{\label{trdef} The total relic density contoured as a function of $(m_E,m_F)$, where the dark matter stable regime is also included, according to the several choices of $w,V$.}	
\end{center}
\end{figure}

For numerical investigation, we use the following parameter values throughout this work,  
$u=v\simeq 174\ \mathrm{GeV},\ s^2_W\simeq 0.231,\ g=\sqrt{4\pi\alpha}/s_W,\ m_{Z_1}=91.187\ \mathrm{GeV}$. Additionally, the atomic numbers of Xenon are $Z=54$ and $A=131$.

Let us investigate the case that the total relic density (\ref{tdmef8}) varies as a function of dark matter masses $(m_E,m_F)$ for $m_E+m_F<m_{W_{34}}$, satisfying the experimental bound $\Omega_{\mathrm{DM}} h^2<0.12$ \cite{Ade:2015xua}. In Figure~\ref{trdef}, we show the viable dark matter mass regime as the overlap of the two colored regions according to the relic density and the stability condition, respectively. Note that the condition for $m_E+m_F<m_{\mathcal{H}_{6}}$ is easily evaded by imposing an appropriate $\la_{16}$ value, which is not taken into account. Moreover, the selections of the new physics scales $w,V$ always satisfy the constraints from the $\rho$-parameter, $Z\bar{f} f$ couplings, FCNCs, and collider bounds, as studied in \cite{VanLoi:2019xud}.      

\begin{figure}[!h]
\begin{center}
	\includegraphics[scale=0.25]{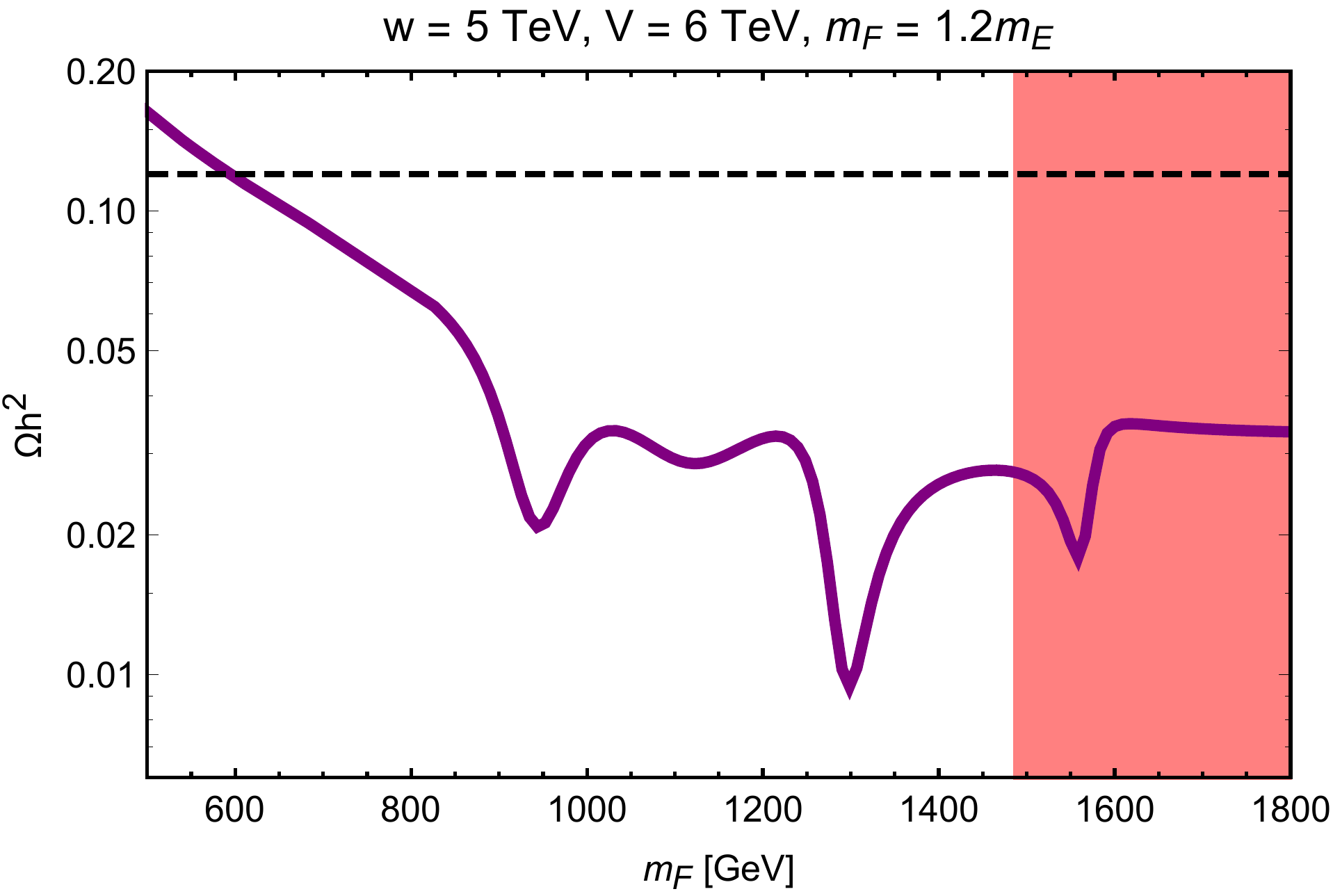}
	\includegraphics[scale=0.25]{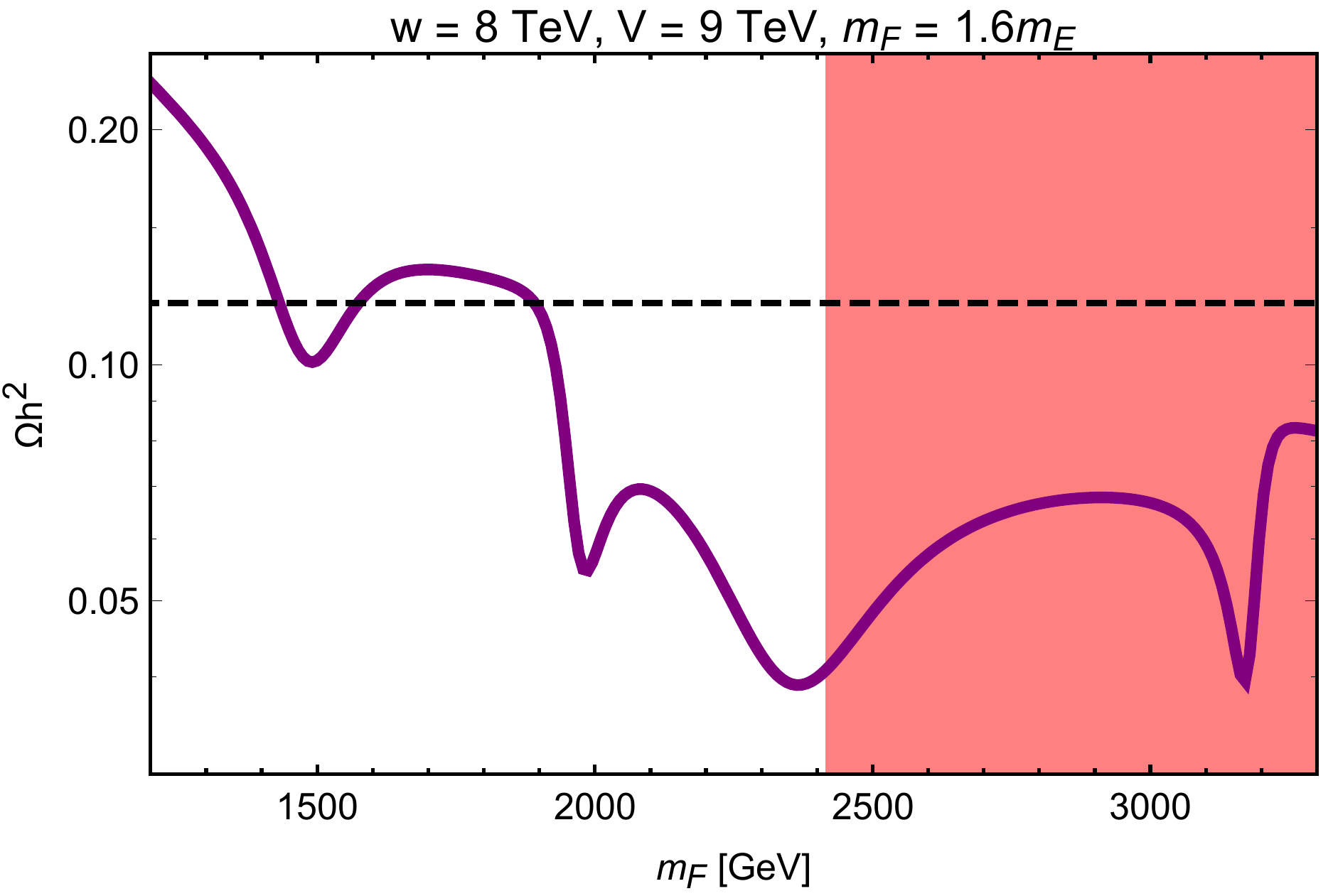}
	\includegraphics[scale=0.25]{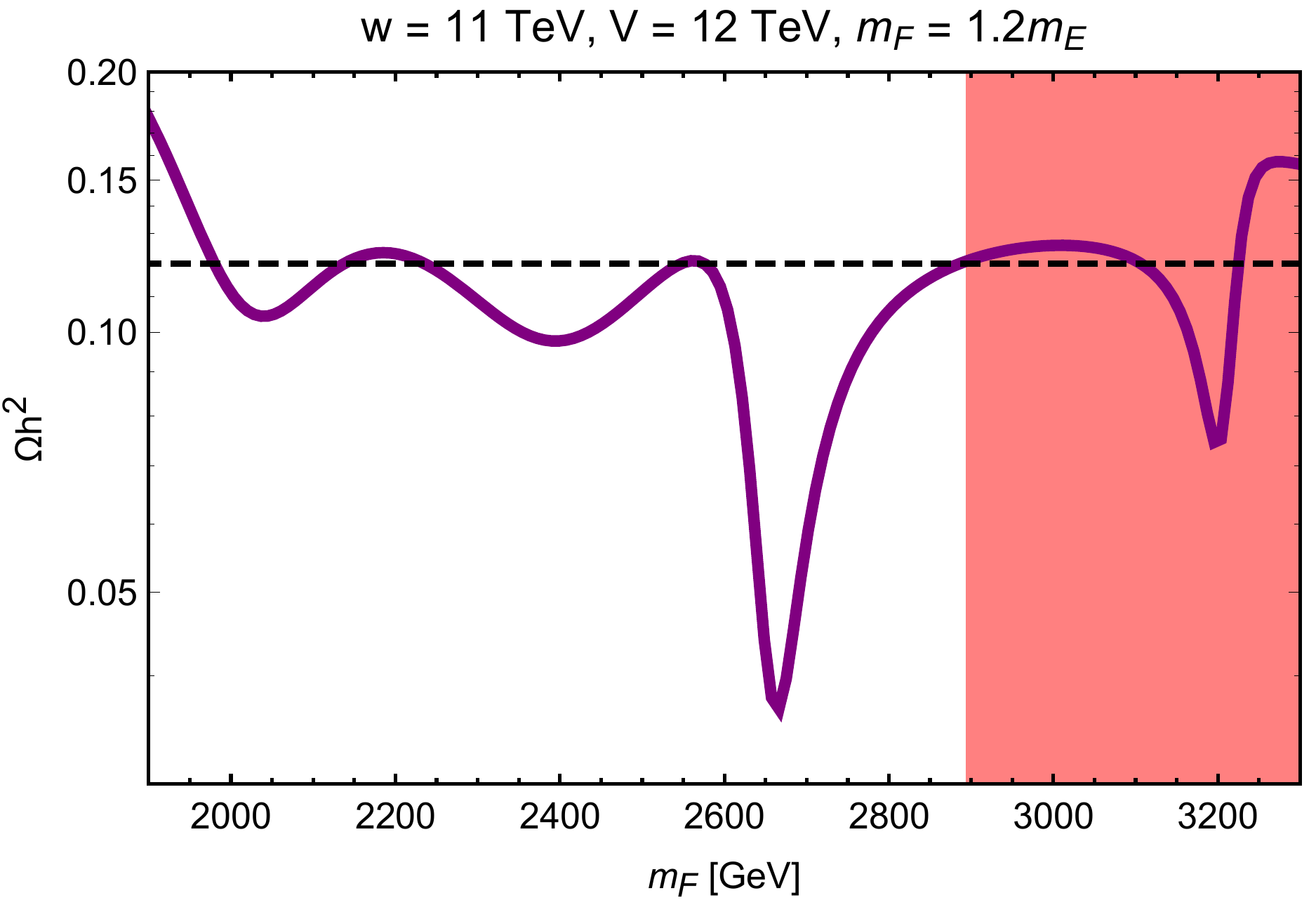}
	\caption{\label{rlef}The total relic density of two-component fermion dark matter as a function of dark matter masses for the case $m_F > m_E$.}	
\end{center}
\end{figure}
To see the physical effect of each dark matter component that contributes to the relic abundance, in Figure~\ref{rlef} we depict the total relic density as a function of $m_F$ for several choices of $w,V$ and $m_E$ as related to $m_F$, which are viable from the above contours. Typically, we determine four resonances in each density curve, corresponding to $m_F=\fr{m_{Z_2}}{2}$, $m_F=\fr{m_{Z_3}}{2}$, and the two others given by $m_E=\fr{m_{Z_2}}{2}$ and $m_E=\fr{m_{Z_3}}{2}$, which are translated to $m_F$ as located at $m_F=\fr{m_F}{m_E}\fr{m_{Z_2}}{2}$ and $m_F=\fr{m_F}{m_E}\fr{m_{Z_3}}{2}$, respectively. It is noted that we always have $m_{Z_3}>m_{Z_2}$ for the mediators and the details of the resonances can be seen in the next figure. Further, due to the contributions of both $E$ and $F$, the total density does not vanish at the resonances. In this case, the viable dark matter mass regime is given below the correct abundance $\Omega h^2< 0.12$ and before the dark matter unstable regime (red) according to $m_E+m_F>m_{W_{34}}$.  

\begin{figure}[!h]
\begin{center}
	\includegraphics[scale=0.25]{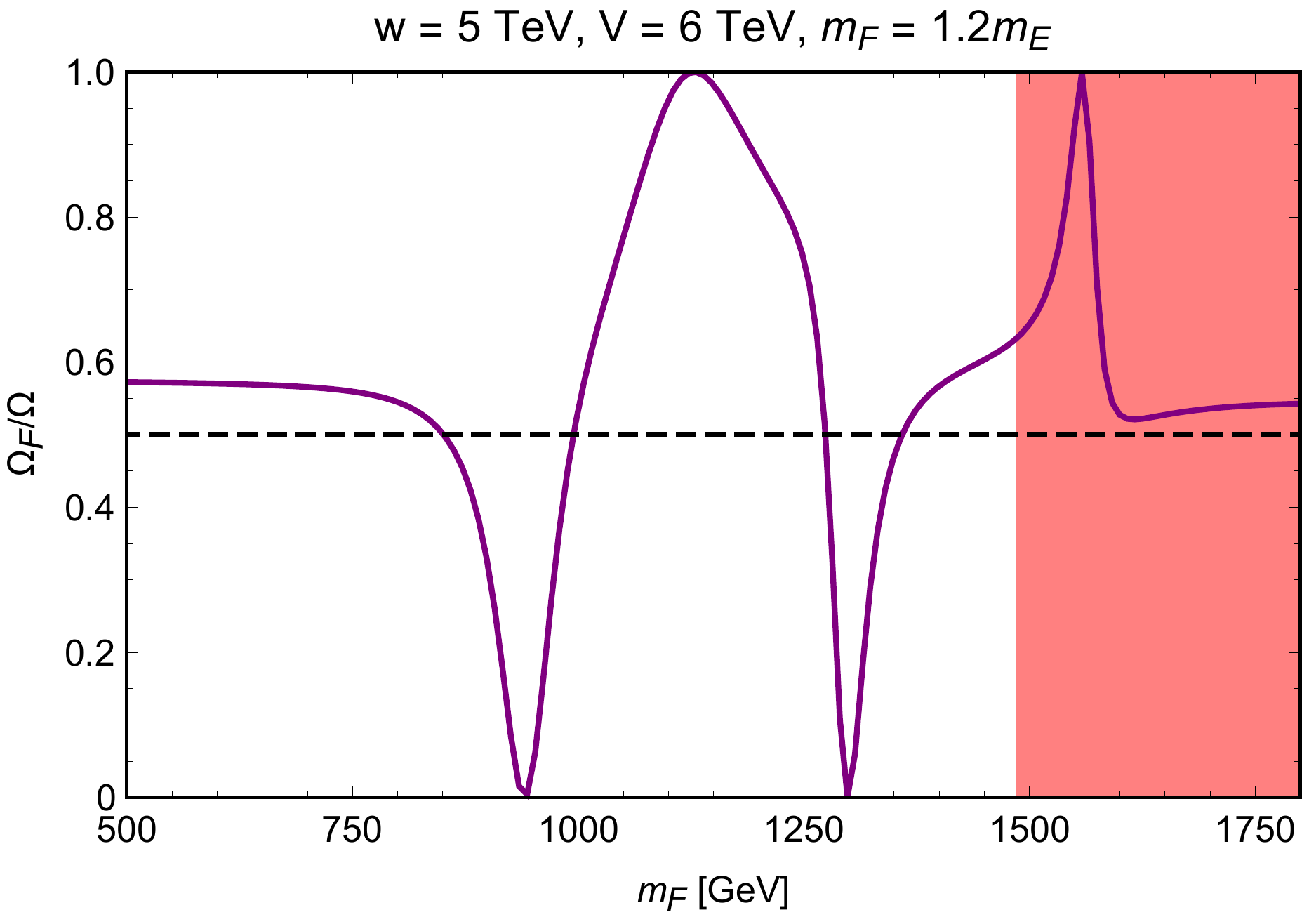}
	\includegraphics[scale=0.25]{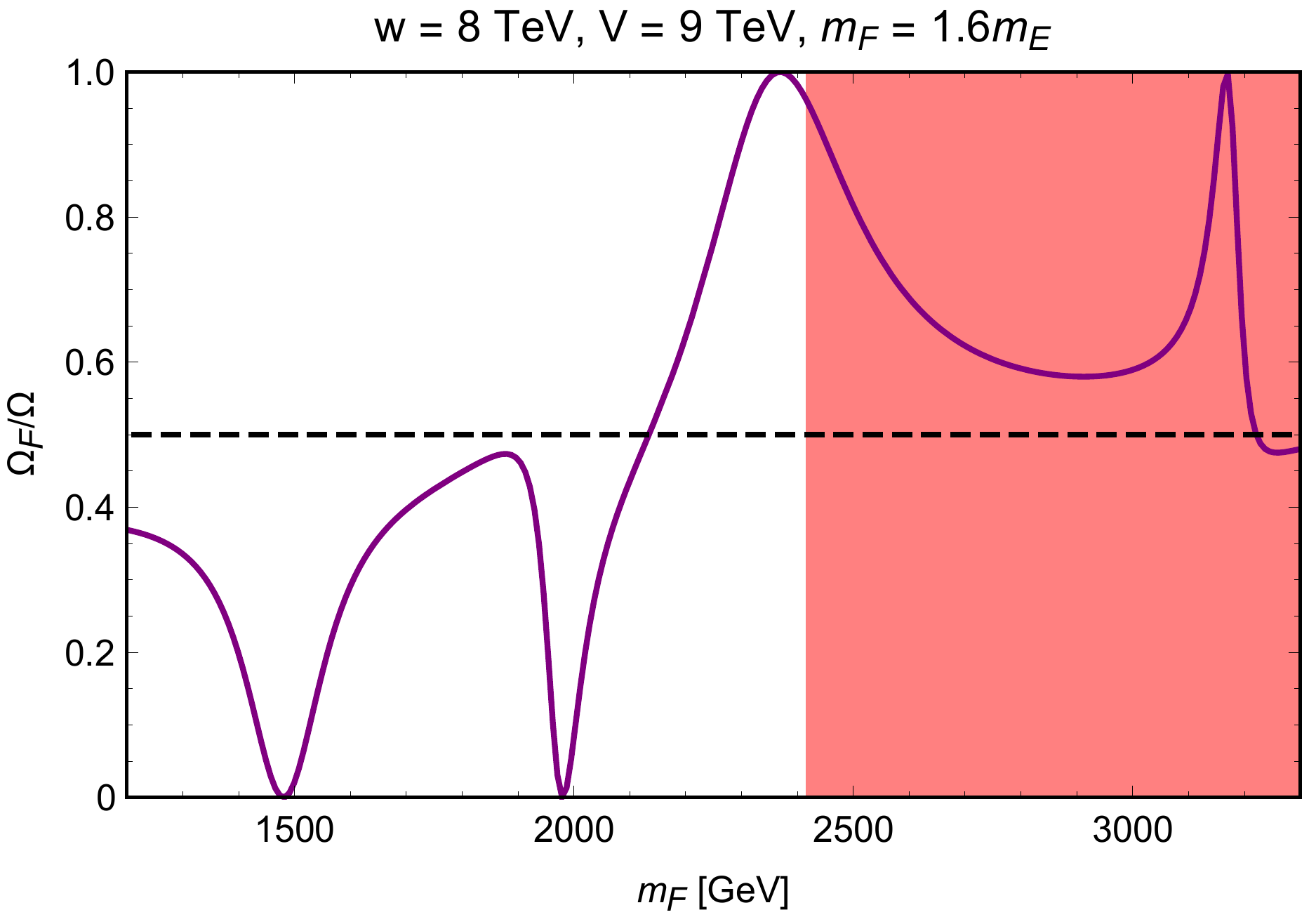}
	\includegraphics[scale=0.25]{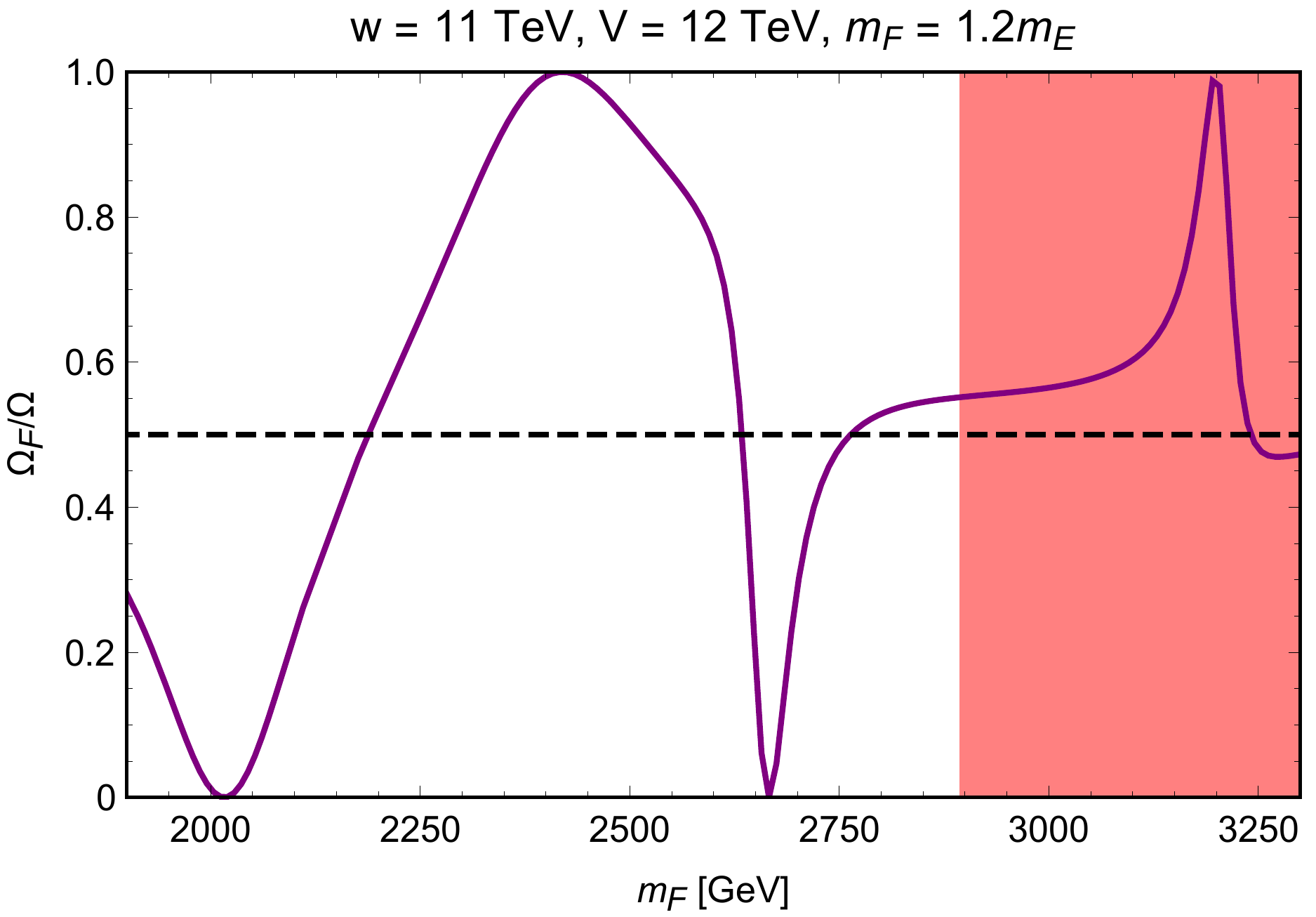}
	\caption{\label{rlef1}The contribution ratio of fermion dark matter components to the density as a function of dark matter masses for the case $m_F > m_E$.}	
\end{center}
\end{figure}
Correspondingly, in Figure \ref{rlef1} we make a comparison between the partial relic densities of dark matter components with the choices of the new physics scales $w,V$ and $m_E$ via $m_F$, as mentioned. It is noteworthy that the region above the line $\Om_F/\Om=0.5$, the candidate $F$ dominantly contributes to the density, and the two peaks at which are due to the $m_E$ resonances. Whereas, below the line $\Om_F/\Om=0.5$, $E$ dominates the density, where the two resonances correspond to the $m_F$ ones. The dark matter unstable region (red) according to $m_E+m_F>m_{W_{34}}$ is also included for completion.           

The above analysis is relevant to $m_F>m_E$. It is noted that the dark matter phenomenologies happen similarly to the case with $m_F<m_E$, thus to the whole dark matter mass regime which is viable from Figure \ref{trdef}. 

We study the direct detection for the dark matter components in our model through their spin-independent (SI) scattering on nuclei. First, let us write the effective Lagrangian describing dark matter--nucleon interaction at the fundamental level through the exchange of the new neutral gauge bosons $Z_{2,3}$ as
\bea
\mathcal{L}^{\text{eff}}_E&=&\sum_{i=2,3}\frac{g^2}{4m^2_{Z_i}}\bar{E}\gamma^\mu\left[g^{Z_i}_V(E)-g^{Z_i}_A(E)\gamma_5\right]E\bar{q}\gamma_\mu\left[g^{Z_i}_V(q)-g^{Z_i}_A(q)\gamma_5\right]q,\\
\mathcal{L}^{\text{eff}}_F&=&\mathcal{L}^{\text{eff}}_E\left(E\leftrightarrow F\right).
\eea

From this effective Lagrangian, one can obtain the SI cross-section for the scattering of the dark matter components on a target nucleus $N$ as
\bea
\sigma^{\text{SI}}_{EN}&=&\sum_{i=2,3}\frac{g^4m^2_{EN}}{16\pi m^4_{Z_i}}\Big|g^{Z_i}_V(E)g^{Z_i}_V(u)(Z+A)+g^{Z_i}_V(E)g^{Z_i}_V(d)(2A-Z)\Big|^2,\\
\sigma^{\text{SI}}_{FN}&=&\sigma^{\text{SI}}_{EN}\left(E\leftrightarrow F\right),
\eea
where $m_{EN}=\frac{m_Em_N}{m_E+m_N}\simeq m_N$ is the dark matter--nucleon reduced mass, $Z$ and $A$ are the atomic number and atomic mass of the nucleus $N$, respectively. However, in the two-component dark matter scenario, the SI cross-section for each dark matter component is calculated as follows
\bea
\sigma^{\text{SI}}_{\text{eff}}(E)&=&\frac{\Omega_Eh^2}{\Omega_{\text{DM}}h^2}\sigma^{\text{SI}}_{EN},\\
\sigma^{\text{SI}}_{\text{eff}}(F)&=&\frac{\Omega_Fh^2}{\Omega_{\text{DM}}h^2}\sigma^{\text{SI}}_{FN}.
\eea

\begin{figure}[!h]
\begin{center}
	\includegraphics[scale=0.35]{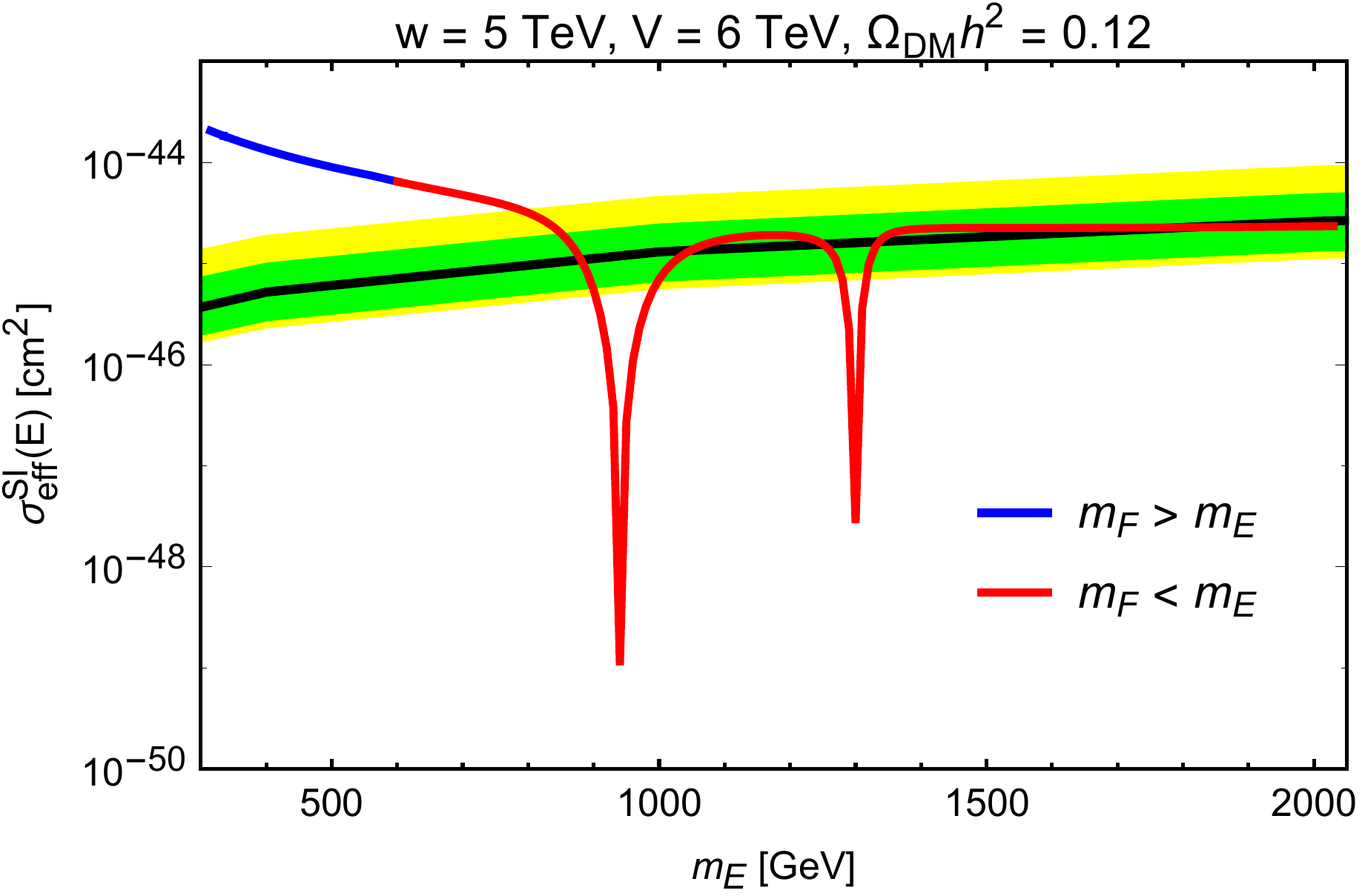}
	\includegraphics[scale=0.35]{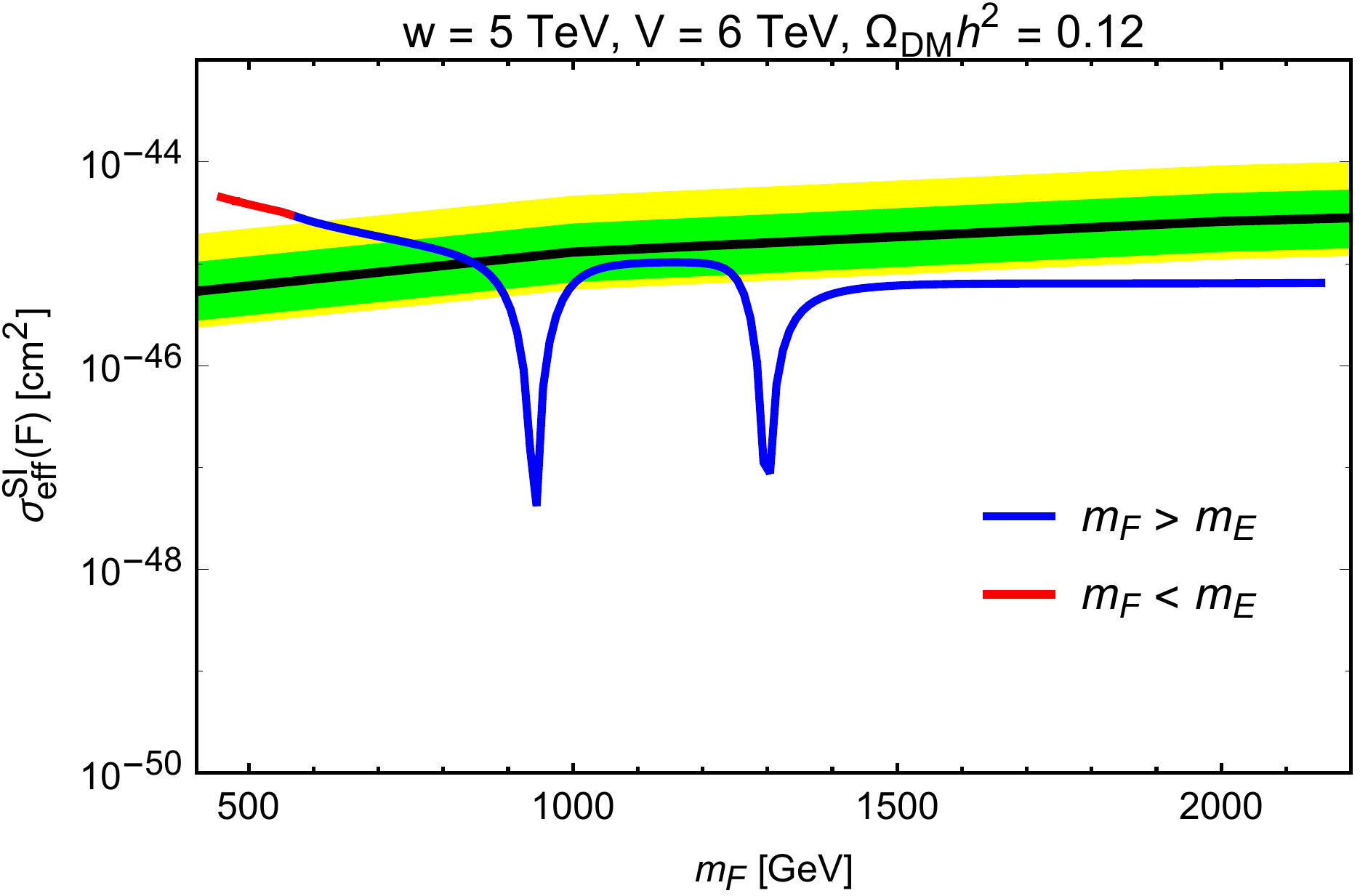}\\
	\includegraphics[scale=0.35]{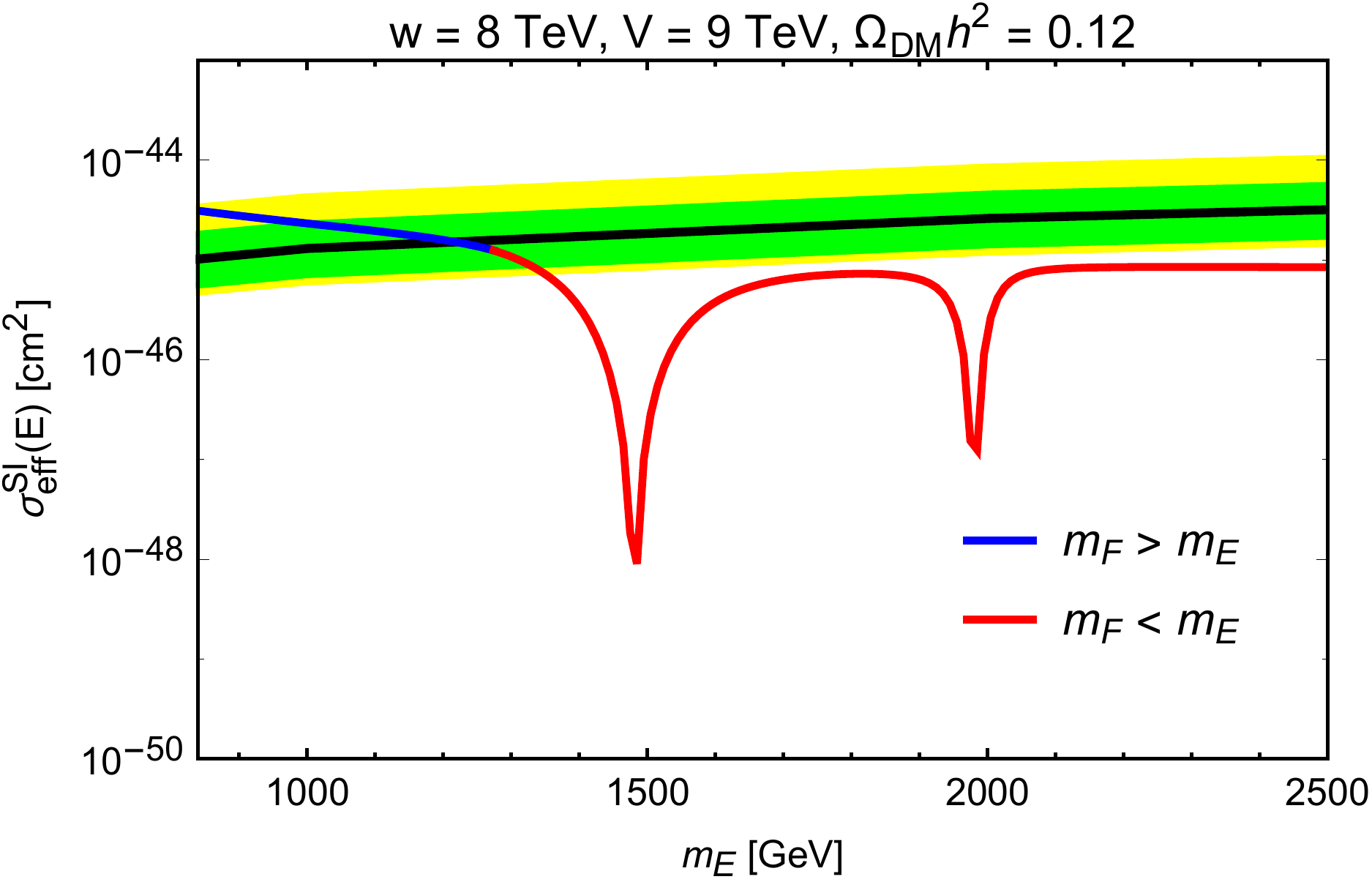}
	\includegraphics[scale=0.35]{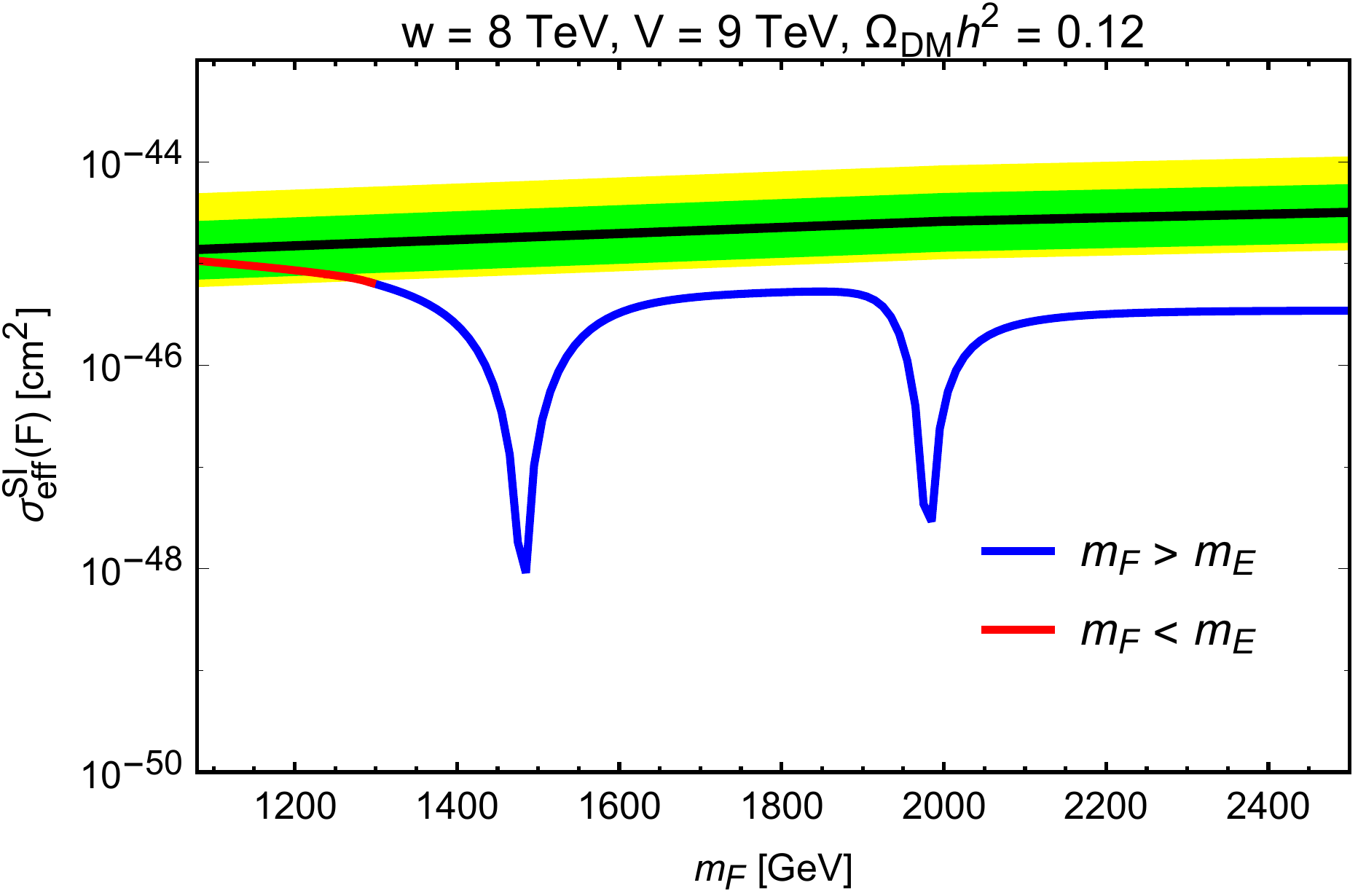}\\
	\includegraphics[scale=0.35]{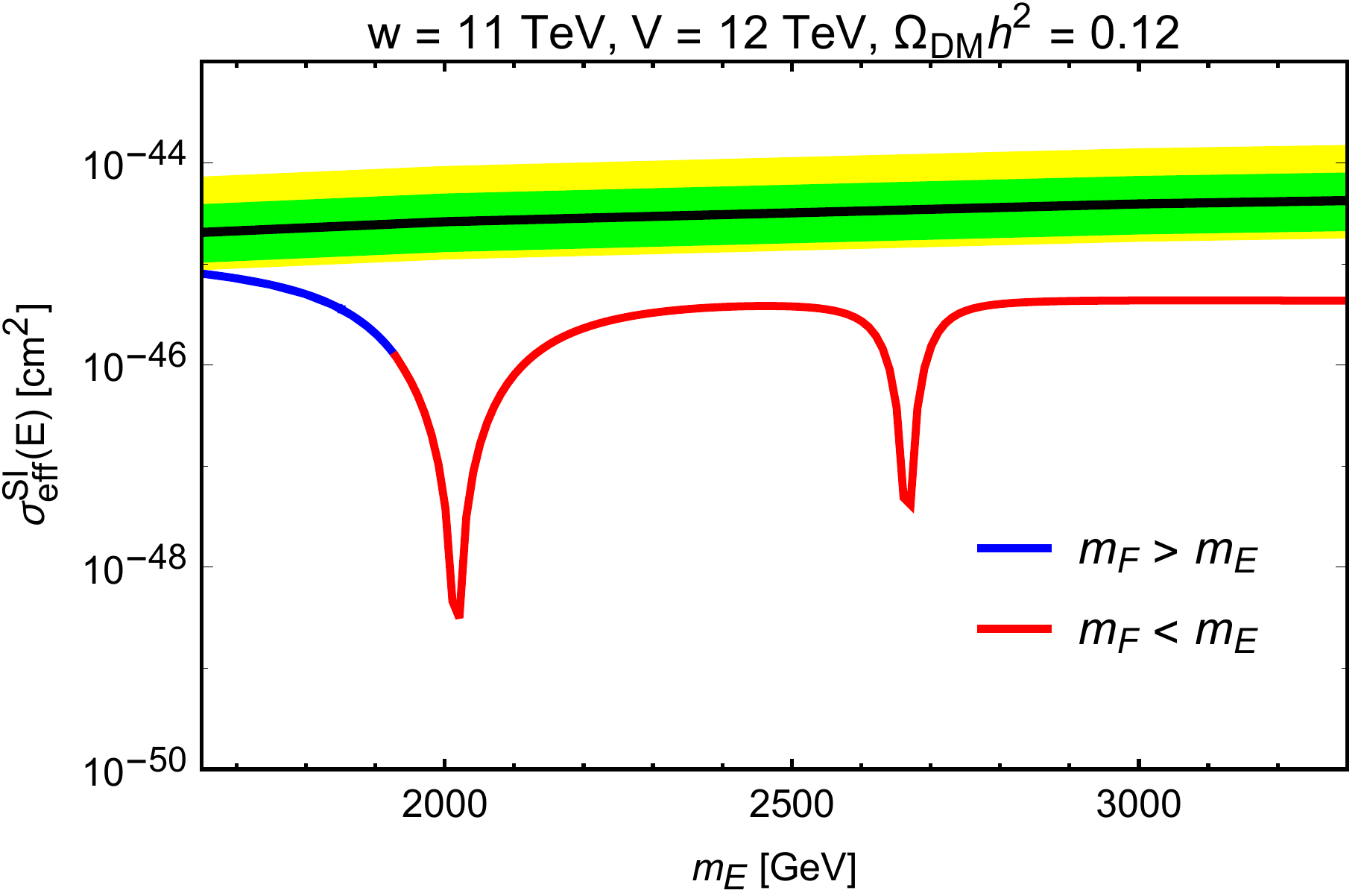}
	\includegraphics[scale=0.35]{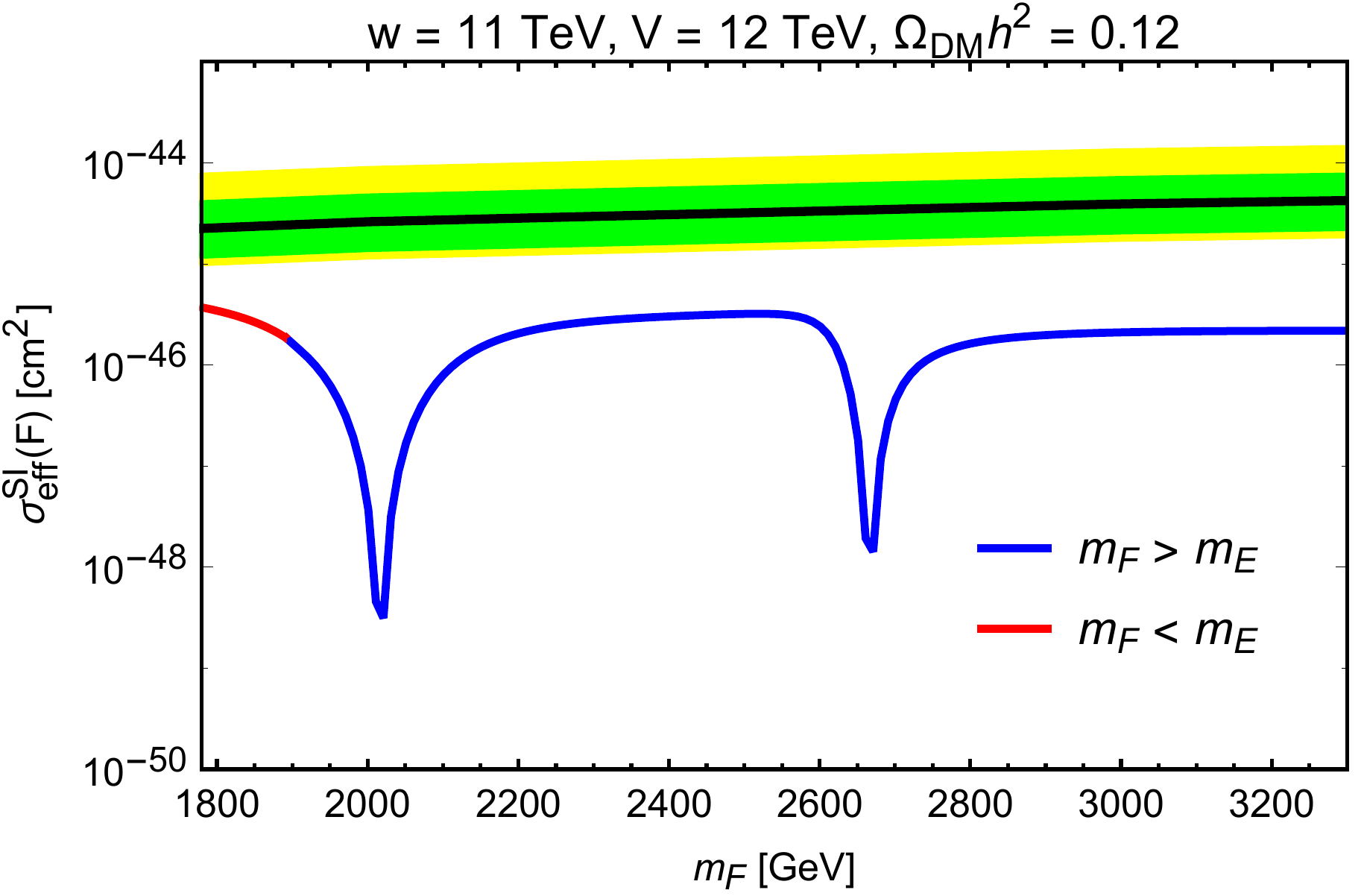}
	\caption{\label{ddef}The spin-independent dark matter-nucleon scattering cross-section limits as a function of dark matter masses according to $(w,V)=(5,6)$, (8,9), and (11,12) TeV, assuming the correct abundance $\Omega_{\mathrm{DM}} h^2=0.12$.}	
\end{center}
\end{figure}
Supposing that the two-component dark matter obtains the correct total density, in Figure \ref{ddef} we plot the SI cross-sections of dark matter components corresponding to the above choices of $(w,V)$ parameters, respectively. Here in each limit curve, we explicitly show which part the case $m_F>m_E$ (blue) and vice versa the opposite case $m_F<m_E$ (red) take place. The experimental bounds \cite{Aprile:2017iyp,Aprile:2018dbl} are also included. That said, the dark matter masses that have been obtained from the relic density and the stable condition also satisfy the direct detection if $(w,V)=(8,9)$ TeV and higher. Correspondingly, both $m_E$ and $m_F$ should be above 1 TeV.    

\subsection{Scenario with two-scalar dark matter}

\begin{figure}[!h]
	\centering
	\includegraphics[scale=0.9]{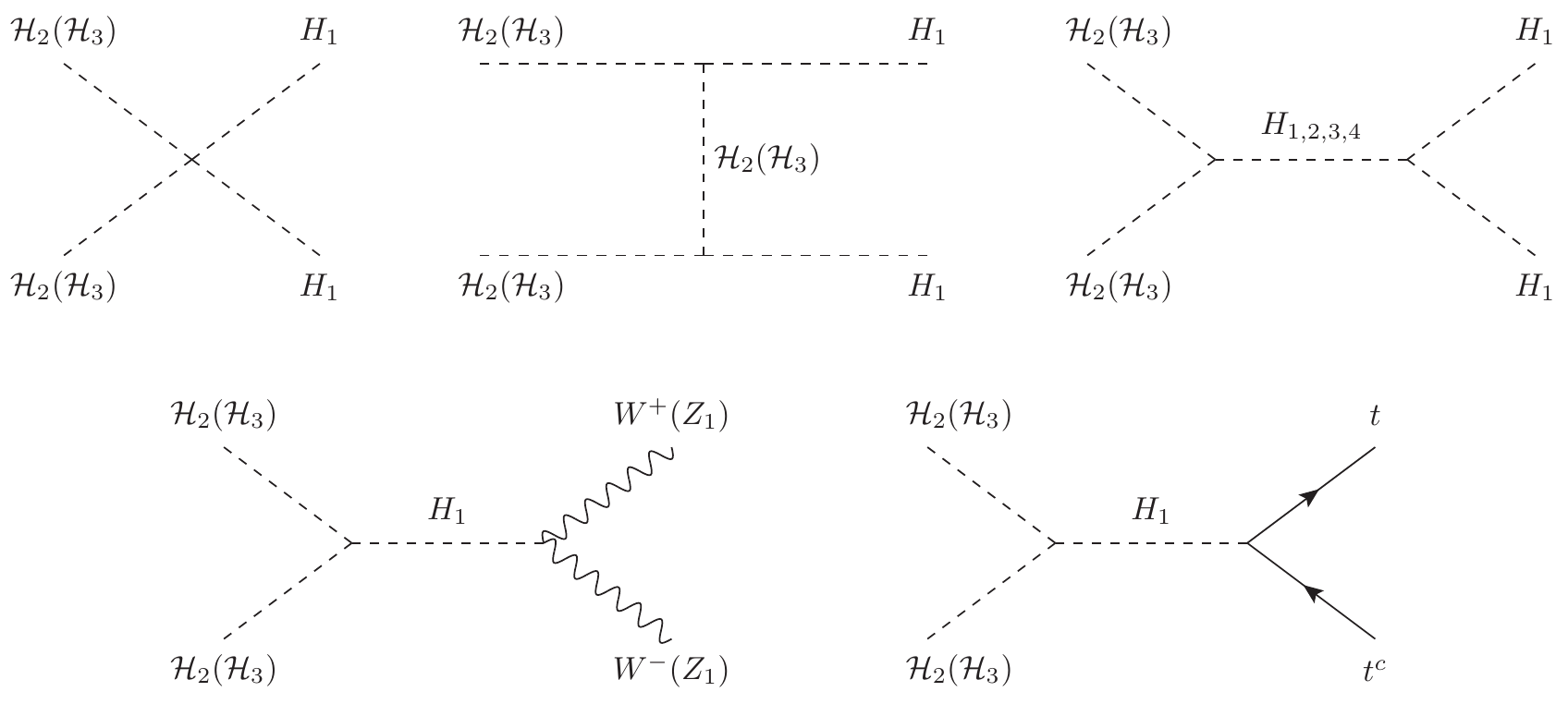}
	\caption{Dominant contributions to annihilation of the two-component scalar dark matter into standard model particles.}	\label{DMSM-2}
\end{figure}

\begin{figure}[!h]
	\centering
	\includegraphics[scale=0.9]{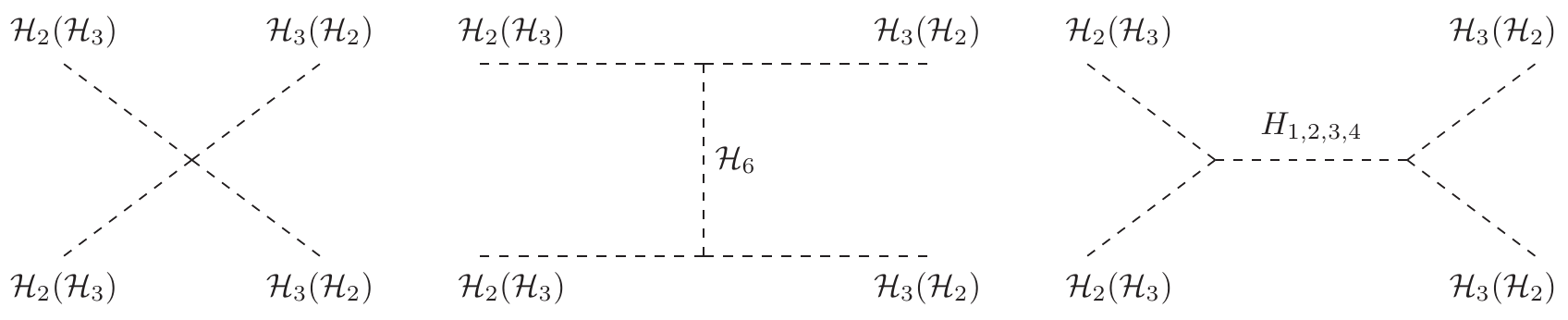}
	\caption{Conversion between two-component scalar dark matter.}	\label{DMDM-2}
\end{figure}

We consider the second case where two-component dark matter contains the scalar particles $\mathcal{H}_2$ and $\mathcal{H}_3$. Here we assume that they are the lightest particles within the classes of singly-wrong particles of the same kind as mentioned, respectively, and they have a net mass smaller than those of $W_{34}$ and $\mathcal{H}_6$. The dark matter pair annihilation into the standard model particles are given by the following dominant channels
\bea 
\mathcal{H}_2\mathcal{H}_2 &\to & H_1H_1,tt^c,W^+W^-,Z_1Z_1,\\
\mathcal{H}_3\mathcal{H}_3 &\to & H_1H_1,tt^c,W^+W^-,Z_1Z_1, 
\eea as presented in Figure \ref{DMSM-2}, while the dark matter conversions are given in Figure \ref{DMDM-2}. Notice that the neutral gauge portals are not available for the channels. Indeed, $Z_1$ does not couple to $\mathcal{H}_2$ and $\mathcal{H}_3$ for $p=q=0$ (cf. Table \ref{1NG2S}). Additionally, $Z_{2,3}$ do not couple to $W^+W^-$ at the effective limit, as aforementioned.

The thermal average annihilation cross-section times the relative velocity for the scalar dark matter components is approximately given by
\bea
\langle\sigma v\rangle_{\mathcal{H}_2\mathcal{H}_2\to\text{SM}\text{SM}}&=&\langle\sigma v\rangle_{\mathcal{H}_2\mathcal{H}_2\to H_1H_1}+\langle\sigma v\rangle_{\mathcal{H}_2\mathcal{H}_2\to tt^c}\crn
&&+\langle\sigma v\rangle_{\mathcal{H}_2\mathcal{H}_2\to W^+W^-}+\langle\sigma v\rangle_{\mathcal{H}_2\mathcal{H}_2\to Z_1Z_1},\\
\langle\sigma v\rangle_{\mathcal{H}_2\mathcal{H}_2\to\mathcal{H}_3\mathcal{H}_3}&=&\frac{\sqrt{m^2_{\mathcal{H}_2}-m^2_{\mathcal{H}_3}}}{16\pi m^3_{\mathcal{H}_2}}\left[\lambda_1+\frac{(\lambda_{12}\omega c_{\alpha_3}+\lambda_{13}Vs_{\alpha_3})^2}{m^2_{\mathcal{H}_2}-m^2_{\mathcal{H}_3}+m^2_{\mathcal{H}_6}}\right.\crn
&&\left. +\frac{(\lambda_6\omega c_{\alpha_1}-\lambda_7Vs_{\alpha_1})^2}{4m^2_{\mathcal{H}_2}-m^2_{H_3}}+\frac{(\lambda_6\omega s_{\alpha_1}+\lambda_7Vc_{\alpha_1})^2}{4m^2_{\mathcal{H}_2}-m^2_{H_4}}
\right]^2,\label{addhh1}\\
\langle\sigma v\rangle_{\mathcal{H}_3\mathcal{H}_3\to\text{SM}\text{SM}}&=&\langle\sigma v\rangle_{\mathcal{H}_2\mathcal{H}_2\to\text{SM}\text{SM}}\left(m_{\mathcal{H}_2}\leftrightarrow m_{\mathcal{H}_3}\right),\\
\langle\sigma v\rangle_{\mathcal{H}_3\mathcal{H}_3\to\mathcal{H}_2\mathcal{H}_2}&=&\langle\sigma v\rangle_{\mathcal{H}_2\mathcal{H}_2\to\mathcal{H}_3\mathcal{H}_3}\left(m_{\mathcal{H}_2}\leftrightarrow m_{\mathcal{H}_3}\right), \label{addhh2}
\eea
where the remaining annihilation cross-sections are     
\bea
\langle\sigma v\rangle_{\mathcal{H}_2\mathcal{H}_2\to H_1H_1}&=&\frac{1}{16\pi m^2_{\mathcal{H}_2}}\left\{2\lambda_1c^2_{\alpha_2}+\lambda_5s^2_{\alpha_2}\right.\crn
&&\left.-\left[\frac{(\lambda_6+\lambda_7)c^2_{\alpha_2}+(\lambda_7+\lambda_9)s^2_{\alpha_2}}{4m^2_{\mathcal{H}_2}-m^2_{H_3}}\right.\right.\crn
&&\left.\left.\times(\omega c_{\alpha_1}-Vs_{\alpha_1})(\lambda_6\omega c_{\alpha_1}-\lambda_7Vs_{\alpha_1})\right.\right.\crn
&&\left.\left.+\left(m_{H_3}\leftrightarrow m_{H_4},c_{\alpha_1}\leftrightarrow s_{\alpha_1},s_{\alpha_1}\leftrightarrow-c_{\alpha_1}\right)\right]\right\}^2,\\
\langle\sigma v\rangle_{\mathcal{H}_2\mathcal{H}_2\to tt^c}&=&\frac{3\left[(2\lambda_1c_{\alpha_2}u+\lambda_5s_{\alpha_2}v)m_tc_{\alpha_2}\right]^2}{16\pi u^2m^4_{\mathcal{H}_2}},\\
\langle\sigma v\rangle_{\mathcal{H}_2\mathcal{H}_2\to W^+W^-}&=&\frac{(2\lambda_1c_{\alpha_2}u+\lambda_5s_{\alpha_2}v)^2}{8\pi(u^2+v^2)m^2_{\mathcal{H}_2}},\\
\langle\sigma v\rangle_{\mathcal{H}_2\mathcal{H}_2\to Z_1Z_1}&=&\frac{(2\lambda_1c_{\alpha_2}u+\lambda_5s_{\alpha_2}v)^2}{16\pi(u^2+v^2)m^2_{\mathcal{H}_2}}.
\eea Above, the annihilation (\ref{addhh1}) happens for $m_{\mathcal{H}_2}>m_{\mathcal{H}_3}$, whereas the annihilation (\ref{addhh2}) exists for $m_{\mathcal{H}_3}>m_{\mathcal{H}_2}$.

\begin{figure}[!h]
\begin{center}
	\includegraphics[scale=0.25]{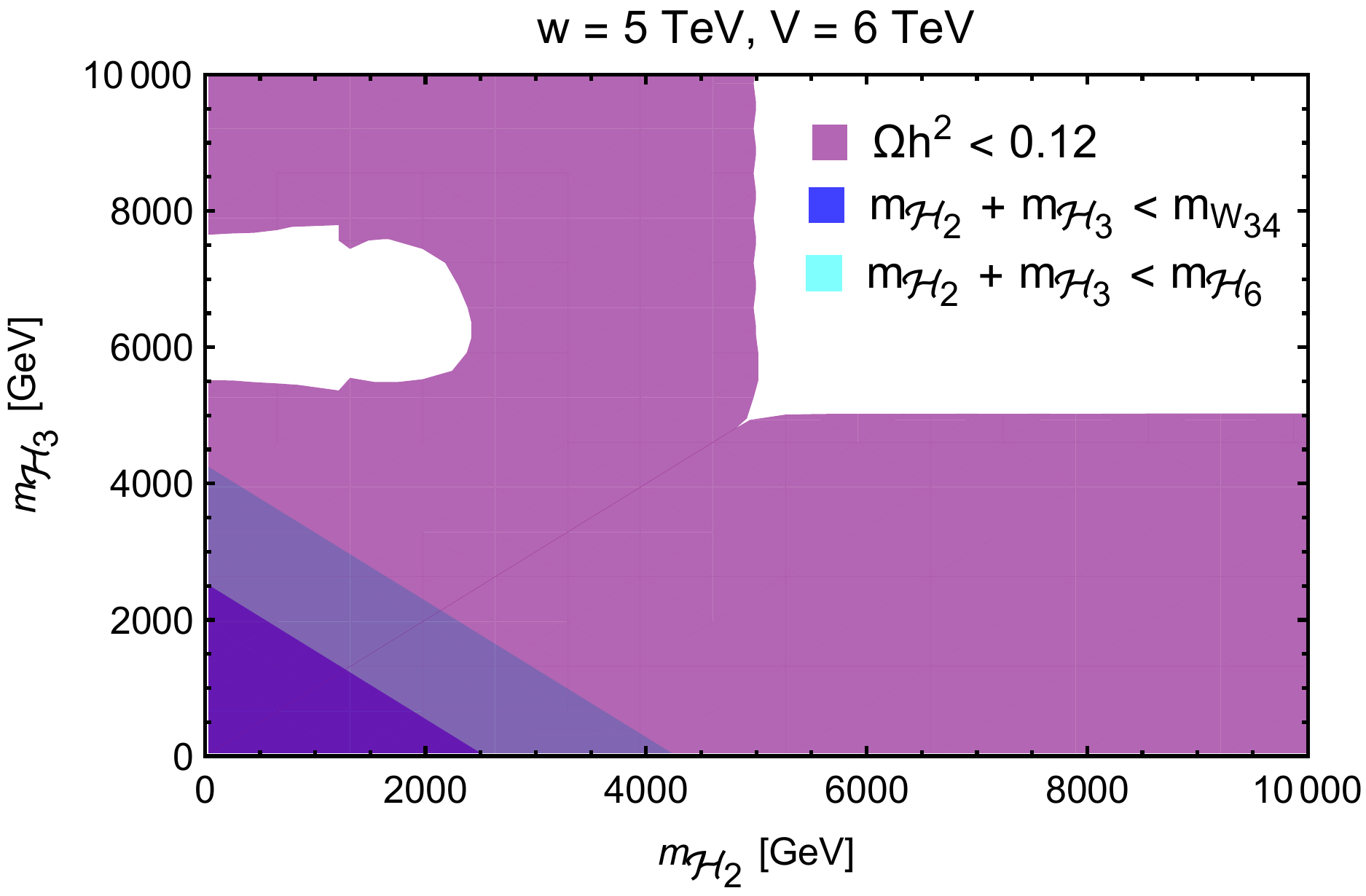}
	\includegraphics[scale=0.25]{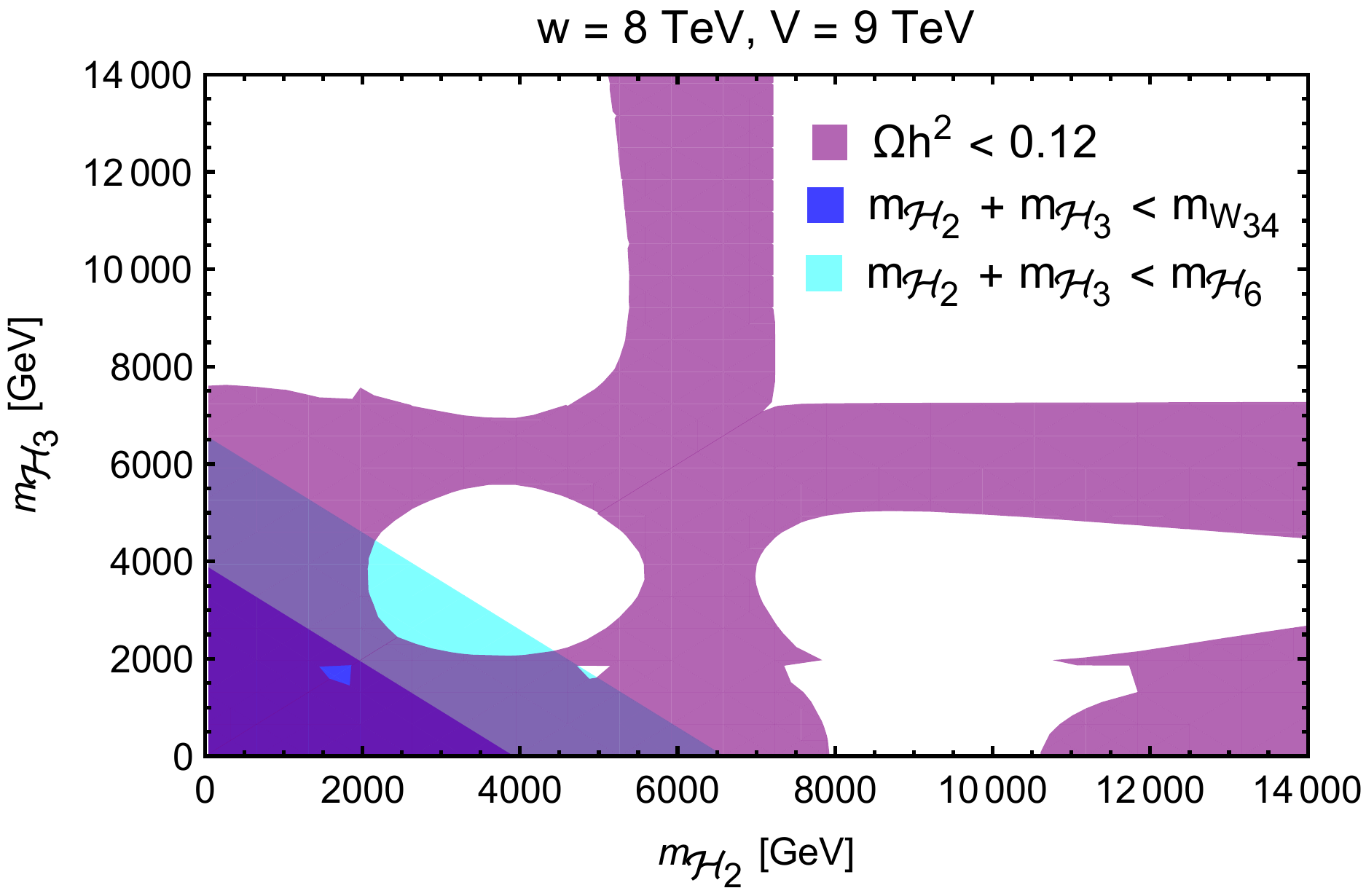}
	\includegraphics[scale=0.25]{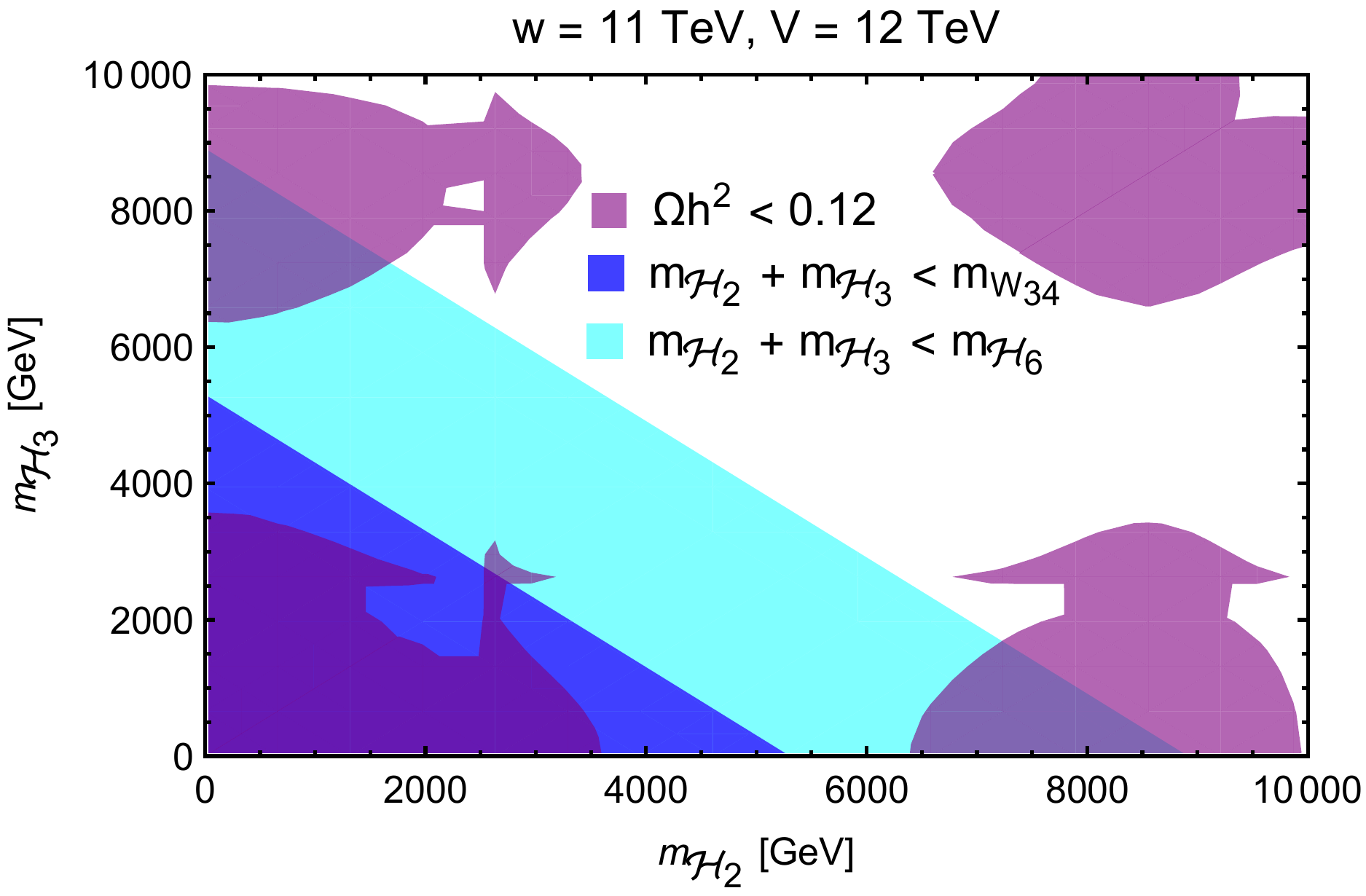}
	\caption{\label{trdh2h3} The total relic density contoured as a function of $(m_{\mathcal{H}_2},m_{\mathcal{H}_3})$, where the dark matter stable regime is also included, according to the several choices of $w,V$.}	
\end{center}
\end{figure}

For numerical computation, we take the following values of parameters into account, 
\bea && m_t\simeq 173.1\ \mathrm{GeV},\ m_{H_1}\simeq 125.3\ \mathrm{GeV},\crn 
&& \la_1=0.1,\ \la_{3,4,9,16}=0.6,\ \la_{5}=-0.15,\ \la_{6,10}=0.5,\ \la_7=0.7,\ \la_{17}=-0.1, \eea throughout this section. Note that the values of potential parameters chosen must satisfy the vacuum stability conditions and positive squared scalar masses.

Corresponding to each choice of $(w,V)$, we make a contour of the total relic density $\Omega_{\mathrm{DM}}h^2<0.12$ due to the contributions of both scalar dark matter candidates to be a function of their masses as in Figure \ref{trdh2h3}, where the regions constrained by the dark matter stable conditions $m_{\mathcal{H}_2}+m_{\mathcal{H}_3}<m_{\mathcal{H}_6}$ and $m_{\mathcal{H}_2}+m_{\mathcal{H}_3}<m_{W_{34}}$ are also indicated. The viable dark matter mass regime is the overlap of the three regions corresponding to the relic density and the stable conditions. 

\begin{figure}[!h]
\begin{center}
	\includegraphics[scale=0.25]{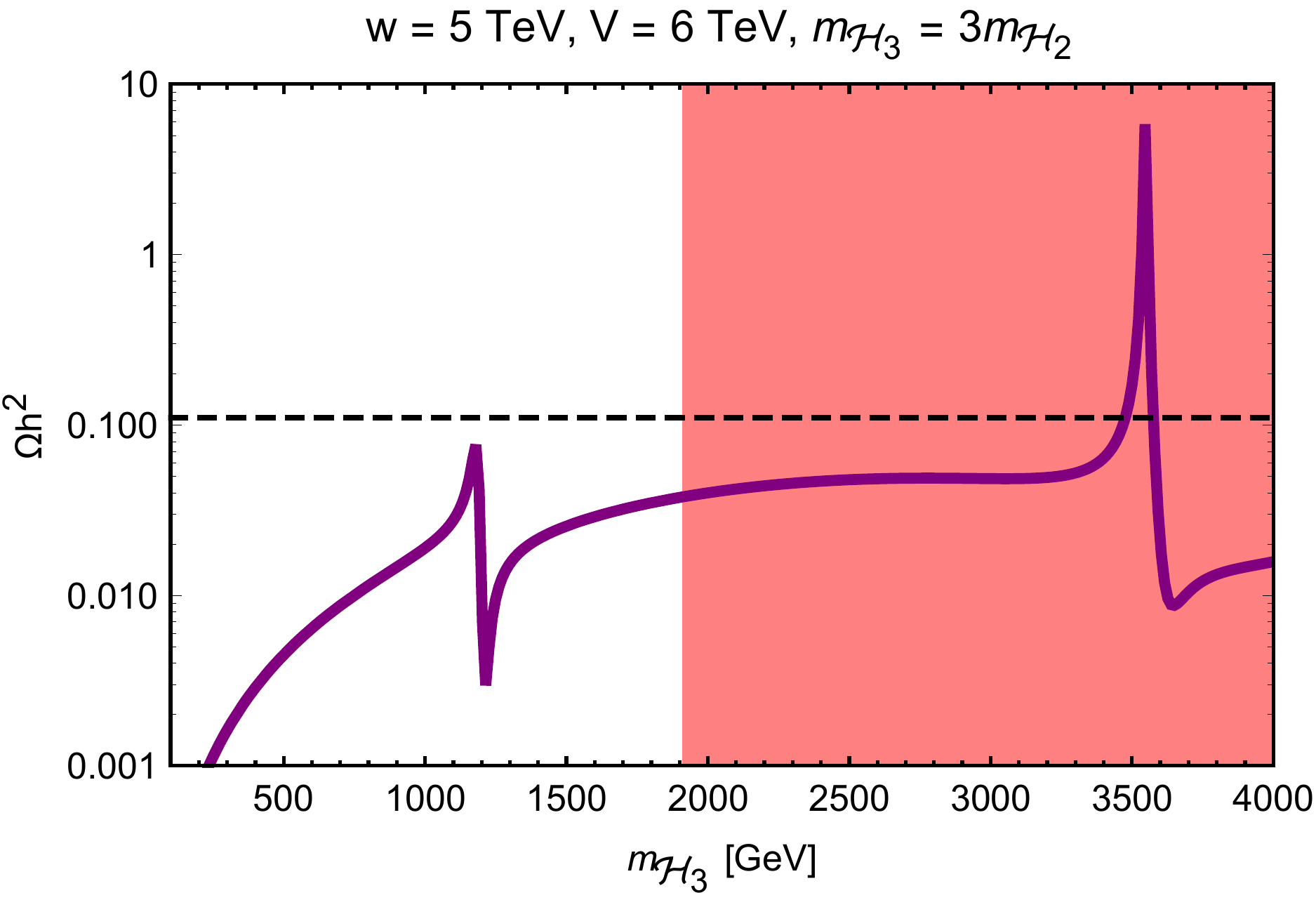}
	\includegraphics[scale=0.25]{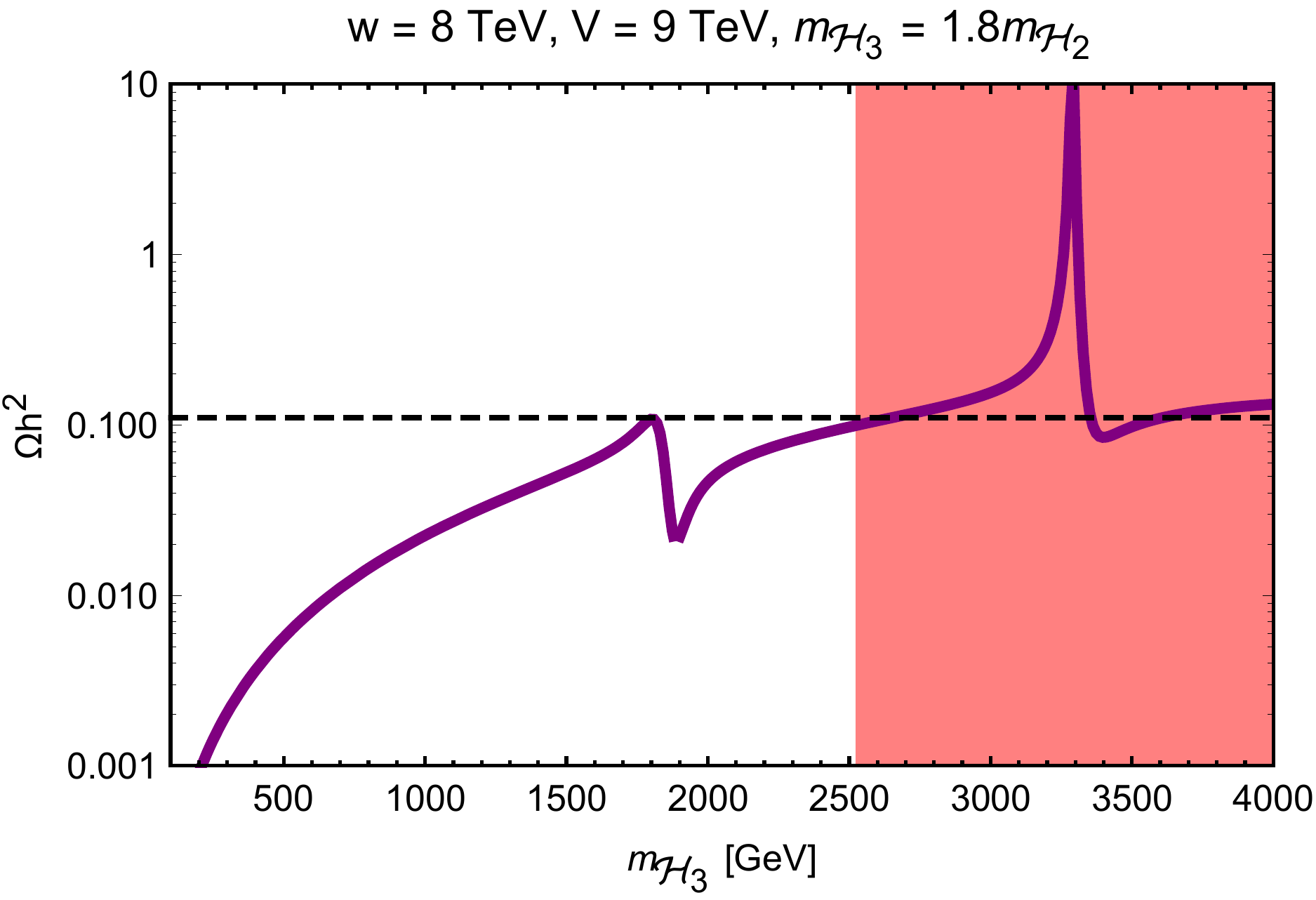}
	\includegraphics[scale=0.25]{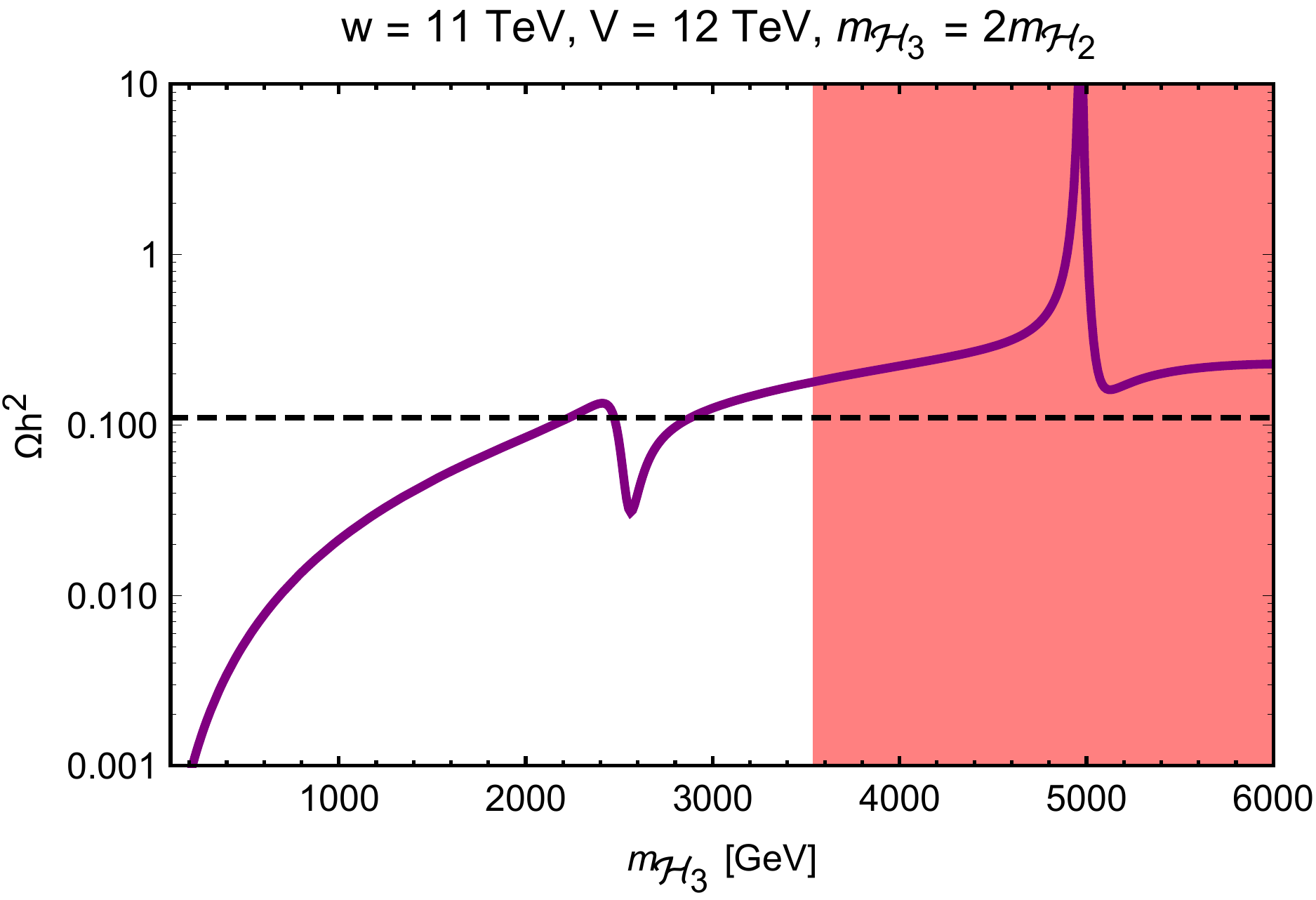}
	\caption{\label{sdm1} The relic density of two-component scalar dark matter as a function of dark matter masses for the case $m_{\mathcal{H}_3} > m_{\mathcal{H}_2}$, where the dark matter unstable regime (red) is also included.}	
\end{center}
\end{figure}
To examine the contribution effects of each scalar dark matter component, we consider the case $m_{\mathcal{H}_3}>m_{\mathcal{H}_2}$. The total relic density is depicted in Figure \ref{sdm1} as a function of $m_{\mathcal{H}_3}$ for different selections of $m_{\mathcal{H}_2}$ and $w,V$, which are viable from the above contours. Each density curve contains four resonances, set by the new neutral Higgs bosons $H_{3,4}$, at which $m_{\mathcal{H}_3}=\fr{m_{H_3}}{2}$, $m_{\mathcal{H}_3}=\fr{m_{H_4}}{2}$, and the two others at $m_{\mathcal{H}_3}=(m_{\mathcal{H}_3}/m_{\mathcal{H}_2})\fr{m_{H_3}}{2}$ and $m_{\mathcal{H}_3}=(m_{\mathcal{H}_3}/m_{\mathcal{H}_2})\fr{m_{H_4}}{2}$, which result from the $\mathcal{H}_2$ resonances, $m_{\mathcal{H}_2}=\fr 1 2 m_{H_3}$ and $m_{\mathcal{H}_2}=\fr 1 2 m_{H_4}$, respectively. The viable mass regime is given below the correct abundance and before the dark matter unstable regime. The density resonance phenomena happen analogously to the case of two-fermion dark matter, but it is now played by the new Higgs portal $H_{3,4}$ instead. Additionally, since the scalar dark matter masses are proportional to $w,V$ beyond the electroweak scale, the standard model Higgs portal $H_1$ negligibly contributes to the relic density. Furthermore, the new Higgs portal $H_2$ gives negligible contributions because it weakly couples to the dark matter components.      

One can consider the total relic density for the case $m_{\mathcal{H}_2}>m_{\mathcal{H}_3}$ as a function of $m_{\mathcal{H}_2}$ with the several values of $w,V$ and $\mathcal{H}_{2,3}$ mass relation. The process happens analogous to the case $m_{\mathcal{H}_2}>m_{\mathcal{H}_3}$.
Hence, the common remark for both cases is that the correct density and stability condition require the scalar dark matter masses to be not too large, limited below several TeVs. Additionally, some $H_{3,4}$ resonances at  the high mass region are already excluded by the stability condition.   

The effective Lagrangian describing dark matter--nucleon interaction in the limit of zero-momentum transfer through the exchange of the Higgs boson $H_1$ is given as 
\bea
\mathcal{L}^{\text{eff}}_{\mathcal{H}_2}&=&\frac{C_qm_q}{m^2_{H_1}}{\mathcal{H}_2}{\mathcal{H}_2}\bar{q}q,\crn
\mathcal{L}^{\text{eff}}_{\mathcal{H}_3}&=&\mathcal{L}^{\text{eff}}_{\mathcal{H}_2}\left(\mathcal{H}_2\leftrightarrow\mathcal{H}_3\right),
\eea
where
\bea
C_u&=&C_c=C_b=\frac{2\sqrt{2}s_{\alpha_2}}{v}(2\lambda_1c_{\alpha_2}u+\lambda_5s_{\alpha_2}v),\crn
C_d&=&C_s=C_t=\frac{2\sqrt{2}c_{\alpha_2}}{u}(2\lambda_1c_{\alpha_2}u+\lambda_5s_{\alpha_2}v).
\eea
Note that $H_{2,3,4}$ give smaller contributions, as neglected. 

Then, the SI cross-section for the scattering of each dark matter component on a target nucleus $N$ is expressed as
\bea
\sigma^{\text{SI}}_{\text{eff}}(\mathcal{H}_2)&=&\frac{\Omega_{\mathcal{H}_2}h^2}{\Omega_{\text{DM}}h^2}\sigma^{\text{SI}}_{\mathcal{H}_2N},\\
\sigma^{\text{SI}}_{\text{eff}}(\mathcal{H}_3)&=&\frac{\Omega_{\mathcal{H}_3}h^2}{\Omega_{\text{DM}}h^2}\sigma^{\text{SI}}_{\mathcal{H}_3N},
\eea
where $\sigma^{\text{SI}}_{\mathcal{H}_{2(3)}N}$ is given by
\bea
\sigma^{\text{SI}}_{\mathcal{H}_{2(3)}N}&=&\left(\frac{2m_{\mathcal{H}_{2(3)}N}}{m^2_{H_1}}\frac{m_p}{m_{\mathcal{H}_{2(3)}}}C_N\right)^2,
\eea 
with $m_{\mathcal{H}_{2(3)}N}=\frac{m_{\mathcal{H}_{2(3)}}m_N}{m_{\mathcal{H}_{2(3)}}+m_N}\simeq m_N$ to be the dark matter--nucleon reduced mass, and the nucleus factor $C_N$ is given by
\bea
C_N=\frac{2}{27}\sum_{q=c,b,t}AC_qf^p_{Tg}+\sum_{q=u,d,s}C_q[Zf^p_{Tq}+(A-Z)f^n_{Tq}],
\eea
with
\bea
f^{p(n)}_{Tu}&\approx&0.020(0.014),\ \ \ \ f^{p(n)}_{Td}\approx0.026(0.036),\crn
f^{p(n)}_{Ts}&\approx&0.118(0.118),\ \ \ \ f^p_{Tg}=1-\sum_{q=u,d,s}f^p_{Tq}.
\eea

\begin{figure}[!h]
\begin{center}
	\includegraphics[scale=0.35]{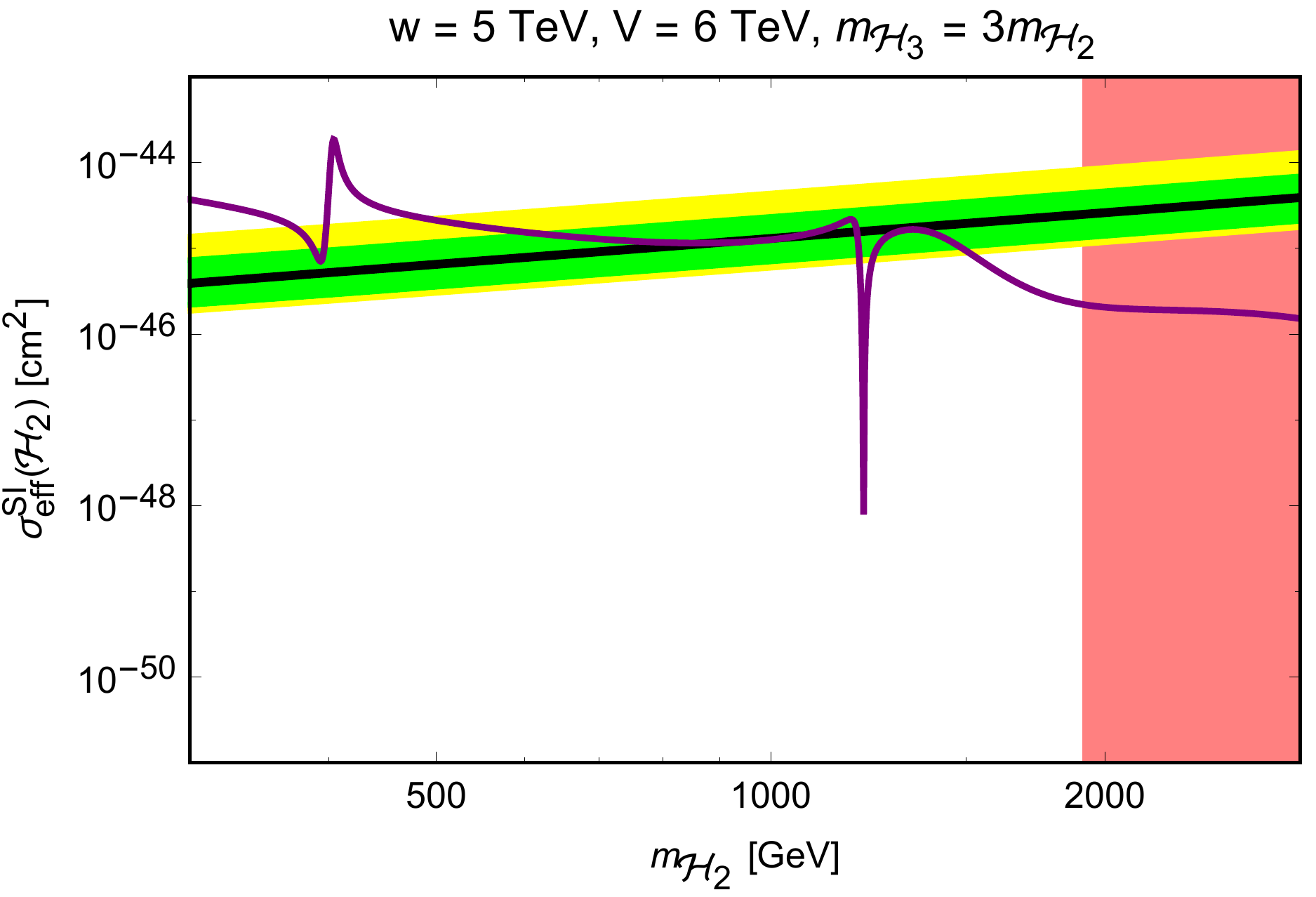}
	\includegraphics[scale=0.35]{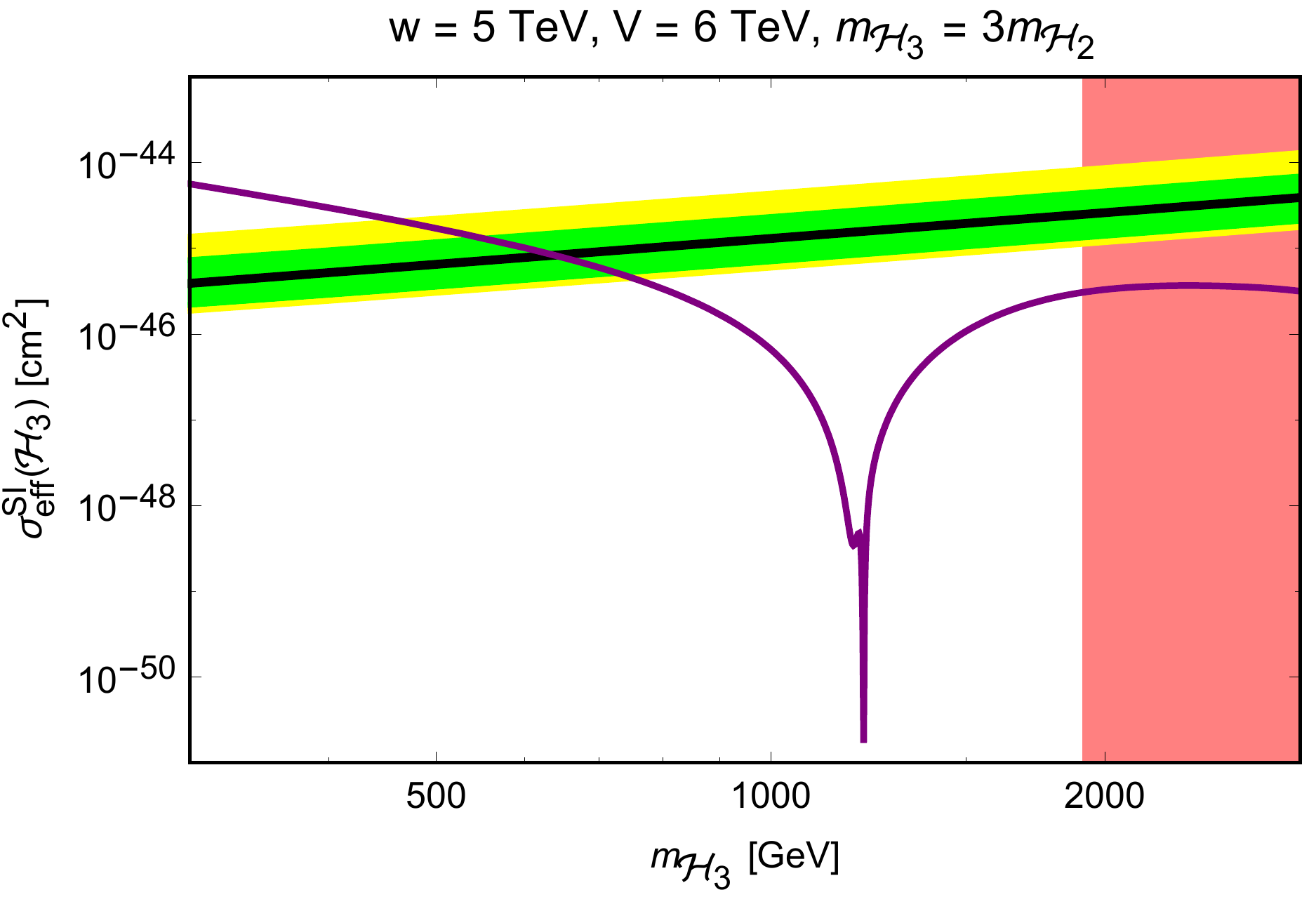}\\
	\includegraphics[scale=0.35]{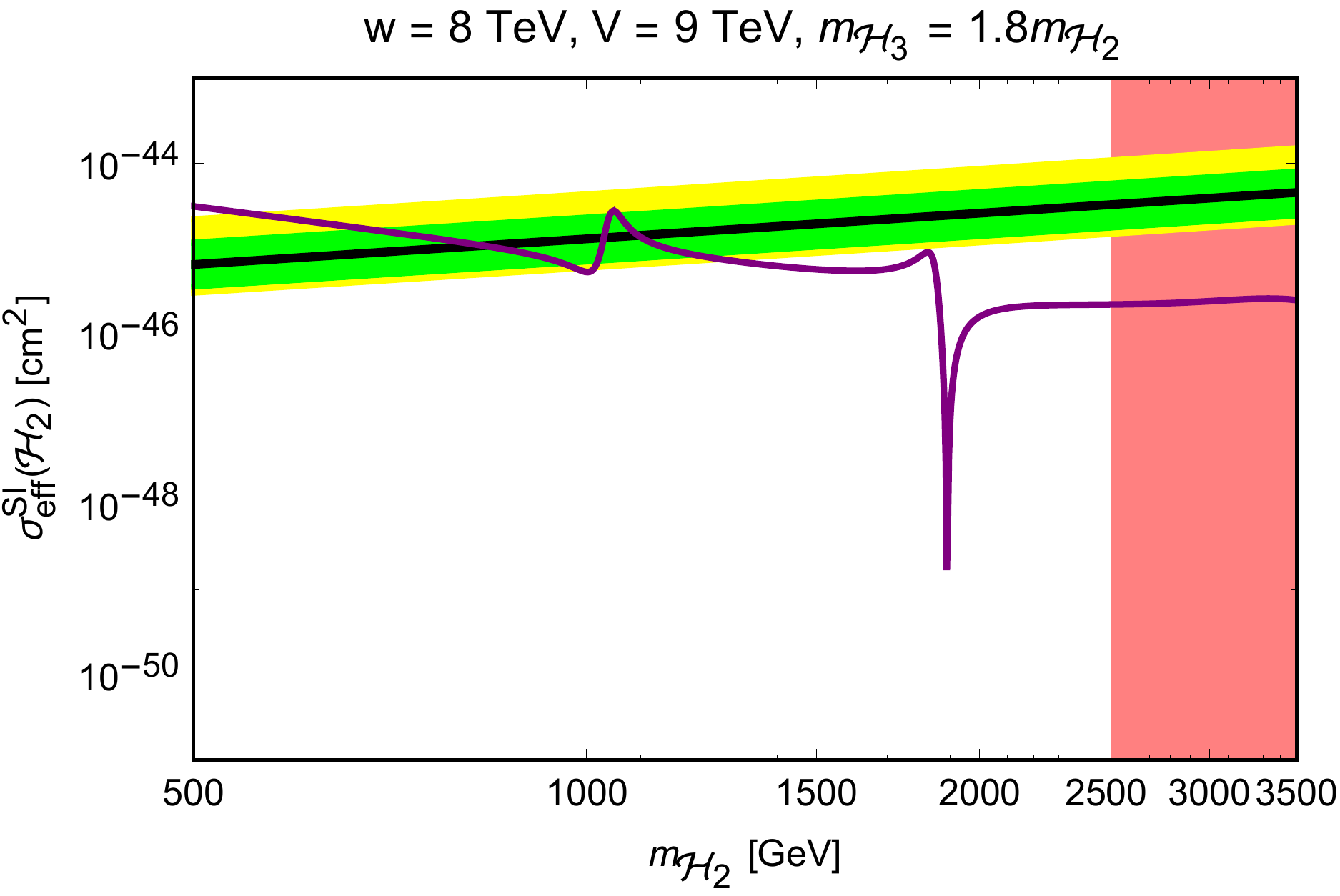}
	\includegraphics[scale=0.35]{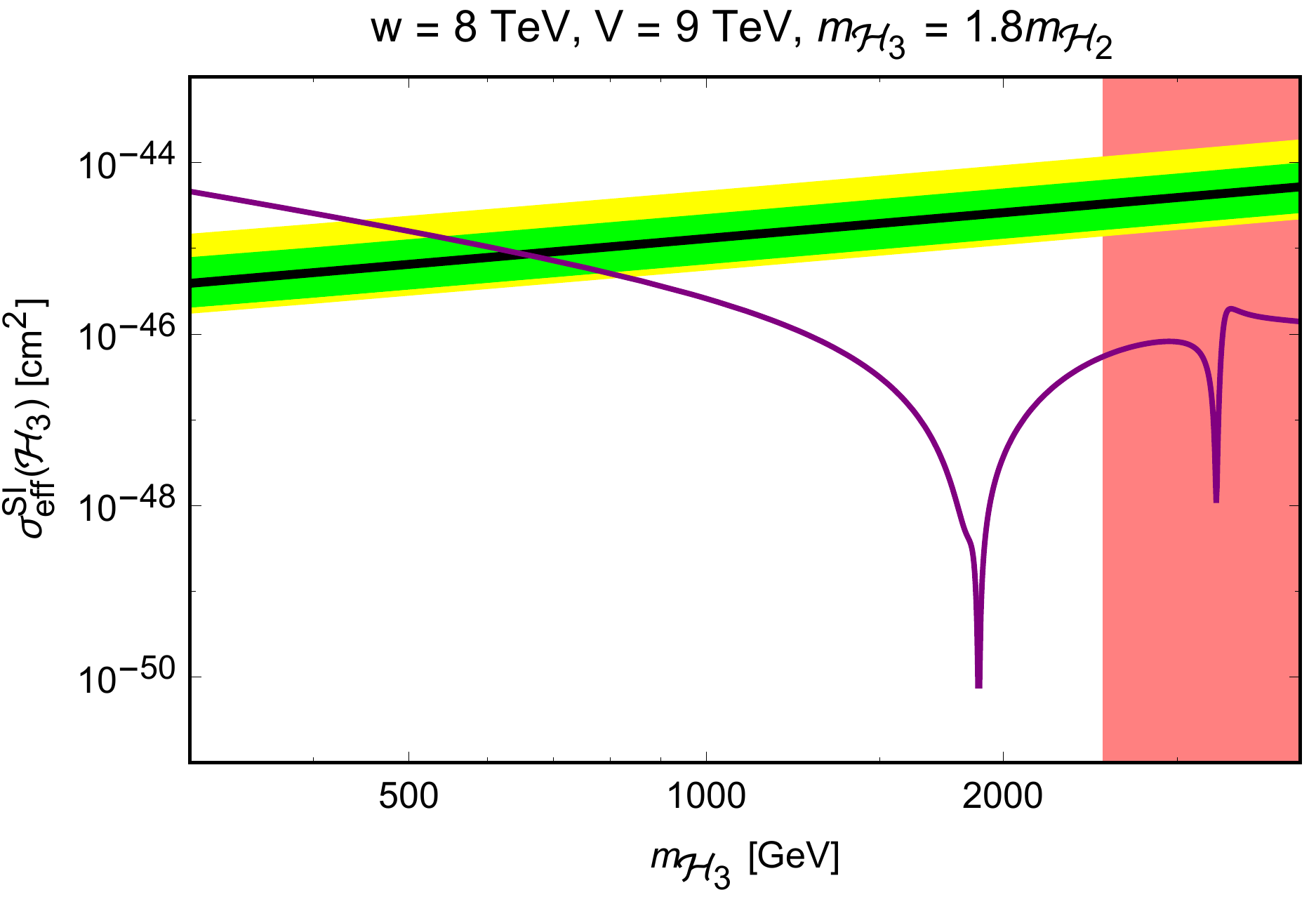}\\
	\includegraphics[scale=0.35]{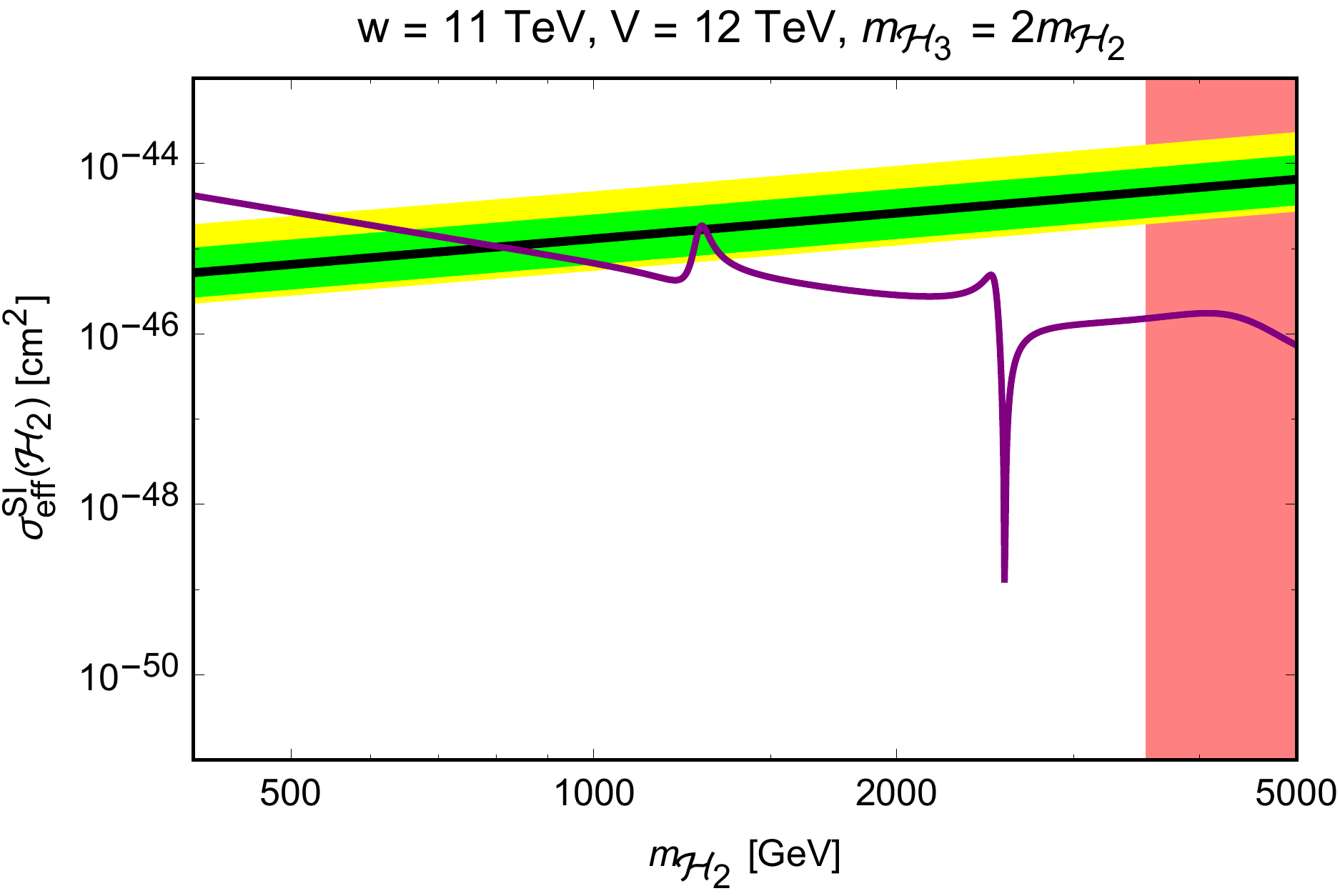}
	\includegraphics[scale=0.35]{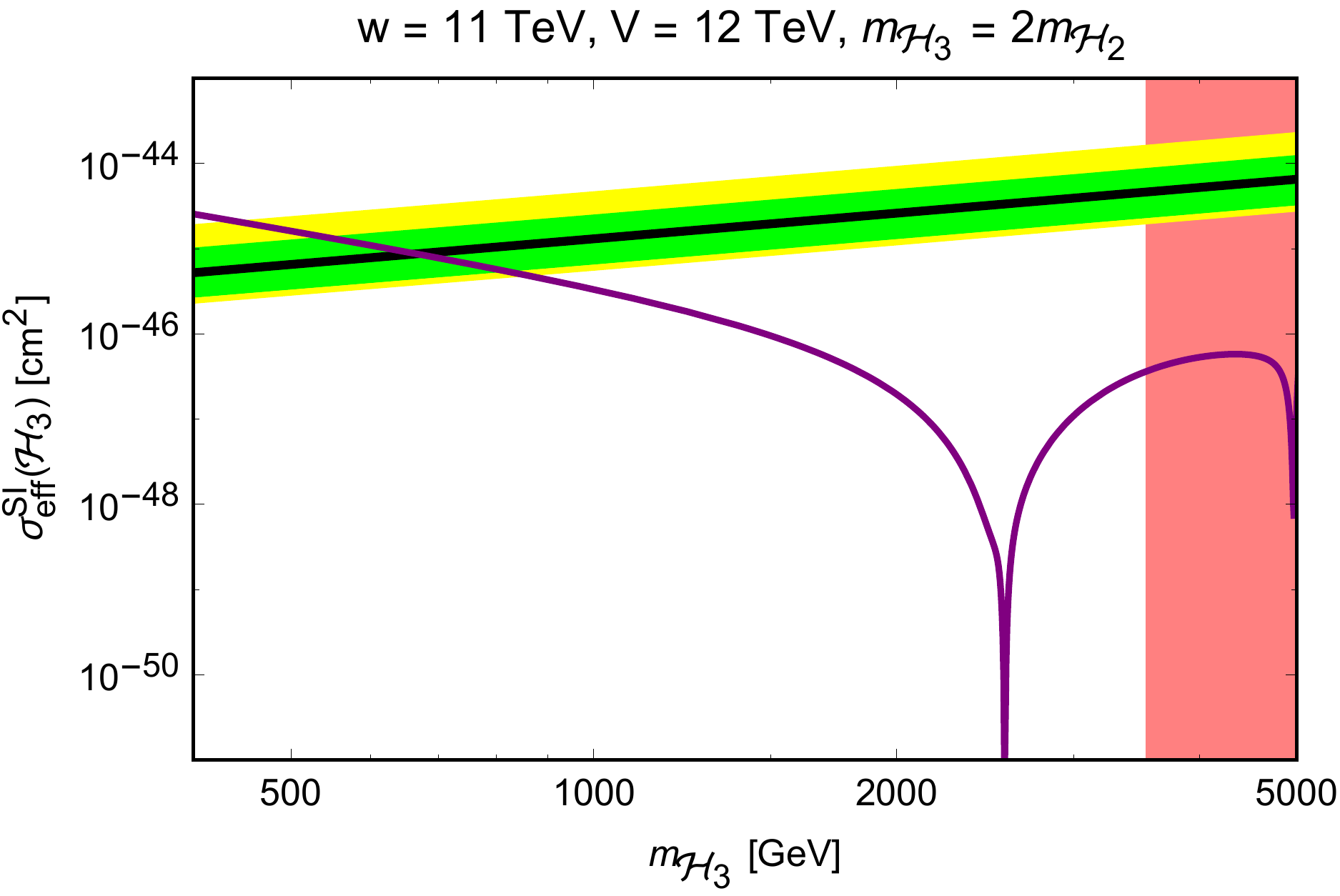}
	\caption{\label{loiadd1}The spin-independent dark matter-nucleon scattering cross-section limits as a function of dark matter masses according to each choice of $w,V$ and $m_{\mathcal{H}_{2,3}}$ relation, where the dark matter unstable regime is input as red.}	
\end{center}
\end{figure}

According to the above choices of $w,V$ and the dark matter mass relations, we plot the SI dark matter nucleon scattering cross-section as the function of the corresponding dark matter mass, where the experimental bounds \cite{Aprile:2017iyp,Aprile:2018dbl} are included. The general remark is that the scalar dark matter masses below around 700 GeV are excluded by the direct detection experiment. Additionally, they with a low mass give small contributions to the abundance as seen from Figure \ref{sdm1}.   

\subsection{Scenario with a fermion and a scalar dark matter}

\begin{figure}[!h]
	\centering
	\includegraphics[scale=0.9]{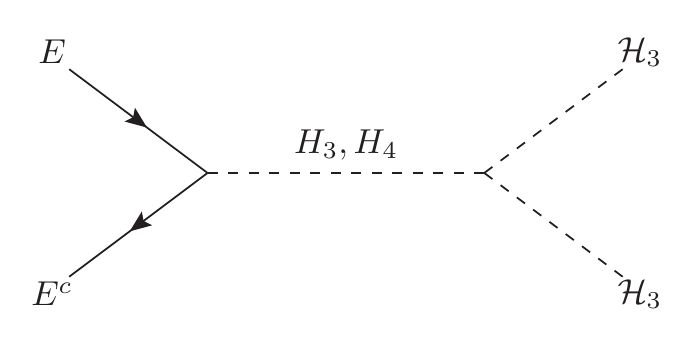}
	\caption{Conversion between fermion and scalar dark matter components.}	\label{DMDM-3}
\end{figure}

In this case we consider $E$ and $\mathcal{H}_3$ to be two-component dark matter candidates, without loss of generality. Their annihilation cross-sections to the standard model particles have been obtained above. Let us examine the fermion and scalar dark matter conversion, as given by the diagram in Figure \ref{DMDM-3}    

The thermal average annihilation cross-section times the relative velocity for the dark matter--dark matter conservation is obtained in the non-relativistic approximation at the leading order as
\bea
\langle\sigma v\rangle_{EE^c\to\mathcal{H}_3\mathcal{H}_3}&\simeq&0,\\
\langle\sigma v\rangle_{\mathcal{H}_3\mathcal{H}_3\to EE^c}&\simeq&\frac{m^2_E}{\pi w^2m^2_{\mathcal{H}_3}}\sqrt{1-\frac{m^2_E}{m^2_{\mathcal{H}_3}}}(m^2_{\mathcal{H}_3}-m^2_E)\crn
&&\times\left[\frac{c_{\alpha_1}(\lambda_6 w c_{\alpha_1}-\lambda_7Vs_{\alpha_1})}{4m^2_{\mathcal{H}_3}-m^2_{H_3}}+\frac{s_{\alpha_1}(\lambda_6 w s_{\alpha_1}+\lambda_7Vc_{\alpha_1})}{4m^2_{\mathcal{H}_3}-m^2_{H_4}}\right]^2.
\eea

\begin{figure}[!h]
\begin{center}
	\includegraphics[scale=0.25]{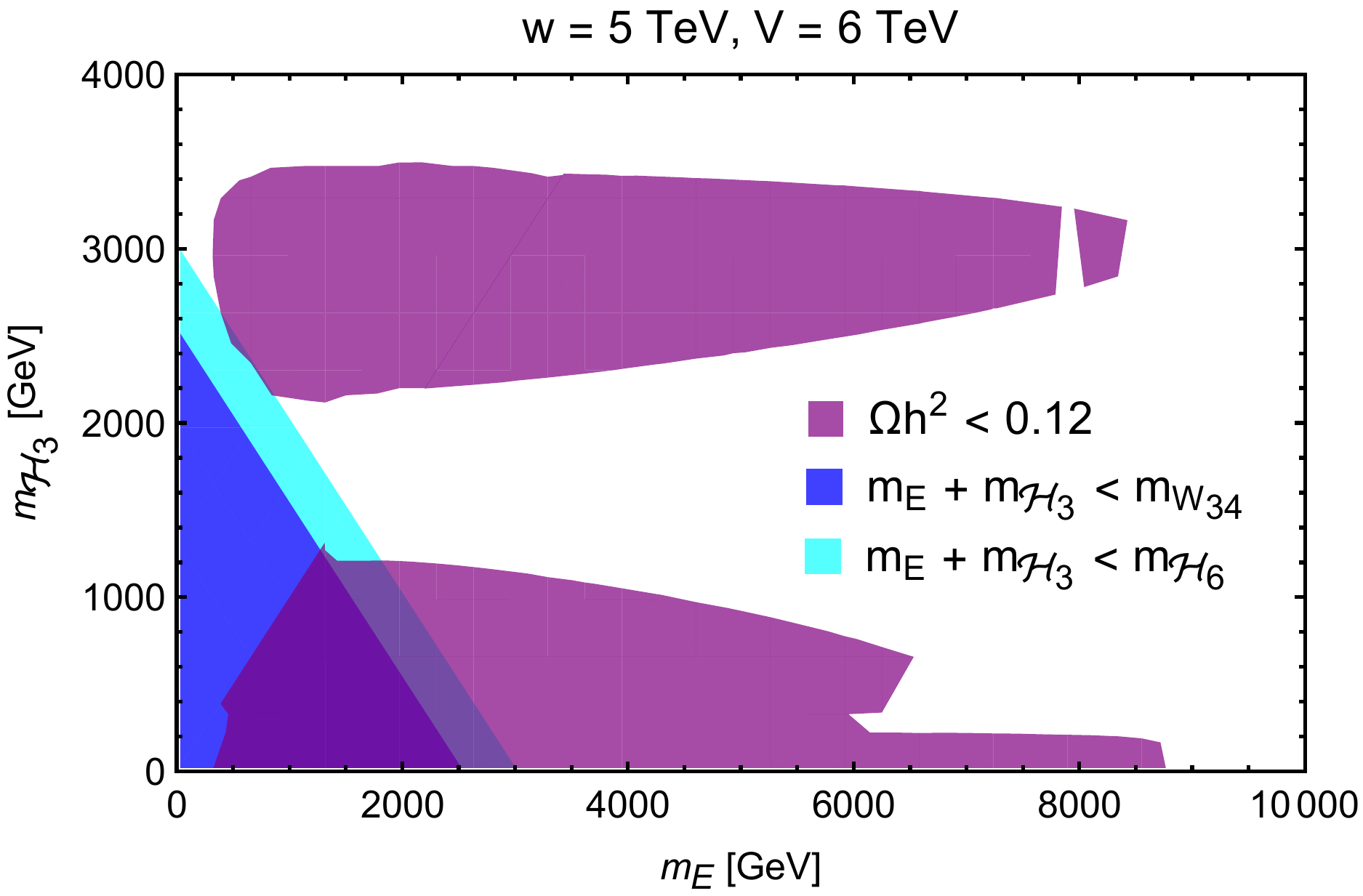}
	\includegraphics[scale=0.25]{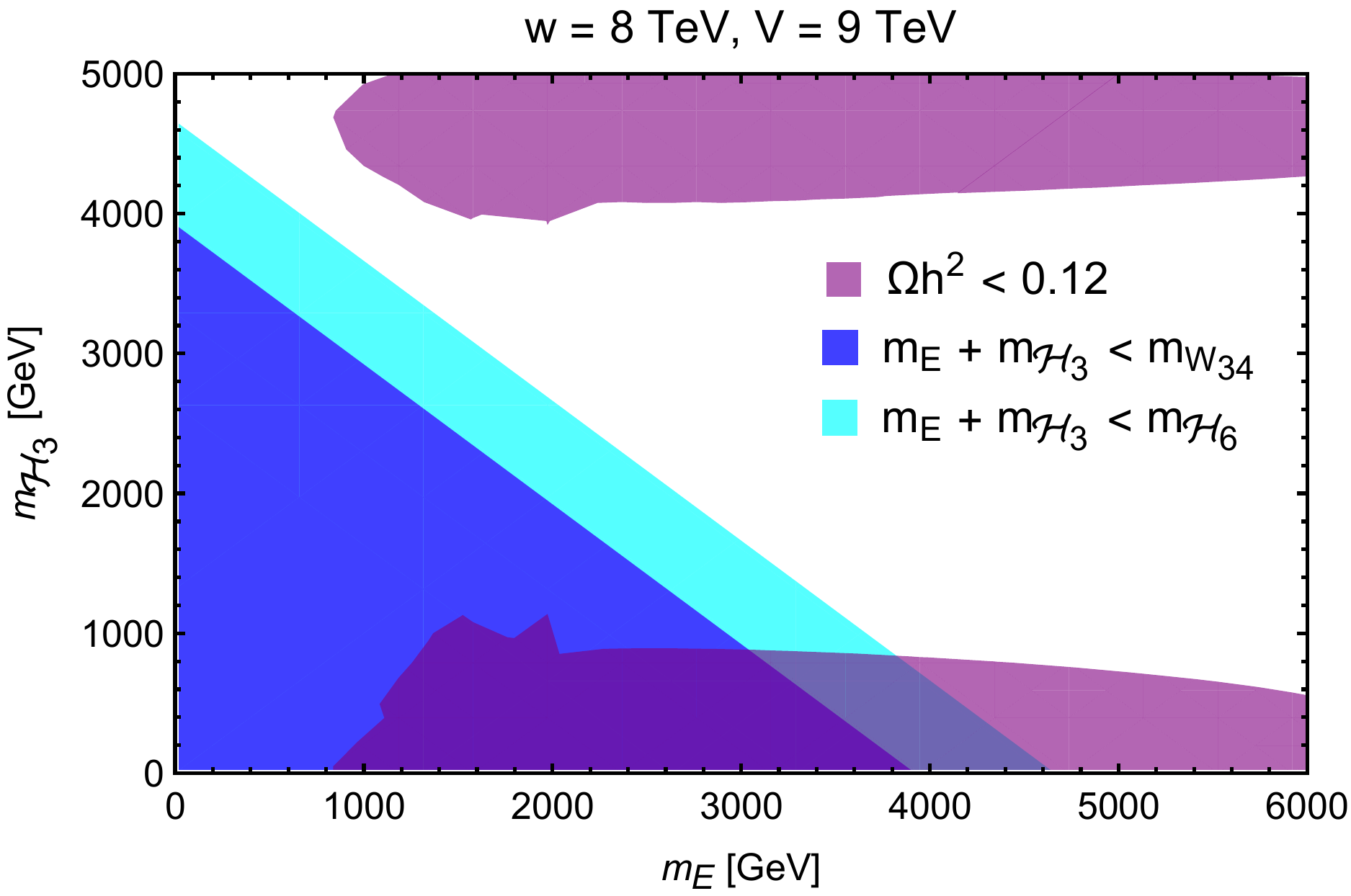}
	\includegraphics[scale=0.25]{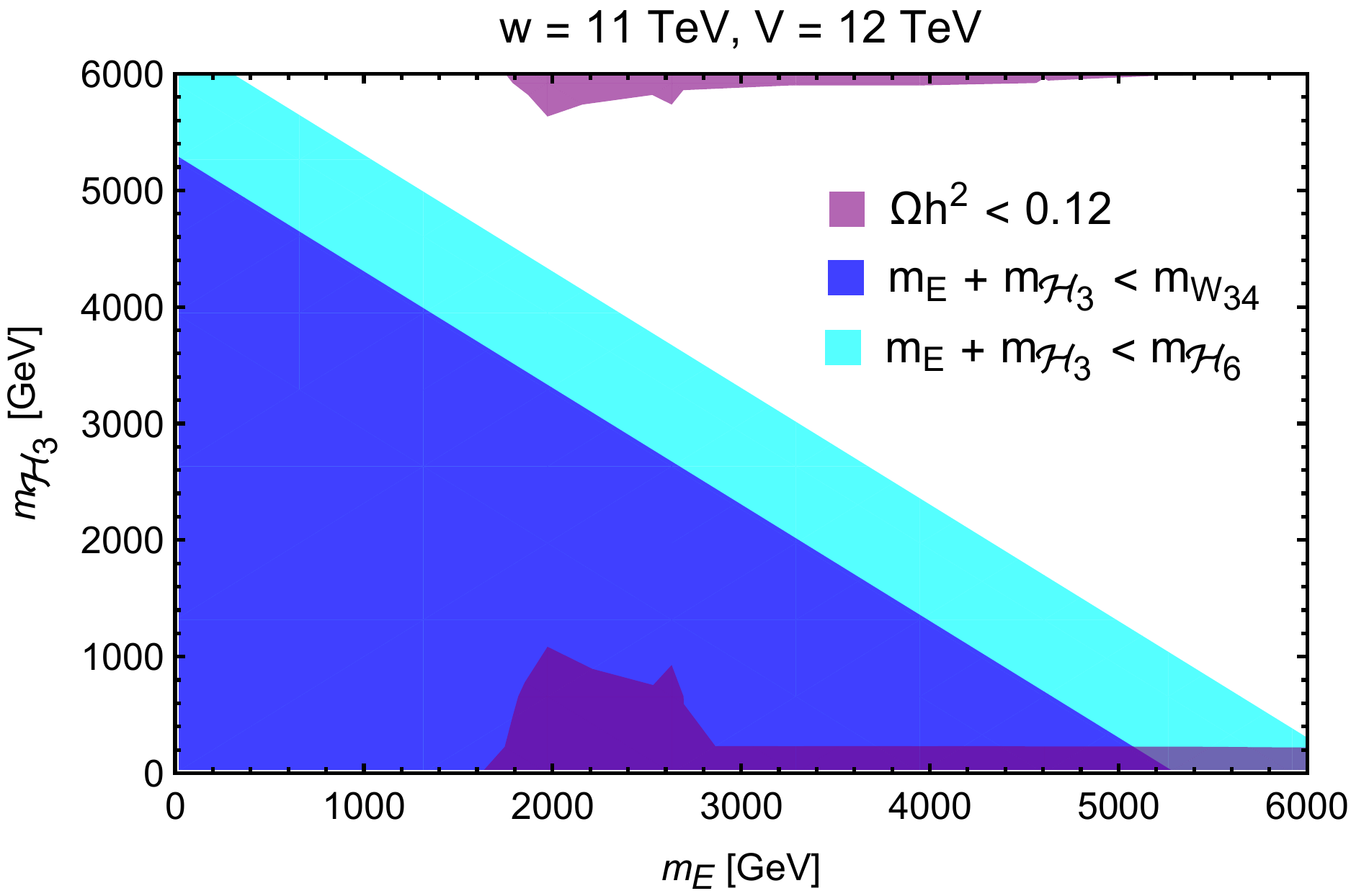}
	\caption{\label{loiadd21} The total abundance contoured as the function of fermion and scalar dark matter masses, where the dark matter stable regimes are also shown, according to each choice of $w,V$.}	
\end{center}
\end{figure}

For numerical calculation, we use the following parameter values,
\be \la_1=0.1,\ \la_{3,4,6,7,9,10}=0.3,\ \la_5=-0.19,\ee throughout this section, where the values of potential parameters chosen must also satisfy the vacuum stability conditions and positive squared scalar masses. Note that some of them differ from the two-scalar dark matter section since the three cases of two-component dark matter are alternative. 

We contour the total relic density as the function of dark matter masses as well as imposing the dark matter stable conditions as displayed in Figure \ref{loiadd21}, corresponding to the fixed values of the new physics scale $w,V$. 

\begin{figure}[!h]
\begin{center}
	\includegraphics[scale=0.25]{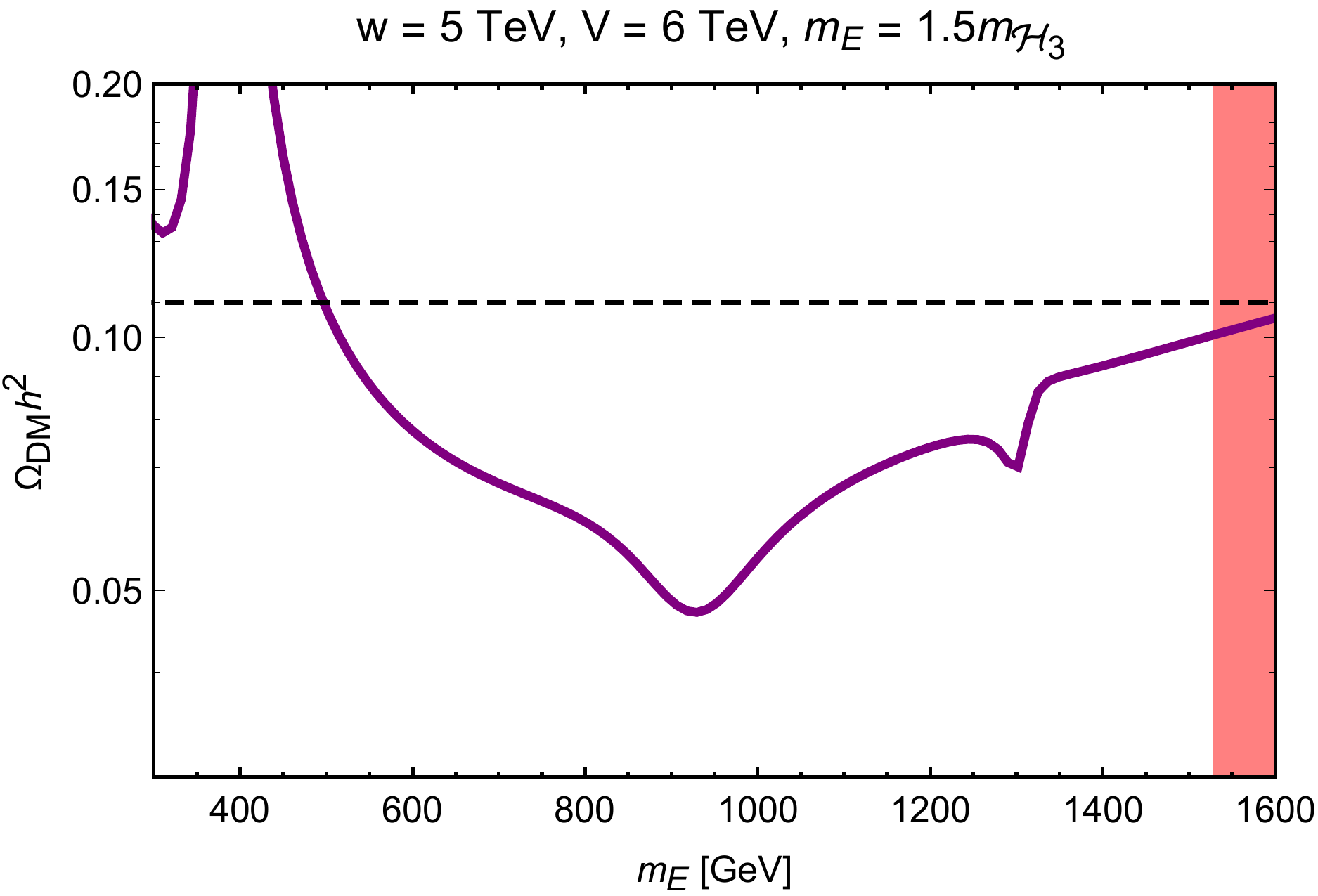}
	\includegraphics[scale=0.25]{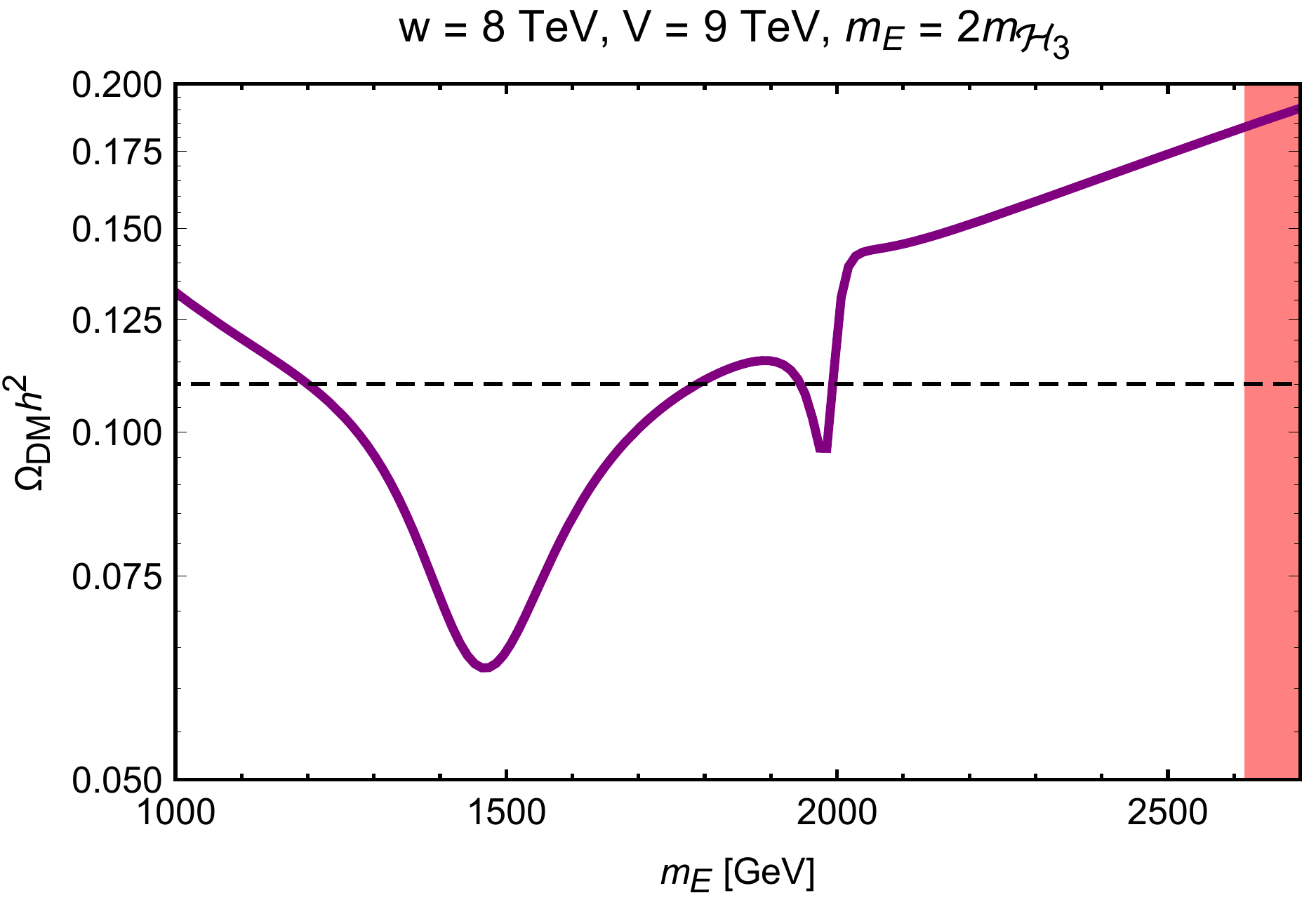}
	\includegraphics[scale=0.25]{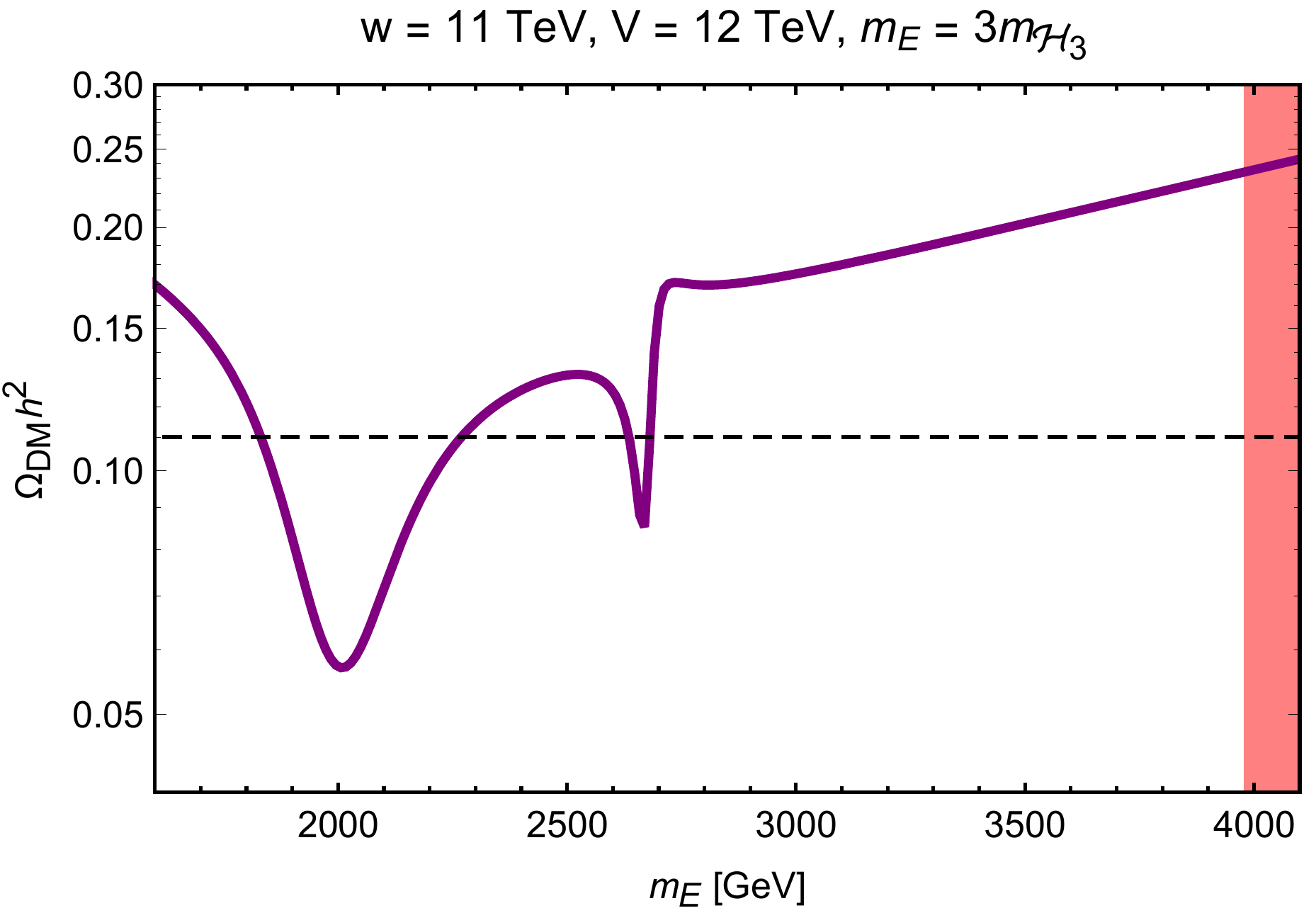}
	\caption{\label{fsf1} The total relic density of fermion and scalar dark matter as the function of the fermion dark matter mass corresponding to the $m_E, m_{\mathcal{H}_3}$ relations and $w, V$ choices.}	
\end{center}
\end{figure}
\begin{figure}[!h]
\begin{center}
	\includegraphics[scale=0.35]{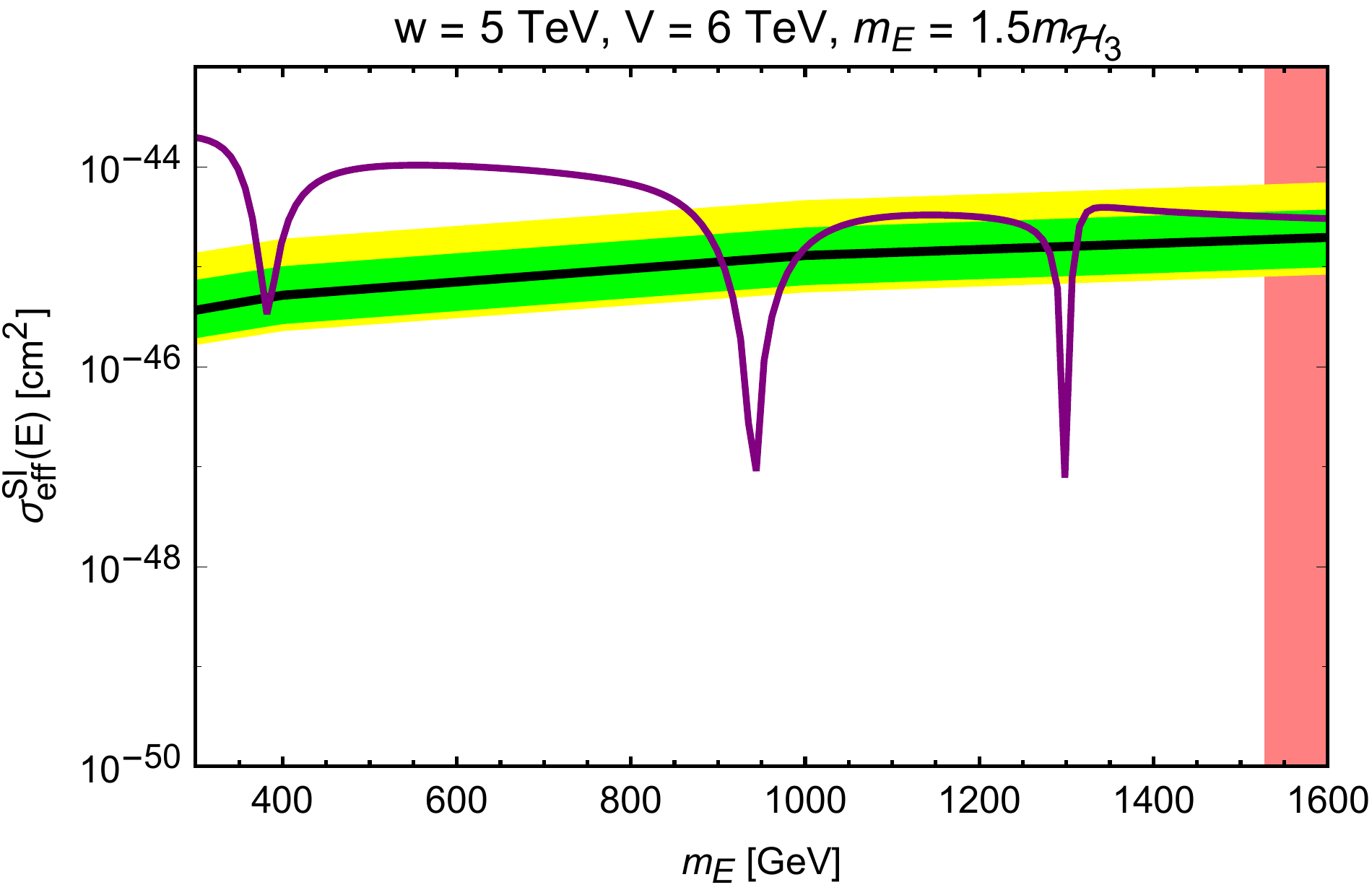}
	\includegraphics[scale=0.35]{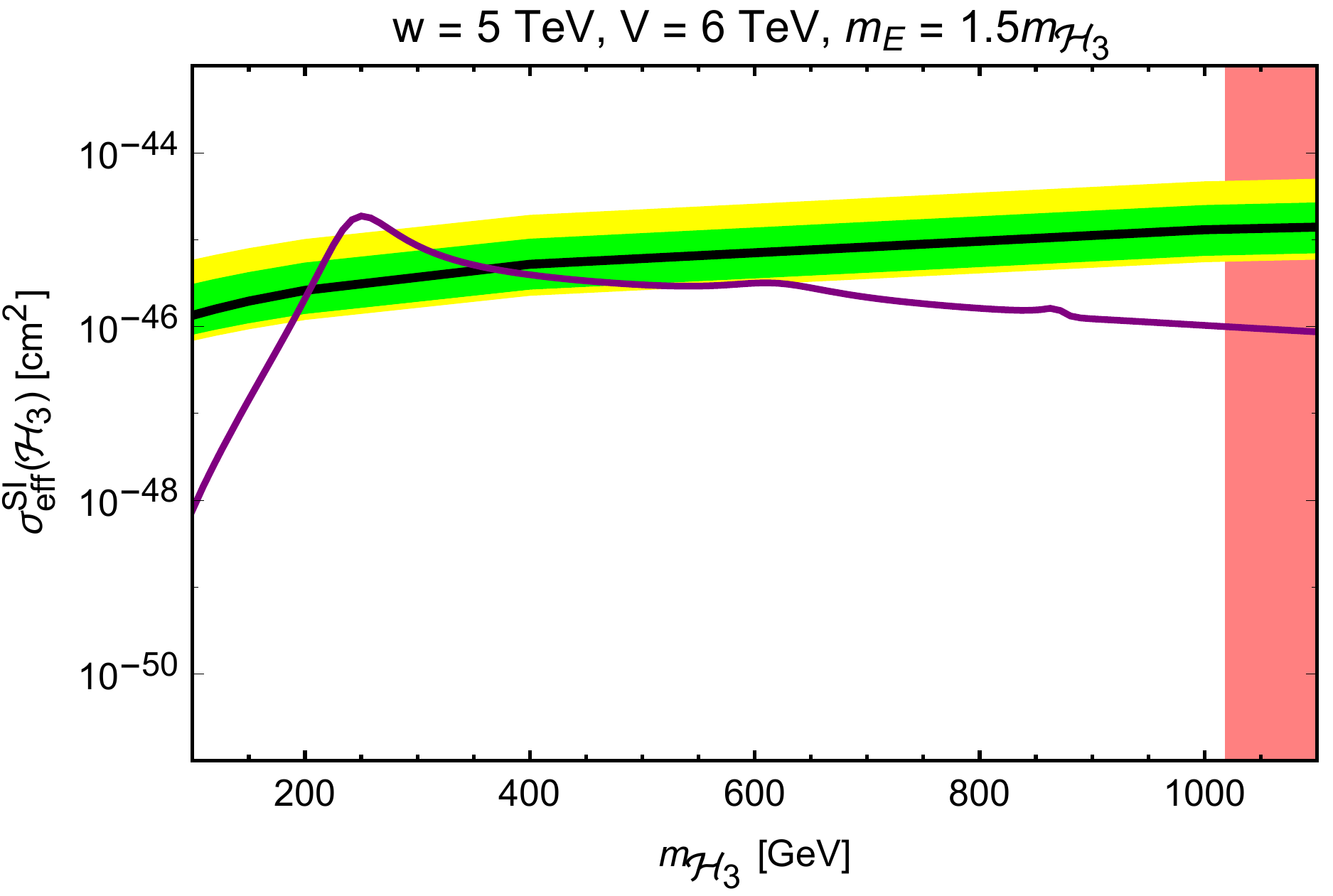}\\
	\includegraphics[scale=0.35]{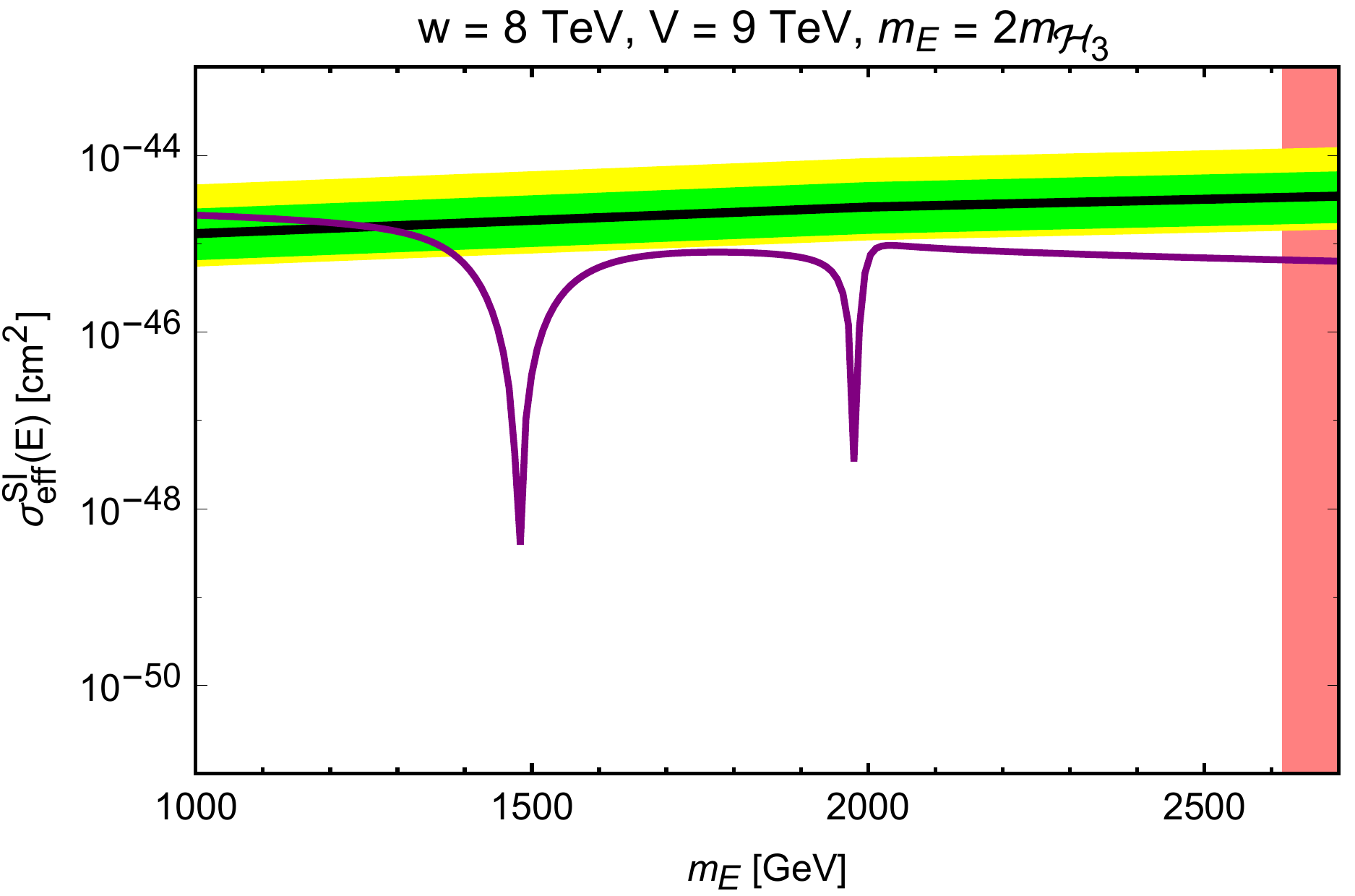}
	\includegraphics[scale=0.35]{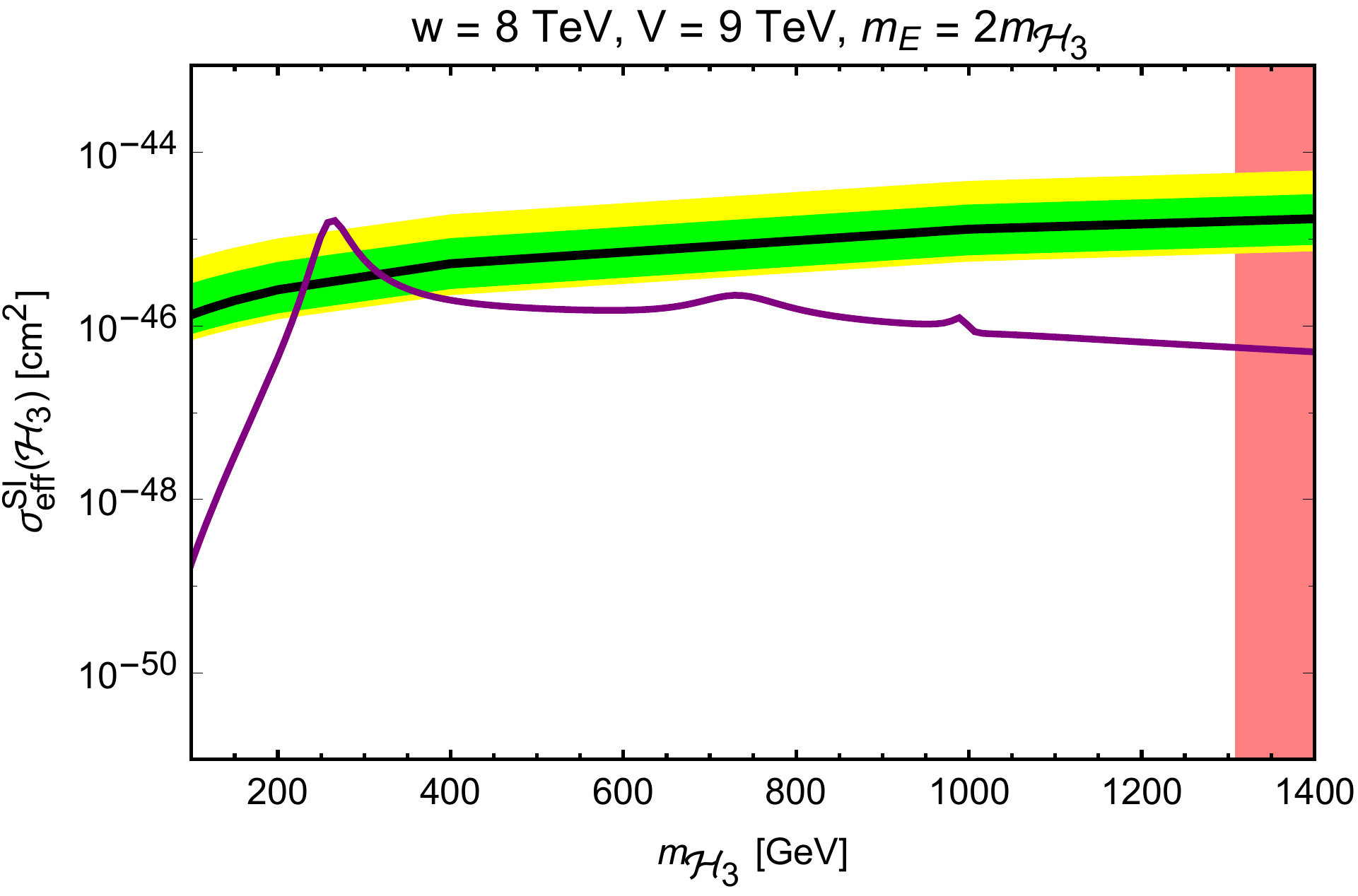}\\
	\includegraphics[scale=0.35]{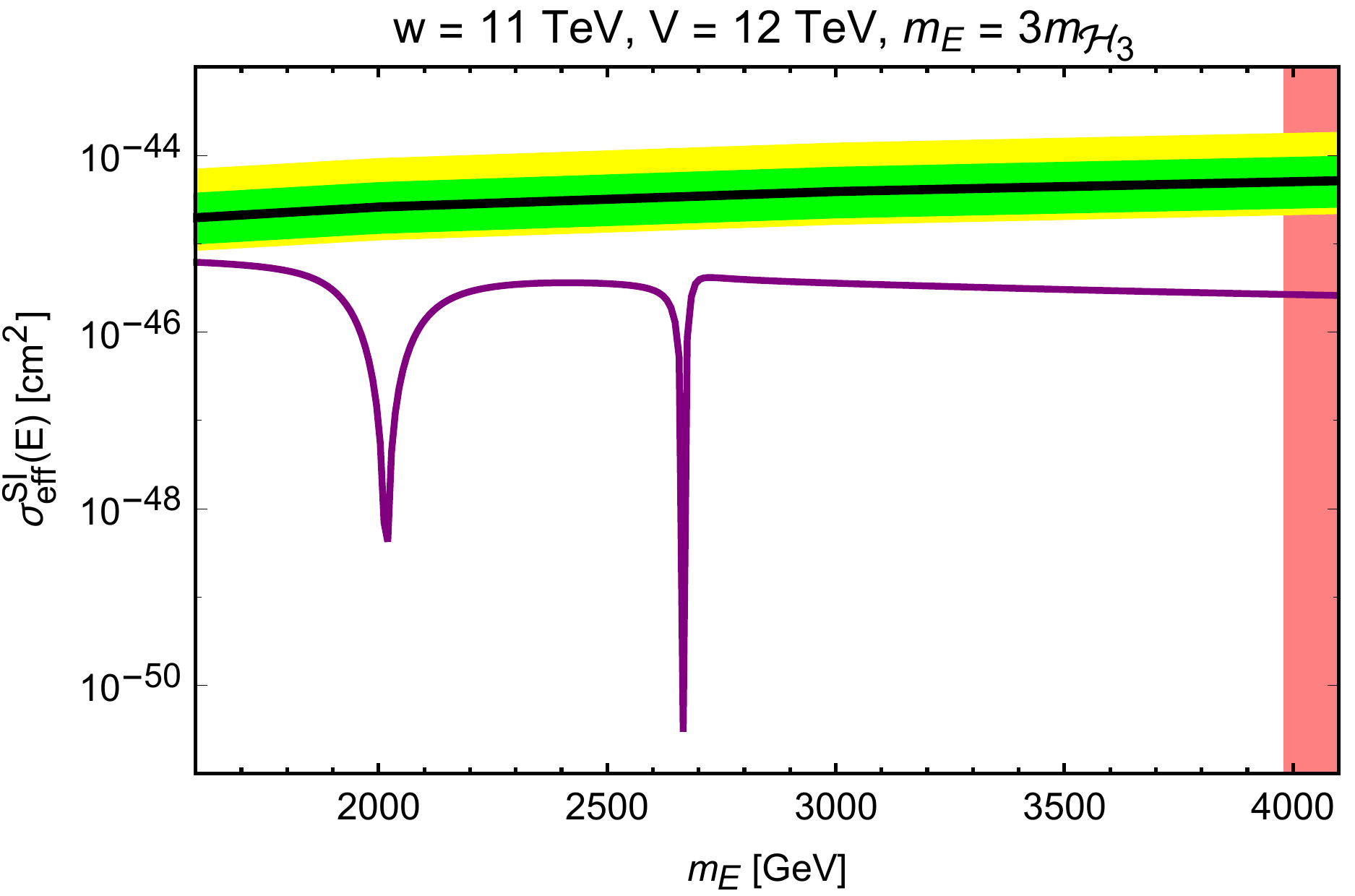}
	\includegraphics[scale=0.35]{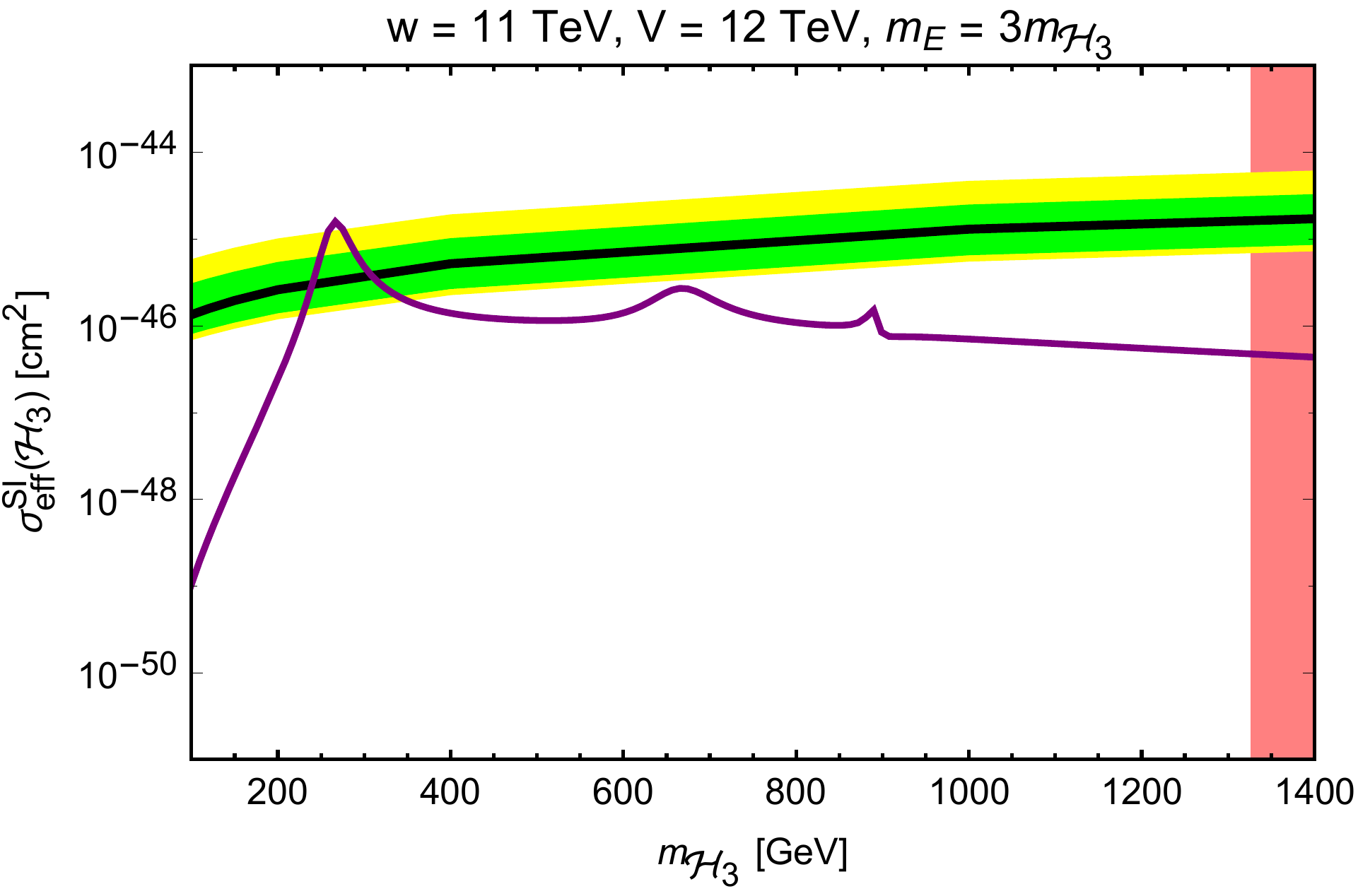}
	\caption{\label{fsf2} Direct detection cross-sections of fermion and scalar dark matter components plotted as the function of the corresponding dark matter mass according to the choices of the dark matter mass relations and $w,V$.}	
\end{center}
\end{figure}
From the density contours according to each pair value of $w,V$, we select the viable dark matter mass relations and plot the total relic density to see the contribution effect of each dark matter component and the direct detection cross-section, which are all given as the functions of a dark matter mass, presented in Figures \ref{fsf1} and \ref{fsf2}, respectively. Here, the dark matter unstable regimes are shown and note that the resonances that are presented in the excluded regimes are omitted. Typically, we obtain the resonance phenomena similar to the two cases above, two resonances for the fermion candidate set by the new gauge portal and other two for the scalar candidate set by the new Higgs portal. With the selection of parameters, in these cases, the resonances are important to govern the mark matter observables. The viable dark matter masses are around one to a few TeV.

\subsection{Remark of three-component dark matter scenario}

Let us remind the reader that with a dark matter stabilizing symmetry as $P=P_n\otimes P_m$---which is basically isomorphic to a $Z_2\otimes Z_2$---there may exist scenarios of three-component dark matter, achieved in a kinematically fortunate situation. As a matter of fact, such a three-component dark matter scheme can be realized in each scenario investigated above. 

Take, for instance, the last scenario. If we assume that the mass relations $m_E+m_{\mathcal{H}_3}>m_{\mathcal{H}_6}$ by contrast and $m_{W_{34}}>m_{\mathcal{H}_6}$ are satisfied, where $E$ and $\mathcal{H}_3$ are the lightest particles within the classes of singly-wrong particles of the same kind, respectively. Then $E$, $\mathcal{H}_3$, and $\mathcal{H}_6$ may be simultaneously stable, providing a scenario of three-component dark matter. In this case, $\mathcal{H}_6$ contributes to the relic density and may own interesting detection aspects. 

Anyway, we have focused the only two-component dark matter scenarios in this work. The prospect for the three-component dark matter schemes are worth exploring to be published elsewhere.  

\section{\label{conl}Conclusion}

We have shown that a gauge theory that includes a higher weak isospin symmetry $SU(P)_L$ must possess a complete gauge symmetry of the form $SU(3)_C\otimes SU(P)_L\otimes U(1)_X\otimes U(1)_N$, where the last two Abelian groups define the electric charge and baryon-minus-lepton charge, respectively. The last charges are unified with the weak charge in the same manner as the electroweak theory. Additionally, the neutrino masses are appropriately induced by the gauge symmetry breaking, supplied in terms of a canonical seesaw mechanism. 

The multiple matter parity $P=\bigotimes^{P-2}_{k=1}P_k$, where each $P_k$ is a $Z_2$, is obtained as a residual gauge symmetry. This parity makes $P-2$ wrong particles stable, providing multicomponent dark matter candidates. The noncommutation of $B-L$ with $SU(P)_L$ yields that the dark matter candidates are nontrivially unified with normal matter in gauge multiplets; in other words, multicomponent dark matter is required to complete the $SU(P)_L$ representations enlarged from the standard model. Therefore, the gauge interactions would govern the dark matter observables. 

The minimal multicomponent dark matter model corresponds to $P=4$, the so-called 3-4-1-1 model. In this case, we have fully diagonalized the scalar and gauge sectors and identified two-component dark matter schemes according to the multiple matter parity $P=P_n\otimes P_m$. All the interactions of fermions and scalars with gauge bosons have been obtained. The 3-4-1-1 model with $q=p=0$ obeys three possibilities of two-component dark matter, including two fermions $(E,F)$, two scalars $(\mathcal{H}_{2,3})$, and a fermion and a scalar e.g. $(E,\mathcal{H}_3)$ candidates, respectively. We have shown the viable parameter space for each scenario/possibility satisfying the relic density and direct detection. Typically, the dark matter masses are obtained in TeV regime. Additionally, there are four resonances in relic density set by the new neutral gauge $Z_{2,3}$ or the new neutral Higgs $H_{3,4}$ portals.

The dark matter scenarios can be discriminated in experiments relying the interacting characters of the dark matter candidates with the standard model particles. For instance, the fermion candidates can interact with leptons, while the scalar candidates do not. Additionally, the fermion candidates scatter inelastically with $Z,H$, while the scalar candidates scatter elastically with $Z,H$ by contrast. All these distinguish the first and the second scenarios. If both the phenomena, elastic and inelastic (or scattering elastically with both leptons and bosons) are observed, it indicates to the third scenario. Another distinction is that the fermion candidates only interact with normal particles via new gauge portals, whereas the scalar candidates do so via Higgs portals, including the standard model one.       

\section*{Acknowledgments}

This research is funded by Vietnam National Foundation for Science and Technology Development (NAFOSTED) under grant number 103.01-2019.353.

\appendix

\section{\label{anomaly} Anomaly cancelation}

The nontrivial anomalies include 
\bea &&[SU(3)_C]^2 U(1)_X,\ [SU(3)_C]^2 U(1)_N,\ 
[SU(P)_L]^2 U(1)_X,\crn 
&&[SU(P)_L]^2 U(1)_N,\ 
[\mathrm{Gravity}]^2 U(1)_X,\ 
[\mathrm{Gravity}]^2 U(1)_N,\\
&& [U(1)_X]^2 U(1)_N,\ 
U(1)_X [U(1)_N]^2,\ 
[U(1)_X]^3,\ [U(1)_N]^3.\nn \eea 

Let us compute each of them, 
\bea [SU(3)_C]^2U(1)_X&\sim& \sum_{\mathrm{quarks}} (X_{q_L}-X_{q_R}) \crn
&=& 2P X_{Q_\al}+PX_{Q_3}-3X_{u_a}-3X_{d_a}-2\sum_{k=1}^{P-2}X_{J_{k\al}}-\sum_{k=1}^{P-2}X_{J_{k3}}\crn
&=&2P\left(\frac{-1}{3}+\frac{1-q}{P}\right)+ P\left(\fr 2 3 +\fr{q-1}{P}\right)-3\times\fr 2 3-3\times\frac {-1}{3}\crn
&&-2\sum_{k=1}^{P-2}\left(-q_k-\frac 1 3\right)-\sum_{k=1}^{P-2}\left(q_k+\frac 2 3\right)= \sum_{k=1}^{P-2}q_k-q = 0. \eea
\bea [SU(3)_C]^2U(1)_N&\sim& \sum_{\mathrm{quarks}} (N_{q_L}-N_{q_R}) \crn
&=& 2P N_{Q_\al}+P N_{Q_3}-3N_{u_a}-3N_{d_a}-2\sum_{k=1}^{P-2}N_{J_{k\al}}-\sum_{k=1}^{P-2}N_{J_{k3}}\crn
&=&2P\left(\frac{-2}{3}+\frac{2-n}{P}\right)+ P\left(\fr 4 3 +\fr{n-2}{P}\right)-3\times\fr 1 3-3\times\fr 1 3\crn
&&-2\sum_{k=1}^{P-2}\left(-n_k-\frac 2 3\right)-\sum_{k=1}^{P-2}\left(n_k+\frac 4 3\right)= \sum_{k=1}^{P-2}n_k-n = 0. \eea
\bea [SU(P)_L]^2 U(1)_X &\sim& \sum_{\mathrm{(anti)P-plets}} X_{F_L}= 3X_{\psi_a}+ 6 X_{Q_\al}+3X_{Q_3} \crn
&=& 3\times\frac{q-1}{P}+6\left(\frac{-1}{3}+\frac{1-q}{P}\right)+3\left(\fr 2 3 +\fr{q-1}{P}\right)=0.  \eea 
\bea [SU(P)_L]^2 U(1)_N &\sim& \sum_{\mathrm{(anti)P-plets}} N_{F_L}= 3N_{\psi_a}+6 N_{Q_\al}+3N_{Q_3} \crn
&=& 3\times\frac{n-2}{P}+6\left(\frac{-2}{3}+\frac{2-n}{P}\right)+3\left(\fr 4 3 +\fr{n-2}{P}\right)=0.  \eea 
\bea [\mathrm{Gravity}]^2U(1)_X&\sim&\sum_{\mathrm{fermions}}(X_{f_L}-X_{f_R})= 3P X_{\psi_a}+6P X_{Q_\al}+3P X_{Q_3}\crn
&&-3X_{\nu_a}-3X_{e_a}-3\sum_{k=1}^{P-2}X_{E_{ka}}-9 X_{u_a}-9 X_{d_a}-6 \sum_{k=1}^{P-2}X_{J_{k\al}}-3\sum_{k=1}^{P-2}X_{J_{k3}}\crn
&=&3P \times\frac{q-1}{P}+6P\left(\frac{-1}{3}+\frac{1-q}{P}\right)+3P\left(\fr 2 3 +\fr{q-1}{P}\right)-3\times 0-3(-1)\crn
&&-3\sum_{k=1}^{P-2}q_k-9\times \fr 2 3-9\times \frac{-1}{3}-6 \sum_{k=1}^{P-2}\left(-q_k-\frac 1 3\right)-3\sum_{k=1}^{P-2}\left(q_k+\frac 2 3\right)\crn
&=& 0.\eea
\bea [\mathrm{Gravity}]^2U(1)_N&\sim&\sum_{\mathrm{fermions}}(N_{f_L}-N_{f_R})= 3P N_{\psi_a}+6P N_{Q_\al}+3P N_{Q_3}\crn
&&-3N_{\nu_a}-3N_{e_a}-3\sum_{k=1}^{P-2}N_{E_{ka}}-9 N_{u_a}-9 N_{d_a}-6 \sum_{k=1}^{P-2}N_{J_{k\al}}-3\sum_{k=1}^{P-2}N_{J_{k3}}\crn
&=&3P \times\frac{n-2}{P}+6P\left(\frac{-2}{3}+\frac{2-n}{P}\right)+3P\left(\fr 4 3 +\fr{n-2}{P}\right)-3 (-1)-3(-1)\crn
&&-3\sum_{k=1}^{P-2}n_k-9\times \fr 1 3-9\times \fr 1 3-6 \sum_{k=1}^{P-2}\left(-n_k-\frac 2 3\right)-3\sum_{k=1}^{P-2}\left(n_k+\frac 4 3\right)\crn
&=& 0.\label{adt5}\eea
\bea [U(1)_X]^2U(1)_N&=&\sum_{\mathrm{fermions}}(X^2_{f_L}N_{f_L}-X^2_{f_R}N_{f_R})= 3P X^2_{\psi_a}N_{\psi_a}+6P X^2_{Q_{\al}}N_{Q_{\al}}+3P X^2_{Q_3} N_{Q_3}\crn 
&&-3X^2_{\nu_a}N_{\nu_a}-3X^2_{e_a} N_{e_a}-3\sum_{k=1}^{P-2}X_{E_{ka}}^2 N_{E_{ka}}-9 X^2_{u_a}N_{u_a}-9 X^2_{d_a}N_{d_a}\crn
&& -6 \sum_{k=1}^{P-2}X_{J_{k\al}}^2 N_{J_{k\al}}-3\sum_{k=1}^{P-2}X_{J_{k3}}^2 N_{J_{k3}}\crn
&=&3P \left(\frac{q-1}{P}\right)^2\left(\frac{n-2}{P}\right)+6P\left(\frac{-1}{3}+\frac{1-q}{P}\right)^2\left(\frac{-2}{3}+\frac{2-n}{P}\right)\crn
&&+3P\left(\fr 2 3 +\fr{q-1}{P}\right)^2\left(\fr 4 3 +\fr{n-2}{P}\right)-3\times 0^2(-1)-3(-1)^2(-1)\crn
&&-3\sum_{k=1}^{P-2}q_k^2n_k-9\left(\fr 2 3\right)^2\left(\fr 1 3\right)-9\left(\frac{-1}{3}\right)^2\left(\fr 1 3\right)\crn
&&-6 \sum_{k=1}^{P-2}\left(-q_k-\frac 1 3\right)^2\left(-n_k-\frac 2 3\right)-3\sum_{k=1}^{P-2}\left(q_k+\frac 2 3\right)^2\left(n_k+\frac 4 3\right)\crn
&=& \fr {2}{3}(n+4q)-\fr {2}{3}\sum_{k=1}^{P-2}(n_k+4q_k) =0.\label{adt4}\eea
 \bea [U(1)_X]U(1)_N^2&=&\sum_{\mathrm{fermions}}(X_{f_L}N^2_{f_L}-X_{f_R}N^2_{f_R})=3P X_{\psi_a}N^2_{\psi_a}+6P X_{Q_{\al}}N^2_{Q_{\al}}+3P X_{Q_3} N^2_{Q_3}\crn 
&&-3X_{\nu_a}N^2_{\nu_a}-3X_{e_a} N^2_{e_a}-3\sum_{k=1}^{P-2}X_{E_{ka}} N^2_{E_{ka}}-9 X_{u_a}N^2_{u_a}-9 X_{d_a}N^2_{d_a}\crn
&& -6 \sum_{k=1}^{P-2}X_{J_{k\al}} N^2_{J_{k\al}}-3\sum_{k=1}^{P-2}X_{J_{k3}} N^2_{J_{k3}}\crn
&=&3P \left(\frac{q-1}{P}\right)\left(\frac{n-2}{P}\right)^2+6P\left(\frac{-1}{3}+\frac{1-q}{P}\right)\left(\frac{-2}{3}+\frac{2-n}{P}\right)^2\crn
&&+3P\left(\fr 2 3 +\fr{q-1}{P}\right)\left(\fr 4 3 +\fr{n-2}{P}\right)^2-3\times 0(-1)^2-3(-1)(-1)^2\crn
&&-3\sum_{k=1}^{P-2}q_kn^2_k-9\left(\fr 2 3\right)\left(\fr 1 3\right)^2-9\left(\frac{-1}{3}\right)\left(\fr 1 3\right)^2\crn
&&-6 \sum_{k=1}^{P-2}\left(-q_k-\frac 1 3\right)\left(-n_k-\frac 2 3\right)^2-3\sum_{k=1}^{P-2}\left(q_k+\frac 2 3\right)\left(n_k+\frac 4 3\right)^2\crn
&=& \fr {8}{3}(n+q)-\fr {8}{3}\sum_{k=1}^{P-2}(n_k+q_k) =0.\label{adt2}\eea
\bea [U(1)_X]^3&=&\sum_{\mathrm{fermions}}(X^3_{f_L}-X^3_{f_R})=3P X^3_{\psi_a}+6P X^3_{Q_{\al}}+3P X^3_{Q_3}-3X^3_{\nu_a}-3X^3_{e_a}\crn 
&&-3\sum_{k=1}^{P-2}X^3_{E_{ka}}-9 X^3_{u_a}-9 X^3_{d_a}-6 \sum_{k=1}^{P-2}X^3_{J_{k\al}}-3\sum_{k=1}^{P-2}X^3_{J_{k3}}\crn
&=&3P \left(\frac{q-1}{P}\right)^3+6P\left(\frac{-1}{3}+\frac{1-q}{P}\right)^3+3P\left(\fr 2 3 +\fr{q-1}{P}\right)^3-3\times 0^3-3(-1)^3\crn
&&-3\sum_{k=1}^{P-2}q^3_k-9\left(\fr 2 3\right)^3-9\left(\frac{-1}{3}\right)^3-6 \sum_{k=1}^{P-2}\left(-q_k-\frac 1 3\right)^3-3\sum_{k=1}^{P-2}\left(q_k+\frac 2 3\right)^3\crn
&=& 2q-2\sum_{k=1}^{P-2}q_k =0.\eea
\bea [U(1)_N]^3&=&\sum_{\mathrm{fermions}}(N^3_{f_L}-N^3_{f_R})=3P N^3_{\psi_a}+6PN^3_{Q_{\al}}+3P N^3_{Q_3}-3N^3_{\nu_a}-3N^3_{e_a}\crn 
&&-3\sum_{k=1}^{P-2}N^3_{E_{ka}}-9N^3_{u_a}-9N^3_{d_a}-6 \sum_{k=1}^{P-2} N^3_{J_{k\al}}-3\sum_{k=1}^{P-2} N^3_{J_{k3}}\crn
&=&3P \left(\frac{n-2}{P}\right)^3+6P\left(\frac{-2}{3}+\frac{2-n}{P}\right)^3+3P\left(\fr 4 3 +\fr{n-2}{P}\right)^3-3(-1)^3-3(-1)^3\crn
&&-3\sum_{k=1}^{P-2}n^3_k-9\left(\fr 1 3\right)^3-9\left(\fr 1 3\right)^3-6 \sum_{k=1}^{P-2}\left(-n_k-\frac 2 3\right)^3-3\sum_{k=1}^{P-2}\left(n_k+\frac 4 3\right)^3\crn
&=& 8n-8\sum_{k=1}^{P-2}n_k =0.\label{adt1}\eea

Hence, all the anomalies are cancelled, independent of $P$ and the $U(1)$'s charge parameters. Additionally, the anomalies (\ref{adt5}), (\ref{adt4}), (\ref{adt2}), and (\ref{adt1}) relevant to $U(1)_N$ vanish with the presence of the right-handed neutrinos. 

\section{\label{fermionmass} Fermion mass}

The Yukawa Lagrangian is given by 
\bea \mathcal{L} &\supset& \fr 1 2 f^\nu_{ab}\bar{\nu}^c_{aR}\phi\nu_{bR} + h^\nu_{ab}\bar{\psi}_{aL}\varphi_1 \nu_{bR}+h^e_{ab}\bar{\psi}_{aL}\varphi_2 e_{bR}+\sum_{k=1}^{P-2}x^k_{ab}\bar{\psi}_{aL}\varphi_{k+2} E_{k bR}\crn
&& + h^u_{3b}\bar{Q}_{3L}\varphi_1 u_{bR}+h^d_{3b}\bar{Q}_{3L}\varphi_2 d_{bR}+\sum_{k=1}^{P-2}y^k_{33}\bar{Q}_{3L}\varphi_{k+2} J_{k3R}
\crn
&& + h^d_{\al b}\bar{Q}_{\al L}\varphi^*_1 d_{bR}+h^u_{\al b}\bar{Q}_{\al L}\varphi^*_2 u_{bR}+\sum_{k=1}^{P-2}y^k_{\al \beta}\bar{Q}_{\al L}\varphi^*_{k+2} J_{k\beta R}\crn
&&+H.c.,\eea where $\varphi_{1,2,3,...,P}$ are $P$ scalar $P$-plets given in two equations (\ref{scalarm1}) and (\ref{scalarm2}), respectively. 

Substituting the vevs $\langle \phi\rangle =\La/\sqrt{2}$ and $\langle \varphi_{i}\rangle_j =v_j \delta_{ij}/\sqrt{2}$ for $i,j=1,2,3,\cdots,P$ as mentioned in the body text, we obtain
\bea && [m_e]_{ab}=-h^e_{ab}\fr{v_2}{\sqrt{2}},\\
&& [m_u]_{\al b} = h^u_{\al b}\fr{v_2}{\sqrt{2}},\hs [m_u]_{3b}=-h^u_{3b}\fr{v_1}{\sqrt{2}},\\ 
&& [m_d]_{\al b} = -h^d_{\al b}\fr{v_1}{\sqrt{2}},\hs [m_d]_{3b}=-h^d_{3b}\fr{v_2}{\sqrt{2}},\\
&& [m_{E_k}]_{ab}=-x^k_{ab}\fr{v_{k+2}}{\sqrt{2}},\hs [m_{J_k}]_{ab}=-y^k_{ab}\fr{v_{k+2}}{\sqrt{2}}. \eea 
Note that $v_1,v_2$ are proportional to the weak scale, since $v^2_1+v^2_2=(246\ \mathrm{GeV})^2$. The ordinary charged leptons and ordinary quarks get appropriate masses, similar to the standard model. Moreover, $E_k$ and $J_k$ get large masses at $v_{3,4,...,P}$ scales in TeV.  

The neutrino mass matrix takes the form,
\bea \mathcal{L}\supset -\fr 1 2 \left(\bar{\nu}_{aL}\ \bar{\nu}^c_{aR}\right)\left(\begin{array}{cc} 
0 & m_{ab} \\
m_{ba} & M_{ab}\end{array}\right)
\left(\begin{array}{c}
\nu^c_{bL}\\
\nu_{bR}\end{array}\right)+H.c.,\eea where $m_{ab}=-h^\nu_{ab}v_1/\sqrt{2}$ and $M_{ab}=-f^\nu_{ab}\La/\sqrt{2}$. Because of $\La\gg v_1$, the seesaw mechanism produces the observed neutrino ($\sim \nu_{aL}$) masses \be m_\nu = -m M^{-1}m^T=h^\nu (f^{\nu})^{-1} (h^\nu)^T\fr{v^2_1}{\sqrt{2}\La}.\ee Whereas, the sterile neutrinos ($\sim \nu_{aR}$) gain heavy masses at $\La$ scale.

\section{\label{gvaf} Vector and axial-vector couplings}

This appendix is devoted to the neutral gauge boson couplings with fermions. 

\begin{table}[!h]
\centering\tabcolsep=0pt\tabulinesep=2pt
\begin{tabu}{XX[2.7,c]X[2.7,c]X[2,c]X[3.5,c]X[1.5,c]}
\hline\hline
$f$ & $g^{Z_1}_V(f)$ & $g^{Z_1}_A(f)$ & $f$ & $g^{Z_1}_V(f)$ & $g^{Z_1}_A(f)$ \\
\hline
$e_a$ & $-\fr 1 2 + 2 s^2_W$ & $-\fr 1 2 $ & $E_a$ & $-2s^2_Wq$ &  $0$ \\
$F_a$ & $-2s^2_Wp$ &  $0$ & $u_a$ & $\fr 1 2 -\fr 4 3 s^2_W$ & $\fr 1 2$ \\
$d_a$ & $-\fr 1 2 + \fr 2 3 s^2_W $ & $-\fr 1 2$ & $J_{\alpha}$ & $2s^2_W (q+\fr 1 3)  $ & $0$ \\
$J_{3}$ & $-2s^2_W (q+\fr 2 3)  $ & $0$ & $K_{\alpha}$ & $2s^2_W (p+\fr 1 3)  $ & $0$ \\
$K_{3}$ & $-2s^2_W (p+\fr 2 3)  $ & $0$ & No data & No data & No data\\
\hline\hline
\end{tabu}
\caption{\label{Z1}The couplings of $Z_1$ with fermions.}  
\end{table}

\begin{table}[!h]
\centering\tabcolsep=0pt\tabulinesep=2pt
\begin{tabu}{XX[40,c]X[27,c]}
\hline\hline
$f$ & $g^{Z_2}_V(f)$ & $g^{Z_2}_A(f)$\\ 
\hline
$e_a$ & $\frac{c_\varphi (1+3\sqrt3\beta t_W^2)}{2\sqrt3\sqrt{1-\beta^2t_W^2}}-\frac{s_\varphi[1+\ga(\ga+\sqrt2\beta+3\sqrt6)t_X^2]}{2\sqrt6\sqrt{1+\ga^2t_X^2}}$ & $\frac{c_\varphi (1-\sqrt3\beta t_W^2)}{2\sqrt3\sqrt{1-\beta^2t_W^2}}-\frac{s_\varphi[1+\ga(\ga+\sqrt2\beta-\sqrt6)t_X^2]}{2\sqrt6\sqrt{1+\ga^2t_X^2}}$ \\
$E_a$ &  $-\frac{c_\varphi (1+2\sqrt3 q\beta t_W^2)}{\sqrt3\sqrt{1-\beta^2t_W^2}}-\frac{s_\varphi[1+\ga(\ga-2\sqrt2\beta-4\sqrt6 q)t_X^2]}{2\sqrt6\sqrt{1+\ga^2t_X^2}}$ & $-\frac{c_\varphi}{\sqrt3\sqrt{1-\beta^2t_W^2}}-\frac{s_\varphi[1+\ga(\ga-2\sqrt2\beta)t_X^2]}{2\sqrt6\sqrt{1+\ga^2t_X^2}}$ \\
$F_a$ &  $-\frac{c_\varphi 2p\beta t_W^2}{\sqrt{1-\beta^2t_W^2}}+\frac{s_\varphi[\sqrt6+\ga(\sqrt6\ga+8p)t_X^2]}{4\sqrt{1+\ga^2t_X^2}}$ & $\frac{s_\varphi\sqrt3\sqrt{1+\ga^2t_X^2}}{2\sqrt2}$ \\
$u_\al$ & $-\frac{c_\varphi (\sqrt3+5\beta t_W^2)}{6\sqrt{1-\beta^2t_W^2}}+\frac{s_\varphi[\sqrt6+\ga(\sqrt6\ga+2\sqrt3\beta+10)t_X^2]}{12\sqrt{1+\ga^2t_X^2}}$ & $-\frac{c_\varphi (1-\sqrt3\beta t_W^2)}{2\sqrt3\sqrt{1-\beta^2t_W^2}}+\frac{s_\varphi[1+\ga(\ga+\sqrt2\beta-\sqrt6)t_X^2]}{2\sqrt6\sqrt{1+\ga^2t_X^2}}$ \\
$u_3$ & $\frac{c_\varphi (\sqrt3-5\beta t_W^2)}{6\sqrt{1-\beta^2t_W^2}}-\frac{s_\varphi[\sqrt6+\ga(\sqrt6\ga+2\sqrt3\beta-10)t_X^2]}{12\sqrt{1+\ga^2t_X^2}}$ & $\frac{c_\varphi (1+\sqrt3\beta t_W^2)}{2\sqrt3\sqrt{1-\beta^2t_W^2}}-\frac{s_\varphi[1+\ga(\ga+\sqrt2\beta+\sqrt6)t_X^2]}{2\sqrt6\sqrt{1+\ga^2t_X^2}}$ \\
$d_\al$ & $-\frac{c_\varphi (\sqrt3-\beta t_W^2)}{6\sqrt{1-\beta^2t_W^2}}+\frac{s_\varphi[\sqrt6+\ga(\sqrt6\ga+2\sqrt3\beta-2)t_X^2]}{12\sqrt{1+\ga^2t_X^2}}$ & $-\frac{c_\varphi (1+\sqrt3\beta t_W^2)}{2\sqrt3\sqrt{1-\beta^2t_W^2}}+\frac{s_\varphi[1+\ga(\ga+\sqrt2\beta+\sqrt6)t_X^2]}{2\sqrt6\sqrt{1+\ga^2t_X^2}}$ \\
$d_3$ & $\frac{c_\varphi (\sqrt3+\beta t_W^2)}{6\sqrt{1-\beta^2t_W^2}}-\frac{s_\varphi[\sqrt6+\ga(\sqrt6\ga+2\sqrt3\beta+2)t_X^2]}{12\sqrt{1+\ga^2t_X^2}}$ & $\frac{c_\varphi (1-\sqrt3\beta t_W^2)}{2\sqrt3\sqrt{1-\beta^2t_W^2}}-\frac{s_\varphi[1+\ga(\ga+\sqrt2\beta-\sqrt6)t_X^2]}{2\sqrt6\sqrt{1+\ga^2t_X^2}}$ \\
$J_{\alpha}$ & $\frac{c_\varphi [\sqrt3-\beta(1+3\sqrt3\beta) t_W^2]}{3\sqrt{1-\beta^2t_W^2}}+\frac{s_\varphi[\sqrt6+\ga(\sqrt6\ga+8\sqrt3\beta+4)t_X^2]}{12\sqrt{1+\ga^2t_X^2}}$ & $\frac{c_\varphi}{\sqrt3\sqrt{1-\beta^2t_W^2}}+\frac{s_\varphi[1+\ga(\ga-2\sqrt2\beta)t_X^2]}{2\sqrt6\sqrt{1+\ga^2t_X^2}}$ \\
$J_{3}$ & $-\frac{c_\varphi [\sqrt3+\beta(1-3\sqrt3\beta) t_W^2]}{3\sqrt{1-\beta^2t_W^2}}-\frac{s_\varphi[\sqrt6+\ga(\sqrt6\ga+8\sqrt3\beta-4)t_X^2]}{12\sqrt{1+\ga^2t_X^2}}$ & $-\frac{c_\varphi}{\sqrt3\sqrt{1-\beta^2t_W^2}}-\frac{s_\varphi[1+\ga(\ga-2\sqrt2\beta)t_X^2]}{2\sqrt6\sqrt{1+\ga^2t_X^2}}$ \\
$K_{\alpha}$ & $\frac{c_\varphi 2(1+3p)\beta t_W^2}{3\sqrt{1-\beta^2t_W^2}}-\frac{s_\varphi[3\sqrt6+\ga(3\sqrt6\ga+24p+8)t_X^2]}{12\sqrt{1+\ga^2t_X^2}}$ & $-\frac{s_\varphi\sqrt3\sqrt{1+\ga^2t_X^2}}{2\sqrt2}$ \\
$K_{3}$ & $-\frac{c_\varphi 2(2+3p)\beta t_W^2}{3\sqrt{1-\beta^2t_W^2}}+\frac{s_\varphi[3\sqrt6+\ga(3\sqrt6\ga+24p+16)t_X^2]}{12\sqrt{1+\ga^2t_X^2}}$ & $\frac{s_\varphi\sqrt3\sqrt{1+\ga^2t_X^2}}{2\sqrt2}$ \\
\hline\hline
\end{tabu}
\caption{\label{Z2}The couplings of $Z_2$ with fermions.}  
\end{table}

\section{\label{sgint} The gauge couplings of scalars}

This appendix is devoted to all the gauge boson and scalar couplings. 

\begin{table}[!h]
\centering\tabcolsep=0pt\tabulinesep=2pt
\begin{tabu}{X[3,l]X[1.2,c]X[4,c]X[1.1,c]}
\hline\hline
Vertex & Coupling & Vertex & Coupling\\ \hline
$W^+_\mu \mathcal{H}_1^-\overlr{\partial}^\mu\mathcal{A}$ & $\frac{1}{2}g$ & $W^+_\mu \mathcal{H}_1^-\overlr{\partial}^\mu H_2$ & $\frac{i}{2}g$\\ 
$W^q_{13\mu}\mathcal{H}_2^{-q}\overlr{\pa}^\mu H_1$ & $-\frac{i}{2}gc_{\al_2}$ &  $W^q_{13\mu}\mathcal{H}_2^{-q}\overlr{\pa}^\mu H_2$ & $-\frac{i}{2}gs_{\al_2}$ \\
$W^q_{13\mu}\mathcal{H}_2^{-q}\overlr{\pa}^\mu \mathcal{A}$ & $\frac{1}{2}gs_{\al_2}$ &  $W^q_{13\mu}\mathcal{H}_4^{-q-1}\overlr{\pa}^\mu \mathcal{H}_1^+$ & $-\frac{i}{\sqrt2}gc_{\al_2}$ \\
$W^p_{14\mu}\mathcal{H}_3^{-p}\overlr{\pa}^\mu H_1$ & $-\frac{i}{2}gc_{\al_2}$ &  $W^p_{14\mu}\mathcal{H}_3^{-p}\overlr{\pa}^\mu H_2$ & $-\frac{i}{2}gs_{\al_2}$ \\
$W^p_{14\mu}\mathcal{H}_3^{-p}\overlr{\pa}^\mu \mathcal{A}$ & $\frac{1}{2}gs_{\al_2}$ &  $W^p_{14\mu}\mathcal{H}_5^{-p-1}\overlr{\pa}^\mu \mathcal{H}_1^+$ & $-\frac{i}{\sqrt2}gc_{\al_2}$ \\ 
$W^{q+1}_{23\mu}\mathcal{H}_4^{-q-1}\overlr{\pa}^\mu H_1$ & $-\frac{i}{2}gs_{\al_2}$ &  $W^{q+1}_{23\mu}\mathcal{H}_4^{-q-1}\overlr{\pa}^\mu H_2$ & $\frac{i}{2}gc_{\al_2}$ \\
$W^{q+1}_{23\mu}\mathcal{H}_4^{-q-1}\overlr{\pa}^\mu \mathcal{A}$ & $\frac{1}{2}gc_{\al_2}$ & $W^{q+1}_{23\mu}\mathcal{H}_2^{-q}\overlr{\pa}^\mu \mathcal{H}_1^-$ & $-\frac{i}{\sqrt2}gs_{\al_2}$ \\ 
$W^{p+1}_{24\mu}\mathcal{H}_5^{-p-1}\overlr{\pa}^\mu H_1$ & $-\frac{i}{2}gs_{\al_2}$ &  $W^{p+1}_{24\mu}\mathcal{H}_5^{-p-1}\overlr{\pa}^\mu H_2$ & $\frac{i}{2}gc_{\al_2}$ \\
$W^{p+1}_{24\mu}\mathcal{H}_5^{-p-1}\overlr{\pa}^\mu \mathcal{A}$ & $\frac{1}{2}gc_{\al_2}$ & $W^{p+1}_{24\mu}\mathcal{H}_3^{-p}\overlr{\pa}^\mu \mathcal{H}_1^-$ & $-\frac{i}{\sqrt2}gs_{\al_2}$ \\ 
$W^{q-p}_{34\mu}\mathcal{H}_6^{p-q}\overlr{\pa}^\mu H_3$ & $\frac{i}{2}gc_{(\al_1-\al_3)}$ &  $W^{q-p}_{34\mu}\mathcal{H}_6^{p-q}\overlr{\pa}^\mu H_4$ & $\frac{i}{2}gs_{(\al_1-\al_3)}$ \\
$W^{q-p}_{34\mu}\mathcal{H}_3^{p}\overlr{\pa}^\mu \mathcal{H}_2^{-q}$ & $\frac{i}{\sqrt2}g$ &  $W^{q-p}_{34\mu}\mathcal{H}_5^{p+1}\overlr{\pa}^\mu \mathcal{H}_4^{-q-1}$ & $\frac{i}{\sqrt2}g$ \\
\hline\hline
\end{tabu}
\caption{\label{1CG2S}The interactions of a charged gauge boson with two scalars.}  
\end{table}

\begin{table}[!h]
\centering\tabcolsep=0pt\tabulinesep=2pt
\begin{tabu}{X[3.3,l]X[2.8,c]X[4.0,c]X[3,c]}
\hline\hline
Vertex & Coupling & Vertex & Coupling\\ \hline
$A_\mu \mathcal{H}_1^-\overlr{\partial}^\mu \mathcal{H}_1^+$ & $-igs_W$ & $A_\mu \mathcal{H}_2^{-q}\overlr{\partial}^\mu \mathcal{H}_2^q$ & $-igs_W q$ \\
$A_\mu \mathcal{H}_3^{-p}\overlr{\partial}^\mu \mathcal{H}_3^p$ & $-igs_W p$ & $A_\mu \mathcal{H}_4^{-q-1}\overlr{\partial}^\mu \mathcal{H}_4^{q+1}$ & $-igs_W (q+1)$ \\
$A_\mu \mathcal{H}_5^{-p-1}\overlr{\partial}^\mu \mathcal{H}_5^{p+1}$ & $-igs_W (p+1)$ & $A_\mu \mathcal{H}_6^{p-q}\overlr{\partial}^\mu \mathcal{H}_6^{q-p}$ & $igs_W (p-q)$ \\
$Z_{1\mu}\mathcal{H}_1^-\overlr{\partial}^\mu \mathcal{H}_1^+$ & $-\frac{i}{2c_W}gc_{2W}$ & $Z_{1\mu} \mathcal{H}_2^{-q}\overlr{\partial}^\mu \mathcal{H}_2^q$ & $igs_Wt_W q$ \\
$Z_{1\mu}\mathcal{H}_3^{-p}\overlr{\partial}^\mu \mathcal{H}_3^p$ & $igs_Wt_Wp$ & $Z_{1\mu} \mathcal{H}_4^{-q-1}\overlr{\partial}^\mu \mathcal{H}_4^{q+1}$ & $igs_Wt_W (q+1)$ \\
$Z_{1\mu}\mathcal{H}_5^{-p-1}\overlr{\partial}^\mu \mathcal{H}_5^{p+1}$ & $igs_Wt_W (p+1)$ & $Z_{1\mu} \mathcal{H}_6^{p-q}\overlr{\partial}^\mu \mathcal{H}_6^{q-p}$ & $igs_Wt_W (q-p)$ \\
$Z_{1\mu}H_2\overlr{\partial}^\mu \mathcal{A}$ & $\frac{1}{2c_W}g$ & No data & No data\\
\hline
\end{tabu}
\begin{tabu}{X[1,l]X[3.1,c]}
Vertex & Coupling\\\hline
$Z_{2\mu} \mathcal{H}_1^{-}\overlr{\partial}^\mu \mathcal{H}_1^+$ & $\frac{i}{2\sqrt3(u^2+v^2)}g [c_\varphi (v^2\beta_1-u^2\beta_2)-s_\varphi(v^2\ga_1-u^2\ga_2)]$\\
$Z_{2\mu} \mathcal{H}_2^{-q}\overlr{\partial}^\mu \mathcal{H}_2^q$ & $\frac{i}{2\sqrt3}g\{c_\varphi [(\beta_2+\beta_1-q(\beta_2-\beta_1)]+s_\varphi\ga_1\}$\\
$Z_{2\mu} \mathcal{H}_3^{-p}\overlr{\partial}^\mu \mathcal{H}_3^p$ & $\frac{i}{2\sqrt3}g\{c_\varphi p (\beta_1-\beta_2)+s_\varphi[(q+p+2)(\ga_2-\ga_1)-3\ga_2]\}$\\
$Z_{2\mu} \mathcal{H}_4^{-q-1}\overlr{\partial}^\mu \mathcal{H}_4^{q+1}$ & $\frac{i}{2\sqrt3}g\{c_\varphi[\beta_1+\beta_2-(1+q)(\beta_2-\beta_1)+s_\varphi\ga_2]\}$\\
$Z_{2\mu} \mathcal{H}_5^{-p-1}\overlr{\partial}^\mu \mathcal{H}_5^{p+1}$ & $\frac{i}{2\sqrt3}g\{c_\varphi(1+p)(\beta_1-\beta_2)+s_\varphi[(q+p)(\ga_2-\ga_1)-3\ga_1]\}$\\
$Z_{2\mu} \mathcal{H}_6^{p-q}\overlr{\partial}^\mu \mathcal{H}_6^{q-p}$ & $\frac{ig}{2\sqrt3}\{c_\varphi[s_{\al_3}^2(\beta_2+\beta_1)+(p-q)(\beta_2-\beta_1)] +s_\varphi[\ga_1-p(\ga_2-\ga_1)+c_{\al_3}^2(\ga_2+\ga_1)]\}$\\
$Z_{2\mu} H_1\overlr{\partial}^\mu \mathcal{A}$ & $\frac{1}{2\sqrt3 (u^2+v^2)} guv [c_\varphi (\beta_2+\beta_1)-s_\varphi (\ga_2+\ga_1)]$\\
$Z_{2\mu} H_2\overlr{\partial}^\mu \mathcal{A}$ & $\frac{1}{2\sqrt3(u^2+v^2)}g [c_\varphi (v^2\beta_1-u^2\beta_2)-s_\varphi(v^2\ga_1-u^2\ga_2)]$\\
$Z_{3\mu}\dots\dots$ & $Z_{2\mu}\dots\dots (c_\varphi\to s_\varphi,s_\varphi\to -c_\varphi)$\\
\hline\hline
\end{tabu}
\caption{\label{1NG2S}The interactions of a neutral gauge boson with two scalars.}  
\end{table}

\begin{table}[!h]
\centering\tabcolsep=0pt\tabulinesep=2pt
\begin{tabu}{X[1,l]X[1.8,c]X[1.8,c]X[1.2,c]}
\hline\hline
Vertex & Coupling & Vertex & Coupling\\ \hline
$H_1 W^+ W^-$ & $\frac{1}{2}g^2\sqrt{u^2+v^2}$ & $H_1 W_{13}^q W_{13}^{-q}$ & $\frac{1}{2}g^2uc_{\al_2}$\\ 
$H_1 W_{14}^p W_{14}^{-p}$ & $\frac{1}{2}g^2uc_{\al_2}$ & $H_1 W_{23}^{q+1} W_{23}^{-q-1}$ & $\frac{1}{2}g^2vs_{\al_2}$\\ 
$H_1 W_{24}^{p+1} W_{24}^{-p-1}$ & $\frac{1}{2}g^2vs_{\al_2}$ & $H_2 W_{13}^q W_{13}^{-q}$ & $\frac{1}{2}g^2us_{\al_2}$ \\
$H_2 W_{14}^p W_{14}^{-p}$ & $\frac{1}{2}g^2us_{\al_2}$ & $H_2 W_{23}^{q+1} W_{23}^{-q-1}$ & $-\frac{1}{2}g^2vc_{\al_2}$ \\
$H_2 W_{24}^{p+1} W_{24}^{-p-1}$ & $-\frac{1}{2}g^2vc_{\al_2}$ & $H_3 W_{13}^q W_{13}^{-q}$ & $\frac{1}{2}g^2wc_{\al_1}$ \\
$H_3 W_{14}^p W_{14}^{-p}$ & $-\frac{1}{2}g^2Vs_{\al_1}$ & $H_3 W_{23}^{q+1} W_{23}^{-q-1}$ & $\frac{1}{2}g^2wc_{\al_1}$ \\
$H_3 W_{24}^{p+1} W_{24}^{-p-1}$ & $-\frac{1}{2}g^2Vs_{\al_1}$ & $H_3 W_{34}^{q-p} W_{34}^{p-q}$ & $\frac{1}{2}g^2(wc_{\al_1}-Vs_{\al_1})$ \\
$H_4 W_{13}^q W_{13}^{-q}$ & $\frac{1}{2}g^2ws_{\al_1}$ & $H_4 W_{14}^p W_{14}^{-p}$ & $\frac{1}{2}g^2Vc_{\al_1}$ \\
$H_4 W_{23}^{q+1} W_{23}^{-q-1}$ & $\frac{1}{2}g^2ws_{\al_1}$ & $H_4 W_{24}^{p+1} W_{24}^{-p-1}$ & $\frac{1}{2}g^2Vc_{\al_1}$ \\
$H_4 W_{34}^{q-p} W_{34}^{p-q}$ & $\frac{1}{2}g^2(ws_{\al_1}+Vc_{\al_1})$ & $\mathcal{H}_1^+ W_{13}^{q} W_{23}^{-q-1}$ & $\frac{1}{\sqrt2}g^2us_{\al_2}$ \\
$\mathcal{H}_1^+ W_{14}^{p} W_{24}^{-p-1}$ & $\frac{1}{\sqrt2}g^2us_{\al_2}$ & $\mathcal{H}_2^q W^+ W_{23}^{-q-1}$ & $\frac{1}{2\sqrt2}g^2u$ \\
$\mathcal{H}_2^q W_{14}^{-p} W_{34}^{p-q}$ & $\frac{1}{2\sqrt2}g^2u$ & $\mathcal{H}_3^p W^+ W_{24}^{-p-1}$ & $\frac{1}{2\sqrt2}g^2u$ \\
$\mathcal{H}_3^p W_{13}^{-q} W_{34}^{q-p}$ & $\frac{1}{2\sqrt2}g^2u$ & $\mathcal{H}_4^{q+1} W^- W_{13}^{-q}$ & $\frac{1}{2\sqrt2}g^2v$ \\
$\mathcal{H}_4^{q+1} W_{24}^{-p-1} W_{34}^{p-q}$ & $\frac{1}{2\sqrt2}g^2v$ & $\mathcal{H}_5^{p+1} W^- W_{14}^{-p}$ & $\frac{1}{2\sqrt2}g^2v$ \\
$\mathcal{H}_5^{p+1} W_{23}^{-q-1} W_{34}^{q-p}$ & $\frac{1}{2\sqrt2}g^2v$ & $\mathcal{H}_6^{q-p} W^{-q}_{13} W_{14}^{p}$ & $\frac{1}{\sqrt2}g^2Vs_{\al_3}$ \\
$\mathcal{H}_6^{q-p} W^{-q-1}_{23} W_{24}^{p+1}$ & $\frac{1}{\sqrt2}g^2Vs_{\al_3}$ & No data & No data \\
\hline\hline
\end{tabu}
\caption{\label{1S2CG}The interactions of a scalar with two charged gauge bosons.}  
\end{table}

\begin{table}[!h]
\centering\tabcolsep=0pt\tabulinesep=2pt
\begin{tabu}{X[1,l]X[4,c]}
\hline\hline
Vertex & Coupling \\ \hline
$\mathcal{H}_1^+ Z_2 W^-$ & $\frac{1}{2\sqrt3\sqrt{u^2+v^2}} g^2uv[c_\varphi (\beta_2+\beta_1)-s_\varphi (\ga_2+\ga_1)]$\\
$\mathcal{H}_2^q A W^{-q}_{13}$ & $\frac{1}{2}g^2qus_W$\\
$\mathcal{H}_2^q Z_1 W^{-q}_{13}$ & $\frac{1}{4c_W}g^2u(1-q+qc_{2W})$\\
$\mathcal{H}_2^q Z_2 W^{-q}_{13}$ & $\frac{1}{8\sqrt3} g^2u\{c_\varphi [(2q-1)(\beta_2-\beta_1)-(\beta_2+\beta_1)]-4s_\varphi \ga_1\}$\\
$\mathcal{H}_3^p A W^{-p}_{14}$ & $\frac{1}{2}g^2pus_W$\\
$\mathcal{H}_3^p Z_1 W^{-p}_{14}$ & $\frac{1}{4c_W}g^2u(1-p+pc_{2W})$\\
$\mathcal{H}_3^p Z_2 W^{-p}_{14}$ & $\frac{1}{8\sqrt3} g^2u\{c_\varphi [\beta_1+\beta_2+(2p-1)(\beta_2-\beta_1)]-2s_\varphi [(1+p+q)(\ga_2-\ga_1)-2\ga_2]\}$\\
$\mathcal{H}_4^{q+1} A W^{-q-1}_{23}$ & $\frac{1}{2}g^2(q+1)vs_W$\\
$\mathcal{H}_4^{q+1} Z_1 W^{-q-1}_{23}$ & $-\frac{1}{4}g^2v[c_W+(3+2q)s_Wt_W]$\\
$\mathcal{H}_4^{q+1} Z_2 W^{-q-1}_{23}$ & $\frac{1}{8\sqrt3}g^2 v\{c_\varphi [(3+2q)(\beta_2-\beta_1)]-(\beta_2+\beta_1)-4s_\varphi\ga_2\}$\\
$\mathcal{H}_5^{p+1} A W^{-p-1}_{24}$ & $\frac{1}{2}g^2(p+1)vs_W$\\
$\mathcal{H}_5^{p+1} Z_1 W^{-p-1}_{24}$ & $-\frac{1}{4}g^2v[c_W+(3+2p)s_Wt_W]$\\
$\mathcal{H}_5^{p+1} Z_2 W^{-p-1}_{24}$ & $\frac{1}{8\sqrt3} g^2v\{c_\varphi [\beta_1+\beta_2+(3+2p)(\beta_2-\beta_1)]-2s_\varphi [(1+p+q)(\ga_2-\ga_1)-2\ga_1]\}$\\
$\mathcal{H}_6^{q-p} Z_2 W^{p-q}_{34}$ & $-\frac{1}{2\sqrt3\sqrt{w^2+V^2}} g^2wV[c_\varphi (\beta_2+\beta_1)-s_\varphi (\ga_2+\ga_1)]$\\
$\dots Z_{3}\dots$ & $\dots Z_{2}\dots (c_\varphi\to s_\varphi,s_\varphi\to -c_\varphi)$\\
\hline\hline
\end{tabu}
\caption{\label{1S1CG1NG}The interactions of a scalar with a neutral and a charged gauge boson.}  
\end{table}

\begin{table}[!h]
\centering\tabcolsep=0pt\tabulinesep=2pt
\begin{tabu}{X[1,l]X[5,c]}
\hline\hline
Vertex & Coupling \\ \hline
$H_1 Z_1 Z_1$ & $\frac{1}{4c_W^2}g^2\sqrt{u^2+v^2}$\\
$H_1 Z_1 Z_2$ & $\frac{1}{2\sqrt3 c_W\sqrt{u^2+v^2}}g^2[u^2(c_\varphi\beta_1-s_\varphi\ga_1)-v^2(c_\varphi\beta_2-s_\varphi\ga_2)]$\\
$H_1 Z_2 Z_2$ & $\frac{1}{12\sqrt{u^2+v^2}}g^2[u^2(c_\varphi\beta_1-s_\varphi\ga_1)^2+v^2(c_\varphi\beta_2-s_\varphi\ga_2)^2]$\\
$H_1 Z_2 Z_3$ & $\frac{1}{6\sqrt{u^2+v^2}}g^2[u^2(c_\varphi\beta_1-s_\varphi\ga_1)(c_\varphi\ga_1+s_\varphi\beta_1)+v^2(c_\varphi\beta_2-s_\varphi\ga_2)(c_\varphi\ga_2+s_\varphi\beta_2)]$\\
$ H_2 Z_1 Z_2$ & $\frac{1}{2\sqrt3 c_W\sqrt{u^2+v^2}}g^2uv[c_\varphi (\beta_1+\beta_2)-s_\varphi (\ga_1+\ga_2)] $\\
$H_2 Z_2 Z_2$ & $\frac{1}{12\sqrt{u^2+v^2}}g^2uv[(c_\varphi\beta_1-s_\varphi\ga_1)^2-(c_\varphi\beta_2-s_\varphi\ga_2)^2]$\\
$H_2 Z_2 Z_3$ & $\frac{1}{6\sqrt{u^2+v^2}}g^2uv[(c_\varphi\beta_1-s_\varphi\ga_1)(c_\varphi\ga_1+s_\varphi\beta_1)-(c_\varphi\beta_2-s_\varphi\ga_2)(c_\varphi\ga_2+s_\varphi\beta_2)]$\\
$H_3 Z_2Z_2$ & $\frac{1}{12}g^2\{ w c_{\al_1}[s_\varphi (\ga_1+q\ga_1-q\ga_2)+c_\varphi (\beta_1+\beta_2)]^2-Vs_{\al_1}s_{\varphi}^2[(q-1)(\ga_1-\ga_2)+3\ga_1]^2\}$\\
$H_3 Z_2 Z_3$ & $\frac{1}{6}g^2\{Vs_{\al_1}s_\varphi c_\varphi [(q-1)(\ga_1-\ga_2)+3\ga_1]^2 - wc_{\al_1}[c_\varphi (\beta_1+\beta_2)+s_\varphi (\ga_1+q\ga_1-q\ga_2)][c_\varphi (\ga_1+q\ga_1-q\ga_2)-s_\varphi (\beta_1+\beta_2)]\}$\\
$H_4 Z_2Z_2$ & $\frac{1}{12}g^2\{ V c_{\al_1}s_\varphi^2 [(q-1)(\ga_1-\ga_2)+3\ga_1]^2+w s_{\al_1}[s_{\varphi}(\ga_1+q\ga_1-q\ga_2)+c_\varphi (\beta_1+\beta_2)]^2\}$\\
$H_4 Z_2 Z_3$ & $\frac{1}{6}g^2\{-Vc_{\al_1}s_\varphi c_\varphi [(q-1)(\ga_1-\ga_2)+3\ga_1]^2 - ws_{\al_1}[c_\varphi (\beta_1+\beta_2)+s_\varphi (\ga_1+q\ga_1-q\ga_2)][c_\varphi (\ga_1+q\ga_1-q\ga_2)-s_\varphi (\beta_1+\beta_2)]\}$\\
$\dots\dots Z_{3}$ & $\dots\dots Z_{2} (c_\varphi\to s_\varphi,s_\varphi\to -c_\varphi)$\\
$\dots Z_{3}\dots$ & $\dots Z_{2}\dots (c_\varphi\to s_\varphi,s_\varphi\to -c_\varphi)$\\
\hline\hline
\end{tabu}
\caption{\label{1S2NG}The interactions of a scalar with two neutral gauge bosons.}  
\end{table}

\begin{table}[!h]
\centering\tabcolsep=0pt\tabulinesep=2pt
\begin{tabu}{X[2,l]X[1.2,c]X[3,c]X[1.1,c]}
\hline\hline
Vertex & Coupling & Vertex & Coupling\\ \hline
$W^+ W^- \mathcal{A} \mathcal{A}$ & $\frac{1}{4}g^2$ & $W^+ W^- \mathcal{H}_1^+ \mathcal{H}_1^-$ & $\frac{1}{2}g^2$\\ 
$W^+ W^- H_1 H_1$ & $\frac{1}{4}g^2$ & $W^+ W^- H_2 H_2$ & $\frac{1}{4}g^2$\\ 
$W^+ H_1 \mathcal{H}_2^q W_{23}^{-q-1}$ & $\frac{1}{2\sqrt2}g^2c_{\al_2}$ & $W^+ H_1 \mathcal{H}_3^p W_{24}^{-p-1}$ & $\frac{1}{2\sqrt2}g^2c_{\al_2}$\\
$W^+ H_1 \mathcal{H}_4^{-q-1} W_{13}^q$ & $\frac{1}{2\sqrt2}g^2s_{\al_2}$ & $W^+ H_1 \mathcal{H}_5^{-p-1} W_{14}^{p}$ & $\frac{1}{2\sqrt2}g^2s_{\al_2}$\\
$W^+ H_2 \mathcal{H}_2^q W_{23}^{-q-1}$ & $\frac{1}{2\sqrt2}g^2s_{\al_2}$ & $W^+ H_2 \mathcal{H}_3^p W_{24}^{-p-1}$ & $\frac{1}{2\sqrt2}g^2s_{\al_2}$\\
$W^+ H_2 \mathcal{H}_4^{-q-1} W_{13}^q$ & $-\frac{1}{2\sqrt2}g^2c_{\al_2}$ & $W^+ H_2 \mathcal{H}_5^{-p-1} W_{14}^p$ & $-\frac{1}{2\sqrt2}g^2c_{\al_2}$\\
$W^+ \mathcal{A} \mathcal{H}_2^q W_{23}^{-q-1}$ & $-\frac{i}{2\sqrt2}g^2s_{\al_2}$ & $W^+ \mathcal{A} \mathcal{H}_3^p W_{24}^{-p-1}$ & $-\frac{i}{2\sqrt2}g^2s_{\al_2}$\\
$W^+ \mathcal{A} \mathcal{H}_4^{-q-1} W_{13}^q$ & $\frac{i}{2\sqrt2}g^2c_{\al_2}$ & $W^+ \mathcal{A} \mathcal{H}_5^{-p-1} W_{14}^p$ & $\frac{i}{2\sqrt2}g^2c_{\al_2}$\\
$W^+ \mathcal{H}_1^- \mathcal{H}_2^{-q} W_{13}^q$ & $\frac{1}{2}g^2s_{\al_2}$ & $W^+ \mathcal{H}_1^- \mathcal{H}_3^{-p} W_{14}^p$ & $\frac{1}{2}g^2s_{\al_2}$\\
$W^+ \mathcal{H}_1^- \mathcal{H}_4^{q+1} W_{23}^{-q-1}$ & $\frac{1}{2}g^2c_{\al_2}$ & $W^+ \mathcal{H}_1^- \mathcal{H}_5^{p+1} W_{24}^{-p-1}$ & $\frac{1}{2}g^2c_{\al_2}$\\
$W^q_{13} W^{-q}_{13} H_1 H_1$ & $\frac{1}{4}g^2c_{\al_2}^2$ & $W^q_{13} W^{-q}_{13} H_1 H_2$ & $\frac{1}{4}g^2 s_{2\al_2}$\\ 
$W^q_{13} W^{-q}_{13} H_2 H_2$ & $\frac{1}{4}g^2 s_{\al_2}^2$ & $W^q_{13} W^{-q}_{13} \mathcal{A} \mathcal{A}$ & $\frac{1}{4}g^2 s_{\al_2}^2$\\ 
$W^q_{13} W^{-q}_{13} H_3 H_3$ & $\frac{1}{4}g^2 c_{\al_1}^2$ & $W^q_{13} W^{-q}_{13} H_3 H_4$ & $\frac{1}{4}g^2 s_{2\al_1}$\\ 
$W^q_{13} W^{-q}_{13} H_4 H_4$ & $\frac{1}{4}g^2 s_{\al_1}^2$ & $W^q_{13} W^{-q}_{13} \mathcal{H}_1^+ \mathcal{H}_1^-$ & $\frac{1}{2}g^2 c_{\al_2}^2$\\ 
$W^q_{13} W^{-q}_{13} \mathcal{H}_2^q \mathcal{H}_2^{-q}$ & $\frac{1}{2}g^2$ & $W^q_{13} W^{-q}_{13} \mathcal{H}_4^{q+1} \mathcal{H}_4^{-q-1}$ & $\frac{1}{2}g^2$\\ 
$W^q_{13} W^{-q}_{13} \mathcal{H}_6^{q-p} \mathcal{H}_6^{p-q}$ & $\frac{1}{2}g^2 s_{\al_3}^2$ & $W^q_{13} H_1 \mathcal{H}_1^+ W_{23}^{-q-1}$ & $\frac{1}{2\sqrt2}g^2 s_{2\al_2}$ \\
$W^q_{13} H_1 \mathcal{H}_3^{-p} W_{34}^{p-q}$ & $\frac{1}{2\sqrt2}g^2 c_{\al_2}$ & $W^q_{13} H_1 \mathcal{H}_4^{-q-1} W^+$ & $\frac{1}{2\sqrt2}g^2 s_{\al_2}$ \\
$W^q_{13} H_2 \mathcal{H}_3^{-p} W_{34}^{p-q}$ & $\frac{1}{2\sqrt2}g^2 s_{\al_2}$ & $W^q_{13} H_2 \mathcal{H}_4^{-q-1} W^+$ & $-\frac{1}{2\sqrt2}g^2 c_{\al_2}$ \\
$W^q_{13} H_3 \mathcal{H}_6^{p-q} W_{14}^{-p}$ & $\frac{1}{2\sqrt2}g^2 c_{(\al_1+\al_3)}$ & $W^q_{13} H_4 \mathcal{H}_6^{p-q} W_{14}^{-p}$ & $\frac{1}{2\sqrt2}g^2 s_{(\al_1+\al_3)}$ \\
$W^q_{13} \mathcal{A} \mathcal{H}_1^+ W_{23}^{-q-1}$ & $-\frac{i}{2\sqrt2}g^2 c_{2\al_2}$ & $W^q_{13} \mathcal{A} \mathcal{H}_3^{-p} W_{34}^{p-q}$ & $\frac{i}{2\sqrt2}g^2 s_{\al_2}$ \\
$W^q_{13} \mathcal{A} \mathcal{H}_4^{-q-1} W^+$ & $\frac{i}{2\sqrt2}g^2 c_{\al_2}$ & $W^q_{13} \mathcal{H}_1^+ H_2 W_{23}^{-q-1}$ & $-\frac{1}{2\sqrt2}g^2 c_{2\al_2}$ \\
$W^q_{13} \mathcal{H}_1^+ \mathcal{H}_5^{-p-1} W_{34}^{p-q}$ & $\frac{1}{2}g^2 c_{\al_2}$ & $W^q_{13} \mathcal{H}_1^- \mathcal{H}_2^{-q} W^+$ & $\frac{1}{2}g^2 s_{\al_2}$ \\
$W^q_{13} \mathcal{H}_2^{-q} \mathcal{H}_3^p W_{14}^{-p}$ & $\frac{1}{2}g^2$ & $W^q_{13} \mathcal{H}_4^{-q-1} \mathcal{H}_5^{p+1} W_{14}^{-p}$ & $\frac{1}{2}g^2$ \\
$W^p_{14} W^{-p}_{14} H_1 H_1$ & $\frac{1}{4}g^2c_{\al_2}^2$ & $W^p_{14} W^{-p}_{14} H_1 H_2$ & $\frac{1}{4}g^2 s_{2\al_2}$ \\
$W^p_{14} W^{-p}_{14} H_2 H_2$ & $\frac{1}{4}g^2 s_{\al_2}^2$ & $W^p_{14} W^{-p}_{14} \mathcal{A} \mathcal{A}$ & $\frac{1}{4}g^2 s_{\al_2}^2$ \\
$W^p_{14} W^{-p}_{14} H_3 H_3$ & $\frac{1}{4}g^2 s_{\al_1}^2$ & $W^p_{14} W^{-p}_{14} H_3 H_4$ & $-\frac{1}{4}g^2 s_{2\al_1}$ \\
$W^p_{14} W^{-p}_{14} H_4 H_4$ & $\frac{1}{4}g^2 c_{\al_1}^2$ & $W^p_{14} W^{-p}_{14} \mathcal{H}_1^+ \mathcal{H}_1^-$ & $\frac{1}{2}g^2 c_{\al_2}^2$ \\
$W^p_{14} W^{-p}_{14} \mathcal{H}_3^p \mathcal{H}_3^{-p}$ & $\frac{1}{2}g^2$ & $W^p_{14} W^{-p}_{14} \mathcal{H}_5^{p+1} \mathcal{H}_5^{-p-1}$ & $\frac{1}{2}g^2$ \\
$W^p_{14} W^{-p}_{14} \mathcal{H}_6^{q-p} \mathcal{H}_6^{p-q}$ & $\frac{1}{2}g^2 c_{\al_3}^2$ & $W^p_{14} H_1 \mathcal{H}_1^+ W_{24}^{-p-1}$ & $\frac{1}{2\sqrt2}g^2 s_{2\al_2}$ \\
$W^p_{14} H_1 \mathcal{H}_2^{-q} W_{34}^{q-p}$ & $\frac{1}{2\sqrt2}g^2 c_{\al_2}$ & $W^p_{14} H_1 \mathcal{H}_5^{-p-1} W^+$ & $\frac{1}{2\sqrt2}g^2 s_{\al_2}$ \\
$W^p_{14} H_2 \mathcal{H}_2^{-q} W_{34}^{q-p}$ & $\frac{1}{2\sqrt2}g^2 s_{\al_2}$ & $W^p_{14} H_2 \mathcal{H}_5^{-p-1} W^+$ & $-\frac{1}{2\sqrt2}g^2 c_{\al_2}$ \\
$W^p_{14} H_3 \mathcal{H}_6^{q-p} W_{13}^{-q}$ & $\frac{1}{2\sqrt2}g^2 c_{(\al_1+\al_3)}$ & $W^p_{14} H_4 \mathcal{H}_6^{q-p} W_{13}^{-q}$ & $\frac{1}{2\sqrt2}g^2 s_{(\al_1+\al_3)}$ \\
$W^p_{14} \mathcal{A} \mathcal{H}_1^+ W_{24}^{-p-1}$ & $-\frac{i}{2\sqrt2}g^2 c_{2\al_2}$ & $W^p_{14} \mathcal{A} \mathcal{H}_2^{-q} W_{34}^{q-p}$ & $\frac{i}{2\sqrt2}g^2 s_{\al_2}$ \\
\hline
\end{tabu}
\caption{\label{2CG2ST1}The interactions of two charged gauge bosons with two scalars (Table continued).}  
\end{table}

\begin{table}[!h]
\centering\tabcolsep=0pt\tabulinesep=2pt
\begin{tabu}{X[2.4,l]X[1.1,c]X[3.0,c]X[1.1,c]}
\hline
Vertex & Coupling & Vertex & Coupling\\ \hline
$W^p_{14} \mathcal{A} \mathcal{H}_5^{-p-1} W^+$ & $\frac{i}{2\sqrt2}g^2 c_{\al_2}$ & $W^p_{14} \mathcal{H}_1^+ H_2 W_{24}^{-p-1}$ & $-\frac{1}{2\sqrt2}g^2 c_{2\al_2}$ \\
$W^p_{14} \mathcal{H}_1^+ \mathcal{H}_4^{-q-1} W_{34}^{q-p}$ & $\frac{1}{2}g^2 c_{\al_2}$ & $W^p_{14} \mathcal{H}_1^- \mathcal{H}_3^{-p} W^+$ & $\frac{1}{2}g^2 s_{\al_2}$ \\
$W^p_{14} \mathcal{H}_2^{q} \mathcal{H}_3^{-p} W_{13}^{-q}$ & $\frac{1}{2}g^2$ & $W^p_{14} \mathcal{H}_4^{q+1} \mathcal{H}_5^{-p-1} W_{13}^{-q}$ & $\frac{1}{2}g^2$ \\
$W^{q+1}_{23} W^{-q-1}_{23} H_1 H_1$ & $\frac{1}{4}g^2s_{\al_2}^2$ & $W^{q+1}_{23} W^{-q-1}_{23} H_1 H_2$ & $-\frac{1}{4}g^2 s_{2\al_2}$ \\
$W^{q+1}_{23} W^{-q-1}_{23} H_2 H_2$ & $\frac{1}{4}g^2 c_{\al_2}^2$ & $W^{q+1}_{23} W^{-q-1}_{23} H_3 H_3$ & $\frac{1}{4}g^2 c_{\al_1}^2$ \\
$W^{q+1}_{23} W^{-q-1}_{23} H_3 H_4$ & $\frac{1}{4}g^2 s_{2\al_1}$ & $W^{q+1}_{23} W^{-q-1}_{23} H_4 H_4$ & $\frac{1}{4}g^2 s_{\al_1}^2$ \\
$W^{q+1}_{23} W^{-q-1}_{23} \mathcal{A} \mathcal{A}$ & $\frac{1}{4}g^2 c_{\al_2}^2$ & $W^{q+1}_{23} W^{-q-1}_{23} \mathcal{H}_1^+ \mathcal{H}_1^-$ & $\frac{1}{2}g^2s_{\al_2}^2$ \\
$W^{q+1}_{23} W^{-q-1}_{23} \mathcal{H}_2^q \mathcal{H}_2^{-q}$ & $\frac{1}{2}g^2$ & $W^{q+1}_{23} W^{-q-1}_{23} \mathcal{H}_4^{q+1} \mathcal{H}_4^{-q-1}$ & $\frac{1}{2}g^2$ \\
$W^{q+1}_{23} W^{-q-1}_{23} \mathcal{H}_6^{q-p} \mathcal{H}_6^{p-q}$ & $\frac{1}{2}g^2 s_{\al_3}^2$ & $W^{q+1}_{23} H_1 \mathcal{H}_1^- W_{13}^{-q}$ & $\frac{1}{2\sqrt2}g^2 s_{2\al_2}$ \\
$W^{q+1}_{23} H_1 \mathcal{H}_2^{-q} W^-$ & $\frac{1}{2\sqrt2}g^2 c_{\al_2}$ & $W^{q+1}_{23} H_1 \mathcal{H}_5^{-p-1} W_{34}^{p-q}$ & $\frac{1}{2\sqrt2}g^2 s_{\al_2}$ \\
$W^{q+1}_{23} H_2 \mathcal{H}_2^{-q} W^-$ & $\frac{1}{2\sqrt2}g^2 s_{\al_2}$ & $W^{q+1}_{23} H_2 \mathcal{H}_5^{-p-1} W_{34}^{p-q}$ & $-\frac{1}{2\sqrt2}g^2 c_{\al_2}$ \\
$W^{q+1}_{23} H_3 \mathcal{H}_6^{p-q} W_{24}^{-p-1}$ & $\frac{1}{2\sqrt2}g^2 c_{(\al_1+\al_3)}$ & $W^{q+1}_{23} H_4 \mathcal{H}_6^{p-q} W_{24}^{-p-1}$ & $\frac{1}{2\sqrt2}g^2 s_{(\al_1+\al_3)}$ \\
$W^{q+1}_{23} \mathcal{A} \mathcal{H}_1^- W_{13}^{-q}$ & $\frac{i}{2\sqrt2}g^2 c_{2\al_2}$ & $W^{q+1}_{23} \mathcal{A} \mathcal{H}_2^{-q} W^-$ & $\frac{i}{2\sqrt2}g^2 s_{\al_2}$ \\
$W^{q+1}_{23} \mathcal{A} \mathcal{H}_5^{-p-1} W_{34}^{p-q}$ & $\frac{i}{2\sqrt2}g^2 c_{\al_2}$ & $W^{q+1}_{23} \mathcal{H}_1^+ \mathcal{H}_4^{-q-1} W^-$ & $\frac{1}{2}g^2 c_{\al_2}$ \\
$W^{q+1}_{23} \mathcal{H}_1^- \mathcal{H}_3^{-p} W_{34}^{p-q}$ & $\frac{1}{2}g^2 s_{\al_2}$ & $W^{q+1}_{23} \mathcal{H}_1^- H_2 W_{13}^{-q}$ & $-\frac{1}{2\sqrt2}g^2 c_{2\al_2}$ \\
$W^{q+1}_{23} \mathcal{H}_2^{-q} \mathcal{H}_3^{p} W_{24}^{-p-1}$ & $\frac{1}{2}g^2$ & $W^{q+1}_{23} \mathcal{H}_4^{-q-1} \mathcal{H}_5^{p+1} W_{24}^{-p-1}$ & $\frac{1}{2}g^2$ \\
$W^{p+1}_{24} W^{-p-1}_{24} H_1 H_1$ & $\frac{1}{4}g^2s_{\al_2}^2$ & $W^{p+1}_{24} W^{-p-1}_{24} H_1 H_2$ & $-\frac{1}{4}g^2 s_{2\al_2}$ \\
$W^{p+1}_{24} W^{-p-1}_{24} H_2 H_2$ & $\frac{1}{4}g^2 c_{\al_2}^2$ & $W^{p+1}_{24} W^{-p-1}_{24} H_3 H_3$ & $\frac{1}{4}g^2 s_{\al_1}^2$ \\
$W^{p+1}_{24} W^{-p-1}_{24} H_3 H_4$ & $-\frac{1}{4}g^2 s_{2\al_1}$ & $W^{p+1}_{24} W^{-p-1}_{24} H_4 H_4$ & $\frac{1}{4}g^2 c_{\al_1}^2$ \\
$W^{p+1}_{24} W^{-p-1}_{24} \mathcal{A} \mathcal{A}$ & $\frac{1}{4}g^2 c_{\al_2}^2$ & $W^{p+1}_{24} W^{-p-1}_{24} \mathcal{H}_1^+ \mathcal{H}_1^-$ & $\frac{1}{2}g^2s_{\al_2}^2$ \\
$W^{p+1}_{24} W^{-p-1}_{24} \mathcal{H}_3^p \mathcal{H}_3^{-p}$ & $\frac{1}{2}g^2$ & $W^{p+1}_{24} W^{-p-1}_{24} \mathcal{H}_5^{p+1} \mathcal{H}_5^{-p-1}$ & $\frac{1}{2}g^2$ \\
$W^{p+1}_{24} W^{-p-1}_{24} \mathcal{H}_6^{q-p} \mathcal{H}_6^{p-q}$ & $\frac{1}{2}g^2 c_{\al_3}^2$ & $W^{p+1}_{24} H_1 \mathcal{H}_1^- W_{14}^{-p}$ & $\frac{1}{2\sqrt2}g^2 s_{2\al_2}$ \\
$W^{p+1}_{24} H_1 \mathcal{H}_3^{-p} W^-$ & $\frac{1}{2\sqrt2}g^2 c_{\al_2}$ & $W^{p+1}_{24} H_1 \mathcal{H}_4^{-q-1} W_{34}^{q-p}$ & $\frac{1}{2\sqrt2}g^2 s_{\al_2}$ \\
$W^{p+1}_{24} H_2 \mathcal{H}_3^{-p} W^-$ & $\frac{1}{2\sqrt2}g^2 s_{\al_2}$ & $W^{p+1}_{24} H_2 \mathcal{H}_4^{-q-1} W_{34}^{q-p}$ & $-\frac{1}{2\sqrt2}g^2 c_{\al_2}$ \\
$W^{p+1}_{24} H_3 \mathcal{H}_6^{q-p} W_{23}^{-q-1}$ & $\frac{1}{2\sqrt2}g^2 c_{(\al_1+\al_3)}$ & $W^{p+1}_{24} H_4 \mathcal{H}_6^{q-p} W_{23}^{-q-1}$ & $\frac{1}{2\sqrt2}g^2 s_{(\al_1+\al_3)}$ \\
$W^{p+1}_{24} \mathcal{A} \mathcal{H}_1^- W_{14}^{-p}$ & $\frac{i}{2\sqrt2}g^2 c_{2\al_2}$ & $W^{p+1}_{24} \mathcal{A} \mathcal{H}_3^{-p} W^-$ & $\frac{i}{2\sqrt2}g^2 s_{\al_2}$ \\
$W^{p+1}_{24} \mathcal{A} \mathcal{H}_4^{-q-1} W_{34}^{q-p}$ & $\frac{i}{2\sqrt2}g^2 c_{\al_2}$ & $W^{p+1}_{24} \mathcal{H}_1^+ \mathcal{H}_5^{-p-1} W^-$ & $\frac{1}{2}g^2 c_{\al_2}$ \\
$W^{p+1}_{24} \mathcal{H}_1^- \mathcal{H}_2^{-q} W_{34}^{q-p}$ & $\frac{1}{2}g^2 s_{\al_2}$ & $W^{p+1}_{24} \mathcal{H}_1^- H_2 W_{14}^{-p}$ & $-\frac{1}{2\sqrt2}g^2 c_{2\al_2}$ \\
$W^{p+1}_{24} \mathcal{H}_2^{q} \mathcal{H}_3^{-p} W_{23}^{-q-1}$ & $\frac{1}{2}g^2$ & $W^{p+1}_{24} \mathcal{H}_4^{q+1} \mathcal{H}_5^{-p-1} W_{23}^{-q-1}$ & $\frac{1}{2}g^2$ \\
$W^{q-p}_{34} W^{p-q}_{34} H_3 H_3$ & $\frac{1}{4}g^2$ & $W^{q-p}_{34} W^{p-q}_{34} H_4 H_4$ & $\frac{1}{4}g^2$ \\
$W^{q-p}_{34} W^{p-q}_{34} \mathcal{H}_2^q \mathcal{H}_2^{-q}$ & $\frac{1}{2}g^2$ & $W^{q-p}_{34} W^{p-q}_{34} \mathcal{H}_3^{p} \mathcal{H}_3^{-p}$ & $\frac{1}{2}g^2$ \\
$W^{q-p}_{34} W^{p-q}_{34} \mathcal{H}_4^{q+1} \mathcal{H}_4^{-q-1}$ & $\frac{1}{2}g^2$ & $W^{q-p}_{34} W^{p-q}_{34} \mathcal{H}_5^{p+1} \mathcal{H}_5^{-p-1}$ & $\frac{1}{2}g^2$ \\
$W^{q-p}_{34} W^{p-q}_{34} \mathcal{H}_6^{q-p} \mathcal{H}_6^{p-q}$ & $\frac{1}{2}g^2$ & $W^{q-p}_{34} H_1 \mathcal{H}_2^{-q} W_{14}^{p}$ & $\frac{1}{2\sqrt2}g^2 c_{\al_2}$\\
\hline
\end{tabu}
\caption{\label{2CG2ST2}The interactions of two charged gauge bosons with two scalars (Continued).}  
\end{table}

\begin{table}[!h]
\centering\tabcolsep=0pt\tabulinesep=2pt
\begin{tabu}{X[3,l]X[1.2,c]X[4,c]X[1.1,c]}
\hline
Vertex & Coupling & Vertex & Coupling\\ \hline
$W^{q-p}_{34} H_1 \mathcal{H}_3^{p} W_{13}^{-q}$ & $\frac{1}{2\sqrt2}g^2 c_{\al_2}$ & $W^{q-p}_{34} H_1 \mathcal{H}_4^{-q-1} W_{24}^{p+1}$ & $\frac{1}{2\sqrt2}g^2 s_{\al_2}$ \\
$W^{q-p}_{34} H_1 \mathcal{H}_5^{p+1} W_{23}^{-q-1}$ & $\frac{1}{2\sqrt2}g^2 s_{\al_2}$ & $W^{q-p}_{34} H_2 \mathcal{H}_2^{-q} W_{14}^{p}$ & $\frac{1}{2\sqrt2}g^2 s_{\al_2}$ \\
$W^{q-p}_{34} H_2 \mathcal{H}_3^{p} W_{13}^{-q}$ & $\frac{1}{2\sqrt2}g^2 s_{\al_2}$ & $W^{q-p}_{34} H_2 \mathcal{H}_4^{-q-1} W_{24}^{p+1}$ & $-\frac{1}{2\sqrt2}g^2 c_{\al_2}$ \\
$W^{q-p}_{34} H_2 \mathcal{H}_5^{p+1} W_{23}^{-q-1}$ & $-\frac{1}{2\sqrt2}g^2 c_{\al_2}$ & $W^{q-p}_{34} \mathcal{A} \mathcal{H}_2^{-q} W_{14}^{p}$ & $\frac{i}{2\sqrt2}g^2 s_{\al_2}$ \\
$W^{q-p}_{34} \mathcal{A} \mathcal{H}_3^{p} W_{13}^{-q}$ & $-\frac{i}{2\sqrt2}g^2 s_{\al_2}$ & $W^{q-p}_{34} \mathcal{A} \mathcal{H}_4^{-q-1} W_{24}^{p+1}$ & $\frac{i}{2\sqrt2}g^2 c_{\al_2}$ \\
$W^{q-p}_{34} \mathcal{A} \mathcal{H}_5^{p+1} W_{23}^{-q-1}$ & $-\frac{i}{2\sqrt2}g^2 c_{\al_2}$ & $W^{q-p}_{34} \mathcal{H}_1^+ \mathcal{H}_3^{p} W_{23}^{-q-1}$ & $\frac{1}{2}g^2 s_{\al_2}$ \\
$W^{q-p}_{34} \mathcal{H}_1^+ \mathcal{H}_4^{-q-1} W_{14}^{p}$ & $\frac{1}{2}g^2 c_{\al_2}$ & $W^{q-p}_{34} \mathcal{H}_1^- \mathcal{H}_2^{-q} W_{24}^{p+1}$ & $\frac{1}{2}g^2 s_{\al_2}$ \\
$W^{q-p}_{34} \mathcal{H}_1^- \mathcal{H}_5^{p+1} W_{13}^{-q}$ & $\frac{1}{2}g^2 c_{\al_2}$ & No data & No data \\
\hline\hline
\end{tabu}
\caption{\label{2CG2ST3}The interactions of two charged gauge bosons with two scalars (Continued).}  
\end{table}

\begin{table}[!h]
\centering\tabcolsep=0pt\tabulinesep=2pt
\begin{tabu}{X[2,l]X[2.2,c]X[2.7,c]X[2.2,c]}
\hline\hline
Vertex & Coupling & Vertex & Coupling\\ \hline
$A W^+ H_2 \mathcal{H}_1^-$ & $-\frac{1}{2}g^2 s_W$ & $A W^+ \mathcal{A} \mathcal{H}_1^-$ & $\frac{i}{2}g^2 s_W$\\
$A W_{13}^q H_1 \mathcal{H}_2^{-q}$ & $\frac{1}{2}g^2 q c_{\al_2} s_W$ & $A W_{13}^q H_2 \mathcal{H}_2^{-q}$ & $\frac{1}{2}g^2 q s_{\al_2} s_W$\\
$A W_{13}^q \mathcal{A} \mathcal{H}_2^{-q}$ & $\frac{i}{2}g^2 q s_{\al_2} s_W$ & $A W_{13}^q \mathcal{H}_1^+ \mathcal{H}_4^{-q-1}$ & $\frac{1}{\sqrt2}g^2 (q+2) c_{\al_2} s_W$\\
$A W_{14}^p H_1 \mathcal{H}_3^{-p}$ & $\frac{1}{2}g^2 p c_{\al_2} s_W$ & $A W_{14}^p H_2 \mathcal{H}_3^{-p}$ & $\frac{1}{2}g^2 p s_{\al_2} s_W$\\
$A W_{14}^p \mathcal{A} \mathcal{H}_3^{-p}$ & $\frac{i}{2}g^2 p s_{\al_2} s_W$ & $A W_{14}^p \mathcal{H}_1^+ \mathcal{H}_5^{-p-1}$ & $\frac{1}{\sqrt2}g^2 (p+2) c_{\al_2} s_W$\\
$A W_{23}^{q+1} H_1 \mathcal{H}_4^{-q-1}$ & $\frac{1}{2}g^2 (q+1) s_{\al_2} s_W$ & $A W_{23}^{q+1} H_2 \mathcal{H}_4^{-q-1}$ & $-\frac{1}{2}g^2 (q+1) c_{\al_2} s_W$\\
$A W_{23}^{q+1} \mathcal{A} \mathcal{H}_4^{-q-1}$ & $\frac{i}{2}g^2 (q+1) c_{\al_2} s_W$ & $A W_{23}^{q+1} \mathcal{H}_1^- \mathcal{H}_2^{-q}$ & $\frac{1}{\sqrt2}g^2 (q-1) s_{\al_2} s_W$\\
$A W_{24}^{p+1} H_1 \mathcal{H}_5^{-p-1}$ & $\frac{1}{2}g^2 (p+1) s_{\al_2} s_W$ & $A W_{24}^{p+1} H_2 \mathcal{H}_5^{-p-1}$ & $-\frac{1}{2}g^2 (p+1) c_{\al_2} s_W$\\
$A W_{24}^{p+1} \mathcal{A} \mathcal{H}_5^{-p-1}$ & $\frac{i}{2}g^2 (p+1) c_{\al_2} s_W$ & $A W_{24}^{p+1} \mathcal{H}_1^- \mathcal{H}_3^{-p}$ & $\frac{1}{\sqrt2}g^2 (p-1) s_{\al_2} s_W$\\
$A W_{34}^{q-p} H_3 \mathcal{H}_6^{p-q}$ & $\frac{1}{2}g^2 (p-q) c_{(\al_1-\al_3)} s_W$ & $A W_{34}^{q-p} H_4 \mathcal{H}_6^{p-q}$ & $\frac{1}{2}g^2 (p-q) s_{(\al_1-\al_3)} s_W$\\
$A W_{34}^{q-p} \mathcal{H}_2^{-q} \mathcal{H}_3^{p}$ & $\frac{1}{\sqrt2}g^2 (p+q) s_W$ & $A W_{34}^{q-p} \mathcal{H}_4^{-q-1} \mathcal{H}_5^{p+1}$ & $\frac{1}{\sqrt2}g^2 (p+q+2) s_W$\\
$Z_1 W^+ H_2 \mathcal{H}_1^-$ & $\frac{1}{2}g^2 s_Wt_W$ & $Z_1 W^+ \mathcal{A} \mathcal{H}_1^-$ & $-\frac{i}{2}g^2 s_Wt_W$ \\
$Z_1 W_{13}^q H_1 \mathcal{H}_2^{-q}$ & $\frac{1}{4 c_W}g^2  (1-2qs_W^2)c_{\al_2}$ & $Z_1 W_{13}^q H_2 \mathcal{H}_2^{-q}$ & $\frac{1}{4 c_W}g^2 (1-2qs_W^2) s_{\al_2}$\\
$Z_1 W_{13}^q \mathcal{A} \mathcal{H}_2^{-q}$ & $\frac{i}{4 c_W}g^2 (1-2qs_W^2) s_{\al_2}$ & $Z_1 W_{13}^q \mathcal{H}_1^+ \mathcal{H}_4^{-q-1}$ & $\frac{g^2[c_W^2-(3+2q)s_W^2]c_{\al_2}}{2\sqrt2 c_W}$\\
$Z_1 W_{14}^p H_1 \mathcal{H}_3^{-p}$ & $\frac{1}{4 c_W}g^2  (1-2ps_W^2)c_{\al_2}$ & $Z_1 W_{14}^p H_2 \mathcal{H}_3^{-p}$ & $\frac{1}{4 c_W}g^2 (1-2ps_W^2) s_{\al_2}$\\
$Z_1 W_{14}^p \mathcal{A} \mathcal{H}_3^{-p}$ & $\frac{i}{4 c_W}g^2 (1-2ps_W^2) s_{\al_2}$ & $Z_1 W_{14}^p \mathcal{H}_1^+ \mathcal{H}_5^{-p-1}$ & $\frac{g^2[c_W^2-(3+2p)s_W^2]c_{\al_2}}{2\sqrt2 c_W}$\\
$Z_1 W_{23}^{q+1} H_1 \mathcal{H}_4^{-q-1}$ & $\frac{1}{2}g^2 (q+1) s_{\al_2} s_W$ & $Z_1 W_{23}^{q+1} H_2 \mathcal{H}_4^{-q-1}$ & $-\frac{1}{2}g^2 (q+1) c_{\al_2} s_W$\\
$Z_1 W_{23}^{q+1} \mathcal{A} \mathcal{H}_4^{-q-1}$ & $\frac{i}{2}g^2 (q+1) c_{\al_2} s_W$ & $Z_1 W_{23}^{q+1} \mathcal{H}_1^- \mathcal{H}_2^{-q}$ & $\frac{1}{\sqrt2}g^2 (q-1) s_{\al_2} s_W$\\
$Z_1 W_{24}^{p+1} H_1 \mathcal{H}_5^{-p-1}$ & $-\frac{g^2[c_W^2+(3+2p)s_W^2]s_{\al_2}}{4 c_W}$ & $Z_1 W_{24}^{p+1} H_2 \mathcal{H}_5^{-p-1}$ & $\frac{g^2[c_W^2+(3+2p)s_W^2]c_{\al_2}}{4 c_W}$\\
$Z_1 W_{24}^{p+1} \mathcal{A} \mathcal{H}_5^{-p-1}$ & $-\frac{ig^2[c_W^2+(3+2p)s_W^2]c_{\al_2}}{4 c_W}$ & $Z_1 W_{24}^{p+1} \mathcal{H}_1^- \mathcal{H}_3^{-p}$ & $-\frac{g^2[c_W^2-(1-2p)s_W^2]s_{\al_2}}{2\sqrt2 c_W}$\\
$Z_1 W_{34}^{q-p} H_3 \mathcal{H}_6^{p-q}$ & $\frac{g^2 (q-p) c_{(\al_1-\al_3)} s_W^2}{2c_W}$ & $Z_1 W_{34}^{q-p} H_4 \mathcal{H}_6^{p-q}$ & $\frac{g^2 (q-p) s_{(\al_1-\al_3)} s_W^2}{2c_W}$\\
$Z_1 W_{34}^{q-p} \mathcal{H}_2^{-q} \mathcal{H}_3^{p}$ & $-\frac{1}{\sqrt2}g^2 (p+q) s_Wt_W$ & $Z_1 W_{34}^{q-p} \mathcal{H}_4^{-q-1} \mathcal{H}_5^{p+1}$ & $-\frac{1}{\sqrt2}g^2 (p+q+2) s_Wt_W$\\
\hline
\end{tabu}
\caption{\label{1NG1CG2ST1}The interactions of a neutral and a charged gauge boson with two scalars (Table continued).}  
\end{table}

\begin{table}[!h]
\centering\tabcolsep=0pt\tabulinesep=2pt
\begin{tabu}{X[1,l]X[3,c]}
\hline
Vertex & Coupling \\ \hline
$Z_2 W^+ H_1 \mathcal{H}_1^-$ & $\frac{1}{2\sqrt3 (u^2+v^2)}g^2uv [c_\varphi (\beta_1+\beta_2)-s_\varphi (\ga_1+\ga_2)]$ \\
$Z_2 W^+ H_2 \mathcal{H}_1^-$ & $\frac{1}{2\sqrt3 (u^2+v^2)}g^2 [c_\varphi (\beta_1 v^2-\beta_2u^2)-s_\varphi (\ga_1v^2-\ga_2u^2)]$ \\
$Z_2 W^+ \mathcal{A} \mathcal{H}_1^-$ & $-\frac{i}{2\sqrt3 (u^2+v^2)}g^2 [c_\varphi (\beta_1 v^2-\beta_2u^2)-s_\varphi (\ga_1v^2-\ga_2u^2)]$ \\
$Z_2 W_{13}^q H_1 \mathcal{H}_2^{-q}$ & $-\frac{1}{8\sqrt3 \sqrt{u^2+v^2}}g^2u \{c_\varphi (\beta_1 +\beta_2)[1+(1-4q^2)t_W^2]+4s_\varphi \ga_1\}$ \\
$Z_2 W_{13}^q H_2 \mathcal{H}_2^{-q}$ & $-\frac{1}{8\sqrt3 \sqrt{u^2+v^2}}g^2v \{c_\varphi (\beta_1 +\beta_2)[1+(1-4q^2)t_W^2]+4s_\varphi \ga_1\}$ \\
$Z_2 W_{13}^q \mathcal{A} \mathcal{H}_2^{-q}$ & $-\frac{i}{8\sqrt3 \sqrt{u^2+v^2}}g^2v \{c_\varphi (\beta_1 +\beta_2)[1+(1-4q^2)t_W^2]+4s_\varphi \ga_1\}$ \\
$Z_2 W_{13}^q \mathcal{H}_1^+ \mathcal{H}_4^{-q-1}$ & $-\frac{1}{4\sqrt6 \sqrt{u^2+v^2}}g^2u \{c_\varphi (\beta_1 +\beta_2)[1-(3+8q+4q^2)t_W^2]+4s_\varphi \ga_2\}$ \\
$Z_2 W_{14}^p H_1 \mathcal{H}_3^{-p}$ & $\frac{1}{4\sqrt3 \sqrt{u^2+v^2}}g^2u \{c_\varphi [\beta_1(1-p) +\beta_2 p]+s_\varphi[(q+p+1)(\ga_1-\ga_2)+2\ga_2]\}$ \\
$Z_2 W_{14}^p H_2 \mathcal{H}_3^{-p}$ & $\frac{1}{4\sqrt3 \sqrt{u^2+v^2}}g^2v \{c_\varphi [\beta_1(1-p) +\beta_2 p]+s_\varphi[(q+p+1)(\ga_1-\ga_2)+2\ga_2]\}$ \\
$Z_2 W_{14}^p \mathcal{A} \mathcal{H}_3^{-p}$ & $\frac{i}{4\sqrt3 \sqrt{u^2+v^2}}g^2v \{c_\varphi [\beta_1(1-p) +\beta_2 p]+s_\varphi[(q+p+1)(\ga_1-\ga_2)+2\ga_2]\}$ \\
$Z_2 W_{14}^p \mathcal{H}_1^+ \mathcal{H}_5^{-p-1}$ & $\frac{1}{2\sqrt6 \sqrt{u^2+v^2}}g^2u \{c_\varphi [\beta_2(2+p)-\beta_1 (1+p)]+s_\varphi[(q+p+1)(\ga_1-\ga_2)+2\ga_1]\}$ \\
$Z_2 W_{23}^{q+1} H_1 \mathcal{H}_4^{-q-1}$ & $-\frac{1}{8\sqrt3 \sqrt{u^2+v^2}}g^2v \{c_\varphi (\beta_1+\beta_2)[1-(3+8q+4q^2)t_W^2]+4s_\varphi\ga_2\}$ \\
$Z_2 W_{23}^{q+1} H_2 \mathcal{H}_4^{-q-1}$ &  $\frac{1}{8\sqrt3 \sqrt{u^2+v^2}}g^2u \{c_\varphi (\beta_1+\beta_2)[1-(3+8q+4q^2)t_W^2]+4s_\varphi\ga_2\}$ \\
$Z_2 W_{23}^{q+1} \mathcal{A} \mathcal{H}_4^{-q-1}$ & $-\frac{i}{8\sqrt3 \sqrt{u^2+v^2}}g^2u \{c_\varphi (\beta_1+\beta_2)[1-(3+8q+4q^2)t_W^2]+4s_\varphi\ga_2\}$ \\
$Z_2 W_{23}^{q+1} \mathcal{H}_1^- \mathcal{H}_2^{-q}$ & $-\frac{1}{4\sqrt6 \sqrt{u^2+v^2}}g^2v \{c_\varphi (\beta_1 +\beta_2)[1+(1-4q^2)t_W^2]+4s_\varphi \ga_1\}$ \\
$Z_2 W_{24}^{p+1} H_1 \mathcal{H}_5^{-p-1}$ & $\frac{1}{4\sqrt3 \sqrt{u^2+v^2}}g^2v \{c_\varphi [\beta_2(2+p)-\beta_1 (1+p)]+s_\varphi[(q+p+1)(\ga_1-\ga_2)+2\ga_1]\}$ \\
$Z_2 W_{24}^{p+1} H_2 \mathcal{H}_5^{-p-1}$ & $-\frac{1}{4\sqrt3 \sqrt{u^2+v^2}}g^2u \{c_\varphi [\beta_2(2+p)-\beta_1 (1+p)]+s_\varphi[(q+p+1)(\ga_1-\ga_2)+2\ga_1]\}$ \\
$Z_2 W_{24}^{p+1} \mathcal{A} \mathcal{H}_5^{-p-1}$ & $\frac{i}{4\sqrt3 \sqrt{u^2+v^2}}g^2u \{c_\varphi [\beta_2(2+p)-\beta_1 (1+p)]+s_\varphi[(q+p+1)(\ga_1-\ga_2)+2\ga_1]\}$ \\
$Z_2 W_{24}^{p+1} \mathcal{H}_1^- \mathcal{H}_3^{-p}$ & $\frac{1}{2\sqrt6 \sqrt{u^2+v^2}}g^2v \{c_\varphi [\beta_1(1-p)+\beta_2 p]+s_\varphi[(q+p+1)(\ga_1-\ga_2)+2\ga_2]\}$ \\
$Z_2 W_{34}^{q-p} H_3 \mathcal{H}_6^{p-q}$ & $\frac{1}{4\sqrt3\sqrt{w^2+V^2}}g^2\{(1+q-p)[ws_{\al_1}(\beta_1c_\varphi-\ga_1s_\varphi)-Vc_{\al_1}(\beta_2c_\varphi-\ga_2s_\varphi)]+(1-q+p)[ws_{\al_1}(\beta_2c_\varphi-\ga_2s_\varphi)-Vc_{\al_1}(\beta_1c_\varphi-\ga_1s_\varphi)]\}$\\
$Z_2 W_{34}^{q-p} H_4 \mathcal{H}_6^{p-q}$ & $\frac{1}{4\sqrt3\sqrt{w^2+V^2}}g^2\{(p-q-1)[wc_{\al_1}(\beta_1c_\varphi-\ga_1s_\varphi)+Vs_{\al_1}(\beta_2c_\varphi-\ga_2s_\varphi)]-(1-q+p)[Vs_{\al_1}(\beta_1c_\varphi-\ga_1s_\varphi)+wc_{\al_1}(\beta_2c_\varphi-\ga_2s_\varphi)]\}$\\
$Z_2 W_{34}^{q-p} \mathcal{H}_2^{-q} \mathcal{H}_3^p$ & $\frac{1}{2\sqrt6}g^2[(1+q+p)(s_\varphi\ga_1-c_\varphi\beta_1)+(1-q-p)(s_\varphi\ga_2-c_\varphi\beta_2)]$\\
$Z_2 W_{34}^{q-p} \mathcal{H}_4^{-q-1} \mathcal{H}_5^{p+1}$ & $\frac{1}{2\sqrt6}g^2[(1+q+p)(c_\varphi\beta_2-s_\varphi\ga_2)-(3+q+p)(c_\varphi\beta_1-s_\varphi\ga_1)]$\\
$Z_{3}\dots\dots$ & $Z_{2}\dots\dots (c_\varphi\to s_\varphi,s_\varphi\to -c_\varphi)$\\
\hline\hline
\end{tabu}
\caption{\label{1NG1CG2ST2}The interactions of a neutral and a charged gauge boson with two scalars (Continued).}  
\end{table}

\begin{table}[!h]
\centering\tabcolsep=0pt\tabulinesep=2pt
\begin{tabu}{X[2,l]X[2.2,c]X[2.7,c]X[2.2,c]}
\hline\hline
Vertex & Coupling & Vertex & Coupling\\ \hline
$A A \mathcal{H}_1^+ \mathcal{H}_1^-$ & $g^2 s_W^2$ & $A A \mathcal{H}_2^q \mathcal{H}_2^{-q}$ & $g^2 q^2 s_W^2$ \\
$A A \mathcal{H}_3^p \mathcal{H}_3^{-p}$ & $g^2 p^2 s_W^2$ & $A A \mathcal{H}_4^{q+1} \mathcal{H}_4^{-q-1}$ & $g^2 (1+q)^2 s_W^2$ \\
$A A \mathcal{H}_5^{p+1} \mathcal{H}_5^{-p-1}$ & $g^2 (1+p)^2 s_W^2$ & $A A \mathcal{H}_6^{q-p} \mathcal{H}_6^{p-q}$ & $g^2 (p-q)^2 s_W^2$ \\
$A Z_1 \mathcal{H}_1^+ \mathcal{H}_1^-$ & $g^2 (s_{2W}-t_W)$ & $A Z_1 \mathcal{H}_2^q \mathcal{H}_2^{-q}$ & $-2g^2 q^2 s_W^2t_W$ \\
$A Z_1 \mathcal{H}_3^p \mathcal{H}_3^{-p}$ & $-2g^2 p^2 s_W^2t_W$ & $A Z_1 \mathcal{H}_4^{q+1} \mathcal{H}_4^{-q-1}$ & $-2g^2 (1+q)^2 s_W^2t_W$ \\
$A Z_1 \mathcal{H}_5^{p+1} \mathcal{H}_5^{-p-1}$ & $-2g^2 (1+p)^2 s_W^2t_W$ & $A Z_1 \mathcal{H}_6^{q-p} \mathcal{H}_6^{p-q}$ & $-2g^2 (p-q)^2 s_W^2t_W$ \\
$Z_1 Z_1 \mathcal{H}_1^+ \mathcal{H}_1^-$ & $\frac{1}{4c_W^2} g^2 c_{2W}^2$ & $Z_1 Z_1 \mathcal{H}_2^q \mathcal{H}_2^{-q}$ & $g^2 q^2 s_W^2t_W^2$ \\
$Z_1 Z_1 \mathcal{H}_3^p \mathcal{H}_3^{-p}$ & $g^2 p^2 s_W^2t_W^2$ & $Z_1 Z_1 \mathcal{H}_4^{q+1} \mathcal{H}_4^{-q-1}$ & $g^2 (1+q)^2 s_W^2t_W^2$ \\
$Z_1 Z_1 \mathcal{H}_5^{p+1} \mathcal{H}_5^{-p-1}$ & $g^2 (1+p)^2 s_W^2t_W^2$ & $Z_1 Z_1 \mathcal{H}_6^{q-p} \mathcal{H}_6^{p-q}$ & $g^2 (p-q)^2 s_W^2t_W^2$ \\
\hline
\end{tabu}
\begin{tabu}{X[1,l]X[3,c]}
Vertex & Coupling \\ \hline
$A Z_2 \mathcal{H}_1^+ \mathcal{H}_1^-$ & $\frac{1}{\sqrt3(u^2+v^2)} g^2 s_W [c_\varphi(u^2\beta_2-v^2\beta_1)-s_\varphi(u^2\ga_2-v^2\ga_1)]$ \\
$A Z_2 \mathcal{H}_2^q \mathcal{H}_2^{-q}$ & $\frac{1}{\sqrt3} g^2 s_W q \{c_\varphi[q(\beta_2-\beta_1)-(\beta_2+\beta_1)]-s_\varphi\ga_1\}$ \\
$A Z_2 \mathcal{H}_3^p \mathcal{H}_3^{-p}$ & $\frac{1}{\sqrt3} g^2 s_W p \{c_\varphi p(\beta_2-\beta_1)-s_\varphi[(q+p+2)(\ga_2-\ga_1)-3\ga_2\}$ \\
$A Z_2 \mathcal{H}_4^{q+1} \mathcal{H}_4^{-q-1}$ & $\frac{1}{\sqrt3} g^2 s_W (1+q) \{c_\varphi[q(\beta_2-\beta_1)-2\beta_1]-s_\varphi\ga_2\}$ \\
$A Z_2 \mathcal{H}_5^{p+1} \mathcal{H}_5^{-p-1}$ & $\frac{1}{\sqrt3} g^2 s_W (1+p) \{c_\varphi (1+p)(\beta_2-\beta_1)-s_\varphi[(q+p)(\ga_2-\ga_1)-3\ga_1\}$ \\
$A Z_2 \mathcal{H}_6^{q-p} \mathcal{H}_6^{p-q}$ & $\frac{1}{\sqrt3} g^2 s_W (p-q) \{c_\varphi [s_{\al_3}^2(\beta_2+\beta_1)+(p-q)(\beta_2-\beta_1)]+s_\varphi[c_{\al_3}^2(\ga_2+\ga_1)-p(\ga_2-\ga_1)+\ga_1]\}$ \\
$A Z_3 \dots$ & $A Z_2 \dots (c_\varphi\to s_\varphi, s_\varphi\to -c_\varphi) $\\
$Z_1 Z_2 H_1 H_2$ & $\frac{1}{2\sqrt3 c_W (u^2+v^2)}g^2 u v [c_\varphi (\beta_2+\beta_1)-s_\varphi (\ga_2+\ga_1)]$\\
$Z_1 Z_2 \mathcal{H}_1^+ \mathcal{H}_1^-$ & $\frac{1}{2\sqrt3(u^2+v^2)c_W} g^2 c_{2W} [c_\varphi(u^2\beta_2-v^2\beta_1)-s_\varphi(u^2\ga_2-v^2\ga_1)]$ \\
$Z_1 Z_2 \mathcal{H}_2^q \mathcal{H}_2^{-q}$ & $-\frac{1}{\sqrt3} g^2 s_Wt_W q \{c_\varphi[q(\beta_2-\beta_1)-(\beta_2+\beta_1)]-s_\varphi\ga_1\}$ \\
$Z_1 Z_2 \mathcal{H}_3^p \mathcal{H}_3^{-p}$ & $-\frac{1}{\sqrt3} g^2 s_Wt_W p \{c_\varphi p(\beta_2-\beta_1)-s_\varphi[(q+p+2)(\ga_2-\ga_1)-3\ga_2]\}$ \\
$Z_1 Z_2 \mathcal{H}_4^{q+1} \mathcal{H}_4^{-q-1}$ & $-\frac{1}{\sqrt3} g^2 s_Wt_W (1+q) \{c_\varphi[q(\beta_2-\beta_1)-2\beta_1]-s_\varphi\ga_2\}$ \\
$Z_1 Z_2 \mathcal{H}_5^{p+1} \mathcal{H}_5^{-p-1}$ & $-\frac{1}{\sqrt3} g^2 s_Wt_W (1+p) \{c_\varphi (1+p)(\beta_2-\beta_1)-s_\varphi[(p+q)(\ga_2-\ga_1)-3\ga_1]\}$ \\
$Z_1 Z_2 \mathcal{H}_6^{q-p} \mathcal{H}_6^{p-q}$ & $-\frac{1}{\sqrt3} g^2 s_Wt_W (p-q) \{c_\varphi [s_{\al_3}^2(\beta_2+\beta_1)+(p-q)(\beta_2-\beta_1)]+s_\varphi[c_{\al_3}^2(\ga_2+\ga_1)-p(\ga_2-\ga_1)+\ga_1]\}$ \\
$Z_1 Z_3 \dots$ & $Z_1 Z_2 \dots (c_\varphi\to s_\varphi, s_\varphi\to -c_\varphi) $\\
$Z_2 Z_2 H_1 H_1$ & $\frac{1}{24(u^2+v^2)}g^2 [u^2(c_\varphi\beta_1-s_\varphi\ga_1)^2+v^2(c_\varphi\beta_2-s_\varphi\ga_2)^2]$\\
$Z_2 Z_2 H_1 H_2$ & $\frac{1}{12(u^2+v^2)}g^2 u v [(c_\varphi\beta_1-s_\varphi\ga_1)^2-(c_\varphi\beta_2-s_\varphi\ga_2)^2]$\\
$Z_2 Z_2 H_2 H_2$ & $\frac{1}{24(u^2+v^2)}g^2 [v^2(c_\varphi\beta_1-s_\varphi\ga_1)^2+u^2(c_\varphi\beta_2-s_\varphi\ga_2)^2]$\\
$Z_2 Z_2 H_3 H_3$ & $\frac{g^2}{24}  \{c_{\al_1}^2[c_\varphi(\beta_2+\beta_1)+s_\varphi(\ga_1+q\ga_1-q\ga_2)]^2+s_{\al_1}^2s_\varphi^2[q(\ga_2-\ga_1)-\ga_2-2\ga_1]^2\}$ \\
\hline
\end{tabu}
\caption{\label{2NG2ST1}The interactions of two neutral gauge bosons with two scalars (Table Continued).}  
\end{table}

\begin{table}[!h]
\centering\tabcolsep=0pt\tabulinesep=2pt
\begin{tabu}{X[1,l]X[3,c]}
\hline
Vertex & Coupling \\ \hline
$Z_2 Z_2 H_3 H_4$ & $\frac{1}{24} g^2 s_{2\al_1}\{[c_\varphi(\beta_2+\beta_1)+s_\varphi(\ga_1+q\ga_1-q\ga_2)]^2-s_\varphi^2[q(\ga_2-\ga_1)-\ga_2-2\ga_1]^2\}$ \\
$Z_2 Z_2 H_4 H_4$ & $\frac{g^2}{24}  \{s_{\al_1}^2[c_\varphi(\beta_2+\beta_1)+s_\varphi(\ga_1+q\ga_1-q\ga_2)]^2+c_{\al_1}^2s_\varphi^2[q(\ga_2-\ga_1)-\ga_2-2\ga_1]^2\}$ \\
$Z_2 Z_2 \mathcal{A} \mathcal{A}$ & $\frac{1}{24(u^2+v^2)}g^2 [v^2(c_\varphi\beta_1-s_\varphi\ga_1)^2+u^2(c_\varphi\beta_2-s_\varphi\ga_2)^2]$\\
$Z_2 Z_2 \mathcal{H}_1^+ \mathcal{H}_1^-$ & $\frac{1}{12(u^2+v^2)}g^2 [v^2(c_\varphi\beta_1-s_\varphi\ga_1)^2+u^2(c_\varphi\beta_2-s_\varphi\ga_2)^2]$\\
$Z_2 Z_2 \mathcal{H}_2^q \mathcal{H}_2^{-q}$ & $\frac{1}{12} g^2 \{c_\varphi[q(\beta_2-\beta_1)-(\beta_2+\beta_1)]-s_\varphi\ga_1\}^2$ \\
$Z_2 Z_2 \mathcal{H}_3^p \mathcal{H}_3^{-p}$ & $\frac{1}{12} g^2 \{c_\varphi p(\beta_2-\beta_1)-s_\varphi[(q+p+2)(\ga_2-\ga_1)-3\ga_2]\}^2$ \\
$Z_2 Z_2 \mathcal{H}_4^{q+1} \mathcal{H}_4^{-q-1}$ & $\frac{1}{12} g^2 \{c_\varphi[q(\beta_2-\beta_1)-2\beta_1]-s_\varphi\ga_2\}^2$ \\
$Z_2 Z_2 \mathcal{H}_5^{p+1} \mathcal{H}_5^{-p-1}$ & $\frac{1}{12} g^2\{c_\varphi (1+p)(\beta_2-\beta_1)-s_\varphi[(p+q)(\ga_2-\ga_1)-3\ga_1]\}^2$ \\
$Z_2 Z_2 \mathcal{H}_6^{q-p} \mathcal{H}_6^{p-q}$ & $\frac{1}{12(w^2+V^2)} g^2 \{V^2[c_\varphi(p-q)(\beta_2-\beta_1)-s_\varphi (p\ga_2-p\ga_1-\ga_2-2\ga_1)]^2+w^2[c_\varphi((p-q)(\beta_2-\beta_1)+\beta_2+\beta_1)-s_\varphi (p\ga_2-p\ga_1-\ga_1)]^2\}$ \\
$Z_2 Z_3 H_1 H_1$ & $\frac{g^2}{12(u^2+v^2)} [u^2(c_\varphi\beta_1-s_\varphi\ga_1)(c_\varphi\ga_1+s_\varphi\beta_1)+v^2(c_\varphi\beta_2-s_\varphi\ga_2)(c_\varphi\ga_2+s_\varphi\beta_2)]$\\
$Z_2 Z_3 H_1 H_2$ & $\frac{1}{6(u^2+v^2)}g^2 u v [(c_\varphi\beta_1-s_\varphi\ga_1)(c_\varphi\ga_1+s_\varphi\beta_1)-(c_\varphi\beta_2-s_\varphi\ga_2)(c_\varphi\ga_2+s_\varphi\beta_2)]$\\
$Z_2 Z_3 H_2 H_2$ & $\frac{g^2}{12(u^2+v^2)} [v^2(c_\varphi\beta_1-s_\varphi\ga_1)(c_\varphi\ga_1+s_\varphi\beta_1)+u^2(c_\varphi\beta_2-s_\varphi\ga_2)(c_\varphi\ga_2+s_\varphi\beta_2)]$\\
$Z_2 Z_3 H_3 H_3$ & $-\frac{1}{12}g^2\{ c_{\al_1}^2[c_\varphi(\beta_2+\beta_1)+s_\varphi(\ga_1+q\ga_1-q\ga_2)][c_\varphi(\ga_1+q\ga_1-q\ga_2)-s_\varphi(\beta_2+\beta_1)]+s_{\al_1}^2s_\varphi c_\varphi[(1-q)(\ga_2-\ga_1)+3\ga_1]^2\}$ \\
$Z_2 Z_3 H_4 H_3$ & $-\frac{1}{12}g^2s_{2\al_1}\{ [c_\varphi(\beta_2+\beta_1)+s_\varphi(\ga_1+q\ga_1-q\ga_2)][c_\varphi(\ga_1+q\ga_1-q\ga_2)-s_\varphi(\beta_2+\beta_1)]-s_\varphi c_\varphi[(1-q)(\ga_2-\ga_1)+3\ga_1]^2\}$ \\
$Z_2 Z_3 H_4 H_4$ & $-\frac{1}{12}g^2\{ s_{\al_1}^2[c_\varphi(\beta_2+\beta_1)+s_\varphi(\ga_1+q\ga_1-q\ga_2)][c_\varphi(\ga_1+q\ga_1-q\ga_2)-s_\varphi(\beta_2+\beta_1)]+c_{\al_1}^2s_\varphi c_\varphi[(1-q)(\ga_2-\ga_1)+3\ga_1]^2\}$ \\
$Z_2 Z_3 \mathcal{A} \mathcal{A}$ & $\frac{g^2}{12(u^2+v^2)} [v^2(c_\varphi\beta_1-s_\varphi\ga_1)(c_\varphi\ga_1+s_\varphi\beta_1)+u^2(c_\varphi\beta_2-s_\varphi\ga_2)(c_\varphi\ga_2+s_\varphi\beta_2)]$\\
$Z_2 Z_3 \mathcal{H}_1^+ \mathcal{H}_1^-$ & $\frac{g^2}{6(u^2+v^2)} [v^2(c_\varphi\beta_1-s_\varphi\ga_1)(c_\varphi\ga_1+s_\varphi\beta_1)+u^2(c_\varphi\beta_2-s_\varphi\ga_2)(c_\varphi\ga_2+s_\varphi\beta_2)]$\\
$Z_2 Z_3 \mathcal{H}_2^q \mathcal{H}_2^{-q}$ & $\frac{g^2}{6} \{c_\varphi[\beta_2+\beta_1-q(\beta_2-\beta_1)]+s_\varphi\ga_1\}\{s_\varphi[\beta_2+\beta_1-q(\beta_2-\beta_1)]-c_\varphi\ga_1\}$\\
$Z_2 Z_3 \mathcal{H}_3^p \mathcal{H}_3^{-p}$ & $\frac{1}{6}g^2 \{c_\varphi p(\beta_2-\beta_1)-s_\varphi[(p+q+2)(\ga_2-\ga_1)-3\ga_2]\}\{s_\varphi p(\beta_2-\beta_1)+c_\varphi[(p+q+2)(\ga_2-\ga_1)-3\ga_2]\}$\\
$Z_2 Z_3 \mathcal{H}_4^{q+1} \mathcal{H}_4^{-q-1}$ & $\frac{1}{6}g^2 \{c_\varphi[q(\beta_2-\beta_1)-2\beta_1]-s_\varphi\ga_2\}\{s_\varphi[q(\beta_2-\beta_1)-2\beta_1]+c_\varphi\ga_2\}$\\
$Z_2 Z_3 \mathcal{H}_5^{p+1} \mathcal{H}_5^{-p-1}$ & $\frac{1}{6}g^2 \{c_\varphi (1+p)(\beta_2-\beta_1)-s_\varphi[(p+q)(\ga_2-\ga_1)-3\ga_1]\}\{s_\varphi (1+p)(\beta_2-\beta_1)+c_\varphi[(p+q)(\ga_2-\ga_1)-3\ga_1]\}$\\
$Z_2 Z_3 \mathcal{H}_6^{q-p} \mathcal{H}_6^{p-q}$ & $\frac{g^2c_{2\varphi} }{6(w^2+V^2)}\{V^2(p-q)[(p-1)(\ga_2-\ga_1)-3\ga_1](\beta_2-\beta_1)-w^2[\ga_1-p(\ga_2-\ga_1)][\beta_2+\beta_1+(p-q)(\beta_2-\beta_1)]\}+\frac{g^2s_{2\varphi} }{12(w^2+V^2)}\{V^2[(p-q)^2(\beta_2-\beta_1)^2-[(p-1)(\ga_2-\ga_1)-3\ga_1]^2]+w^2[[\ga_1-p(\ga_2-\ga_1)]^2-[\beta_2+\beta_1+(p-q)(\beta_2-\beta_1)]^2]\}$\\
$Z_3 Z_3 \dots$ & $Z_2 Z_2 \dots (c_\varphi\to s_\varphi, s_\varphi\to -c_\varphi) $\\
\hline
\end{tabu}
\caption{\label{2NG2ST2}The interactions of two neutral gauge bosons with two scalars (Continued).}  
\end{table}

\bibliographystyle{JHEP}

\bibliography{combine}

\end{document}